\documentclass[11pt]{article}

\usepackage{amsfonts}
\usepackage{amsmath}
\usepackage{mathrsfs}
\usepackage{fancyhdr}
\usepackage{epsfig,morefloats}

\usepackage[authoryear]{natbib}

\usepackage[figuresright]{rotating}
\usepackage{amsthm}
\usepackage{amssymb}
\usepackage{color}
\usepackage{lscape}
\usepackage{rotating}
\usepackage{caption}
\usepackage{subfigure}
\usepackage{graphicx}
\usepackage[compact]{titlesec}
\usepackage{lineno}
\usepackage{cancel}
\usepackage{multirow}
\usepackage{enumerate}
\usepackage{url} % not crucial - just used below for the URL
\usepackage{chngcntr}
\usepackage{enumitem, lscape, bbding,pifont}
\usepackage{array, booktabs, multirow, pdflscape, graphicx, epstopdf}
\usepackage{amsmath, amsthm, amssymb, amsfonts, fancyhdr}
\usepackage{bbm, setspace, color, chngcntr, listings}
\usepackage{algorithmic,algorithm, chngcntr}
\usepackage{setspace}

\definecolor{mygray}{gray}{0.6}

\newcommand{\Date}[1]{\def\@Date{#1}}
\def\today{\number\day~\ifcase\month\or
 January\or February\or March\or April\or May\or June\or
 July\or August\or September\or October\or November\or December\fi~\number\year}

\makeatletter
\newcommand{\figcaption}{\def\@captype{figure}\caption}
\newcommand{\tabcaption}{\def\@captype{table}\caption}
\makeatother

\numberwithin{equation}{section}
\newcommand{\E}{\mathbb{E}}
\newcommand{\bP}{\mathbb{P}}

\newcommand{\bS}{\mathbb{S}}
\newcommand{\Cov}{\mbox{Cov}}
\newcommand{\Var}{\mbox{Var}}
\newcommand{\md}{\mbox{d}}
\newcommand{\cF}{\mathcal{F}}

\newcommand{\cG}{\mathcal{G}}
\newcommand{\cI}{\mathcal{I}}

\newcommand{\cB}{\mathcal{B}}
\newcommand{\cS}{\mathcal{S}}
\newcommand{\cW}{\mathcal{W}}
\newcommand{\cX}{\mathcal{X}}
\newcommand{\cY}{\mathcal{Y}}
\newcommand{\cZ}{\mathcal{Z}}

\newcommand{\cE}{\mathcal{E}}

\newcommand{\bzero}{\boldsymbol{0}}
\newcommand{\R}{\mathbb{R}}

\newcommand{\bR}{\mathbf R}
\newcommand{\bA}{\mathbf A}
\newcommand{\bH}{\mathbf H}
\newcommand{\bu}{\mathbf u}
\newcommand{\bv}{\mathbf v}
\newcommand{\bx}{{\mathbf x}}
\newcommand{\by}{\mathbf y}
\newcommand{\bz}{\mathbf z}
\newcommand{\bw}{\mathbf w}
\newcommand{\bg}{\mathbf g}
\newcommand{\bh}{\mathbf h}

\newcommand{\bs}{\mathbf s}
\newcommand{\ba}{{\mathbf a}}
\newcommand{\bb}{{\mathbf b}}
\newcommand{\bM}{{\mathbf M}}

\newcommand{\be}{\boldsymbol \eta}
\newcommand{\bdel}{\boldsymbol \delta}
\newcommand{\btheta}{\boldsymbol \theta}
\newcommand{\bmu}{\boldsymbol \mu}
\newcommand{\bpsi}{\boldsymbol \psi}
\newcommand{\bxi}{\boldsymbol \xi}

\newcommand{\T}{{\scriptscriptstyle {\rm \top}}}
\newcommand{\p}{{\rm p}}

\newcommand{\calN}{{\mathcal{N}}}
\newcommand{\calK}{{\mathcal{K}}}
\newcommand{\bphi}{\boldsymbol{\phi}}
\newcommand{\bgamma}{\boldsymbol{\gamma}}
\newcommand{\bXi}{\boldsymbol{\Xi}}
\newcommand{\bSigma}{\boldsymbol{\Sigma}}
\newcommand{\bOmega}{\boldsymbol{\Omega}}
\newcommand{\Z}{\mathbb{Z}}
\newtheorem{theorem}{Theorem}

\newtheorem{proposition}{Proposition}
\newtheorem{remark}{Remark}

\newtheorem{condition}{Condition}
\newcommand{\blind}{1}

\allowdisplaybreaks

\counterwithout{equation}{section}
\counterwithout{remark}{section}

\usepackage{geometry}
\usepackage{authblk}

\begin{document}

% \def\spacingset#1{\renewcommand{\baselinestretch}%
% {#1}\small\normalsize} \spacingset{1}

%\renewcommand{\baselinestretch}{1.0}

%%%%%%%%%%%%%%%%%%%%%%%%%%%%%%%%%%%%%%%%%%%%%%%%%%%%%%%%%%%%%%%%%%%%%%%%%%%%%%

\if1\blind
{
  \title{\bf Testing the Martingale Difference Hypothesis in High Dimension
  	\thanks{
  The authors equally contributed to the paper.
 Chang and Jiang were supported in part by the National Natural Science Foundation of China (grant nos.~71991472, 72125008, 11871401 and 12001442). Chang was also supported by the Center of Statistical Research at Southwestern University of Finance and Economics.}}
% Jiang was supported in part by
% the National Natural Science Foundation
% of China (grant nos. 71991472 and 12001442).
% We would like to thank the editor, the associate editor and two reviewers for their constructive suggestions which led to substantial improvements.

 %Corresponding author: Xiaofeng Shao. Email addresses:  changjinyuan@swufe.edu.cn  (J. Chang),  jiangqing@bnu.edu.cn (Q. Jiang), xshao@illinois.edu (X. Shao). }
%\hspace{.2cm}\\}

\author[a]{Jinyuan Chang}
\author[b,a]{Qing Jiang}%\thanks{ Corresponding author. Email addresses: \url{ changjinyuan@swufe.edu.cn } (J.~Chang),  \url{zhentao.shi@cuhk.edu.hk} (Z.~Shi), \url{zhangjia@swufe.edu.cn} (J.~Zhang). }
\author[c]{Xiaofeng Shao}
%\affil[a]{\it \small Guizhou Key Laboratory of Big Data Statistical Analysis, Guizhou University of Finance and Economics, Guiyang, Guizhou  Province, China}
\affil[a]{\it \small Joint Laboratory of Data Science and Business Intelligence, Southwestern University of Finance and Economics, Chengdu, Sichuan Province, China}
\affil[b]{\it \small Center for Statistics and Data Science, Beijing Normal University, Zhuhai, Guangdong Province, China}
\affil[c]{\it \small Department of Statistics, University of Illinois at Urbana-Champaign, Champaign, IL, U.S.A. }

\date{}

  \maketitle
} \fi

\if0\blind
{
  \bigskip
  \bigskip
  \bigskip
  \begin{center}
    {\LARGE\bf Testing the Martingale Difference Hypothesis in High Dimension}
\end{center}
  \medskip
} \fi

%\maketitle

\bigskip

\begin{abstract}

In this paper, we consider testing the martingale difference hypothesis for high-dimensional time series.
 Our test is built on the sum of squares of the element-wise max-norm of the proposed matrix-valued nonlinear dependence measure at different lags. To conduct the inference, we approximate the null distribution of our test statistic by Gaussian approximation and provide a simulation-based approach to generate critical values. The asymptotic behavior of the test statistic under the alternative is also studied. Our approach is nonparametric as the null hypothesis only assumes the time series concerned is martingale difference without specifying any parametric forms of its conditional moments. As an advantage of Gaussian approximation, our test is robust to the cross-series dependence of unknown magnitude. To the best of our knowledge, this is the first  valid test for the martingale difference hypothesis that not only allows for large dimension but also captures  nonlinear serial dependence. The practical usefulness of our test is illustrated via simulation and  a real data analysis. The test is implemented in a user-friendly R-function.

\end{abstract}

\bigskip \bigskip
\noindent%

\bigskip
%\begin{quote}
%{\sl MSC2010 subject classifications}: Primary 62G99; secondary 62F40
%\end{quote}

%\vfill

%\begin{quote}
%\noindent
%{\sl Keywords}: Empirical likelihood; Estimating equations; High-dimensional statistical methods; Misspecification; Moment selection; Penalized likelihood.
%\end{quote}

%\thispagestyle{empty}
%\pagenumbering{gobble}

%\newpage
%\tableofcontents

%\newpage
%\pagenumbering{arabic}
%\setcounter{page}{1}

%\linenumbers

\newpage
% \spacingset{1.1} % DON'T change the spacing!
\onehalfspacing

\section{Introduction}

Testing the martingale difference hypothesis is a fundamental problem in econometrics and time series analysis. The concept of  martingale difference plays an important role in many areas of economics and finance. Several economic and financial theories such as the efficient markets hypothesis \citep{Fama1970,Fama1991,LeRoy1989,Lo1997}, rational expectations \citep{Hall1978} and optimal asset pricing \citep{Cochrane2005,Fama2013}, yield such dependence restrictions on the underlying economic and financial variables. More formally,	let $\{\bx_t\}$ be a $p$-dimensional time series with $\E(\bx_t)=\bzero$ for any $t\in\Z$. Write $\bx_t=(x_{t,1},\ldots,x_{t,p})^\T$ and denote by $\mathscr{F}_{s}$ the $\sigma$-field generated by $\{\bx_t\}_{t\leqslant s}$. We call $\{\bx_t\}_{t\in\Z}$ a martingale difference sequence (MDS) if and only if $\E(\bx_t\,|\,\mathscr{F}_{t-1})=\bzero$ for any $t\in\Z$. Given the observations $\{\bx_t\}_{t=1}^{n}$, we are interested in the hypothesis testing problem:
\begin{align}\label{H0}
H_0: \{\bx_t\}_{t\in \Z} \mbox{ is a MDS} ~~~~\mbox{versus}~~~~ H_1:\{\bx_t\}_{t\in \Z} \mbox{ is not a MDS.}
\end{align}
The MDS hypothesis implies that the past information does not help to improve the prediction of future values of a MDS, so the best nonlinear predictor of the future values of a MDS given the current information set is just its unconditional expectation. The theme of the lack of predictability
is of central interest in economics and finance and has stimulated a huge literature in both econometrics and time series analysis. %Below we provide a brief literature review.

%\subsection{Literature review}

So far most of the work on MDS testing is restricted to the univariate case, i.e., $p=1$.
In one strand of literature, the MDS testing problem is reduced to testing the uncorrelatedness in either time domain or spectral domain. See \cite{BP1970}, \cite{LB1978}, \cite{Durlauf1991}, \cite{Hong1996}, \cite{Deo2000}, \cite{LNS2001}, and  \cite{Shao2011a,Shao2011b}, among others.
These tests target on serial correlation but are unable to capture nonlinear serial  dependence.
There are examples of uncorrelated processes that are not MDS such as certain bilinear processes and nonlinear moving average processes, see \cite{DL2003} for specific examples. Hence, it is important to develop tests that can go beyond linear serial dependence. In the specification testing literature, the exponential function based approach,  pioneered by \cite{Bierens1984, Bierens1990}, \cite{DeJong1996} and \cite{BP1997}, is capable of detecting nonlinear serial dependence. Using the characteristic function,  \cite{Hong1999} proposed the generalized spectral density as a new tool for specification testing in a nonlinear time series framework; see \cite{HL2003} and \cite{HL2005} for further developments. As an interesting extension of \cite{Hong1999}, \cite{EV2006a} developed a MDS test based on the generalized spectral distribution function to capture nonlinear serial dependence at  all lags. Parallel to the exponential/characteristic function based approach, the indicator/distribution function based approach has been taken by \cite{Stute1997}, \cite{KS1999},
\cite{DL2003}, and \cite{PW2005} among others. We refer to \cite{EL2009} for a comprehensive review.

For the multivariate time series, i.e., $p>1$, the literature for the MDS testing is scarce.
Although it is expected that most of the above-mentioned tests can be extended to relatively low dimensional case,
the theoretical and empirical properties of these tests are unknown. Recently, \cite{HLZ2017} proposed a multivariate extension of the classical univariate variance ratio test \citep{LM1988,PS1988,CD2006} to test a weak form of the efficient markets hypothesis, i.e., uncorrelatedness of $\bx_t$.
As argued in \cite{HLZ2017}, the rationale to consider the MDS test for multivariate time series is that even if
the MDS hypothesis holds for each component series $\{x_{t,j}\}_{t\in\Z}$, the MDS hypothesis could be violated at the
multivariate level. In particular, the current return on the $i$th asset may be predicted by past observations of the $j$th asset. A univariate test may fail to detect this kind of cross-serial dependence, which can be captured by  a multivariate test. %As an important theoretical contribution, \cite{HLZ2017} allow the dimension $p$ to grow relative to the sample size in their asymptotic theory. However,
Since it is well known that the variance ratio test  only targets on serial correlation, the test of \cite{HLZ2017} is unable to capture  nonlinear serial dependence.

Nowadays, time series of moderate or high dimension are routinely collected or generated owing to the advance in science and technology. For example, S\&P 500 index measures the stock performance of 500 large companies listed on stock exchanges in the United States, and it is tempting to ask whether the stock returns of the 500 companies are predictable at the daily or weekly frequency for a given time period (say, 5 years).  The same question can be asked for the stocks within the same sector,
such as those in the real estate sector (see Section~\ref{sec:real} for data illustration). This naturally leads us to the regime where the dimension  $p$ is comparable to  or exceeds the sample size $n$. To the best of our knowledge, there is no MDS testing procedure available in the literature that allows the dimension $p$ to exceed the sample size $n$. Most of the aforementioned tests developed in the univariate setting require nontrivial modification to accommodate the high-dimensionality. The multivariate variance ratio test in \cite{HLZ2017} allows for growing dimension $p$ in their theory (i.e., $1/p+p/n=o(1)$) but is quite limited since their test cannot be implemented  when $p>n$ and may encounter computational problems when $p$ is large (say, $p>120$); see Section~\ref{sec:numerical} for more details.

To fill this gap, we introduce a  new test for the MDS hypothesis of multivariate and possibly high-dimensional time series. We first use the element-wise max-norm of a  sample-based matrix to characterize the nonlinear dependence of underlying $p$-dimensional time series $\{\bx_t\}$ at a given lag $j\geqslant1$, and then combine such information at different lags to propose our test statistic. Owing to the high-dimensionality and unknown temporal and cross-series dependence, the limiting null distribution of our test statistic is hard to derive, and it  may  not even  have a closed form. To circumvent such difficulty, we employ the celebrated Gaussian approximation technique \citep{CCK2013}, which has undergone a rapid development recently,  to establish the asymptotic equivalence between the null distribution of our test statistic
and that of a certain function of a multivariate Gaussian random vector. Our theoretical analysis shows that our proposed test works even if $p$ grows exponentially with respect to the sample size $n$, provided that some suitable regularity assumptions hold. To facilitate feasible inference, we propose a simulation-based approach to generate critical values. We also investigate the power behavior of our test under some local alternatives.

%\subsection{Our connection and contribution to the literature}
Since the seminal contribution of \cite{CCK2013}, the literature on Gaussian approximation  in the high-dimensional setting has been growing rapidly.  For the sample mean of independent random vectors,
we mention \cite{CCK2013, CCK2017},  \cite{DZ2020}, \cite{FK2020},  \cite{KMB2020},  \cite{CCKK2019}, and \cite{CCK2020}.
 For high-dimensional $U$-statistics and $U$-processes,  see \cite{Chen2018} and \cite{CK2019} for recent developments. The applicability of Gaussian approximation has also been extended to high-dimensional time series setting by
\cite{ZW2017}, \cite{ZC2018}, \cite{CCK2019} and \cite{CCW2020}. Also see \cite{CYZ2017,CZZZ2017,CZZW2017,CQYZ2018}, and \cite{YC2020} among others for the use of Gaussian approximation or variants in high-dimensional statistical inference.

\cite{ZW2017} and \cite{ZC2018} considered the Gaussian approximation for $$\max_{1\leqslant j\leqslant p}\frac{1}{\sqrt{n}}\sum_{t=1}^nx_{t,j}$$ with the physical dependence measure \citep{Wu2005} imposed on $\{\bx_t\}$, and \cite{CCK2019} considered the same problem when $\{\bx_t\}$ is a $\beta$-mixing sequence. \cite{CCW2020} studied the Gaussian approximations for $\mathbb{P}(n^{-1/2}\sum_{t=1}^n\bx_t\in A)$ over some general classes of the set $A$ (hyper-rectangles, simple convex sets and sparsely convex sets) under three different dependency framework ($\alpha$-mixing, $m$-dependent, and physical dependence measure), which include the results obtained in \cite{ZW2017}, \cite{ZC2018} and \cite{CCK2019} as special cases. Compared to the use of Gaussian approximation results for high-dimensional time series in the existing works, %\cite{ZW2017}, \cite{ZC2018}, \cite{CCK2019} and \cite{CCW2020},
our test statistic is considerably more involved and motivates  us to develop new techniques for establishing the asymptotic equivalence between the null distribution of our test statistic
and that of a certain function of a multivariate Gaussian random vector.
More specifically, the theoretical analysis in this paper targets on the Gaussian approximation for some function of the high-dimensional vector $(n-K)^{-1/2}\sum_{t=1}^{n-K}\be_t$, where $\be_t$ is a newly defined vector based on $\{\bx_t,\bx_{t+1},\ldots,\bx_{t+K}\}$ and $K$ is the number of lags involved in our test statistic. %In addition, we only need to  assume $\{\bx_t\}$ is $\alpha$-mixing,  which is weaker than the $\beta$-mixing assumption in \cite{CCK2019}.
Since $K$ is allowed to grow with the sample size $n$ in our setting, the dependence structure among $\{\be_t\}$ will vary with $K$ which cannot be covered in the frameworks of above mentioned works, and the existing Gaussian approximation results cannot be applied here. Some nontrivial technical challenges need to be addressed in our theoretical analysis.

From a methodological and practical viewpoint, we highlight a few appealing features of our proposed test:

(a) Our approach is nonparametric as the null hypothesis only assumes the time series concerned is martingale difference without specifying any parametric forms of its conditional moments. Hence, it is robust to second-order and higher-order conditional moments of unknown forms, including conditional heteroscedasticity,  a prominent feature of many financial time series. 

(b) It allows the dimension $p$ to grow exponentially with respect to the sample size $n$,  and works well for a broad range of dimension $p$ even at a medium sample size (e.g., $n=300$) as shown in our simulation studies. We have developed an R-function {\verb"MartG_test"} in the package {\verb"HDTSA"} which implements the test in an automatic manner.

(c) There is no particular requirement on the strength of cross-series dependence in our theory, so our test is applicable to time series with cross-series dependence of unknown magnitude. Strong cross-series dependence has been commonly observed in many real high-dimensional time series data. %usually implied by the static factor model  \citep{FLM2013}.
%For example, \cite{FFX2016} used the static factor model to model the S\&P 500 Equity Index constituents where the strong dependence within and across the sectors can be well captured.

%\subsection{Organization and notation}
The rest of this paper is organized as follows. The methodology and theoretical analysis are given in Sections \ref{SecMed} and \ref{SecThe}, respectively. Section \ref{sec:general} extends the proposed test to more general settings.
Section \ref{sec:numerical} studies the finite sample performance of our proposed test. A real data analysis is presented in Section \ref{sec:real}. Section \ref{sec:conc} concludes the paper. Section \ref{sec:pfs} includes the mathematical proofs of our main results. Some additional technical arguments and numerical studies are given in the supplementary material.
At the end of this section, we  introduce some notation that is used throughout the paper. For any positive integer $q\geqslant2$, we write $[q]=\{1,\ldots,q\}$ and denote by $\bS^{q-1}$ the $q$-dimensional unit sphere. For any $q_1\times q_2$ matrix $\bM=(m_{i,j})_{q_1\times q_2}$, let $|\bM|_{\infty}=\max_{i\in[q_1],j\in[q_2]}|m_{i,j}|$ and $|\bM|_0=\sum_{i=1}^{q_1}\sum_{j=1}^{q_2}I(m_{i,j}\neq0)$, where $I(\cdot)$ denotes the indicator function. Specifically, if $q_2=1$, we use $|\bM|_\infty=\max_{i\in[q_1]}|m_{i,1}|$ and $|\bM|_0=\sum_{i=1}^{q_1}I(m_{i,1}\neq 0)$ to denote the $L_\infty$-norm and $L_0$-norm of the  $q_1$-dimensional vector $\bM$, respectively. For any $q$-dimensional vector $\ba=(a_1,\ldots,a_q)^\T$, write $\psi(\ba)$ as the $q$-dimensional vector $\{\psi(a_1),\ldots,\psi(a_q)\}^\T$ for given function $\psi:\mathbb{R}\rightarrow\mathbb{R}$, and denote by $\ba_{\mathcal{L}}$ the subvector of $\ba$ collecting the components indexed by a given index set $\mathcal{L}\subset[q]$.

\section{Methodology}\label{SecMed}
\subsection{Test statistic and the associated critical values}	
	Let $\{\bx_t\}$ be a $p$-dimensional time series with $\E(\bx_t)=\bzero$ for any $t$. Given the observations $\{\bx_t\}_{t=1}^{n}$, we shall develop a martingale difference hypothesis test that can capture certain nonlinear dependence between $\bx_t$ and $\bx_{t+j}$, for $j\in \mathbb N_+$. To this end, we let $\bphi(\cdot): \R^p\rightarrow \R^d$ represent a map that is provided by the user. For example, $\bphi(\bx)=\bx$ is the linear identity map; $\bphi(\bx)=\{\bx^{\T},(\bx^2)^{\T}\}^{\T}$ includes both linear and quadratic terms, where $\bx^2=(x_{1}^2, \ldots, x_{p}^2)^{\T}$ with $\bx=(x_1,\ldots,x_p)^\T$;
$\bphi(\bx)=\cos(\bx)$ captures certain type of nonlinear dependence, where $\cos(\bx)=\{\cos(x_1),\ldots,\cos(x_p)\}^\T$ with $\bx=(x_1,\ldots,x_p)^\T$.
	
%	Denote by ${\cal F}_{s}$ the $\sigma$-field generated by $\{\bx_t\}_{t\leqslant s}$.
%	We call $\{\bx_t\}_{t=1}^n$ a martingale difference sequence (MDS) if and only if $\E(\bx_t\,|\,\calF_{t-1})=\bzero$ for any $2\leqslant t\leqslant n$. We are interested in the hypothesis testing problem:
%\begin{align}\label{H0}
%  H_0': \{\bx_t\}_{t=1}^n \mbox{ is a MDS} ~~~~\mbox{versus}~~~~ H_1:\{\bx_t\}_{t=1}^n %\mbox{ is not a MDS.}
%\end{align}
%There are many tests for the univariate case, i.e., $p=1$. For multivariate case with fixed $p$, see \cite{EV2006a}, \cite{Shao2011}, \cite{HLZ2017}. In this work, we address the high-dimensional problem, where $p$ can grow with respect to sample size $n$ in our asymptotics, thus we include the case $p>n$.

Denote $\bgamma_j=(n-j)^{-1}\sum_{t=1}^{n-j} \E[{\rm vec}\{\bphi(\bx_{t})\bx_{t+j}^{\T}\}]$ for each $j\geqslant1$.
Our proposal for testing the martingale difference hypothesis consists in checking all the pairwise covariance between $\bphi(\bx_t)$ and $\bx_{t+j}$, namely, our null hypothesis is now
\begin{align}\label{H0_1}
  H_0': \bgamma_j=\bzero \mbox{ for all }j\geqslant 1\,.
\end{align}
It is easy to see that $H_0$ in \eqref{H0} implies $H_0'$ in \eqref{H0_1} but not vice versa. In theory, it would be ideal to develop a test that is consistent with any violation of $H_0$ but this is very challenging in a model free setting, since the alternative we target is huge owing to the high-dimensionality and nonlinear serial dependence at all lags. As argued in \cite{PJ2014}, ``{\it Typically, the information set includes the infinite past history of the series,.... If a finite number of lagged values is included in the conditioning set, some dependence structure in the process may be missed due to omitted lags. However, tests that are designed to cope with the infinite lag case may have very low power (e.g., \citealp{DeJong1996}) and may not be feasible in empirical applications.}''
Thus even in the low-dimensional setting, it is not clear whether there is a practical benefit for a test that is consistent with all alternatives. This motivates us to relax the null hypothesis $H_0$ and focus on the directional alternatives encoded by the function $\bphi(\cdot)$, which is pre-specified by the user and can incorporate some prior information.

Note that if the time series $\{\bx_t\}$ is strictly stationary, then $\bgamma_j=\E[{\rm vec}\{\bphi(\bx_{0})\bx_{j}^{\T}\}]$ which represents the population-level nonlinear dependence measure at lag $j$. In our asymptotic theory, no stationarity assumption needs to be imposed.
%\xs{Do we need to assume stationarity for convenience?} \jc{Actually, we do not need the stationary assumption here.}
To test $H_0'$,  it is natural to consider a test statistic with the following form
\begin{align}\label{TestStat}
T_n=n\sum_{j=1}^{K}|\hat{\bgamma}_j|_{\infty}^2 \,,
\end{align}
where
$\hat{\bgamma}_j=(n-j)^{-1}\sum_{t=1}^{n-j} {\rm vec}\{\bphi(\bx_{t})\bx_{t+j}^{\T}\}$ is the estimator of $\bgamma_j$. Here $K=o(n)$ is a truncation lag and is allowed to grow with respect to the sample size $n$. This flexibility is important when there exists nonlinear serial dependence at large lags.

Intuitively, a large value of $T_n$ provides evidence against $H_0'$ in \eqref{H0_1} and then we can reject $H_0$ in \eqref{H0} if
\begin{align}\label{eq:testcv}
T_n>{\rm cv}_\alpha\,,
\end{align}
where ${\rm cv}_\alpha>0$ is the critical value at the significance level $\alpha\in(0,1)$. To determine ${\rm cv}_\alpha$, we need to derive the distribution of $T_n$ under $H_0$. Write $\hat{\bgamma}=(\hat{\bgamma}_1^\T,\ldots,\hat{\bgamma}_K^\T)^\T$ and $\bgamma=(\bgamma_1^\T,\ldots,\bgamma_K^\T)^\T$.
 For fixed $(p,d,K)$ and under suitable moment and weak dependence conditions, it follows from the central limit theorem that $\sqrt{n}(\hat{\bgamma}-\bgamma)\rightarrow_d\mathcal{N}(\bzero,\mathring{\bSigma}_{K})$ as $n\rightarrow\infty$ for some positive definite matrix $\mathring{\bSigma}_{K}\in\mathbb{R}^{(Kpd)\times(Kpd)}$. Let $\mathring{\bg}:=(\mathring{g}_1,\ldots,\mathring{g}_{Kpd})^\T\sim \calN(\bzero,\mathring{\bSigma}_K)$. By the continuous mapping theorem, the distribution of $T_n$ under $H_0$ can be approximated by that of its Gaussian analogue $\mathring{G}_K = \sum_{j=1}^{K}  |\mathring{\bg}_{\mathcal{L}_j}|_\infty^2$ in the scenario with fixed $(p,d,K)$, where $\mathcal{L}_j=\{(j-1)pd+1,\ldots,jpd\}$. Write $\tilde{n}=n-K$ and
let
\begin{align}\label{eq:ft}
\be_t=([{\rm vec} \{\bphi(\bx_t)\bx_{t+1}^{\T}\}]^\T,\ldots, [{\rm vec} \{ \bphi(\bx_t)\bx_{t+K}^{\T}\}]^\T )^{\T}
\end{align}
for any $t\in[\tilde{n}]$. Define
\begin{align}\label{eq:longruncov}
\bSigma_{n,K}={\rm Cov} \bigg(\frac{1}{\sqrt{\tilde{n}}}\sum_{t=1}^{\tilde{n}}\be_t\bigg)\,,
\end{align}
which is the long-run covariance matrix of the sequence $\{\be_t\}_{t=1}^{\tilde{n}}$. For fixed $(p,d,K)$, the asymptotic covariance $\mathring{\bSigma}_K$ of 	$\sqrt{n}(\hat{\bgamma}-\bgamma)$ is essentially the limit of $\bSigma_{n,K}$ specified in \eqref{eq:longruncov} as $n\rightarrow\infty$. In the high-dimensional scenarios, i.e.,  when $(p,d,K)$ is diverging with respect to $n$, Proposition \ref{prop.H0}  indicates that such approximation for the null distribution of $T_n$ is still valid even when $p$ and $d$ grow exponentially with respect to the sample size $n$.

\begin{proposition}\label{prop.H0}
Assume Conditions {\rm \ref{cond.tail}--\ref{cond.m}} in Section {\rm \ref{SecThe}} hold and $G_K = \sum_{j=1}^{K}   |\bg_{\mathcal{L}_j}|_\infty^2$, where $\bg=(g_1,\ldots,g_{Kpd})^\T \\\sim\mathcal{N}(\bzero,\bSigma_{n,K})$ and $\mathcal{L}_j=\{(j-1)pd+1,\ldots,jpd\}$. Let $K=O(n^\delta)$ for some constant $0\leqslant \delta<f_1(\tau_1,\tau_2)$ with $f_1(\tau_1,\tau_2)$ defined as \eqref{eq:f1} in Section {\rm \ref{SecThe}}.
Then it holds that
$
  \sup_{x>0}|\bP_{H_0}( T_n >x) - \bP(G_K > x)|=o(1)$
as $n\rightarrow\infty$, provided that $\log(pd)=o(n^{c})$ for some constant $c>0$ only depending on $(\tau_1,\tau_2,\delta)$.
\end{proposition}

Proposition \ref{prop.H0} reveals that the Kolmogorov-Smirnov distance between the null distribution of the proposed test statistic $T_n$  and the distribution of $G_K$  converges to zero, even when $p$ and $d$ diverge at some exponential rate of $n$. Letting
\begin{align}\label{eq:cvalpha}
{\rm cv}_\alpha=\inf\{x>0:\bP(G_K\leqslant x) \geqslant 1-\alpha\}
\end{align}
in \eqref{eq:testcv}, Proposition \ref{prop.H0} yields that $\bP_{H_0}(T_n>{\rm cv}_\alpha)\to \alpha$  as $n\to\infty$. Since the long-run covariance matrix $\bSigma_{n,K}$ is usually unknown in practice, we need to replace it by some estimate $\widehat{\bSigma}_{n,K}$ and then use $\hat{{\rm cv}}_\alpha$ defined below to approximate the desired critical value ${\rm cv}_{\alpha}$ specified in \eqref{eq:cvalpha}:
\begin{align}\label{eq:hatcv}
\hat{{\rm cv}}_{\alpha}=\inf\{x>0:\mathbb{P}(\hat{G}_K\leqslant x\,|\,\mathcal{X}_n)\geqslant1-\alpha\}\,,
\end{align}
where $\mathcal{X}_n=\{\bx_1,\ldots,\bx_n\}$ and  $\hat{G}_K = \sum_{j=1}^{K} |\hat{\bg}_{\mathcal{L}_j}|_\infty^2$ with $\hat{\bg}:=(\hat{g}_1,\ldots,\hat{g}_{Kpd})^\T\sim \calN(\bzero,\widehat\bSigma_{n,K})$ and $\mathcal{L}_j=\{(j-1)pd+1,\ldots,jpd\}$. Then we reject the null hypothesis $H_0$ specified in \eqref{H0} if
\begin{align}\label{eq:testproc}
T_n>\hat{{\rm cv}}_{\alpha}\,.
\end{align}
We defer the details of $\widehat{\bSigma}_{n,K}$ to Section \ref{sec:covest}.

\begin{remark}{\rm
If we select the function $\bphi(\bx)=\bx$, the test statistic $T_n$ defined in \eqref{TestStat} can also be applied to test the high-dimensional white noise hypothesis, i.e., $H_0:\{\bx_t\}_{t\in\mathbb{Z}}$ is white noise versus $H_1:\{\bx_t\}_{t\in\mathbb{Z}}$ is not white noise.  \cite{CYZ2017} considered this hypothesis testing problem with $L_\infty$-type test statistic using the maximum absolute autocorrelations and cross-correlations of the component series in $\bx_t$ over all lags $k\in[K]$. It is well known that the $L_\infty$-type test statistic is powerful against the sparse alternatives, that is,   only a small fraction of the elements in $\bgamma=(\bgamma_1^\T,\ldots,\bgamma_K^\T)^\T$ are nonzero, while it can be powerless for the dense but faint alternatives, i.e., when  most elements in $\bgamma=(\bgamma_1^\T,\ldots,\bgamma_K^\T)^\T$ are nonzero but with very small magnitudes. To remedy such weakness, our proposed $T_n$ in \eqref{TestStat} combines the signals from different lags together using the sum of squares and  is expected to improve the power performance in case of  dense but faint alternatives. On the technical side,  constructing the Gaussian approximation to the null distribution of $T_n$ defined in \eqref{TestStat} is more challenging than that for the $L_\infty$-type statistic used in \cite{CYZ2017}. \cite{CYZ2017} only considered the case with fixed $K$ under the $\beta$-mixing assumption. The null distribution of their test statistic can be  easily obtained by the associated Gaussian approximation results developed in \cite{CCK2019}. In this paper, we only impose  the $\alpha$-mixing assumption on $\{\bx_t\}$ and the corresponding $\alpha$-mixing coefficients of $\{\be_t\}$ become a triangular array  owing to the divergence of  $K$. To the best of our knowledge, our paper is the first attempt to derive the Gaussian approximation results in such a complex setting.}
\end{remark}

\begin{remark}{\rm
As we mentioned earlier, the only paper that allows growing dimension for the martingale difference hypothesis testing is \cite{HLZ2017}, which generalized the variance ratio test to multivariate time series. In their asymptotic theory, they considered both finite/fixed horizon (i.e.,  fixed $K$) and increasing horizon (i.e., $K\rightarrow\infty$ but $K^2/n\rightarrow 0$), which is
also allowed in our theory. In their Theorem 7, they presented the limiting null distribution of a particular test statistic $Zd_{\rm tr}$ under the restriction that the dimension $p$ grows but $p/n\rightarrow 0$. Their another two test statistics $Z_{\rm tr}$ and $Z_{\rm det}$ for the setting of fixed $p$ cannot be implemented in practice when $p>\sqrt{n}$. By contrast, our test statistic can work for a much broader range of $p$, including the case $p\gg n$,  and thus is advantageous in dealing with the martingale difference hypothesis testing for high-dimensional time series. In addition, we can capture nonlinear serial dependence owing to the flexibility of user-chosen $\bphi(\cdot)$, which yields a nonlinear dependence measure. In practice, we need to set the lags $K$ and the user-chosen map $\bphi(\cdot)$, which can incorporate some prior information we have. For example, if the time series is expected to exhibit seasonal dependence, then $K$ should be large enough to include some seasonal lags. If we are dealing with stock return data, then including quadratic terms in $\bphi(\cdot)$ might help to capture potential nonlinear dependence.
%We provide an in-depth theoretical study of the behavior of our test statistic $T_n$ under both the null and alternatives in Section~\ref{SecThe}.
}
\end{remark}

\begin{remark}\label{rk:3}
{\rm If the time series $\{\bx_t\}$ is strictly stationary, we know the transformed data $\{\be_t\}$ is also strictly stationary and our test statistic $T_n$ given in \eqref{TestStat} essentially converts the MDS testing problem for $\bx_t$ to testing zero mean for the transformed data $\be_t$. There are indeed several papers in the literature of Gaussian approximation that tackle the mean testing problem for high-dimensional time series; see \cite{ZW2017}, \cite{ZC2018}, \cite{CCK2019} and \cite{CCW2020}. \cite{ZW2017} and \cite{ZC2018} considered the Gaussian approximation theory in the framework that assumes the physical dependence \citep{Wu2005} among $\{\be_t\}$. \cite{CCK2019} and \cite{CCW2020} considered the Gaussian approximation theory, respectively, in the frameworks that assume the $\beta$-mixing assumption and $\alpha$-mixing assumption for $\{\be_t\}$. Notice that the dependence structure among $\{\be_t\}$ will vary with $K$. The dependence framework for $\{\be_t\}$ assumed in these existing works do not cover our current setting, thus the existing Gaussian approximation results cannot be used for approximating the null distribution of our proposed test statistic $T_n$.
 }
\end{remark}

\subsection{Estimation of long-run covariance matrix}\label{sec:covest}

In the low-dimensional setting, long-run covariance matrix estimation (or heteroscedastic-autocorrelation-consistent estimation) is a classic problem in econometrics and time series analysis and there is a rich literature. We refer the readers to two foundational papers by \cite{NW1987} and \cite{Andrews1991}.
In the high-dimensional setting, the estimator proposed in the low-dimensional environment can still be used, but establishing the proper probabilistic bounds for the difference is very challenging. Recall $\tilde{n}=n-K$.
Following \cite{CYZ2017}, we adopt  the following estimate for the long-run covariance matrix $\bSigma_{n,K}$:
\begin{align}\label{Jn}
  \widehat \bSigma_{n,K} = \sum_{j=-\tilde{n}+1}^{\tilde{n}-1}\mathcal{K}\bigg(\frac{j}{b_n}\bigg) \widehat \bH_j\,,
\end{align}
where $\widehat \bH_j=\tilde{n}^{-1}\sum_{t=j+1}^{\tilde{n}}(\be_t-\bar{\be})(\be_{t-j}-\bar{\be})^{\T}$ if $j\geqslant0$
and  $\widehat \bH_j=\tilde{n}^{-1}\sum_{t=-j+1}^{\tilde{n}}({\be}_{t+j}-\bar{\be})({\be}_{t}-\bar{\be})^{\T}$ otherwise, with $\bar{{\be}}=\tilde{n}^{-1}\sum_{t=1}^{\tilde{n}} {\be}_t$. Here $\mathcal{K}(\cdot)$ is a symmetric kernel function that is continuous at 0, and $b_n$ is the bandwidth diverging with $n$. The theoretical property of $\widehat{\bSigma}_{n,K}$ defined as \eqref{Jn} is summarized in Proposition \ref{prop.sigma} in Section \ref{SecThe}. As indicated in \cite{Andrews1991}, to make $\widehat{\bSigma}_{n,K}$ given in \eqref{Jn} be positive semi-definite, we can require the kernel function $\mathcal{K}(\cdot)$ to satisfy $\int_{-\infty}^\infty \mathcal{K}(x)e^{-{\rm i}x\lambda}\,{\rm d}x\geqslant0$ for any $\lambda\in\mathbb{R}$, where ${\rm i}=\sqrt{-1}$. The Bartlett kernel, Parzen kernel and Quadratic Spectral kernel all satisfy this requirement. See Section \ref{sec:numerical} for the explicit forms of these kernels.

Given $\widehat{\bSigma}_{n,K}$, to compute $\hat{{\rm cv}}_{\alpha}$ given in \eqref{eq:hatcv}, we need to generate $\hat{\bg}:=(\hat{g}_1,\ldots,\hat{g}_{Kpd})^\T\sim \calN(\bzero,\widehat\bSigma_{n,K})$. Notice that $\widehat{\bSigma}_{n,K}$ is a $(Kpd)\times(Kpd)$ matrix.  The standard procedure is  based on the Cholesky decomposition of $\widehat{\bSigma}_{n,K}$ but generating $\hat{\bg}$ is a computationally $(nK^2p^2d^2+K^3p^3d^3)$-hard problem that requires a large storage space for $\widehat{\bSigma}_{n,K}$. In practice, $p$ and $d$ can be quite large. As suggested in \cite{CYZ2017}, we can generate $\hat{\bg}$ as follows:

\begin{algorithm}[H] \caption{Procedure for generating $\hat{\bg}$} \label{alg1}
\vspace{0.1in}
\noindent{\bf Step 1.} Let $\boldsymbol{\Theta}$ be a $\tilde{n}\times\tilde{n}$ matrix with $(i,j)$th element $\mathcal K\{(i-j)/b_n\}$. \\[0.5em]
\noindent{\bf Step 2.} Generate $\boldsymbol\xi=(\xi_1,\ldots,\xi_{\tilde{n}})^{\T} \sim \calN(\bzero,\boldsymbol{\Theta})$ independent of $\mathcal{X}_n$. \\[0.5em]
\noindent{\bf Step 3.} Let
$  \hat{\bg}=(\hat{g}_{1},\ldots,\hat{g}_{Kpd})^\T=\tilde{n}^{-1/2}\sum_{t=1}^{\tilde{n}}\xi_t ({\be}_t-\bar{{\be}})$.
\vspace{0.1in}
\end{algorithm}
\noindent We can show that $\hat{\bg}$ obtained in Algorithm \ref{alg1} satisfies $\hat{\bg} \,|\, \cX_n\sim \mathcal{N}(\bzero,\widehat{\bSigma}_{n,K})$. The computational complexity of Step 2 in Algorithm \ref{alg1} is just $O(n^3)$ which is independent of $(p,d)$. When $p$ and $d$ are large, the required storage space of Algorithm \ref{alg1} is also much smaller than that of the standard procedure since it only requires to store $\{\be_t\}_{t=1}^{\tilde{n}}$ and $\bar{\be}$ rather than $\widehat{\bSigma}_{n,K}$. In practice, we can draw $\hat{\bg}_1,\ldots,\hat{\bg}_B$ independently by Algorithm \ref{alg1} for some large integer $B$ and then take the $\lfloor B\alpha\rfloor$th largest value among $\hat{G}_{K,1},\ldots,\hat{G}_{K,B}$ to approximate $\hat{\rm cv}_\alpha$ defined as \eqref{eq:hatcv}, where $\hat{G}_{K,i} = \sum_{j=1}^K |\hat{\bg}_{i,\mathcal{L}_j}|_\infty^2$ with $\hat{\bg}_i=(\hat{g}_{i,1},\ldots,\hat{g}_{i,Kpd})^\T$ and $\mathcal{L}_j=\{(j-1)pd+1,\ldots,jpd\}$.

\section{Theoretical property}\label{SecThe}

Recall $T_n=n\sum_{j=1}^{K}|\hat{\bgamma}_j|_{\infty}^2$. Since the distribution of  $\hat{\bgamma}=(\hat{\bgamma}_1^\T,\ldots,\hat{\bgamma}_K^\T)^\T$ can be well approximated by that of $\bar{\be}=\tilde{n}^{-1}\sum_{t=1}^{\tilde{n}}\be_t$ with $\tilde{n}=n-K$, the difference between the distributions of $T_n$ and $\tilde{T}_n:=\tilde{n}\sum_{j=1}^K|\bar{\be}_{\mathcal{L}_j}|_\infty^2$ is expected to be asymptotically negligible. See Lemma L2 in Section \ref{sec:pfs}. The key step in our theoretical analysis is to approximate the null distribution of $\tilde{T}_n$ by Gaussian approximation.

For any $j_1,\ldots,j_K\in[pd]$ and $x>0$, let $\mathcal A_{j_1,\ldots,j_K}(x)=\{\bb \in\R^{Kpd}: \bb_{S_{j_1,\ldots,j_K}}^{\T} \bb_{S_{j_1,\ldots,j_K}}\leqslant x\}$ with  $S_{j_1,\ldots,j_K}=\{j_1,j_2+pd,\ldots,j_K+(K-1)pd\}$. Define $\mathcal A(x;K) = \bigcap_{j_1=1}^{pd} \cdots \bigcap_{j_K=1}^{pd} \mathcal A_{j_1,\ldots,j_K}(x)$. We then have $\{\tilde T_n\leqslant x\}=\{\tilde{n}^{1/2}\bar{\be}\in\mathcal A(x;K)\}$. Note that the set $\mathcal A_{j_1,\ldots,j_K}(x)$ is convex that only depends on the components in $S_{j_1,\ldots,j_K}$. We can reformulate $\mathcal A_{j_1,\ldots,j_K}(x)$ as follows:
\begin{align*}
  \mathcal A_{j_1,\ldots,j_K}(x)=\bigcap_{\ba\in\bS^{Kpd-1}:\,\ba_{S_{j_1,\ldots,j_K}}\in\bS^{K-1}} \{\bb\in\R^{Kpd}:\ba^{\T} \bb \leqslant x^{1/2}\} \,.
\end{align*}
Define $\cF
=\bigcup_{j_1=1}^{pd}\cdots\bigcup_{j_K=1}^{pd}\{\ba\in\bS^{Kpd-1}:\ba_{S_{j_1,\ldots,j_K}}\in\bS^{K-1}\}$. Then
$
  \mathcal A(x;K)=\bigcap_{\ba\in\cF}\{\bb\in\R^{Kpd}: \ba^{\T} \bb\leqslant x^{1/2}\}$ and
\begin{align}\label{eq:tildeTx}
\{\tilde{T}_n\leqslant x\}=\bigg\{\frac{1}{\sqrt{\tilde{n}}}\sum_{t=1}^{\tilde{n}}\ba^\T\be_t\leqslant x^{1/2}~\textrm{for any}~\ba\in\mathcal{F}\bigg\}
\end{align}
for any $x>0$. As indicated in \eqref{eq:tildeTx}, to construct the Gaussian approximation of $\mathbb{P}_{H_0}(\tilde{T}_n\leqslant x)$, we need to impose the following assumption on the tail behavior of $\ba^\T\be_t$. See also \cite{CCK2017} and \cite{CCW2020}.

\begin{condition}\label{cond.tail}
There exist some universal constants $C_1>1$, $C_2>0$ and $\tau_1\in(0,1]$ independent of $(K,p,d,n)$ such that
\begin{align*}
   \sup_{t\in[n]}\sup_{\ba\in\cal F}\bP(|\ba^{\T}\be_t|>x) \leqslant C_1\exp(-C_2 x^{\tau_1})
\end{align*}
for any $x>0$.
\end{condition}

Condition \ref{cond.tail} is stronger than necessary for the theoretical justification of our proposed method, % which is just for simplifying our technical arguments
and it can be weakened at the expense of much lengthier proofs. For example, Condition \ref{cond.tail} can be replaced by the assumption:
\begin{equation}\label{eq:newcond1}
\max_{t\in[n]}\max_{\ell\in[Kpd]}\mathbb{P}(|\eta_{t,\ell}|>x)\leqslant C_1\exp(-C_2x^{\tau_1})
\end{equation}
for any $x>0$. Recall $\eta_{t,\ell}=\phi_{l_1}(\bx_t)x_{t+k,l_2}$ for some $l_1\in[d], l_2\in[p]$ and $k\in[K]$. If $\bphi(\cdot)$ is selected as some bounded functions, then \eqref{eq:newcond1} holds provided that $\max_{t\in[n]}\max_{l_2\in[p]}\bP(|x_{t,l_2}|>x) \leqslant C_*\exp(-C_{**}x^{\tau_1})$ for any $x>0$. If $\bphi(\cdot)$ and $\bx_t$ satisfy $\max_{t\in[n]}\max_{l_1\in[d]}\bP\{|\phi_{l_1}(\bx_t)|>x\} \leqslant C_*\exp(-C_{**}x^{\tau_*})$ and $\max_{t\in[n]}\max_{l_2\in[p]}\bP(|x_{t,l_2}|>x) \leqslant C_*\exp(-C_{**}x^{\tau_{**}})$ for any $x>0$, by Lemma 2 of \cite{CTW2013}, we know \eqref{eq:newcond1} holds with $\tau_1=\tau_*\tau_{**}/(\tau_*+\tau_{**})$.  For any $\ba\in\cal F$, there exists $(j_1,\ldots,j_K)\in[pd]^K$ such that $\sum_{\ell=1}^K a_{j_\ell+(\ell-1)pd}^2=1$ and $a_{j}=0$ for $j\notin S_{j_1,\ldots,j_K}$, which implies $\sum_{\ell=1}^K|a_{j_\ell+(\ell-1)pd}|\leqslant \sqrt{K}$. By Bonferroni inequality and \eqref{eq:newcond1}, for any given $\ba\in\cal F$, it holds that
\begin{align}
  \bP(|\ba^{\T}\be_t|>x)
  \leqslant&~
  \sum_{\ell=1}^K \bP\bigg\{|\eta_{t,j_{\ell}+(\ell-1)pd}| >\frac{x}{\sum_{\ell=1}^K|a_{j_\ell+(\ell-1)pd}| } \bigg\}  \notag\\
  \leqslant&~
  \sum_{\ell=1}^K \bP\bigg\{|\eta_{t,j_{\ell}+(\ell-1)pd}| >\frac{x}{\sqrt{K}} \bigg\}  \leqslant C_*K\exp(-C_{**}K^{-\tau_1/2}x^{\tau_1})   \label{eq:inqcond1}
\end{align}
for any $x>0$, which provides a rough upper bound for $\max_{t\in[n]}\sup_{\ba\in\mathcal{F}}\bP(|\ba^{\T}\be_t|>x)$. When $K$ is a fixed positive integer, by \eqref{eq:inqcond1}, we know Condition \ref{cond.tail} is satisfied provided that \eqref{eq:newcond1} holds. If we only assume \eqref{eq:newcond1}, we can still establish the associated Gaussian approximation results based on \eqref{eq:inqcond1} rather than Condition \ref{cond.tail} but the associated arguments will be quite cumbersome.

\begin{condition}\label{cond.mix}
 Assume that $\{\bx_t\}$ is $\alpha$-mixing in the sense that
      \begin{align*}
        \alpha(k):=\sup_t\sup_{(A,B)\in \mathscr{F}_{-\infty}^t \times \mathscr{F}_{t+k}^{+\infty}} |\bP(A\cap B) - \bP(A)\bP(B)| \to 0 ~~\mbox{as } k\to \infty \,,
      \end{align*}
      where $\mathscr{F}_{-\infty}^u$ and $\mathscr{F}_{u+k}^{+\infty}$ are the $\sigma$-fields generated respectively by $\{\bx_t\}_{t\leqslant u}$ and $\{\bx_t\}_{t\geqslant u+k}$. Furthermore, there exist some universal constants $C_3>1$, $C_4>0$ and $\tau_2\in(0,1]$ independent of $(K,p,d,n)$ such that
$
        \alpha(k) \leqslant C_3\exp(-C_4 k^{\tau_2})$ for all $k\geqslant 1$.
\end{condition}

The $\alpha$-mixing assumption in Condition \ref{cond.mix} is weaker than the $\beta$-mixing assumption considered in \cite{CCK2019}. Restricting $\tau_2\in(0,1]$ is just to simplify the presentation. If the $\alpha$-mixing coefficients satisfy Condition \ref{cond.mix} with some constant $\tau_2>1$, then Condition \ref{cond.mix} will be satisfied automatically with $\tau_2=1$.
Under certain conditions, VAR processes, multivariate ARCH processes, and multivariate GARCH processes all satisfy Condition \ref{cond.mix} with $\tau_2=1$; see \cite{Hafner:Preminger:2009}, \cite{Boussama_etal:2011} and \cite{Wong_etal:2020}.
In addition, if we only require $\sup_{t\in[n]}\sup_{\ba\in\cal F} \bP(|\ba^{\T}\be_t|>x)=O\{x^{-(\nu+\epsilon)}\}$ for any $x>0$ in Condition 1 and $\alpha(k)=O\{k^{-\nu(\nu+\epsilon)/(2\epsilon)}\}$ for all $k\geqslant 1$ in Condition 2 with some constants $\nu>2$ and $\epsilon>0$, we can also apply the Fuk-Nagaev-type inequalities to construct the upper bounds for the tail probabilities of certain statistics for which our testing procedure still works for $Kpd$ diverging at some polynomial rate of $n$. We refer to Section 3.2 of \cite{CGY2018} for the implementation of the Fuk-Nagaev-type inequalities in such a scenario.

\begin{condition}\label{cond.m}
There exists a universal constant $C_5>0$ independent of $(K,p,d,n)$ such that
\begin{align*}
  \inf_{\ba\in\cal F} {\rm Var}\bigg(\frac{1}{\sqrt{\tilde{n}}}\sum_{t=1}^{\tilde{n}} \ba^{\T}\be_t \bigg)
\geqslant C_5\,.
\end{align*}
\end{condition}

\begin{condition}\label{cond.kern}
The kernel function $\mathcal K(\cdot)$ is continuously differentiable with bounded derivatives on $\R$ satisfying {\rm(i)} $\mathcal K(0)=1$, {\rm(ii)} $\mathcal K(x)=\mathcal K(-x)$ for any $x\in\R$,
%{\rm(iii)} $\int_{-\infty}^{\infty}|\mathcal K(x)|\,{\rm d}x <\infty$,
and {\rm(iii)} $|\mathcal K(x)|\leqslant C_6|x|^{-\vartheta}$ as $|x|\to\infty$ for some universal constants $C_6>0$ and $\vartheta>1$.
\end{condition}

Condition \ref{cond.m} is a mild technical assumption for the validity of the Gaussian approximation which requires the long-run variance of the sequence $\{\ba^\T\be_t\}$ to be  non-degenerate. Note that there are no explicit requirements on the cross-series dependence, and both weak and strong cross-series dependence are allowed in our theory.
%{\color{red} This condition is used for Nazarov's inequality and the proof of Theorem \ref{thm.H1}, while it is not necessary to require the variance of $(n-K)^{-1/2}\sum_{t=1}^{n-K}\ba^{\T}\be_t$ bounded away from 0. This restriction could be relaxed as the variance  of $(n-K)^{-1/2}\sum_{t=1}^{n-K}\ba^{\T}\be_t$ tends to 0 at some rate by using its recent generalization, e.g. Theorem 10 in \cite{DZ2020}. If so, the arguments of the proof are almost identical, except that some convergence rates need to be specified and the relationship between $p$ and $n$ may  change. Thus for the presentation issue, we just display Condition \ref{cond.m} in this paper.}
Condition \ref{cond.kern} is commonly used for the nonparametric estimation of the long-run covariance matrix; see \cite{NW1987} and \cite{Andrews1991}. For the kernel functions with bounded support such as Parzen kernel and Bartlett kernel, we have $\vartheta=\infty$ in Condition \ref{cond.kern}.

%\xs{More discussion about the conditions \ref{cond.tail} and \ref{cond.m} are needed! can they be verified for models with strong panel dependence?}

For $\tau_1$ and $\tau_2$ specified in Conditions \ref{cond.tail} and \ref{cond.mix}, we define
\begin{align}
 f_1(\tau_1,\tau_2)=\min\bigg( \frac{1}{15}\,,\frac{7\tau_1\tau_2}{18\tau_1+18\tau_2-3\tau_1\tau_2}\,,
  \frac{\tau_2}{9-3\tau_2}\bigg)\,.\label{eq:f1}
  \end{align}
Such defined $f_1(\tau_1,\tau_2)$ is used to control the divergence rate of $K$ which is determined from the technical proofs of Gaussian approximation theory. See Proposition \ref{prop.H0} in Section \ref{SecMed}. Notice that $\tau_1, \tau_2\in(0,1]$. When $\tau_1=\tau_2=1$, then $f_1(\tau_1,\tau_2)=1/15$.

Assume that the bandwidth $b_n$ involved in \eqref{Jn} satisfies $b_n\asymp n^\rho$ for some constant $0<\rho< (\vartheta-1)/(3\vartheta-2)$ with $\vartheta$ specified in Condition \ref{cond.kern}.  Let
  \begin{align}
 f_2(\rho,\vartheta)=\min\bigg(\frac{\rho}{5}\,,\frac{2\rho+\vartheta-1-3\rho\vartheta}{6\vartheta-3} \bigg)\,. \label{eq:f2}
\end{align}
Such defined $f_2(\rho,\vartheta)$ is also used to control the divergence rate of $K$ which is obtained from the estimation of long-run covariance matrix $\bSigma_{n,K}$. See Proposition \ref{prop.sigma} below. For given kernel function $\mathcal{K}(\cdot)$, the parameter $\vartheta$ is determined. Since $\vartheta=\infty$ if $\mathcal{K}(\cdot)$ is selected as the kernel functions with bounded support such as Parzen kernel and Bartlett kernel, then $f_2(\rho,\infty)=\min\{\rho/5,(1-3\rho)/6\}$. For given $\vartheta>1$, the optimal selection of $\rho$ that maximizes $f_2(\rho,\vartheta)$ with respect to $\rho$ is $(5\vartheta-5)/(21\vartheta-13)$ and the associated $f_2(\rho,\vartheta)=(\vartheta-1)/(21\vartheta-13)$.

\begin{proposition}\label{prop.sigma}
Assume that Conditions {\rm\ref{cond.tail}}, {\rm\ref{cond.mix}} and {\rm\ref{cond.kern}} hold. Let $b_n\asymp n^\rho$ for some constant $0<\rho<(\vartheta-1)/(3\vartheta-2)$, and $K=O(n^\delta)$ for some constant $0\leqslant \delta<f_2(\rho,\vartheta)$ with  $f_2(\rho,\vartheta)$ defined as  \eqref{eq:f2}. Then 
$
|\widehat{\bSigma}_{n,K}-\bSigma_{n,K}|_\infty
=o_\p[K^{-3}\{\log(npd)\}^{-2}]$
provided that $\log(pd)=o(n^{c})$ for some constant $c>0$ only depending on $(\tau_1,\tau_2,\rho,\vartheta,\delta)$.
\end{proposition}

%\jc{Proposition \ref{prop.sigma} presents a general result for the high-dimensional long-run covariance matrix estimation that allows the $\alpha$-mixing coefficients varying with the sample size $n$. The term $O(K^2n^{-\rho})$ gives the decaying rate of the bias in $\widehat{\bSigma}_{n,K}-\bSigma_{n,K}$. If $\mathcal{K}(\cdot)$ has bounded support, the convergence rate stated in Proposition \ref{prop.sigma} can be simplified as $O(K^2n^{-\rho})+O_\p[n^{(3\rho-1)/2}\{\log(npd)\}^{1/2}]+O_\p[n^{2\rho-1}\{\log(npd)\}^{2/\tau_1}]+O_\p[n^{\rho-1}\{\log(npd)\}^{1/\tau_*}]$. To make the estimate $\widehat{\bSigma}_{n,K}$ be consistent under the maximum element-wise loss in this setting, we need to require $\rho<1/3$ and $K=o(n^{\rho/2})$. Furthermore, with fixed $K$, the optimal selection of $\rho$ is $1/5$ and then $|\widehat{\bSigma}_{n,K}-\bSigma_{n,K}|_\infty=O_\p[n^{-1/5}\{\log(npd)\}^{1/2}]+O_\p[n^{-3/5}\{\log(npd)\}^{2/\tau_1}]+O_\p[n^{-4/5}\{\log(npd)\}^{1/\tau_*}]$.

%}

%\xs{can we provide some discussion regarding the 4 error bounds, can we explain the source of these errors?
%	this may get a bit technical, but could help the reader understand. also can be discuss some special cases, like when
%	$(p,d,K)$ fixed, or $p=d=n^{\kappa}$? }

Different from the existing literature of high-dimensional covariance matrix estimation, our procedure does not require $\widehat{\bSigma}_{n,K}$ to be consistent under the matrix $L_2$-operator norm and therefore it can work without imposing any structural assumptions on the underlying long-run covariance matrix $\bSigma_{n,K}$. More specifically, our procedure only requires $|\widehat{\bSigma}_{n,K}-\bSigma_{n,K}|_\infty=o_\p[K^{-3}\{\log(npd)\}^{-2}]$, which is a quite mild requirement and our proposed $\widehat{\bSigma}_{n,K}$ in Section \ref{sec:covest} satisfies this even when $p$ and $d$ grow exponentially with $n$. Now  we are ready to present the theoretical guarantees of the testing procedure \eqref{eq:testproc}.

%The consistency of the testing procedures is established in the following theorems.

\begin{theorem}\label{thm.H0}
Assume Conditions {\rm \ref{cond.tail}--\ref{cond.kern}} hold. Let $b_n\asymp n^\rho$ for some constant $0<\rho<(\vartheta-1)/(3\vartheta-2)$.
Select $K=O(n^\delta)$ for some constant $0\leqslant \delta<\min\{f_1(\tau_1,\tau_2)\,,f_2(\rho,\vartheta)\}$ with $f_1(\tau_1,\tau_2)$ and $f_2(\rho,\vartheta)$ defined as \eqref{eq:f1} and \eqref{eq:f2}, respectively.
Then 
$  \bP_{H_0}( T_n > \hat{\rm cv}_\alpha) \to \alpha$ as $n\rightarrow\infty$, provided that $\log(pd)=o(n^{c})$ for some constant $c>0$ only depending on $(\tau_1,\tau_2,\rho,\vartheta,\delta)$.
\end{theorem}

Theorem \ref{thm.H0} reveals the validity of our proposed test in the sense that the testing procedure maintains the nominal significance level asymptotically under the null hypothesis, where $pd$ is allowed to diverge exponentially with respect to the sample size $n$.
In Theorem \ref{thm.H1}, the asymptotic power of the proposed tests is analyzed.

\begin{theorem}\label{thm.H1}
Assume the conditions of Theorem {\rm\ref{thm.H0}} hold.
Let $\varrho$ be the largest element in the main diagonal of $\bSigma_{n,K}$, and write $\lambda(K,p,d,\alpha)=\{2\log(pd)\}^{1/2} +\{2\log(4K/\alpha)\}^{1/2}$. If
$
  \sum_{j=1}^{K}|\bgamma_j|_\infty^2
\geqslant n^{-1}K\varrho\lambda^2(K,p,d,\alpha)(1+\epsilon_n)^2$ under the alternative hypothesis
for some $\epsilon_n>0$ satisfying $\epsilon_n\rightarrow 0$ and $\varrho\lambda^2(K,p,d,\alpha)K^{-1}(\log K)^{-1}\epsilon_n^2\rightarrow\infty$,
then
$  \bP_{H_1}(T_n>\hat{\rm cv}_\alpha)\to 1$ as $n\to\infty$.
\end{theorem}

%\xs{a little bit discussion here would be helpful. Suppose $K$ is fixed and $\varrho$ is $O(1)$, what is the rate of alternative we can detect?}

Theorem \ref{thm.H1} shows that our proposed test is consistent under local alternatives. Recall $\bgamma=(\bgamma_1^{\T},\ldots,\bgamma_K^{\T})^{\T}$ with each $\bgamma_j\in\R^{pd}$. When $K$ is fixed and $\varrho=O(1)$, the latter of which holds under suitable assumptions on the data generating process, the condition that $|\bgamma|_\infty \geqslant Cn^{-1/2}\{\log(Kpd)\}^{1/2}$ for some positive constant $C$, is sufficient  for $\sum_{j=1}^K|\bgamma_j|^2_\infty \geqslant n^{-1}K\varrho\lambda^2(K,p,d,\alpha)(1+\epsilon_n)^2$. As we have discussed in Remark \ref{rk:3}, if the time series $\{\bx_t\}$ is strictly stationary, we know the transformed data $\{\be_t\}$ is also strictly stationary and the proposed test statistic $T_n$ given in \eqref{TestStat} essentially tests whether $\bgamma=\mathbb{E}(\be_t)=\bzero$ or not. As shown in Theorem 3 of \cite{CLX2013}, $n^{-1/2}\{\log(Kpd)\}^{1/2}$ is the minimax optimal separation rate of any tests for the $(Kpd)$-dimensional mean vector  hypothesis testing problem $H_0:\bgamma=\bzero$ versus $H_1:\bgamma\neq\bzero$ based on the data $\{\be_t\}_{t=1}^n$ if the smallest eigenvalues of ${\rm Var}(\be_t)$ are uniformly bounded away from zero.
That is, for any $\alpha,\, \beta>0$ satisfying $\alpha+\beta<1$, there exists a constant $\delta_0>0$ such that $\inf_{\bgamma\in\mathcal{M}(\delta_0)}\sup_{\xi_\alpha\in\mathcal{T}_\alpha}\bP_{H_1}(\mbox{reject $H_0$ based on $\xi_\alpha$})\leqslant 1-\beta$ for all sufficiently large $n$, $p$ and $d$, where $\mathcal{M}(\delta_0)=\{\bgamma\in\mathbb{R}^{Kpd}:|\bgamma|_\infty\geqslant \delta_0 n^{-1/2}\{\log(Kpd)\}^{1/2}\}$, and $\mathcal{T}_\alpha$ is the set of all $\alpha$-level tests for the test $H_0:\bgamma=\bzero$ versus $H_1:\bgamma\neq\bzero$. Hence, if the time series $\{\bx_t\}$ is strictly stationary, our proposed testing procedure with fixed $K$ will share some minimax optimal property.

\section{General martingale difference hypothesis and specification testing}\label{sec:general}
Our test procedure can also be extended to a more general martingale difference hypothesis, that is
\begin{align}\label{eq:newnull}
    H_0: \E(\bx_t\,|\,\mathscr{F}_{t-1})=\bmu_x\mbox{ for any } t\in\mathbb{Z}\,,
\end{align}
where $\bmu_x\in\R^p$ is an unknown vector. In this scenario, we can consider the test statistic
\begin{align}\label{eq:Tnnew}
T_n^{\rm new} = n\sum_{j=1}^K|\hat\bgamma_j^{\rm new}|_{\infty}^2 \,,
\end{align}
where $\hat{\bgamma}_j^{\rm new}
=(n-j)^{-1}\sum_{t=1}^{n-j}{\rm vec}\{\bphi(\bx_t)(\bx_{t+j}-\bar{\bx})^{\T}\}$ with $\bar{\bx}=n^{-1}\sum_{t=1}^n\bx_t$. Write $\mathring{\bx}_t=\bx_t-\bmu_x$. In comparison to $T_n$ given in \eqref{TestStat}, we replace $\bx_{t+j}$ there by its mean-centered
version $\bx_{t+j}-\bar{\bx}$ in $T_n^{\rm new}$. Notice that
\begin{align*}
\hat{\bgamma}_j^{\rm new}
%=&~ \frac{1}{n-j}\sum_{t=1}^{n-j}{\rm vec}\{\bphi(\bx_t)(\bx_{t+j}-\bar{\bx})^{\T}\} \\
%=& {\rm vec}\bigg\{\frac{1}{n-j}\sum_{t=1}^{n-j}\bphi(\bx_t)\bigg(\mathring{\bx}_{t+j}-\frac{1}{n}\sum_{t=1}^n\mathring{\bx}_t\bigg)^{\T}\bigg\} \\
%=&~ {\rm vec}\bigg\{\frac{1}{n-j}\sum_{t=1}^{n-j}\bphi(\bx_t)\mathring{\bx}_{t+j}^{\T} \bigg\} - {\rm vec}\bigg(\bigg[\frac{1}{n-j}\sum_{t=1}^{n-j}\E\{\bphi(\bx_t)\}\bigg]\bigg(\frac{1}{n}\sum_{t=1}^n\mathring{\bx}_t\bigg)^{\T}\bigg)  \\
%&~ - {\rm vec}\bigg\{\bigg(\frac{1}{n-j}\sum_{t=1}^{n-j}[\bphi(\bx_t)-\E\{\bphi(\bx_t)\}]\bigg)\bigg(\frac{1}{n}\sum_{t=1}^n\mathring{\bx}_t\bigg)^{\T}\bigg\} \\
=&~ \underbrace{\frac{1}{n-j}\sum_{t=1}^{n-j}{\rm vec}\bigg(\bphi(\bx_t)\mathring{\bx}_{t+j}^{\T} -\bigg[\frac{1}{n-j}\sum_{s=1}^{n-j}\E\{\bphi(\bx_s)\}\bigg]\mathring{\bx}_t^{\T}\bigg)}_{{\rm I}_j} \\
&~ + \underbrace{{\rm vec}\bigg(\bigg[\frac{1}{n-j}\sum_{t=1}^{n-j}\E\{\bphi(\bx_t)\}\bigg] \bigg(\frac{1}{n-j}\sum_{t=1}^{n-j}\mathring{\bx}_t - \frac{1}{n}\sum_{t=1}^n\mathring{\bx}_t\bigg)^{\T}\bigg)}_{{\rm II}_j} \\
&~ - \underbrace{{\rm vec}\bigg\{\bigg(\frac{1}{n-j}\sum_{t=1}^{n-j}[\bphi(\bx_t)-\E\{\bphi(\bx_t)\}]\bigg)\bigg(\frac{1}{n}\sum_{t=1}^n\mathring{\bx}_t\bigg)^{\T}\bigg\}}_{{\rm III}_j}\,.
%\\
%=&~ \underbrace{{\rm vec}\bigg\{\frac{1}{n-j}\sum_{t=1}^{n-j}\bphi(\bx_t)\mathring{\bx}_{t+j}^{\T}\bigg\}}_{\rm I}
%-\underbrace{{\rm vec} \bigg[ \bigg\{\frac{1}{n-j}\sum_{t=1}^{n-j}\bphi(\bx_{t})\bigg\}\bigg(\frac{1}{n}\sum_{t=1}^{n}\mathring{\bx}_t\bigg)^{\T}\bigg]}_{\rm II}
\end{align*}
Since $K=o(n)$ and $j\in[K]$, ${\rm I}_j$ is the leading term of $\hat{\bgamma}_j^{\rm new}$, and ${\rm II}_j$ and ${\rm III}_j$ are the negligible terms in comparison to ${\rm I}_j$.
Define
\begin{align*}
  \be_t^{\rm new}=
  \left(
    \begin{array}{c}
      {\rm vec} (\bphi(\bx_t)\mathring{\bx}_{t+1}^{\T} -[(n-1)^{-1}\sum_{s=1}^{n-1}\E\{\bphi(\bx_{s})\}]\mathring{\bx}_t^{\T}) \\
      \vdots \\
      {\rm vec} (\bphi(\bx_t)\mathring{\bx}_{t+K}^{\T} -[(n-K)^{-1}\sum_{s=1}^{n-K}\E\{\bphi(\bx_{s})\}]\mathring{\bx}_t^{\T}) \\
    \end{array}
  \right) \,.
\end{align*}
Write $\tilde{n}=n-K$. If Conditions \ref{cond.tail} and \ref{cond.m} hold for $\be_t^{\rm new}$, together with Condition \ref{cond.mix}, we know the null distribution of $T_n^{\rm new}$ can be approximated by that of its Gaussian analogue $G_K^{\rm new} = \sum_{j=1}^{K}   |\bg^{\rm new}_{\mathcal{L}_j}|_\infty^2$, where  $\mathcal{L}_j=\{(j-1)pd+1,\ldots,jpd\}$ and $\bg^{\rm new}=(g_1^{\rm new},\ldots,g_{Kpd}^{\rm new})^\T \sim\mathcal{N}(\bzero,\bSigma_{n,K}^{\rm new})$ with $\bSigma_{n,K}^{\rm new}={\rm Cov} (\tilde{n}^{-1/2}\sum_{t=1}^{\tilde{n}}\be_t^{\rm new})$. Write $\hat{\mathring{\bx}}_t=\bx_t-\bar{\bx}$ and
\begin{align*}
  \hat{\be}_t^{\rm new}=
  \left(
    \begin{array}{c}
      {\rm vec}[\bphi(\bx_t)\hat{\mathring{\bx}}_{t+1}^{\T} -\{(n-1)^{-1}\sum_{s=1}^{n-1}\bphi(\bx_{s})\}\hat{\mathring{\bx}}_t^{\T}] \\
      \vdots \\
      {\rm vec}[\bphi(\bx_t)\hat{\mathring{\bx}}_{t+K}^{\T} -\{(n-K)^{-1}\sum_{s=1}^{n-K}\bphi(\bx_{s})\}\hat{\mathring{\bx}}_t^{\T}] \\
    \end{array}
  \right) \,.
\end{align*}
Identical to \eqref{Jn}, we can adopt the following estimate for $\bSigma_{n,K}^{\rm new}$:
\begin{align*}
    \widehat{\bSigma}_{n,K}^{\rm new} = \sum_{j=-\tilde{n}+1}^{\tilde{n}-1}\calK\bigg(\frac{j}{b_n}\bigg) \widehat\bH_j^{\rm new}\,,
\end{align*}
where $\widehat{\bH}_j^{\rm new}=\tilde{n}^{-1}\sum_{t=j+1}^{\tilde{n}}(\hat\be_t^{\rm new}-\bar{\hat{\be}}^{\rm new})(\hat\be_{t-j}^{\rm new}-\bar{\hat{\be}}^{\rm new})$ if $j\geqslant 0$ and $\widehat{\bH}_j^{\rm new} =\tilde{n}^{-1} \sum_{t=-j+1}^{\tilde{n}} (\hat\be_{t+j}^{\rm new}-\bar{\hat{\be}}^{\rm new})(\hat\be_{t}^{\rm new} -\bar{\hat{\be}}^{\rm new})$ otherwise, with $\bar{\hat{\be}}^{\rm new} =\tilde{n}^{-1}\sum_{t=1}^{\tilde{n}}\hat\be_t^{\rm new}$. Algorithm  \ref{alg3} states how to implement the proposed general martingale difference hypothesis test in practice.

\begin{algorithm}[H] \caption{Testing procedure for general martingale difference hypothesis} \label{alg3}
\vspace{0.3em}
\begin{algorithmic}
\STATE\hspace{-1.3em}
{\bf Step 1.} Compute the test statistic $T_n^{\rm new}$ as in \eqref{eq:Tnnew}, and let $\boldsymbol{\Theta}$ be a $\tilde{n}\times\tilde{n}$ matrix with $(i,j)$th element \STATE\hspace{2.7em} $\mathcal K\{(i-j)/b_n\}$.
\vspace{0.5em}
\STATE\hspace{-1.3em}
{\bf Step 2.} Generate $\boldsymbol\xi=(\xi_1,\ldots,\xi_{\tilde{n}})^{\T} \sim \calN(\bzero,\boldsymbol{\Theta})$ independent of $\mathcal{X}_n$, and let
$  \hat{\bg}^{\rm new}=\tilde{n}^{-1/2}\sum_{t=1}^{\tilde{n}}\xi_t (\hat{\be}_t^{\rm new}-$
\STATE\hspace{2.71em} $\bar{\hat{\be}}^{\rm new})$.
\vspace{0.5em}
\STATE\hspace{-1.3em}
{\bf Step 3.} Draw $\hat\bg_1^{\rm new},\ldots,\hat\bg_B^{\rm new}$ independently by Step 2 for some large integer $B$.
\vspace{0.5em}
\STATE\hspace{-1.3em}
{\bf Step 4.} For given significance level $\alpha\in(0,1)$, take the $\lfloor B\alpha\rfloor$th largest value among $\hat{G}_{K,1}^{\rm new},\ldots,\hat{G}_{K,B}^{\rm new}$
\STATE\hspace{2.7em}
as the critical value $\hat{\rm cv}_{\alpha}$, where
$   \hat{G}_{K,i}^{\rm new} =\sum_{j=1}^{K} |\hat\bg_{i,\mathcal{L}_j}^{\rm new}|_\infty^2 $ with $\hat\bg_i^{\rm new}=(\hat{g}_{i,1}^{\rm new},\ldots,\hat{g}_{i,Kpd}^{\rm new})^{\T}$ and
\STATE\hspace{2.7em}
$\mathcal{L}_j=\{(j-1)pd+1,\ldots,jpd\}$.
\vspace{0.5em}
\STATE\hspace{-1.3em}
{\bf Step 5.} We reject $H_0$ defined as \eqref{eq:newnull} if $T_n^{\rm new}>\hat{\rm cv}_\alpha$.
\end{algorithmic}
\vspace{0.04in}
\end{algorithm}

Below we shall  provide some detailed discussion about potential extension of our test to the specification testing framework. Let $\by_t$ and $\bu_t$ be observable $p$-dimensional and $q$-dimensional time series, respectively. Consider the time series model
\begin{equation}\label{eq:model}
\by_t=\bh(\bu_t;\btheta_0)+\bx_t\,,\end{equation}
where $\bx_t$ is the error process, and $\bh(\cdot;\cdot)\in\mathbb{R}^p$ is a known link function with unknown truth $\btheta_0\in\mathbb{R}^m$. Without loss of generality, we assume $\mathbb{E}(\bx_t\,|\,\bu_t)=\bzero$. Model \eqref{eq:model} is quite general for our analysis where we can select $\bu_t$ as $\by_{t-1},\ldots,\by_{t-\ell}$ for some integer $\ell\geqslant1$. For the model diagnosis, we are interested in the hypothesis testing problem:
\begin{equation}\label{eq:nullhypo}
H_0: \{\bx_t\}_{t\in\mathbb{Z}}~\mbox{is a MDS}~~~~~\mbox{versus}~~~~~H_1:\{\bx_t\}_{t\in\mathbb{Z}}~\mbox{is not a MDS}.
\end{equation}
Based on the conditional moment restrictions $\mathbb{E}(\bx_t\,|\,\bu_t)=\bzero$, for given basis functions $\bpsi(\cdot):\mathbb{R}^{q}\rightarrow\mathbb{R}^l$ with $pl\geqslant m$, we can identify the unknown truth $\btheta_0$ by the $pl$ unconditional moment restrictions \begin{equation*}%\label{eq:uncond}
\mathbb{E}[\{\by_t-\bh(\bu_t;\btheta_0)\}\otimes\bpsi(\bu_t)]=\bzero\,,\end{equation*}
where $\otimes$ denotes the Kronecker product. %Notice that there are $pl$ moment restrictions used to identify $\btheta_0$.

{\it Case 1}. If $m$ is fixed or diverges slowly with the sample size $n$, applying the estimation procedure suggested in \cite{CCC2015}, we can obtain a consistent estimator $\hat{\btheta}_n$ for $\btheta_0$ and it admits the following asymptotic expansion:
\begin{equation}\label{eq:expan}
\hat{\btheta}_n-\btheta_0=\frac{1}{n}\sum_{t=1}^n\bw(\by_t,\bu_t)+\mbox{high order term}\,,
\end{equation}
where $\bw(\cdot)$ is the influence function such that $\mathbb{E}\{\bw(\by_t,\bu_t)\}=\bzero$.
Write $\hat{\bx}_t=\by_t-\bh(\bu_t;\hat{\btheta}_n)$. Together with \eqref{eq:expan}, it holds that
\begin{equation*}
\hat{\bx}_t=\bx_t-\nabla_{\btheta} \bh(\bu_t;\btheta_0)\cdot\frac{1}{n}\sum_{s=1}^n\bw(\by_s,\bu_s)+\mbox{high order term}\,.
\end{equation*}
Based on obtained $\{\hat{\bx}_t\}_{t=1}^n$, we can propose the following test statistic for \eqref{eq:nullhypo}:
\begin{equation}\label{eq:newtest}
T_n^{\natural}=n\sum_{j=1}^K|\bgamma_j^{\natural}|_\infty^2\,,
\end{equation}
where $\bgamma_j^{\natural}=(n-j)^{-1}\sum_{t=1}^{n-j}{\rm vec}\{\bphi(\hat{\bx}_t)\hat{\bx}_{t+j}^{\T}\}$. In comparison to the original test statistic $T_n$ given in (3) based on observed $\{\bx_t\}_{t=1}^n$, we replace $\bx_t$ there by its estimate $\hat{\bx}_t$.
By Taylor expansion, under some regularity conditions, it holds that
\begin{align*}
\bgamma_j^{\natural}
=&~ \frac{1}{n-j} \sum_{t=1}^{n-j}{\rm vec}\{\bphi(\bx_t)\bx_{t+j}^{\T}\}- \frac{1}{n}\sum_{t=1}^n\bA_j\bw(\by_t,\bu_t) + \mbox{high order term} \,,
\end{align*}
where $\bA_j=(n-j)^{-1}\sum_{t=1}^{n-j}\mathbb{E}\{\bx_{t+j}\otimes[\nabla_{\bx}\bphi(\bx_t)\nabla_{\btheta}\bh(\bu_t;\btheta_0)] +[\nabla_{\btheta}\bh(\bu_{t+j};\btheta_0)]\otimes\bphi(\bx_t)\}$. Define
\begin{align*}
  \be_t^{\natural}=
  \left(
    \begin{array}{c}
      {\rm vec} \{\bphi(\bx_t){\bx}_{t+1}^{\T} \}-\bA_1\bw(\by_t,\bu_t) \\
      \vdots \\
      {\rm vec} \{\bphi(\bx_t){\bx}_{t+K}^{\T} \}-\bA_K\bw(\by_t,\bu_t) \\
    \end{array}
  \right) \,.
\end{align*}
Recall $\tilde{n}=n-K$. Following the same arguments in Section 2.1, the null distribution of $T_n^{\natural}$ can be approximated by that of its Gaussian analogue $ G_K ^{\natural}= \sum_{j=1}^{K}   |\bg_{\mathcal{L}_j}^{\natural}|_\infty^2$, where  $\mathcal{L}_j=\{(j-1)pd+1,\ldots,jpd\}$ and $\bg^{\natural}=( g_1^{\natural},\ldots, g_{Kpd}^{\natural})^\T \sim\mathcal{N}(\bzero,\bSigma_{n,K}^{\natural})$ with $\bSigma_{n,K}^{\natural}={\rm Cov} (\tilde{n}^{-1/2}\sum_{t=1}^{\tilde{n}}\be_t^{\natural})$. The key challenge here is to construct a valid estimate $\widehat{\bSigma}_{n,K}^{\natural}$ satisfying $|\widehat{\bSigma}_{n,K}^{\natural}-{\bSigma}_{n,K}^{\natural}|_\infty=o_{\p}[K^{-3}\{\log(npd)\}^{-2}]$ with unknown $\bA_1,\ldots,\bA_K$ and unobserved $\{\bx_t\}$.

{\it Case 2}. If $m\gg n$, we need to assume the unknown truth $\btheta_0=(\theta_{0,1},\ldots,\theta_{0,m})^{\T}$ in \eqref{eq:model} is sparse. Let $\mathcal{S}=\{k\in[m]:\theta_{0,k}\neq0\}$. Using the penalized estimation procedure, for example, \cite{CTW2018}, we can obtain a sparse estimate $\hat{\btheta}_n$ for $\btheta_0$ satisfying the oracle property:  (i) $\mathbb{P}(\hat{\btheta}_{n,\mathcal{S}^{\rm c}}=\bzero)\rightarrow1$ as $n\rightarrow\infty$, and (ii) $\hat{\btheta}_{n,\mathcal{S}}$ follows the asymptotic expansion:
\begin{equation}\label{eq:asypexp}
\hat{\btheta}_{n,\mathcal{S}}-\btheta_{0,\mathcal{S}}-\bxi_n=\frac{1}{n}\sum_{t=1}^n\tilde{\bw}(\by_t,\bu_t)+\mbox{high order term}\,,
\end{equation}
where $\tilde{\bw}(\cdot)$ is the influence function such that $\mathbb{E}\{\tilde{\bw}(\by_t,\bu_t)\}=\bzero$, and $\bxi_n$ is the asymptotic bias satisfying $|\bxi_n|_\infty=O_{\p}(\delta_n)$ for some $\delta_n=o(1)$ but $\delta_n\gg n^{-1/2}$. To propose the testing procedure in the setting with $m\gg n$, we need to do the next three steps first: (a) identify the index set $\mathcal{S}$, (b) estimate the asymptotic bias $\bxi_n$, (c) obtain the bias-corrected estimate $\tilde{\btheta}_n$ for $\btheta_0$ based on the estimate of $\bxi_n$. Write $\hat{\bx}_t=\by_t-\bh(\bu_t;\tilde{\btheta}_n)$. We can still use the test statistic $T_n^{\natural}$ given in \eqref{eq:newtest} in current setting. To determine the associated critical value, we only need to replace $\bw(\cdot)$ and $\nabla_{\btheta}\bh(\cdot;\btheta_0)$ by $\tilde{\bw}(\cdot)$ and $\nabla_{\btheta_{\mathcal{S}}}\bh(\cdot;\btheta_0)$, respectively, in the procedure for the setting with fixed or slowly diverging $m$. However, as commented in \cite{CCTW2021}, if $\bh(\cdot;\btheta)$ is a nonlinear function of $\btheta$, the asymptotic bias $\bxi_n$ may include some unknown information which makes the estimation of $\bxi_n$ extremely difficult (if not impossible). How to address this problem requires further study.

\section{Simulation studies}\label{sec:numerical}
%\subsection{Preliminary}
In this section, we examine the finite sample performance  of our proposed test in comparison with the ones proposed by \cite{HLZ2017}. % for both low-dimensional and high-dimensional settings.
%when $p$ is low-dimension.
%In the second case, we investigate the finite sample performance of our test and Hong's test in high-dimensional scenarios.
All tests in our simulation are implemented at the $5\%$ significance level using $4000$ Monte Carlo replications, and the number of bootstrap replications used to determine the critical value $\hat{{\rm cv}}_\alpha$ in our procedure is chosen as $B=2000$. We
set the sample size $n\in\{100,300\}$ and lags $K\in\{2,4,6,8\}$. The dimension $p$ is set according to the ratio   $p/n \in \{0.04, 0.08, 0.15, 0.4, 1.2\}$, which covers low-, moderate- and high-dimensional scenarios.
Two types of maps are considered, i.e.,
(i) linear function ($d=p$), $\bphi(\bx_t)=\bx_t$;
(ii) both linear and quadratic functions ($d=2p$), $\bphi(\bx_t)=\{\bx_t^{\T}, (\bx_t^2)^{\T}\}^{\T}$.
Furthermore, we use three kernel functions for the estimation of long-run covariance matrix $\bSigma_{n,K}$, i.e.,
\begin{itemize}
  \item[(a)]Quadratic Spectral (QS) kernel: $\mathcal{K}_{\rm{QS}}(x)=25(12\pi^2x^2)^{-1}\{ (6\pi x/5)^{-1}\sin(6\pi x/5) - \cos(6\pi x/5)\}$.
  \item[(b)]Parzen (PR) kernel: $\mathcal{K}_{\rm{PR}}(x)=(1-6x^2+6|x|^3)I(0\leqslant|x|\leqslant 1/2)+2(1-|x|)^3I(1/2 < |x|\leqslant 1)$.

  \item[(c)]Bartlett (BT) kernel: $\mathcal{K}_{\rm{BT}}(x)=(1-|x|)I(|x|\leqslant 1)$.
  \end{itemize}

Recall $\tilde{n}=n-K$. We use the data-driven bandwidth formulas developed in \cite{Andrews1991} to determine the associated bandwidth $b_n$ involved in these three kernel functions, that is,
$
b_{{\rm QS}}=1.3221\{\hat a(2) \tilde{n}\}^{1/5}$,
$b_{{\rm PR}}=2.6614\{\hat a(2) \tilde{n}\}^{1/5}$ and
$b_{{\rm BT}}=1.1447\{\hat a(1) \tilde{n}\}^{1/3}$,
where
$\hat a(2)=\{\sum_{\ell=1}^{Kpd} 4\hat\rho_\ell^2 \hat\sigma_\ell^4 (1-\hat\rho_\ell)^{-8}\}\{\sum_{\ell=1}^{Kpd}\hat\sigma_\ell^4(1-\hat\rho_\ell)^{-4}\}^{-1}$
and $\hat a(1)=\{\sum_{\ell=1}^{Kpd} 4\hat\rho_\ell^2 \hat\sigma_\ell^4 (1-\hat\rho_\ell)^{-6}(1+\hat\rho_\ell)^{-2}\} \{\sum_{\ell=1}^{Kpd}\hat\sigma_\ell^4(1-\hat\rho_\ell)^{-4}\}^{-1}$,
with $\hat\rho_\ell$ and $\hat\sigma_\ell^2$ being, respectively, the estimated autoregressive coefficient and innovation variance from fitting an AR(1) model to time series $\{\eta_{t,\ell}\}_{t=1}^{\tilde{n}}$, the $\ell$th component sequence of $\{\be_t\}_{t=1}^{\tilde{n}}$ defined in \eqref{eq:ft}.
Denote the test statistics based on the three kernels with linear map by $T_{{\rm{QS}}}^{l}$, $T_{{\rm{PR}}}^{l}$ and $T_{{\rm{BT}}}^{l}$, respectively,
and denote the ones with both linear and quadratic map by $T_{{\rm{QS}}}^{q}$, $T_{{\rm{PR}}}^{q}$ and $T_{{\rm{BT}}}^{q}$, respectively. Note that the data-driven formulas by \cite{Andrews1991} are based on AR(1) model assumption and also deliver an estimation-optimal bandwidth in the low-dimensional setting. Here we apply it to determine the associated bandwidth $b_n$ in both moderate- and high-dimensional settings since there are no other known formulas and the numerical studies in \cite{CYZ2017} show such formula seems to work well when the dimension is large.
We also include three tests proposed by  \cite{HLZ2017} in our simulation comparison, i.e., the trace-based test $Z_{\rm{tr}}$, the determinant-based test  $Z_{\rm{det}}$, and the large-dimensional test $Zd_{\rm tr}$. Note that  \cite{HLZ2017} only examined the finite sample performance of $Z_{\rm{tr}}$ and $Z_{\rm{det}}$, which cannot be implemented when $p>\sqrt{n}$, whereas $Zd_{\rm tr}$ is shown to be valid under the assumption $p/n\rightarrow 0$ and its implementation becomes infeasible when $p>n$. The tests of \cite{HLZ2017} require the matrix normalization which is computationally prohibitive in the high-dimensional setting. See Section \ref{sec:cost} in the supplementary material for the comparison of computational cost between our test and the tests of \cite{HLZ2017}.

\subsection{Empirical size}
To examine the empirical size, we consider the following models:

\begin{itemize}[leftmargin=1.8cm]
  \item[Model 1.~] i.i.d. normal sequence: $\bx_t\overset{{\rm i.i.d.}}{\sim} \mathcal{N}(\bzero, \bA)$ where $\bA=(a_{kl})_{p\times p}$ with $a_{kl}=0.995^{|k-l|}$ for any $k,l\in [p]$.
   \item[Model 2.~] Stochastic volatility model: $\bx_t=\boldsymbol{\varepsilon}_t\exp(\boldsymbol{\sigma}_t)$
  with $\boldsymbol{\sigma}_t=0.25\boldsymbol{\sigma}_{t-1}+0.05{\bf u}_t$,
       $\boldsymbol{\varepsilon}_t\overset{{\rm i.i.d.}}{\sim}\calN(\bzero,\bOmega_\varepsilon)$ and ${\bf u}_t\overset{{\rm i.i.d.}}{\sim}\calN(\bzero,\bOmega_u)$, where $\bOmega_\varepsilon=(\omega_{\varepsilon,kl})_{p\times p}$ and $\bOmega_{u}=(\omega_{u,kl})_{p\times p}$ with $\omega_{\varepsilon,kl}=I(k=l)+0.4I(k\neq l)$ and $\omega_{u,kl}=0.9^{|k-l|}$ for any $k,l\in[p]$.
  \item[Model 3.~] Bivariate constant conditional correlation GARCH(1,1) model:
  $
    \bx_t ={\bf b}_t^{1/2}\circ \boldsymbol{\varepsilon}_t$ with ${\bf b}_t = {\ba}_0 + \bA_1{\bf b}_{t-1} + \bA_2\bx_{t-1}^2$ and
    $\boldsymbol{\varepsilon}_t \overset{{\rm i.i.d.}}{\sim} \calN(\bzero,\bOmega_\varepsilon)$,
  where $\circ$ denotes the Hadamard product, $\ba_0=(0.2,0.1\,{\bf 1}_{p-1}^{\T})^{\T}$, $\bA_1=0.9\,{\bf I}_p$,
  $\bA_2={\rm diag}(0.05,0.08,0.03\, {\bf 1}_{p-2}^{\T})$,
  and $\bOmega_\varepsilon=(\omega_{\varepsilon,kl})_{p\times p}$ with $\omega_{\varepsilon,kl}=I(k=l)+0.5I(k\neq l)$ for any $k,l\in[p]$. Here ${\bf 1}_q$ and ${\bf I}_q$ denote, respectively, the $q$-dimensional vector with all components being $1$ and $q$-dimensional identity matrix for any given integer $q$.
\end{itemize}
A few comments are in order. Model 1 was used by \cite{CYZ2017} in their simulation for high-dimensional white noise testing problem. Model 2 is the multivariate extension of the univariate stochastic volatility model considered in \cite{EV2006a} for the univariate martingale difference hypothesis testing problem.
Model 3 is motivated from \cite{HLZ2017}, which reduces to the bivariate GARCH model considered in \cite{HLZ2017} when $p=2$.

As seen from Table \ref{tab:M1-M3}, our tests have quite accurate size when the dimension $p$ is low for all models. For a fixed sample size $n$, the rejection rates tend to decrease as the dimension $p$ increases, showing the impact on the bootstrap-based approximation from the dimension $p$. For a fixed dimension $p$,  enlarging sample size from $n=100$ to $n=300$ helps to bring down the size distortion to some extent for most kernels and maps, e.g., the empirical sizes for Model 1--3 are undersized when $n=100$ and $p/n=1.2~(p=120)$, and  the empirical sizes increase and become much closer to the 5\% nominal level when $n=300$ and $p/n=0.4~(p=120)$.  Overall our tests show reasonably good size control and the undersize phenomenon for the moderate- and high-dimensional scenarios could be due to the bandwidth choice, which is always a difficult issue in practice. The three tests of \cite{HLZ2017} also show quite accurate size for Models 1 and 2, and there is some noticeable over-rejection   for Model 3 when $n=100$. When $n=300$ and $p/n=0.4$, we are unable to implement the test $Zd_{\rm tr}$  even though $p<n$. The reason is that the computation of $Zd_{\rm tr}$ requires to store  five $120^2\times 120^2$ matrices,  and product of three $120^2\times 120^2$ matrices during the calculation, which results in running out of the memory (RAM: 8158 MB). This indicates the difficulty of implementing their tests for $p=120$ and beyond.

In order to investigate the influence of the data-driven bandwidth used in our simulation,
we examine the sensitivity of our size and power results by replacing the data-driven bandwidth $b_n$ by its scaled version $c\cdot b_n$ with $c\in\{2^{-3},2^{-2},2^{-1},2^1,2^2,2^3\}$. Simulation results for Bartlett kernel are displayed in Tables \ref{tab:size_cbn} and  \ref{tab:power_cbn}. Simulation results for Quadratic Spectral kernel and Parzen kernel are reported in the supplementary material.
For different multiplies $c$, the sizes and powers are relatively robust. In addition, we find that the results for $c<1$ perform a little better than these for $c>1$ in general, but not by much. Therefore, the choice of $c=1$ in our simulation is reasonable.

%Then we also record 'NA' in the table of simulation results, which means that Hong's method for high dimensional case needs more memory than ours. Thus our proposed method is more applicable in practice.

%It can be seen that our tests under-reject for all cases. For Model $M_1$ and $M_3$, the size distortion gets worse as we increase $p$ for a fixed sample size $n$. For a fixed $p$,

%For the model $M_2$, when $n=300$, the size is quite close to the nominal level $5\%$ for most combinations of kernel and map, and seems quite satisfactory overall. Although Hong's method performs sightly better than our method when $p$ is small,  we can see that the sizes for Hong's method inflate as the increase of $K$ for a fixed sample size $n$ and dimension $p$, which means our method is more stable.

%We did not include Hong's test in the table, as the only feasible case for their tests is when $(n,p)=(300,15)$, which yields empirical rejection rates $0.203\,(K=4)$, $0.384\,(K=6)$ and $0.536\,(K=8)$, respectively.
%This is not surprising in view of the results presented earlier.

%\begin{landscape}
\begin{sidewaystable}[htp]
\scriptsize
  \centering
  \caption{Empirical sizes ($\%$) of the tests $T_{\rm QS}^l$, $T_{\rm PR}^l$, $T_{\rm BT}^l$, $T_{\rm QS}^q$, $T_{\rm PR}^q$, $T_{\rm BT}^q$, $Z_{\rm tr}$, $Z_{\rm det}$ and $Zd_{\rm tr}$ for Models 1--3 at the 5\% nominal level. }
  \resizebox{!}{6.3cm}{
    \begin{tabular}{ccc|ccccccccc| ccccccccc| ccccccccc}
          &       &       & \multicolumn{9}{c|}{Model 1}                                   & \multicolumn{9}{c|}{Model 2}                                     & \multicolumn{9}{c}{Model 3} \\[0.6em]
    $n$     & $p/n$   & $K$    & $T_{\rm QS}^l$  & $T_{\rm PR}^l$  & $T_{\rm BT}^l$  & $T_{\rm QS}^q$  & $T_{\rm PR}^q$  & $T_{\rm BT}^q$    & $Z_{\rm tr}$   & $Z_{\rm det}$  & $Zd_{\rm tr}$ & $T_{\rm QS}^l$  & $T_{\rm PR}^l$  & $T_{\rm BT}^l$  & $T_{\rm QS}^q$  & $T_{\rm PR}^q$  & $T_{\rm BT}^q$    & $Z_{\rm tr}$   & $Z_{\rm det}$  & $Zd_{\rm tr}$  & $T_{\rm QS}^l$  & $T_{\rm PR}^l$  & $T_{\rm BT}^l$  & $T_{\rm QS}^q$  & $T_{\rm PR}^q$  & $T_{\rm BT}^q$    & $Z_{\rm tr}$   & $Z_{\rm det}$  & $Zd_{\rm tr}$ \\[0.5em]
    \hline
    100   & 0.04  & 2     & 4.2   & 4.5   & 4.5   & 4.3   & 4.3   & 4.4   & 5.2   & 4.4   & 5.5   & 4.2   & 4.4   & 4.7   & 2.2   & 2.4   & 2.5   & 5.2   & 4.7   & 4.8   & 3.5   & 3.6   & 4.2   & 2.9   & 2.8   & 3.2   & 6.7   & 6.3   & 5.2  \\
          &       & 4     & 5.1   & 5.0   & 5.2   & 4.6   & 4.3   & 4.5   & 5.2   & 5.2   & 5.4   & 3.1   & 3.3   & 3.5   & 2.9   & 2.8   & 3.2   & 4.3   & 4.9   & 6.3   & 3.2   & 3.2   & 3.5   & 3.0   & 3.1   & 3.4   & 6.9   & 6.9   & 6.3  \\
          &       & 6     & 4.6   & 4.4   & 4.8   & 4.5   & 4.5   & 4.6   & 5.0   & 4.7   & 6.0   & 3.0   & 2.9   & 3.5   & 2.8   & 2.7   & 2.9   & 5.4   & 4.4   & 6.0   & 3.1   & 3.1   & 3.6   & 3.9   & 4.0   & 4.1   & 6.5   & 6.8   & 6.1  \\
          &       & 8     & 4.4   & 4.4   & 4.7   & 5.0   & 4.9   & 5.1   & 5.6   & 4.9   & 6.2   & 2.8   & 2.8   & 3.3   & 2.9   & 2.8   & 3.1   & 5.4   & 4.5   & 6.5   & 3.3   & 3.3   & 4.0   & 4.5   & 4.4   & 4.9   & 6.7   & 7.9   & 6.9  \\[0.3em]
          & 0.08  & 2     & 4.1   & 4.1   & 4.1   & 3.1   & 3.2   & 3.4   & 4.8   & 4.9   & 5.2   & 3.8   & 3.9   & 4.0   & 1.9   & 1.8   & 1.8   & 4.4   & 4.9   & 5.2   & 3.3   & 3.2   & 3.5   & 2.7   & 2.5   & 2.7   & 5.9   & 8.0   & 5.2  \\
          &       & 4     & 3.9   & 3.8   & 4.0   & 4.4   & 4.3   & 4.7   & 6.0   & 4.8   & 5.2   & 3.0   & 3.0   & 3.5   & 1.8   & 1.9   & 2.1   & 5.4   & 5.1   & 6.0   & 2.6   & 2.5   & 3.0   & 2.7   & 2.7   & 2.7   & 7.1   & 8.7   & 5.2  \\
          &       & 6     & 4.6   & 4.4   & 4.6   & 4.4   & 4.4   & 4.7   & 6.7   & 6.3   & 6.2   & 2.2   & 2.1   & 2.5   & 2.2   & 2.2   & 2.4   & 5.5   & 5.9   & 5.6   & 2.2   & 2.3   & 2.8   & 4.4   & 4.2   & 4.5   & 7.5   & 8.2   & 7.1  \\
          &       & 8     & 4.8   & 4.9   & 5.2   & 4.0   & 3.9   & 4.2   & 7.4   & 5.7   & 5.4   & 2.3   & 2.0   & 3.0   & 2.1   & 2.2   & 2.4   & 7.2   & 5.5   & 6.9   & 2.8   & 2.8   & 3.3   & 4.4   & 4.4   & 4.9   & 8.2   & 8.9   & 7.6  \\[0.3em]
          & 0.15  & 2     & 4.2   & 4.4   & 4.4   & 4.0   & 3.8   & 4.2   & NA    & NA    & 4.6   & 3.4   & 3.4   & 3.7   & 1.9   & 1.9   & 2.3   & NA    & NA    & 4.8   & 3.4   & 3.4   & 4.1   & 1.9   & 1.9   & 1.9   & NA    & NA    & 5.1  \\
          &       & 4     & 4.3   & 4.2   & 4.6   & 3.2   & 3.2   & 3.5   & NA    & NA    & 4.3   & 2.5   & 2.5   & 2.7   & 1.7   & 1.8   & 2.1   & NA    & NA    & 5.9   & 2.2   & 2.2   & 2.7   & 2.5   & 2.4   & 2.7   & NA    & NA    & 5.8  \\
          &       & 6     & 4.2   & 4.0   & 4.4   & 3.8   & 3.8   & 4.0   & NA    & NA    & 5.2   & 2.7   & 2.9   & 3.3   & 1.7   & 1.6   & 1.9   & NA    & NA    & 5.7   & 2.2   & 2.1   & 2.8   & 2.8   & 2.7   & 3.0   & NA    & NA    & 7.4  \\
          &       & 8     & 4.0   & 4.1   & 4.5   & 4.4   & 4.5   & 4.8   & NA    & NA    & 6.3   & 2.7   & 2.6   & 3.1   & 2.1   & 2.2   & 2.4   & NA    & NA    & 6.4   & 1.8   & 1.8   & 2.6   & 3.4   & 3.6   & 3.6   & NA    & NA    & 8.3  \\[0.3em]
          & 0.40   & 2     & 3.8   & 4.0   & 4.1   & 2.1   & 2.3   & 2.6   & NA    & NA    & 4.8   & 3.4   & 3.5   & 3.9   & 2.0   & 2.3   & 2.5   & NA    & NA    & 4.7   & 2.7   & 2.5   & 3.0   & 1.8   & 1.8   & 1.9   & NA    & NA    & 5.5  \\
          &       & 4     & 2.8   & 2.9   & 3.2   & 2.5   & 2.6   & 2.6   & NA    & NA    & 5.3   & 3.2   & 3.2   & 3.6   & 2.1   & 2.1   & 2.3   & NA    & NA    & 5.4   & 1.7   & 1.7   & 2.1   & 2.0   & 2.2   & 1.9   & NA    & NA    & 6.6  \\
          &       & 6     & 2.9   & 3.0   & 3.4   & 2.8   & 2.8   & 3.2   & NA    & NA    & 5.6   & 2.6   & 2.6   & 3.1   & 2.1   & 2.0   & 2.1   & NA    & NA    & 6.0   & 1.1   & 1.1   & 1.7   & 2.4   & 2.6   & 2.3   & NA    & NA    & 7.9  \\
          &       & 8     & 3.2   & 3.0   & 3.4   & 3.2   & 3.1   & 3.4   & NA    & NA    & 5.8   & 2.8   & 2.8   & 3.2   & 2.6   & 2.5   & 2.8   & NA    & NA    & 5.2   & 1.5   & 1.4   & 2.2   & 3.3   & 3.6   & 3.4   & NA    & NA    & 9.2  \\[0.3em]
          & 1.20   & 2     & 2.2   & 2.3   & 2.5   & 1.1   & 1.2   & 1.3   & NA    & NA    & NA    & 3.8   & 3.8   & 4.3   & 2.6   & 2.8   & 2.7   & NA    & NA    & NA    & 1.6   & 1.6   & 2.0   & 2.9   & 3.3   & 2.5   & NA    & NA    & NA \\
          &       & 4     & 2.0   & 2.1   & 2.6   & 1.1   & 1.2   & 1.2   & NA    & NA    & NA    & 2.9   & 3.1   & 3.2   & 2.5   & 2.5   & 2.7   & NA    & NA    & NA    & 1.1   & 1.1   & 1.8   & 3.9   & 4.1   & 3.1   & NA    & NA    & NA \\
          &       & 6     & 2.0   & 2.1   & 2.4   & 1.1   & 1.1   & 1.3   & NA    & NA    & NA    & 3.3   & 3.3   & 3.9   & 2.3   & 2.4   & 2.4   & NA    & NA    & NA    & 1.1   & 1.1   & 1.5   & 4.9   & 5.3   & 3.9   & NA    & NA    & NA \\
          &       & 8     & 1.4   & 1.7   & 2.0   & 1.2   & 1.2   & 1.4   & NA    & NA    & NA    & 3.1   & 3.2   & 3.8   & 2.9   & 3.1   & 3.2   & NA    & NA    & NA    & 1.1   & 1.0   & 1.5   & 6.3   & 6.7   & 5.5   & NA    & NA    & NA \\[0.3em]
    \hline
    300   & 0.04  & 2     & 5.6   & 5.5   & 5.8   & 4.2   & 4.1   & 4.4   & 5.1   & 5.5   & 5.8   & 4.1   & 4.1   & 4.2   & 3.8   & 3.7   & 3.9   & 4.9   & 5.6   & 4.7   & 4.0   & 4.0   & 4.0   & 3.6   & 3.4   & 3.8   & 6.2   & 7.5   & 5.2  \\
          &       & 4     & 3.9   & 4.2   & 4.5   & 4.7   & 4.6   & 5.0   & 5.9   & 5.4   & 5.0   & 3.7   & 3.9   & 4.2   & 2.9   & 2.8   & 3.2   & 6.1   & 5.9   & 5.6   & 3.8   & 3.7   & 4.1   & 3.3   & 3.4   & 3.9   & 6.2   & 6.6   & 5.5  \\
          &       & 6     & 4.2   & 4.1   & 4.2   & 5.5   & 5.2   & 5.5   & 6.4   & 6.7   & 5.6   & 3.9   & 3.6   & 3.9   & 3.9   & 4.0   & 4.2   & 6.6   & 6.8   & 6.4   & 3.7   & 3.7   & 4.1   & 4.7   & 4.7   & 4.9   & 7.3   & 7.9   & 5.6  \\
          &       & 8     & 4.7   & 4.8   & 5.0   & 6.0   & 6.0   & 6.3   & 7.1   & 6.9   & 5.8   & 3.7   & 3.8   & 4.0   & 4.1   & 4.0   & 4.3   & 7.1   & 6.4   & 5.1   & 3.2   & 3.0   & 3.4   & 4.4   & 4.4   & 4.8   & 8.6   & 8.0   & 6.3  \\[0.3em]
          & 0.08  & 2     & 4.8   & 4.8   & 5.0   & 4.0   & 4.0   & 4.1   & NA    & NA    & 5.5   & 4.2   & 4.3   & 4.4   & 3.5   & 3.6   & 3.8   & NA    & NA    & 4.8   & 3.6   & 3.5   & 3.8   & 3.2   & 3.2   & 3.2   & NA    & NA    & 5.6  \\
          &       & 4     & 3.8   & 3.8   & 3.9   & 4.1   & 4.0   & 4.2   & NA    & NA    & 5.2   & 3.8   & 3.5   & 3.8   & 3.7   & 3.6   & 3.9   & NA    & NA    & 5.0   & 3.7   & 3.5   & 4.0   & 3.2   & 3.0   & 3.4   & NA    & NA    & 5.4  \\
          &       & 6     & 4.6   & 4.4   & 5.0   & 5.0   & 5.0   & 5.4   & NA    & NA    & 5.0   & 3.6   & 3.3   & 3.8   & 4.1   & 4.2   & 4.4   & NA    & NA    & 5.4   & 3.3   & 3.2   & 3.7   & 3.4   & 3.3   & 3.7   & NA    & NA    & 5.3  \\
          &       & 8     & 3.9   & 4.2   & 4.3   & 5.7   & 5.9   & 6.1   & NA    & NA    & 5.4   & 3.7   & 3.6   & 4.1   & 3.8   & 3.7   & 4.0   & NA    & NA    & 5.4   & 3.0   & 3.0   & 3.4   & 3.6   & 3.7   & 4.1   & NA    & NA    & 5.9  \\[0.3em]
          & 0.15  & 2     & 4.7   & 4.6   & 4.8   & 3.8   & 3.9   & 4.3   & NA    & NA    & 6.0   & 4.4   & 4.4   & 4.7   & 3.7   & 3.6   & 3.9   & NA    & NA    & 4.6   & 3.9   & 4.0   & 4.2   & 3.1   & 2.9   & 3.3   & NA    & NA    & 5.0  \\
          &       & 4     & 4.4   & 4.4   & 4.6   & 4.4   & 4.3   & 4.6   & NA    & NA    & 4.6   & 3.7   & 3.9   & 3.9   & 4.2   & 4.2   & 4.4   & NA    & NA    & 5.1   & 3.2   & 3.2   & 3.6   & 3.2   & 3.0   & 3.3   & NA    & NA    & 5.3  \\
          &       & 6     & 3.9   & 3.9   & 4.1   & 4.6   & 4.4   & 4.8   & NA    & NA    & 5.1   & 3.5   & 3.5   & 3.8   & 3.4   & 3.7   & 3.8   & NA    & NA    & 5.3   & 3.2   & 3.0   & 3.4   & 3.6   & 3.5   & 3.8   & NA    & NA    & 6.8  \\
          &       & 8     & 3.9   & 4.0   & 4.2   & 4.4   & 4.4   & 4.7   & NA    & NA    & 5.6   & 3.5   & 3.5   & 3.7   & 4.2   & 4.3   & 4.4   & NA    & NA    & 5.6   & 3.0   & 3.0   & 3.4   & 3.6   & 3.4   & 4.1   & NA    & NA    & 5.9  \\[0.3em]
          & 0.40   & 2     & 4.2   & 4.2   & 4.3   & 2.6   & 2.6   & 2.8   & NA    & NA    & NA    & 4.5   & 4.6   & 4.8   & 3.1   & 3.0   & 3.4   & NA    & NA    & NA    & 3.8   & 3.8   & 4.1   & 2.8   & 2.7   & 3.1   & NA    & NA    & NA \\
          &       & 4     & 3.5   & 3.5   & 3.6   & 3.2   & 3.2   & 3.5   & NA    & NA    & NA    & 4.2   & 4.3   & 4.5   & 3.8   & 3.7   & 4.0   & NA    & NA    & NA    & 3.1   & 3.1   & 3.5   & 2.7   & 2.6   & 3.0   & NA    & NA    & NA \\
          &       & 6     & 3.7   & 3.9   & 4.2   & 4.1   & 4.1   & 4.7   & NA    & NA    & NA    & 4.1   & 4.0   & 4.4   & 3.9   & 4.0   & 4.2   & NA    & NA    & NA    & 2.7   & 2.6   & 3.0   & 2.4   & 2.3   & 2.7   & NA    & NA    & NA \\
          &       & 8     & 3.2   & 3.3   & 3.8   & 4.1   & 4.0   & 4.6   & NA    & NA    & NA    & 4.1   & 4.1   & 4.2   & 4.2   & 4.2   & 4.4   & NA    & NA    & NA    & 2.3   & 2.3   & 2.8   & 2.8   & 2.9   & 3.1   & NA    & NA    & NA \\[0.3em]
          & 1.20   & 2     & 3.1   & 3.0   & 3.4   & 1.8   & 1.7   & 2.0   & NA    & NA    & NA    & 4.0   & 4.2   & 4.2   & 3.9   & 3.9   & 4.0   & NA    & NA    & NA    & 3.8   & 3.8   & 4.0   & 2.4   & 2.4   & 2.8   & NA    & NA    & NA \\
          &       & 4     & 2.3   & 2.2   & 2.4   & 1.8   & 1.8   & 2.0   & NA    & NA    & NA    & 3.8   & 3.9   & 4.0   & 3.9   & 3.9   & 4.2   & NA    & NA    & NA    & 2.7   & 2.7   & 3.1   & 1.8   & 1.9   & 2.0   & NA    & NA    & NA \\
          &       & 6     & 1.3   & 1.2   & 1.7   & 1.7   & 1.7   & 1.9   & NA    & NA    & NA    & 3.8   & 3.5   & 3.9   & 4.2   & 4.4   & 4.7   & NA    & NA    & NA    & 2.1   & 2.2   & 2.6   & 2.1   & 2.1   & 2.2   & NA    & NA    & NA \\
          &       & 8     & 1.1   & 1.3   & 1.8   & 1.7   & 1.8   & 2.1   & NA    & NA    & NA    & 4.2   & 4.4   & 4.6   & 4.0   & 3.9   & 4.2   & NA    & NA    & NA    & 1.8   & 1.7   & 2.4   & 2.0   & 2.0   & 2.5   & NA    & NA    & NA \\
    \end{tabular}%
    }
  \label{tab:M1-M3}%
\end{sidewaystable}%
%\end{landscape}

%\begin{landscape}
\begin{sidewaystable}[htp]
\renewcommand\arraystretch{1.2}
\scriptsize
  \centering
  \caption{Empirical sizes ($\%$) of the tests $T_{\rm BT}^l$ and $T_{\rm BT}^q$ for Models 1--3 at the 5\% nominal level, where $c$ represents the constant which is multiplied by Andrews' bandwidth. }
  \resizebox{!}{5.7cm}{
    \begin{tabular}{ccc|ccccccc|ccccccc|ccccccc|ccccccc|ccccccc|ccccccc}
          &       &       & \multicolumn{7}{c|}{Model 1 with $T_{\rm BT}^l$}            & \multicolumn{7}{c|}{Model 1 with $T_{\rm BT}^q$}            & \multicolumn{7}{c|}{Model 2 with $T_{\rm BT}^l$}            & \multicolumn{7}{c|}{Model 2 with $T_{\rm BT}^q$}            & \multicolumn{7}{c|}{Model 3 with $T_{\rm BT}^l$}            & \multicolumn{7}{c}{Model 3 with $T_{\rm BT}^q$} \\[0.4em]
    \hline
          &       &       & \multicolumn{7}{c|}{$c$}   & \multicolumn{7}{c|}{$c$}   & \multicolumn{7}{c|}{$c$}        & \multicolumn{7}{c|}{$c$}   & \multicolumn{7}{c|}{$c$}    & \multicolumn{7}{c}{$c$} \\[0.2em]
    $n$     & $p/n$   & $K$     & $2^{-3}$ & $2^{-2}$ & $2^{-1}$ & $2^0$  & $2^1$  & $2^2$  & $2^3$  & $2^{-3}$ & $2^{-2}$ & $2^{-1}$ & $2^0$  & $2^1$  & $2^2$  & $2^3$  & $2^{-3}$ & $2^{-2}$ & $2^{-1}$ & $2^0$  & $2^1$  & $2^2$  & $2^3$  & $2^{-3}$ & $2^{-2}$ & $2^{-1}$ & $2^0$  & $2^1$  & $2^2$  & $2^3$  & $2^{-3}$ & $2^{-2}$ & $2^{-1}$ & $2^0$  & $2^1$  & $2^2$  & $2^3$  & $2^{-3}$ & $2^{-2}$ & $2^{-1}$ & $2^0$  & $2^1$  & $2^2$  & $2^3$ \\[0.3em]
    \hline
    100   & 0.04  & 2     & 4.6   & 4.9   & 4.7   & 4.3   & 5.3   & 6.1   & 9.4   & 4.2   & 4.3   & 4.2   & 4.0   & 4.3   & 4.5   & 7.9   & 4.2   & 4.4   & 4.1   & 4.0   & 4.0   & 4.1   & 5.5   & 3.0   & 3.0   & 2.8   & 2.5   & 2.4   & 2.8   & 4.3   & 4.1   & 4.0   & 3.6   & 3.6   & 3.7   & 3.9   & 5.0   & 3.9   & 3.7   & 2.8   & 2.8   & 2.9   & 3.0   & 3.9  \\
          &       & 4     & 5.1   & 5.0   & 4.8   & 4.5   & 4.3   & 5.3   & 6.6   & 5.0   & 4.2   & 4.4   & 4.6   & 3.8   & 4.8   & 6.4   & 4.3   & 4.3   & 3.4   & 3.8   & 2.9   & 2.4   & 3.4   & 2.8   & 3.5   & 3.1   & 3.4   & 2.3   & 2.2   & 2.7   & 4.4   & 4.4   & 3.8   & 3.5   & 2.6   & 2.5   & 2.7   & 3.9   & 4.0   & 4.0   & 3.2   & 2.9   & 3.0   & 3.4  \\
          &       & 6     & 5.1   & 5.0   & 4.1   & 4.8   & 4.5   & 4.1   & 6.0   & 5.3   & 5.0   & 4.9   & 5.1   & 5.1   & 4.5   & 5.8   & 3.8   & 4.0   & 3.2   & 2.8   & 2.0   & 2.1   & 2.1   & 3.1   & 3.2   & 3.5   & 3.0   & 2.9   & 2.3   & 2.7   & 4.2   & 4.3   & 4.0   & 3.8   & 2.8   & 1.9   & 2.2   & 4.4   & 4.6   & 4.5   & 4.2   & 3.7   & 3.6   & 3.6  \\
          &       & 8     & 5.7   & 5.3   & 4.9   & 5.4   & 4.5   & 4.1   & 4.0   & 5.3   & 5.6   & 5.0   & 6.1   & 4.8   & 5.5   & 5.1   & 3.5   & 3.6   & 4.1   & 3.2   & 2.3   & 2.0   & 1.5   & 3.7   & 3.3   & 3.5   & 3.7   & 3.3   & 3.1   & 3.2   & 3.8   & 4.1   & 4.7   & 3.9   & 2.7   & 1.9   & 1.5   & 6.0   & 5.8   & 5.3   & 5.2   & 4.0   & 4.3   & 4.2  \\[0.3em]
          & 0.08  & 2     & 4.9   & 4.2   & 4.3   & 4.6   & 4.9   & 5.9   & 9.2   & 4.2   & 4.3   & 4.1   & 4.2   & 4.3   & 5.2   & 7.2   & 4.3   & 4.0   & 3.4   & 4.1   & 3.5   & 4.1   & 5.7   & 2.5   & 2.6   & 2.2   & 2.2   & 2.3   & 2.2   & 3.7   & 4.0   & 3.8   & 4.2   & 3.4   & 3.2   & 4.0   & 4.5   & 3.1   & 3.3   & 3.0   & 2.7   & 3.0   & 2.5   & 4.3  \\
          &       & 4     & 5.4   & 4.8   & 5.1   & 4.2   & 3.7   & 4.6   & 5.0   & 4.5   & 4.0   & 4.4   & 4.6   & 3.8   & 3.9   & 5.7   & 3.9   & 3.5   & 3.5   & 3.0   & 2.3   & 2.1   & 2.5   & 2.8   & 2.7   & 1.8   & 2.0   & 2.2   & 1.9   & 2.4   & 3.7   & 3.7   & 3.7   & 3.1   & 2.6   & 2.2   & 2.4   & 3.6   & 3.8   & 3.2   & 3.3   & 2.8   & 2.3   & 3.1  \\
          &       & 6     & 4.7   & 4.7   & 5.1   & 4.1   & 3.7   & 4.5   & 4.7   & 4.9   & 5.0   & 5.5   & 4.6   & 4.5   & 3.8   & 5.3   & 3.3   & 3.6   & 3.7   & 3.1   & 2.7   & 1.3   & 2.1   & 2.5   & 2.5   & 2.6   & 2.2   & 2.2   & 1.7   & 2.2   & 4.3   & 3.9   & 3.7   & 2.9   & 2.2   & 1.5   & 1.4   & 4.3   & 4.5   & 3.4   & 4.0   & 3.4   & 2.3   & 2.7  \\
          &       & 8     & 4.8   & 5.2   & 4.6   & 4.0   & 4.5   & 3.3   & 4.3   & 4.8   & 5.0   & 5.3   & 5.2   & 5.0   & 4.2   & 4.4   & 3.2   & 3.0   & 3.4   & 2.6   & 2.1   & 1.5   & 0.8   & 2.5   & 2.9   & 2.7   & 2.1   & 2.0   & 2.3   & 2.1   & 4.5   & 4.0   & 3.7   & 3.4   & 2.0   & 1.3   & 1.2   & 5.0   & 5.5   & 4.8   & 4.4   & 4.1   & 3.4   & 4.0  \\[0.3em]
          & 0.15  & 2     & 5.3   & 4.5   & 4.4   & 4.5   & 4.5   & 5.6   & 8.9   & 4.2   & 3.2   & 3.3   & 4.0   & 3.7   & 4.1   & 6.1   & 4.3   & 4.4   & 4.3   & 3.3   & 3.9   & 4.8   & 5.1   & 2.5   & 2.2   & 2.1   & 2.4   & 2.2   & 2.0   & 2.7   & 4.2   & 3.8   & 3.6   & 4.0   & 3.1   & 2.9   & 3.9   & 3.3   & 3.0   & 2.5   & 2.4   & 1.8   & 2.4   & 3.1  \\
          &       & 4     & 5.0   & 4.7   & 4.3   & 3.9   & 3.7   & 3.8   & 5.6   & 4.3   & 4.0   & 4.0   & 3.8   & 3.2   & 3.5   & 4.6   & 3.8   & 3.9   & 3.6   & 3.0   & 2.4   & 2.6   & 2.7   & 2.4   & 1.9   & 2.2   & 1.9   & 2.0   & 1.9   & 2.2   & 2.9   & 3.3   & 3.5   & 2.8   & 1.9   & 1.6   & 1.8   & 3.4   & 3.6   & 2.9   & 2.5   & 2.3   & 2.5   & 2.6  \\
          &       & 6     & 4.0   & 4.5   & 4.7   & 3.9   & 4.3   & 3.6   & 3.6   & 3.7   & 4.1   & 3.9   & 4.1   & 3.9   & 3.9   & 3.8   & 3.2   & 3.6   & 3.3   & 2.9   & 2.3   & 1.4   & 1.6   & 2.3   & 2.0   & 2.2   & 1.6   & 2.6   & 1.8   & 2.0   & 3.2   & 3.6   & 3.5   & 3.2   & 1.7   & 1.3   & 1.0   & 3.8   & 3.7   & 2.6   & 3.5   & 2.4   & 3.1   & 3.2  \\
          &       & 8     & 4.8   & 4.7   & 4.5   & 3.9   & 3.9   & 3.0   & 3.1   & 4.5   & 4.9   & 4.5   & 4.8   & 4.3   & 3.7   & 4.1   & 3.3   & 3.5   & 3.1   & 3.0   & 2.0   & 1.0   & 0.9   & 2.4   & 2.9   & 2.4   & 2.1   & 2.0   & 1.8   & 2.0   & 3.0   & 3.6   & 3.2   & 2.3   & 1.5   & 0.9   & 0.7   & 3.9   & 4.1   & 3.6   & 3.9   & 3.4   & 3.0   & 3.4  \\[0.3em]
          & 0.40  & 2     & 4.6   & 4.1   & 4.2   & 4.2   & 4.7   & 4.5   & 6.4   & 3.1   & 3.1   & 2.7   & 2.6   & 2.7   & 2.9   & 5.1   & 4.1   & 3.8   & 4.0   & 3.6   & 3.8   & 4.0   & 5.9   & 2.4   & 2.6   & 2.2   & 2.6   & 2.0   & 2.9   & 3.8   & 3.5   & 3.9   & 3.0   & 3.1   & 2.7   & 2.2   & 2.3   & 1.8   & 2.6   & 2.3   & 1.5   & 2.2   & 2.4   & 3.2  \\
          &       & 4     & 3.5   & 3.6   & 3.8   & 3.2   & 3.4   & 2.6   & 3.5   & 2.7   & 2.9   & 3.1   & 2.9   & 3.0   & 2.7   & 3.1   & 4.0   & 3.5   & 4.1   & 3.6   & 2.9   & 2.5   & 3.3   & 2.5   & 2.7   & 2.5   & 2.3   & 2.2   & 2.1   & 2.5   & 3.0   & 2.8   & 2.6   & 2.2   & 1.4   & 1.0   & 0.9   & 2.3   & 2.2   & 2.4   & 2.0   & 2.1   & 2.4   & 3.1  \\
          &       & 6     & 4.0   & 4.2   & 4.1   & 3.8   & 2.8   & 2.2   & 2.8   & 3.0   & 3.7   & 2.9   & 3.2   & 2.2   & 2.3   & 3.2   & 3.9   & 4.0   & 3.2   & 3.4   & 2.8   & 2.0   & 2.2   & 2.5   & 2.9   & 2.4   & 2.4   & 2.1   & 2.0   & 2.2   & 3.2   & 3.2   & 2.9   & 2.0   & 1.2   & 0.6   & 0.5   & 2.9   & 2.8   & 3.2   & 2.6   & 2.7   & 2.9   & 4.3  \\
          &       & 8     & 3.8   & 3.7   & 4.0   & 3.8   & 3.1   & 2.3   & 1.7   & 3.5   & 3.2   & 3.3   & 3.5   & 3.0   & 2.8   & 2.5   & 3.5   & 3.9   & 3.7   & 3.2   & 2.1   & 1.9   & 1.4   & 2.8   & 2.9   & 3.2   & 2.1   & 2.1   & 2.5   & 2.5   & 2.9   & 3.1   & 2.3   & 1.9   & 1.3   & 0.7   & 0.6   & 3.0   & 3.1   & 3.7   & 3.4   & 4.0   & 4.0   & 4.8  \\[0.3em]
          & 1.20  & 2     & 3.9   & 4.4   & 3.1   & 3.1   & 2.9   & 3.1   & 4.1   & 1.8   & 1.8   & 1.4   & 1.4   & 1.1   & 1.4   & 2.2   & 4.5   & 4.0   & 3.9   & 4.1   & 4.3   & 5.1   & 6.5   & 3.1   & 2.4   & 2.7   & 2.6   & 2.5   & 2.9   & 4.7   & 3.1   & 3.4   & 2.6   & 2.4   & 1.8   & 1.8   & 1.5   & 1.8   & 2.1   & 2.1   & 2.1   & 3.0   & 4.6   & 5.0  \\
          &       & 4     & 2.9   & 3.0   & 2.5   & 2.4   & 1.4   & 1.3   & 2.1   & 1.8   & 1.7   & 1.6   & 1.3   & 1.2   & 1.2   & 1.6   & 4.5   & 3.7   & 4.2   & 3.6   & 3.8   & 4.0   & 4.1   & 2.7   & 2.8   & 3.0   & 2.8   & 2.4   & 2.4   & 3.3   & 2.3   & 2.1   & 1.7   & 1.4   & 0.8   & 0.6   & 0.4   & 1.4   & 2.0   & 2.4   & 2.9   & 4.2   & 5.4   & 6.3  \\
          &       & 6     & 2.8   & 3.0   & 2.3   & 2.2   & 1.6   & 0.9   & 0.9   & 1.8   & 1.8   & 1.8   & 1.7   & 1.2   & 1.4   & 1.6   & 3.8   & 4.2   & 3.8   & 3.2   & 3.0   & 3.0   & 3.2   & 2.7   & 3.3   & 2.9   & 3.1   & 2.7   & 2.2   & 2.7   & 2.0   & 1.7   & 1.6   & 1.2   & 0.7   & 0.5   & 0.5   & 1.9   & 2.4   & 2.9   & 3.9   & 4.7   & 6.5   & 8.7  \\
          &       & 8     & 2.9   & 2.8   & 2.4   & 2.3   & 1.4   & 0.8   & 0.5   & 1.9   & 1.9   & 2.0   & 1.7   & 1.4   & 1.4   & 1.1   & 4.0   & 4.1   & 4.1   & 3.7   & 2.9   & 2.4   & 2.4   & 3.2   & 3.2   & 3.2   & 2.9   & 3.2   & 2.7   & 2.8   & 1.7   & 2.0   & 1.7   & 1.4   & 0.6   & 0.5   & 0.4   & 2.3   & 3.0   & 3.7   & 5.3   & 6.9   & 8.4   & 10.6  \\[0.3em]
    \hline
    300   & 0.04  & 2     & 5.2   & 4.8   & 5.2   & 4.7   & 4.9   & 4.6   & 5.6   & 4.7   & 4.6   & 5.0   & 4.4   & 4.8   & 4.8   & 5.6   & 4.4   & 4.9   & 4.0   & 4.0   & 3.8   & 4.4   & 4.3   & 3.7   & 3.9   & 3.5   & 3.6   & 3.2   & 3.7   & 3.4   & 4.2   & 5.0   & 4.4   & 4.4   & 3.8   & 4.1   & 3.9   & 4.0   & 4.1   & 3.8   & 3.4   & 2.9   & 3.0   & 3.0  \\
          &       & 4     & 4.5   & 4.6   & 5.2   & 4.3   & 4.8   & 4.6   & 4.6   & 5.6   & 5.4   & 4.3   & 4.9   & 4.8   & 4.6   & 4.1   & 4.2   & 4.0   & 3.8   & 3.7   & 3.3   & 3.2   & 3.0   & 4.2   & 3.6   & 4.6   & 4.0   & 3.4   & 2.9   & 2.9   & 4.5   & 4.1   & 4.6   & 3.5   & 3.3   & 2.7   & 2.4   & 4.5   & 4.1   & 4.6   & 4.0   & 3.1   & 2.3   & 2.4  \\
          &       & 6     & 4.9   & 4.6   & 4.3   & 4.7   & 4.6   & 3.4   & 3.6   & 5.7   & 5.7   & 5.1   & 5.0   & 4.3   & 4.5   & 3.9   & 4.9   & 3.5   & 4.0   & 4.0   & 3.3   & 2.3   & 1.8   & 3.9   & 4.1   & 4.7   & 4.2   & 3.9   & 3.8   & 2.4   & 4.3   & 5.0   & 3.8   & 3.5   & 2.6   & 2.1   & 1.7   & 5.0   & 5.7   & 4.9   & 4.3   & 3.2   & 2.5   & 2.2  \\
          &       & 8     & 5.1   & 5.0   & 5.6   & 4.3   & 3.7   & 3.5   & 3.2   & 6.8   & 6.5   & 6.3   & 5.6   & 5.4   & 5.4   & 4.7   & 4.4   & 4.5   & 4.1   & 3.5   & 3.3   & 1.7   & 1.7   & 5.3   & 4.6   & 4.4   & 4.4   & 3.9   & 3.1   & 3.4   & 3.7   & 4.2   & 3.8   & 3.3   & 3.2   & 1.7   & 1.2   & 5.2   & 5.7   & 5.0   & 4.9   & 3.5   & 2.9   & 2.1  \\[0.3em]
          & 0.08  & 2     & 4.8   & 4.7   & 4.1   & 4.9   & 4.2   & 4.6   & 5.1   & 4.4   & 4.8   & 4.4   & 4.3   & 4.5   & 4.5   & 4.3   & 4.6   & 4.9   & 4.9   & 4.4   & 4.5   & 4.4   & 4.8   & 4.0   & 4.1   & 3.8   & 3.1   & 3.3   & 3.3   & 3.7   & 4.5   & 4.7   & 4.2   & 4.3   & 3.4   & 3.8   & 2.9   & 3.9   & 3.2   & 3.5   & 3.2   & 3.6   & 3.0   & 2.6  \\
          &       & 4     & 5.2   & 4.6   & 4.6   & 4.1   & 4.3   & 3.7   & 4.3   & 5.1   & 5.1   & 5.0   & 4.8   & 4.4   & 3.8   & 3.7   & 4.2   & 4.1   & 3.9   & 4.4   & 3.7   & 3.1   & 3.1   & 4.8   & 3.9   & 4.4   & 4.2   & 3.6   & 3.1   & 2.9   & 4.7   & 4.5   & 3.9   & 3.9   & 3.3   & 2.4   & 2.3   & 4.2   & 4.6   & 4.2   & 3.2   & 3.3   & 2.3   & 2.0  \\
          &       & 6     & 4.3   & 4.4   & 5.1   & 4.6   & 3.9   & 3.5   & 2.7   & 6.2   & 5.8   & 5.6   & 5.6   & 5.1   & 4.4   & 3.1   & 4.7   & 4.2   & 4.1   & 4.2   & 3.0   & 2.7   & 2.4   & 4.9   & 4.8   & 4.2   & 4.3   & 3.5   & 3.2   & 2.7   & 4.2   & 3.6   & 4.0   & 4.0   & 3.0   & 1.9   & 1.5   & 4.5   & 4.8   & 4.3   & 4.1   & 3.4   & 2.4   & 2.1  \\
          &       & 8     & 5.1   & 4.8   & 5.0   & 4.1   & 3.8   & 2.7   & 2.0   & 6.2   & 5.7   & 5.3   & 5.4   & 4.5   & 4.7   & 4.0   & 4.0   & 4.3   & 4.4   & 3.6   & 3.4   & 2.1   & 1.6   & 4.8   & 4.4   & 4.5   & 3.6   & 4.5   & 3.3   & 3.1   & 4.1   & 4.2   & 3.9   & 2.7   & 2.5   & 1.6   & 0.7   & 5.6   & 5.2   & 5.0   & 4.0   & 3.5   & 3.4   & 1.7  \\[0.3em]
          & 0.15  & 2     & 4.3   & 4.3   & 5.4   & 5.0   & 4.4   & 5.1   & 5.0   & 4.5   & 4.2   & 4.0   & 3.6   & 3.6   & 3.8   & 4.3   & 5.0   & 4.0   & 4.4   & 4.4   & 4.6   & 4.8   & 4.7   & 3.7   & 4.2   & 3.8   & 3.4   & 3.2   & 3.6   & 3.8   & 4.0   & 4.8   & 3.8   & 4.0   & 4.0   & 2.9   & 2.7   & 4.4   & 3.6   & 3.7   & 2.9   & 2.6   & 2.6   & 2.7  \\
          &       & 4     & 4.6   & 4.3   & 4.4   & 4.5   & 3.3   & 3.1   & 3.4   & 4.4   & 4.4   & 4.9   & 4.2   & 4.3   & 3.5   & 3.5   & 4.8   & 4.8   & 4.7   & 4.2   & 3.6   & 4.3   & 3.6   & 4.4   & 4.1   & 3.7   & 3.9   & 3.1   & 2.8   & 2.9   & 4.1   & 4.0   & 4.0   & 3.5   & 2.9   & 2.4   & 1.7   & 4.0   & 3.5   & 3.2   & 3.3   & 2.1   & 2.2   & 1.6  \\
          &       & 6     & 5.2   & 4.3   & 4.1   & 4.6   & 3.7   & 2.9   & 2.9   & 5.1   & 5.0   & 5.0   & 5.0   & 4.0   & 3.5   & 2.8   & 4.6   & 4.8   & 4.8   & 3.9   & 3.3   & 2.9   & 2.8   & 5.5   & 4.3   & 4.6   & 3.6   & 3.9   & 3.1   & 3.1   & 4.2   & 3.7   & 3.6   & 3.5   & 2.0   & 1.6   & 1.0   & 4.5   & 4.3   & 4.3   & 4.3   & 2.8   & 2.1   & 1.5  \\
          &       & 8     & 4.3   & 4.1   & 4.2   & 3.7   & 2.7   & 2.5   & 2.2   & 5.3   & 5.0   & 5.7   & 5.5   & 4.9   & 3.5   & 3.1   & 3.8   & 4.2   & 4.3   & 4.8   & 3.4   & 2.7   & 1.7   & 4.3   & 4.5   & 4.7   & 4.5   & 3.9   & 3.7   & 3.0   & 4.0   & 3.7   & 3.7   & 3.1   & 2.2   & 1.4   & 0.4   & 4.9   & 5.0   & 4.2   & 3.9   & 3.6   & 2.2   & 1.4  \\[0.3em]
          & 0.40  & 2     & 4.2   & 4.2   & 4.7   & 3.8   & 3.7   & 4.2   & 3.5   & 3.8   & 3.8   & 2.9   & 3.2   & 3.1   & 2.7   & 3.0   & 5.2   & 4.5   & 5.1   & 4.8   & 5.0   & 4.9   & 5.3   & 3.5   & 4.2   & 4.4   & 3.5   & 3.3   & 3.6   & 3.6   & 4.3   & 4.5   & 3.8   & 3.8   & 4.1   & 3.0   & 2.8   & 3.9   & 3.4   & 3.5   & 2.8   & 2.5   & 2.0   & 2.0  \\
          &       & 4     & 3.8   & 3.8   & 3.9   & 3.9   & 3.4   & 3.0   & 1.7   & 4.2   & 4.6   & 3.2   & 3.2   & 2.7   & 2.4   & 2.1   & 4.9   & 4.4   & 4.3   & 4.5   & 4.6   & 3.7   & 3.7   & 4.9   & 4.6   & 4.6   & 4.2   & 4.1   & 3.4   & 2.6   & 4.0   & 3.9   & 3.6   & 2.6   & 3.3   & 2.1   & 1.3   & 3.8   & 3.4   & 3.5   & 2.9   & 2.4   & 1.7   & 1.6  \\
          &       & 6     & 3.8   & 4.1   & 3.1   & 3.4   & 2.9   & 2.1   & 1.4   & 3.7   & 4.1   & 4.4   & 3.3   & 2.9   & 2.9   & 1.7   & 4.6   & 4.5   & 4.2   & 4.4   & 4.0   & 3.2   & 3.1   & 4.9   & 4.1   & 4.4   & 4.2   & 4.1   & 3.2   & 3.4   & 3.6   & 3.9   & 3.6   & 3.4   & 2.0   & 1.4   & 0.9   & 3.9   & 3.7   & 3.1   & 3.1   & 2.6   & 1.8   & 1.5  \\
          &       & 8     & 3.8   & 3.2   & 3.8   & 3.4   & 2.6   & 1.5   & 0.7   & 3.7   & 4.0   & 4.3   & 3.8   & 3.5   & 2.6   & 1.9   & 4.2   & 4.7   & 4.3   & 3.7   & 3.5   & 2.8   & 2.4   & 5.8   & 5.1   & 4.6   & 4.4   & 4.2   & 4.0   & 3.6   & 3.4   & 3.4   & 3.1   & 2.6   & 1.8   & 1.0   & 0.4   & 4.5   & 3.9   & 3.9   & 3.0   & 2.4   & 1.6   & 1.3  \\[0.3em]
          & 1.20  & 2     & 3.0   & 3.6   & 3.4   & 3.4   & 3.2   & 2.4   & 1.9   & 2.0   & 1.9   & 1.8   & 1.8   & 1.3   & 1.5   & 1.2   & 4.6   & 5.4   & 4.9   & 5.2   & 4.8   & 4.4   & 5.7   & 4.6   & 4.4   & 4.6   & 4.3   & 4.0   & 4.2   & 4.5   & 4.3   & 4.5   & 3.9   & 3.5   & 2.7   & 3.2   & 2.1   & 3.1   & 2.7   & 2.8   & 2.3   & 2.2   & 2.0   & 1.6  \\
          &       & 4     & 3.6   & 2.9   & 2.9   & 3.5   & 2.2   & 1.2   & 0.8   & 2.0   & 1.9   & 2.2   & 1.6   & 1.4   & 1.0   & 0.6   & 4.6   & 4.8   & 4.5   & 4.1   & 3.8   & 3.7   & 3.7   & 4.1   & 4.3   & 4.2   & 4.1   & 3.4   & 3.8   & 3.8   & 3.2   & 3.6   & 4.0   & 2.8   & 2.2   & 1.3   & 0.9   & 2.8   & 2.8   & 2.5   & 2.3   & 1.7   & 1.3   & 0.9  \\
          &       & 6     & 2.6   & 2.4   & 2.6   & 2.1   & 1.5   & 0.6   & 0.2   & 2.3   & 2.3   & 1.8   & 1.6   & 1.4   & 0.9   & 0.7   & 4.7   & 4.1   & 4.4   & 4.4   & 3.6   & 3.5   & 3.5   & 5.4   & 4.4   & 4.4   & 4.3   & 4.2   & 3.9   & 3.7   & 3.2   & 2.9   & 2.7   & 2.6   & 1.7   & 1.0   & 0.3   & 3.2   & 3.2   & 2.7   & 2.3   & 2.2   & 1.3   & 1.3  \\
          &       & 8     & 2.1   & 2.7   & 2.6   & 1.6   & 1.2   & 0.6   & 0.1   & 2.2   & 2.0   & 2.1   & 2.2   & 1.9   & 0.8   & 0.7   & 4.5   & 3.6   & 4.0   & 4.5   & 4.3   & 3.6   & 2.5   & 5.1   & 4.5   & 5.0   & 4.5   & 4.6   & 3.8   & 3.7   & 2.9   & 2.8   & 2.8   & 2.1   & 1.4   & 0.8   & 0.2   & 3.5   & 3.6   & 3.0   & 2.6   & 2.3   & 1.6   & 1.0  \\
    \end{tabular}%
    }
  \label{tab:size_cbn}%
\end{sidewaystable}%
%\end{landscape}

\subsection{Empirical power}

To study the empirical power of the proposed method, we consider the following models:
\begin{itemize}[leftmargin=1.8cm]
  \item[Model 4.~] First-order exponential autoregressive model:
       $\bx_t = 0.15\bx_{t-1} + \exp(-2\bx_{t-1}^2) + \boldsymbol{\varepsilon}_t$ with $\boldsymbol{\varepsilon}_t\overset{{\rm i.i.d.}}{\sim}\calN(\bzero,\bOmega_\varepsilon)$, where $\bOmega_\varepsilon=(\omega_{\varepsilon,kl})_{p\times p}$ with $\omega_{\varepsilon,kl}=I(k=l)+0.25I(k\neq l)$ for any $k,l\in[p]$.
  \item[Model 5.~] The sum of a white noise and cosine of the first difference of an autoregressive process:
       $\bx_t= \boldsymbol{\varepsilon}_t+0.8\cos(\bz_t-\bz_{t-1})$ with $\bz_t=0.85\bz_{t-1}+{\mathbf u}_t$, $\boldsymbol{\varepsilon}_t\overset{{\rm i.i.d.}}{\sim} \calN(\bzero,\bOmega_\varepsilon)$ and ${\mathbf u}_t\overset{{\rm i.i.d.}}{\sim}\mathcal{N}(\bzero,\bOmega_u)$, where $\bOmega_\varepsilon=(\omega_{\varepsilon,kl})_{p\times p}$ and $\bOmega_{u}=(\omega_{u,kl})_{p\times p}$ with $\omega_{\varepsilon,kl}=I(k=l)+0.3I(k\neq l)$ and $\omega_{u,kl}=0.7^{|k-l|}$ for any $k,l\in [p]$.
  \item[Model 6.~] Threshold autoregressive model of order one: $\bx_t=(x_{t,1},\ldots,x_{t,p})^\T$ with $x_{t,j}=-0.45x_{t-1,j}I(x_{t-1,j}\\ \geqslant 1)+0.6x_{t-1,j}I(x_{t-1,j}< 1)+\varepsilon_{t,j}$ for each $j\in[p]$, where $\boldsymbol{\varepsilon}_t=(\varepsilon_{t,1},\ldots,\varepsilon_{t,p})^{\T}\overset{{\rm i.i.d.}}{\sim}\calN(\bzero,{\bf I}_p)$.
\end{itemize}

Models 4--6 are the multivariate extensions of the univariate models considered in \cite{EV2006a} (see Models 7--9 there).
Table \ref{tab:M4-M6} shows that for Models 4--6, the powers based on three different kernels are similar for the same map with the use of Bartlett kernel exhibiting slightly more power in most cases. When $n=100$ and for Model 4, using the linear and quadratic map leads to more power when $p/n\leqslant0.15$, but less power when $p/n>0.15$. This can be explained by the impact from the high dimension. The additional nonlinear serial dependence captured by the  quadratic map is apparent when $p\leqslant 15$, but as the dimension $p$ increases to $120$, the signal related to nonlinear dependence is likely dominated by that related to linear dependence and possibly the noise, so using linear map alone yields more power.
Similar phenomena occur for Models 5 and 6. As expected, when we increase the sample size $n$ from $100$ to $300$, we see the appreciation of the power as both linear and  nonlinear serial dependence get strengthened at the sample level.
Overall, the powers of our tests are quite encouraging for the three models, and all combinations of kernel and map under consideration.

By contrast, the three tests of \cite{HLZ2017} mostly fail to reject the martingale difference hypothesis for Models 4 and 5 in all settings. This is presumably due to the inability of their tests to capture nonlinear serial dependence. For Model 6, their tests exhibit great power, which is probably due to the fact that the model implies strong linear serial dependence although it is a nonlinear model per se. Indeed, the sample ACF at lag $1,2,3$ are 0.324, 0.120 and 0.046, respectively, based on our simulation. Again their tests cannot be implemented when $p$ is too large relative to $n$, as their ability of handling the high dimension is quite limited.

%For the model $M_4$ and $M_5$, our method performs much more powerful than Hong's method no matter when $n=100$ or 300. And for the model $M_6$, our method is slightly less powerful than Hong's method when $n=100$, but as $n$ increases, our method is comparable with Hong's method.
%In addition,
%The use of Bartlett kernel seems to deliver slightly more power, especially for the linear map. In general, the power decreases as we increase the dimension for a given kernel, map and sample size.

%with the increase of sample size and distance away from the null hypothesis, the powers of our proposed methods increase. This verifies the consistency of the method.

%\begin{landscape}
  \begin{sidewaystable}[htbp]
  \scriptsize
  \centering
  \caption{Empirical power ($\%$) of the tests $T_{\rm QS}^l$, $T_{\rm PR}^l$, $T_{\rm BT}^l$, $T_{\rm QS}^q$, $T_{\rm PR}^q$, $T_{\rm BT}^q$, $Z_{\rm tr}$, $Z_{\rm det}$ and $Zd_{\rm tr}$  for Models 4--6 at the 5\% nominal level. }
    \resizebox{!}{6.3cm}{
    \begin{tabular}{ccc|ccccccccc| ccccccccc| ccccccccc}
          &       &       & \multicolumn{9}{c|}{Model 4}                           & \multicolumn{9}{c|}{Model 5}                           & \multicolumn{9}{c}{Model 6} \\[0.6em]
    $n$     & $p/n$   & $K$    & $T_{\rm QS}^l$  & $T_{\rm PR}^l$  & $T_{\rm BT}^l$  & $T_{\rm QS}^q$  & $T_{\rm PR}^q$  & $T_{\rm BT}^q$    & $Z_{\rm tr}$   & $Z_{\rm det}$  & $Zd_{\rm tr}$ & $T_{\rm QS}^l$  & $T_{\rm PR}^l$  & $T_{\rm BT}^l$  & $T_{\rm QS}^q$  & $T_{\rm PR}^q$  & $T_{\rm BT}^q$    & $Z_{\rm tr}$   & $Z_{\rm det}$  & $Zd_{\rm tr}$  & $T_{\rm QS}^l$  & $T_{\rm PR}^l$  & $T_{\rm BT}^l$  & $T_{\rm QS}^q$  & $T_{\rm PR}^q$  & $T_{\rm BT}^q$    & $Z_{\rm tr}$   & $Z_{\rm det}$  & $Zd_{\rm tr}$ \\[0.5em]
    \hline
    100   & 0.04  & 2     & 79.0  & 78.1  & 81.2  & 93.5  & 93.5  & 94.5  & 6.1   & 5.5   & 6.0   & 65.7  & 65.2  & 69.1  & 94.2  & 94.3  & 95.5  & 4.5   & 4.7   & 5.6   & 77.0  & 77.4  & 84.4  & 80.5  & 81.3  & 85.4  & 100   & 63.9  & 100  \\
          &       & 4     & 87.8  & 87.5  & 89.2  & 97.8  & 97.8  & 98.3  & 6.7   & 6.0   & 5.9   & 81.5  & 80.3  & 83.4  & 98.3  & 98.4  & 98.6  & 4.5   & 4.6   & 6.6   & 66.7  & 66.5  & 77.7  & 75.7  & 76.2  & 81.8  & 99.7  & 62.4  & 99.4  \\
          &       & 6     & 91.1  & 90.9  & 92.6  & 98.7  & 98.7  & 99.0  & 6.4   & 5.0   & 6.6   & 87.1  & 86.8  & 89.1  & 99.0  & 99.0  & 99.2  & 5.0   & 5.4   & 7.3   & 64.7  & 63.8  & 76.8  & 75.6  & 76.8  & 82.3  & 97.7  & 59.6  & 97.0  \\
          &       & 8     & 93.5  & 93.0  & 94.5  & 99.3  & 99.1  & 99.4  & 4.6   & 5.7   & 6.8   & 90.9  & 90.8  & 92.3  & 99.2  & 99.1  & 99.4  & 4.9   & 5.3   & 7.9   & 66.6  & 65.1  & 78.0  & 77.8  & 77.9  & 84.8  & 94.2  & 52.5  & 91.1  \\[0.3em]
          & 0.08  & 2     & 82.2  & 81.6  & 84.8  & 93.0  & 93.2  & 94.7  & 6.4   & 7.0   & 5.7   & 71.8  & 71.7  & 76.3  & 94.9  & 95.0  & 96.0  & 4.3   & 5.0   & 4.9   & 75.0  & 75.1  & 85.2  & 71.0  & 72.3  & 78.7  & 100   & 51.7  & 100  \\
          &       & 4     & 91.8  & 91.4  & 93.0  & 97.7  & 97.8  & 98.2  & 6.7   & 6.9   & 6.3   & 86.7  & 86.0  & 89.2  & 97.9  & 97.8  & 98.4  & 5.7   & 5.6   & 6.9   & 65.8  & 65.2  & 79.9  & 65.6  & 67.6  & 75.3  & 100   & 63.3  & 100  \\
          &       & 6     & 93.9  & 93.6  & 95.3  & 98.6  & 98.7  & 98.9  & 5.4   & 7.4   & 6.7   & 90.7  & 90.4  & 92.7  & 99.0  & 99.1  & 99.3  & 6.5   & 6.3   & 8.2   & 63.6  & 62.7  & 78.6  & 65.0  & 65.7  & 75.0  & 100   & 66.0  & 99.9  \\
          &       & 8     & 95.3  & 95.0  & 96.3  & 99.1  & 99.2  & 99.3  & 5.8   & 6.7   & 7.0   & 92.6  & 92.6  & 94.9  & 99.0  & 98.9  & 99.2  & 6.6   & 6.1   & 9.1   & 61.6  & 60.4  & 77.8  & 65.8  & 66.5  & 75.9  & 99.7  & 61.4  & 99.7  \\[0.3em]
          & 0.15  & 2     & 84.2  & 84.1  & 87.2  & 90.8  & 91.0  & 92.7  & NA    & NA    & 5.9   & 74.7  & 74.6  & 80.1  & 91.7  & 91.9  & 93.7  & NA    & NA    & 5.3   & 71.3  & 71.6  & 83.3  & 60.3  & 62.4  & 69.9  & NA    & NA    & 100  \\
          &       & 4     & 92.2  & 91.6  & 94.4  & 96.8  & 96.7  & 97.3  & NA    & NA    & 6.3   & 88.1  & 87.5  & 91.3  & 96.8  & 96.8  & 97.6  & NA    & NA    & 6.6   & 58.4  & 58.5  & 76.7  & 53.1  & 54.8  & 64.9  & NA    & NA    & 100  \\
          &       & 6     & 95.2  & 95.0  & 96.6  & 97.7  & 97.7  & 97.9  & NA    & NA    & 7.9   & 91.5  & 91.4  & 94.2  & 97.7  & 97.9  & 98.1  & NA    & NA    & 9.4   & 55.5  & 54.6  & 75.4  & 50.1  & 51.5  & 62.9  & NA    & NA    & 100  \\
          &       & 8     & 96.0  & 95.8  & 97.1  & 98.3  & 98.2  & 98.6  & NA    & NA    & 8.2   & 94.4  & 94.4  & 96.2  & 98.4  & 98.5  & 98.8  & NA    & NA    & 11.2  & 54.3  & 53.2  & 75.3  & 48.9  & 50.8  & 61.7  & NA    & NA    & 100  \\[0.3em]
          & 0.40   & 2     & 85.0  & 84.5  & 88.7  & 78.8  & 79.7  & 82.4  & NA    & NA    & 6.3   & 73.7  & 73.4  & 79.6  & 80.5  & 81.4  & 84.7  & NA    & NA    & 5.9   & 62.0  & 62.8  & 81.7  & 44.5  & 47.7  & 55.3  & NA    & NA    & 100  \\
          &       & 4     & 91.5  & 91.1  & 94.3  & 88.4  & 89.0  & 90.4  & NA    & NA    & 7.0   & 87.6  & 87.0  & 91.4  & 90.0  & 90.7  & 92.6  & NA    & NA    & 9.2   & 47.2  & 47.0  & 72.9  & 34.6  & 37.7  & 44.9  & NA    & NA    & 100  \\
          &       & 6     & 94.6  & 94.2  & 96.9  & 91.7  & 92.2  & 93.4  & NA    & NA    & 6.7   & 91.7  & 91.2  & 94.5  & 91.8  & 92.2  & 93.7  & NA    & NA    & 12.8  & 40.4  & 38.8  & 69.5  & 32.3  & 34.1  & 41.1  & NA    & NA    & 100  \\
          &       & 8     & 95.2  & 94.9  & 97.2  & 92.2  & 92.7  & 93.3  & NA    & NA    & 9.4   & 92.9  & 92.8  & 95.8  & 93.6  & 93.9  & 95.2  & NA    & NA    & 14.5  & 36.3  & 35.1  & 65.4  & 32.0  & 34.1  & 39.8  & NA    & NA    & 100  \\[0.3em]
          & 1.20   & 2     & 78.5  & 78.0  & 83.9  & 44.3  & 45.6  & 48.8  & NA    & NA    & NA    & 67.6  & 68.3  & 77.9  & 52.5  & 55.0  & 58.8  & NA    & NA    & NA    & 49.3  & 49.1  & 77.3  & 43.7  & 46.8  & 45.8  & NA    & NA    & NA \\
          &       & 4     & 87.6  & 86.9  & 91.8  & 57.0  & 58.6  & 61.8  & NA    & NA    & NA    & 82.8  & 82.1  & 88.8  & 64.3  & 66.2  & 70.1  & NA    & NA    & NA    & 30.0  & 29.4  & 60.8  & 46.9  & 49.4  & 43.5  & NA    & NA    & NA \\
          &       & 6     & 89.2  & 88.8  & 93.2  & 58.3  & 60.0  & 62.7  & NA    & NA    & NA    & 87.1  & 86.4  & 92.2  & 68.6  & 71.0  & 73.6  & NA    & NA    & NA    & 21.7  & 20.9  & 52.3  & 49.7  & 53.1  & 44.8  & NA    & NA    & NA \\
          &       & 8     & 90.8  & 90.3  & 94.5  & 62.6  & 64.2  & 66.9  & NA    & NA    & NA    & 88.9  & 88.6  & 92.9  & 69.4  & 71.4  & 74.8  & NA    & NA    & NA    & 17.5  & 16.7  & 44.4  & 53.9  & 56.2  & 47.9  & NA    & NA    & NA \\[0.3em]
    \hline
    300   & 0.04  & 2     & 100   & 100   & 100   & 100   & 100   & 100   & 13.6  & 8.3   & 8.6   & 100   & 100   & 100   & 100   & 100   & 100   & 5.0   & 5.3   & 5.2   & 100   & 100   & 100   & 99.4  & 99.5  & 99.7  & 100   & 99.0  & 100  \\
          &       & 4     & 100   & 100   & 100   & 100   & 100   & 100   & 15.1  & 8.9   & 10.4  & 100   & 100   & 100   & 100   & 100   & 100   & 5.1   & 5.9   & 5.7   & 99.8  & 99.8  & 99.9  & 98.9  & 99.0  & 99.3  & 100   & 98.6  & 100  \\
          &       & 6     & 100   & 100   & 100   & 100   & 100   & 100   & 10.4  & 8.9   & 7.9   & 100   & 100   & 100   & 100   & 100   & 100   & 6.1   & 6.5   & 6.1   & 99.8  & 99.7  & 99.9  & 98.6  & 98.5  & 99.1  & 100   & 97.0  & 100  \\
          &       & 8     & 100   & 100   & 100   & 100   & 100   & 100   & 9.7   & 9.2   & 6.2   & 100   & 100   & 100   & 100   & 100   & 100   & 6.5   & 6.9   & 6.2   & 100   & 100   & 100   & 99.2  & 99.2  & 99.6  & 100   & 95.7  & 100  \\[0.3em]
          & 0.08  & 2     & 100   & 100   & 100   & 100   & 100   & 100   & NA    & NA    & 9.3   & 100   & 100   & 100   & 100   & 100   & 100   & NA    & NA    & 4.8   & 100   & 100   & 100   & 98.6  & 98.6  & 99.3  & NA    & NA    & 100  \\
          &       & 4     & 100   & 100   & 100   & 100   & 100   & 100   & NA    & NA    & 11.7  & 100   & 100   & 100   & 100   & 100   & 100   & NA    & NA    & 5.9   & 99.9  & 99.9  & 100   & 98.2  & 98.2  & 99.1  & NA    & NA    & 100  \\
          &       & 6     & 100   & 100   & 100   & 100   & 100   & 100   & NA    & NA    & 8.3   & 100   & 100   & 100   & 100   & 100   & 100   & NA    & NA    & 6.0   & 99.9  & 99.9  & 100   & 98.3  & 98.4  & 99.1  & NA    & NA    & 100  \\
          &       & 8     & 100   & 100   & 100   & 100   & 100   & 100   & NA    & NA    & 7.3   & 100   & 100   & 100   & 100   & 100   & 100   & NA    & NA    & 7.0   & 100   & 100   & 100   & 98.3  & 98.3  & 99.0  & NA    & NA    & 100  \\[0.3em]
          & 0.15  & 2     & 100   & 100   & 100   & 100   & 100   & 100   & NA    & NA    & 10.2  & 100   & 100   & 100   & 100   & 100   & 100   & NA    & NA    & 5.6   & 100   & 100   & 100   & 98.2  & 98.4  & 98.8  & NA    & NA    & 100  \\
          &       & 4     & 100   & 100   & 100   & 100   & 100   & 100   & NA    & NA    & 11.1  & 100   & 100   & 100   & 100   & 100   & 100   & NA    & NA    & 6.1   & 99.9  & 99.9  & 100   & 97.0  & 97.0  & 98.3  & NA    & NA    & 100  \\
          &       & 6     & 100   & 100   & 100   & 100   & 100   & 100   & NA    & NA    & 9.9   & 100   & 100   & 100   & 100   & 100   & 100   & NA    & NA    & 6.8   & 99.9  & 99.9  & 100   & 97.2  & 97.3  & 98.6  & NA    & NA    & 100  \\
          &       & 8     & 100   & 100   & 100   & 100   & 100   & 100   & NA    & NA    & 7.0   & 100   & 100   & 100   & 100   & 100   & 100   & NA    & NA    & 7.3   & 99.9  & 100   & 100   & 96.8  & 97.0  & 98.6  & NA    & NA    & 100  \\[0.3em]
          & 0.40   & 2     & 100   & 100   & 100   & 100   & 100   & 100   & NA    & NA    & NA    & 100   & 100   & 100   & 100   & 100   & 100   & NA    & NA    & NA    & 100   & 100   & 100   & 96.4  & 96.4  & 98.2  & NA    & NA    & NA \\
          &       & 4     & 100   & 100   & 100   & 100   & 100   & 100   & NA    & NA    & NA    & 100   & 100   & 100   & 100   & 100   & 100   & NA    & NA    & NA    & 100   & 99.9  & 100   & 94.3  & 94.4  & 97.4  & NA    & NA    & NA \\
          &       & 6     & 100   & 100   & 100   & 100   & 100   & 100   & NA    & NA    & NA    & 100   & 100   & 100   & 100   & 100   & 100   & NA    & NA    & NA    & 99.9  & 99.9  & 100   & 92.0  & 92.1  & 96.9  & NA    & NA    & NA \\
          &       & 8     & 100   & 100   & 100   & 100   & 100   & 100   & NA    & NA    & NA    & 100   & 100   & 100   & 100   & 100   & 100   & NA    & NA    & NA    & 100   & 100   & 100   & 92.5  & 92.3  & 96.9  & NA    & NA    & NA \\[0.3em]
          & 1.20   & 2     & 100   & 100   & 100   & 100   & 100   & 100   & NA    & NA    & NA    & 100   & 100   & 100   & 100   & 100   & 100   & NA    & NA    & NA    & 100   & 100   & 100   & 92.9  & 92.7  & 97.0  & NA    & NA    & NA \\
          &       & 4     & 100   & 100   & 100   & 100   & 100   & 100   & NA    & NA    & NA    & 100   & 100   & 100   & 100   & 100   & 100   & NA    & NA    & NA    & 100   & 100   & 100   & 82.9  & 83.2  & 93.0  & NA    & NA    & NA \\
          &       & 6     & 100   & 100   & 100   & 100   & 100   & 100   & NA    & NA    & NA    & 100   & 100   & 100   & 100   & 100   & 100   & NA    & NA    & NA    & 100   & 100   & 100   & 76.0  & 76.1  & 89.6  & NA    & NA    & NA \\
          &       & 8     & 100   & 100   & 100   & 100   & 100   & 100   & NA    & NA    & NA    & 100   & 100   & 100   & 100   & 100   & 100   & NA    & NA    & NA    & 100   & 100   & 100   & 69.3  & 69.3  & 87.1  & NA    & NA    & NA \\
    \end{tabular}%
    }
  \label{tab:M4-M6}%
\end{sidewaystable}%
%\end{landscape}

%%%%%%%%%%%%%%%%%%%%%%%%%%%%%%%%%%%%%%%%%%%%%%%%%%%%%%%%%%%%%%%%%%%%%%%%%%%%%%%%%%%%
%\begin{landscape}
\begin{sidewaystable}[htbp]
\scriptsize
\renewcommand\arraystretch{1.2}
  \centering
  \caption{Empirical powers ($\%$) of the tests $T_{\rm BT}^l$ and $T_{\rm BT}^q$ for Models 4--6 at the 5\% nominal level, where $c$ represents the constant which is multiplied by Andrews' bandwidth. }
  \resizebox{!}{5.2cm}{
    \begin{tabular}{ccc|ccccccc|ccccccc|ccccccc|ccccccc|ccccccc|ccccccc}
          &       &       & \multicolumn{7}{c|}{Model 4 with $T_{\rm BT}^l$}            & \multicolumn{7}{c|}{Model 4 with $T_{\rm BT}^q$}            & \multicolumn{7}{c|}{Model 5 with $T_{\rm BT}^l$}            & \multicolumn{7}{c|}{Model 5 with $T_{\rm BT}^q$}            & \multicolumn{7}{c|}{Model 6 with $T_{\rm BT}^l$}            & \multicolumn{7}{c}{Model 6 with $T_{\rm BT}^q$} \\[0.4em]
    \hline
          &       &       & \multicolumn{7}{c|}{$c$}    & \multicolumn{7}{c|}{$c$}     & \multicolumn{7}{c|}{$c$}    & \multicolumn{7}{c|}{$c$}    & \multicolumn{7}{c|}{$c$}     & \multicolumn{7}{c}{$c$} \\[0.3em]
    $n$     & $p/n$   & $K$     & $2^{-3}$ & $2^{-2}$ & $2^{-1}$ & $2^0$  & $2^1$  & $2^2$  & $2^3$  & $2^{-3}$ & $2^{-2}$ & $2^{-1}$ & $2^0$  & $2^1$  & $2^2$  & $2^3$ & $2^{-3}$ & $2^{-2}$ & $2^{-1}$ & $2^0$  & $2^1$  & $2^2$  & $2^3$  & $2^{-3}$ & $2^{-2}$ & $2^{-1}$ & $2^0$  & $2^1$  & $2^2$  & $2^3$  & $2^{-3}$ & $2^{-2}$ & $2^{-1}$ & $2^0$  & $2^1$  & $2^2$  & $2^3$  & $2^{-3}$ & $2^{-2}$ & $2^{-1}$ & $2^0$  & $2^1$  & $2^2$  & $2^3$ \\[0.4em]
    \hline
    100   & 0.04  & 2     & 81.2  & 81.8  & 80.5  & 78.8  & 76.6  & 73.2  & 72.1  & 95.7  & 95.9  & 95.6  & 93.5  & 93.8  & 92.8  & 92.4  & 73.2  & 73.7  & 72.0  & 69.0  & 64.3  & 61.5  & 61.2  & 96.0  & 96.3  & 96.0  & 95.8  & 95.0  & 94.0  & 93.9  & 99.5  & 98.2  & 93.2  & 84.3  & 79.1  & 79.9  & 88.1  & 99.6  & 96.7  & 90.6  & 84.4  & 84.7  & 87.3  & 95.6  \\
          &       & 4     & 89.9  & 89.8  & 89.3  & 88.6  & 86.7  & 80.6  & 78.3  & 99.1  & 98.8  & 98.9  & 98.0  & 97.7  & 96.6  & 97.0  & 85.9  & 85.6  & 85.9  & 83.3  & 79.6  & 72.9  & 69.7  & 99.0  & 98.9  & 98.8  & 98.8  & 97.9  & 97.5  & 96.7  & 98.5  & 97.7  & 91.6  & 78.6  & 68.2  & 68.0  & 78.2  & 99.4  & 96.9  & 91.2  & 82.8  & 78.8  & 84.0  & 93.1  \\
          &       & 6     & 93.2  & 93.2  & 94.0  & 91.3  & 89.8  & 85.1  & 80.6  & 99.3  & 99.2  & 99.4  & 99.0  & 99.0  & 98.2  & 98.2  & 90.2  & 90.4  & 90.5  & 89.3  & 85.9  & 79.9  & 75.9  & 99.0  & 99.5  & 99.1  & 99.0  & 98.8  & 98.6  & 98.2  & 98.7  & 97.6  & 91.8  & 78.5  & 65.0  & 64.2  & 72.5  & 99.7  & 97.7  & 91.0  & 84.3  & 79.7  & 82.6  & 92.0  \\
          &       & 8     & 94.4  & 94.7  & 94.4  & 93.6  & 91.2  & 88.3  & 82.7  & 99.6  & 99.6  & 99.2  & 99.4  & 99.1  & 99.1  & 98.8  & 93.1  & 93.4  & 92.5  & 91.8  & 89.1  & 84.7  & 78.3  & 99.5  & 99.2  & 99.4  & 99.5  & 99.4  & 98.9  & 98.9  & 98.6  & 97.7  & 91.2  & 78.9  & 65.4  & 61.6  & 70.3  & 99.5  & 98.0  & 91.8  & 85.1  & 80.3  & 83.2  & 92.4  \\[0.3em]
          & 0.08  & 2     & 87.0  & 86.7  & 86.6  & 85.0  & 80.9  & 76.6  & 72.7  & 95.7  & 95.7  & 95.2  & 94.9  & 94.2  & 92.6  & 92.4  & 80.6  & 80.8  & 78.9  & 76.6  & 69.9  & 65.3  & 63.2  & 96.0  & 96.4  & 95.7  & 95.3  & 93.9  & 93.5  & 93.3  & 99.9  & 99.3  & 94.6  & 85.1  & 77.4  & 77.6  & 88.7  & 99.7  & 95.3  & 86.8  & 78.4  & 76.5  & 84.0  & 94.3  \\
          &       & 4     & 93.7  & 93.5  & 93.8  & 92.6  & 89.8  & 83.9  & 77.3  & 98.4  & 98.9  & 98.7  & 98.2  & 97.5  & 97.1  & 96.4  & 90.4  & 90.7  & 91.1  & 88.9  & 84.1  & 76.9  & 70.5  & 98.9  & 98.4  & 98.6  & 98.7  & 97.6  & 96.8  & 96.3  & 99.9  & 99.2  & 94.0  & 78.3  & 66.9  & 66.9  & 79.4  & 99.3  & 96.1  & 85.7  & 74.2  & 71.9  & 79.8  & 92.6  \\
          &       & 6     & 96.7  & 96.2  & 96.1  & 94.6  & 93.5  & 87.2  & 80.4  & 99.2  & 99.2  & 99.3  & 99.0  & 98.5  & 98.2  & 97.9  & 93.9  & 94.3  & 93.8  & 92.6  & 89.9  & 82.0  & 74.6  & 99.1  & 99.2  & 99.3  & 99.0  & 98.6  & 98.1  & 98.1  & 99.7  & 99.3  & 94.0  & 78.5  & 62.2  & 59.3  & 73.7  & 99.3  & 95.8  & 85.9  & 76.4  & 70.0  & 77.7  & 91.8  \\
          &       & 8     & 97.1  & 96.9  & 97.2  & 96.2  & 93.9  & 90.2  & 83.5  & 99.4  & 99.4  & 99.1  & 99.2  & 98.6  & 98.9  & 98.4  & 95.9  & 95.8  & 95.6  & 95.1  & 92.1  & 86.6  & 79.1  & 99.5  & 99.5  & 99.4  & 99.3  & 99.0  & 99.0  & 98.4  & 99.8  & 99.5  & 94.8  & 78.5  & 61.2  & 56.5  & 70.6  & 99.1  & 96.3  & 87.9  & 76.0  & 69.7  & 76.2  & 90.5  \\[0.3em]
          & 0.15  & 2     & 88.4  & 88.5  & 89.5  & 87.0  & 82.2  & 75.4  & 69.6  & 94.4  & 93.6  & 93.7  & 92.8  & 90.5  & 90.3  & 90.3  & 84.7  & 83.8  & 83.9  & 79.0  & 72.5  & 65.9  & 59.6  & 95.1  & 95.4  & 94.4  & 93.2  & 92.1  & 91.1  & 90.5  & 100  & 99.8  & 95.4  & 83.1  & 76.1  & 76.3  & 87.1  & 99.0  & 93.9  & 81.1  & 71.2  & 70.1  & 79.7  & 93.2  \\
          &       & 4     & 95.6  & 95.4  & 94.7  & 94.6  & 91.4  & 83.6  & 74.9  & 97.9  & 97.7  & 97.4  & 97.3  & 97.1  & 95.8  & 94.7  & 93.0  & 92.6  & 92.5  & 91.2  & 86.1  & 75.5  & 69.4  & 97.9  & 97.9  & 98.1  & 97.3  & 96.9  & 95.3  & 94.9  & 99.9  & 99.5  & 94.8  & 76.5  & 61.1  & 60.5  & 77.3  & 98.3  & 93.0  & 78.6  & 63.0  & 59.7  & 71.7  & 90.7  \\
          &       & 6     & 97.2  & 96.8  & 97.2  & 96.9  & 94.2  & 87.8  & 76.9  & 98.3  & 98.4  & 98.3  & 98.0  & 98.2  & 97.0  & 95.7  & 95.6  & 95.6  & 95.5  & 93.8  & 90.7  & 81.4  & 74.8  & 98.9  & 98.6  & 98.5  & 98.2  & 97.8  & 96.6  & 96.8  & 99.9  & 99.4  & 94.6  & 75.3  & 58.6  & 53.0  & 69.7  & 98.1  & 93.3  & 78.7  & 63.0  & 58.3  & 68.4  & 89.5  \\
          &       & 8     & 97.8  & 98.0  & 98.3  & 97.0  & 94.9  & 89.5  & 80.8  & 98.6  & 98.6  & 99.0  & 98.7  & 98.3  & 97.1  & 96.4  & 96.9  & 96.9  & 96.6  & 95.7  & 93.1  & 85.4  & 77.3  & 98.8  & 98.8  & 98.8  & 98.6  & 98.5  & 97.7  & 97.2  & 99.9  & 99.5  & 94.9  & 78.2  & 55.2  & 48.3  & 66.7  & 97.9  & 93.6  & 79.2  & 65.0  & 56.9  & 65.7  & 87.5  \\[0.3em]
          & 0.40  & 2     & 89.7  & 90.9  & 89.5  & 88.2  & 81.8  & 71.2  & 63.0  & 82.9  & 84.6  & 82.3  & 81.4  & 79.2  & 76.2  & 78.4  & 85.7  & 85.3  & 85.3  & 81.4  & 72.6  & 61.7  & 53.4  & 87.1  & 87.3  & 86.9  & 85.2  & 81.8  & 79.7  & 81.6  & 99.9  & 99.5  & 96.2  & 82.6  & 67.3  & 67.3  & 83.4  & 95.6  & 87.2  & 69.5  & 55.1  & 55.0  & 68.6  & 90.4  \\
          &       & 4     & 95.6  & 96.3  & 95.7  & 93.9  & 88.7  & 79.3  & 66.2  & 90.4  & 90.2  & 90.7  & 89.6  & 88.0  & 84.7  & 84.0  & 93.2  & 94.2  & 93.1  & 91.2  & 84.8  & 73.1  & 59.2  & 93.6  & 93.4  & 93.2  & 91.4  & 90.4  & 87.7  & 86.9  & 99.9  & 99.3  & 94.8  & 73.1  & 49.7  & 47.0  & 68.7  & 92.6  & 84.6  & 63.1  & 44.1  & 43.0  & 56.7  & 86.5  \\
          &       & 6     & 97.2  & 97.8  & 97.0  & 96.1  & 92.3  & 82.3  & 68.2  & 92.4  & 92.6  & 92.0  & 92.1  & 91.7  & 88.6  & 87.6  & 95.7  & 96.5  & 96.3  & 93.5  & 88.9  & 76.9  & 64.9  & 94.8  & 94.6  & 94.6  & 93.5  & 93.0  & 89.7  & 89.1  & 99.8  & 99.4  & 95.3  & 69.5  & 41.9  & 36.9  & 60.5  & 89.4  & 80.8  & 58.7  & 40.1  & 41.6  & 52.0  & 83.8  \\
          &       & 8     & 97.8  & 97.8  & 97.9  & 96.7  & 93.0  & 85.7  & 71.0  & 93.3  & 93.1  & 93.7  & 93.1  & 92.4  & 90.7  & 88.7  & 97.5  & 97.4  & 97.0  & 95.6  & 92.3  & 81.1  & 67.4  & 94.9  & 95.1  & 95.2  & 94.1  & 93.6  & 91.3  & 90.2  & 99.8  & 99.4  & 95.0  & 66.2  & 38.8  & 33.1  & 54.8  & 87.6  & 79.8  & 56.9  & 39.6  & 38.1  & 51.9  & 83.1  \\[0.3em]
          & 1.20  & 2     & 86.7  & 86.2  & 88.1  & 83.9  & 76.3  & 62.3  & 49.4  & 51.2  & 49.4  & 49.8  & 48.7  & 45.3  & 43.6  & 46.9  & 84.2  & 84.1  & 84.0  & 78.5  & 65.1  & 52.3  & 41.8  & 64.0  & 63.2  & 63.0  & 58.6  & 55.2  & 51.8  & 58.8  & 99.7  & 99.3  & 97.0  & 77.6  & 54.1  & 54.0  & 75.4  & 78.9  & 67.6  & 51.2  & 46.8  & 50.3  & 64.6  & 89.0  \\
          &       & 4     & 93.6  & 93.6  & 93.0  & 90.4  & 84.4  & 69.1  & 51.9  & 60.3  & 60.9  & 60.7  & 58.4  & 55.2  & 51.9  & 52.5  & 91.4  & 91.5  & 91.2  & 88.3  & 79.5  & 59.5  & 45.3  & 72.4  & 71.0  & 71.7  & 69.7  & 65.4  & 60.1  & 61.8  & 99.1  & 99.0  & 95.2  & 63.0  & 30.7  & 28.2  & 53.1  & 64.1  & 56.3  & 41.5  & 42.9  & 49.9  & 59.9  & 85.0  \\
          &       & 6     & 95.3  & 96.1  & 96.0  & 93.9  & 87.1  & 72.2  & 53.4  & 63.6  & 63.9  & 63.1  & 63.7  & 60.1  & 56.4  & 54.5  & 94.5  & 94.6  & 94.5  & 91.8  & 84.5  & 67.7  & 50.3  & 75.2  & 74.7  & 74.1  & 73.7  & 69.4  & 64.4  & 66.4  & 99.1  & 98.0  & 93.0  & 53.2  & 23.1  & 20.5  & 43.1  & 55.7  & 49.7  & 40.3  & 45.1  & 54.0  & 64.0  & 84.4  \\
          &       & 8     & 96.4  & 96.0  & 96.3  & 94.7  & 89.3  & 75.2  & 55.3  & 63.8  & 65.7  & 65.5  & 65.8  & 61.0  & 58.6  & 57.2  & 96.1  & 95.5  & 95.4  & 93.3  & 86.7  & 73.2  & 54.2  & 75.6  & 74.6  & 74.8  & 74.8  & 70.8  & 68.1  & 67.5  & 98.3  & 96.8  & 92.8  & 48.7  & 20.8  & 18.3  & 40.4  & 50.8  & 46.6  & 41.2  & 48.1  & 57.5  & 68.9  & 85.1  \\[0.4em]
    \hline
    300   & 0.04  & 2     & 100   & 100   & 100   & 100   & 100   & 100   & 100   & 100   & 100   & 100   & 100   & 100   & 100   & 100   & 100   & 100   & 100   & 100   & 100   & 100   & 100   & 100   & 100   & 100   & 100   & 100   & 100   & 100   & 100   & 100   & 100   & 100   & 99.9  & 99.9  & 99.9  & 100   & 99.9  & 99.8  & 99.6  & 99.5  & 99.4  & 99.7  \\
          &       & 4     & 100   & 100   & 100   & 100   & 100   & 100   & 100   & 100   & 100   & 100   & 100   & 100   & 100   & 100   & 100   & 100   & 100   & 100   & 100   & 100   & 100   & 100   & 100   & 100   & 100   & 100   & 100   & 100   & 100   & 100   & 100   & 100   & 99.9  & 99.7  & 99.5  & 100   & 100   & 99.8  & 99.5  & 99.1  & 99.2  & 99.4  \\
          &       & 6     & 100   & 100   & 100   & 100   & 100   & 100   & 100   & 100   & 100   & 100   & 100   & 100   & 100   & 100   & 100   & 100   & 100   & 100   & 100   & 100   & 100   & 100   & 100   & 100   & 100   & 100   & 100   & 100   & 100   & 100   & 100   & 100   & 99.8  & 99.6  & 99.2  & 100   & 99.9  & 99.8  & 99.4  & 99.0  & 99.1  & 99.5  \\
          &       & 8     & 100   & 100   & 100   & 100   & 100   & 100   & 100   & 100   & 100   & 100   & 100   & 100   & 100   & 100   & 100   & 100   & 100   & 100   & 100   & 100   & 100   & 100   & 100   & 100   & 100   & 100   & 100   & 100   & 100   & 100   & 100   & 100   & 99.9  & 99.6  & 99.3  & 100   & 100   & 99.8  & 99.5  & 99.4  & 99.2  & 99.0  \\[0.3em]
          & 0.08  & 2     & 100   & 100   & 100   & 100   & 100   & 100   & 100   & 100   & 100   & 100   & 100   & 100   & 100   & 100   & 100   & 100   & 100   & 100   & 100   & 100   & 99.9  & 100   & 100   & 100   & 100   & 100   & 100   & 100   & 100   & 100   & 100   & 100   & 99.9  & 99.9  & 99.9  & 100   & 99.9  & 99.7  & 99.3  & 99.0  & 99.0  & 99.5  \\
          &       & 4     & 100   & 100   & 100   & 100   & 100   & 100   & 100   & 100   & 100   & 100   & 100   & 100   & 100   & 100   & 100   & 100   & 100   & 100   & 100   & 100   & 100   & 100   & 100   & 100   & 100   & 100   & 100   & 100   & 100   & 100   & 100   & 100   & 99.9  & 99.6  & 99.3  & 100   & 99.9  & 99.5  & 98.9  & 98.9  & 98.5  & 99.1  \\
          &       & 6     & 100   & 100   & 100   & 100   & 100   & 100   & 100   & 100   & 100   & 100   & 100   & 100   & 100   & 100   & 100   & 100   & 100   & 100   & 100   & 100   & 100   & 100   & 100   & 100   & 100   & 100   & 100   & 100   & 100   & 100   & 100   & 100   & 99.9  & 99.7  & 99.3  & 100   & 100   & 99.5  & 98.9  & 98.5  & 98.4  & 99.1  \\
          &       & 8     & 100   & 100   & 100   & 100   & 100   & 100   & 100   & 100   & 100   & 100   & 100   & 100   & 100   & 100   & 100   & 100   & 100   & 100   & 100   & 100   & 100   & 100   & 100   & 100   & 100   & 100   & 100   & 100   & 100   & 100   & 100   & 100   & 100   & 99.6  & 99.2  & 100   & 99.8  & 99.7  & 99.3  & 98.5  & 98.5  & 99.1  \\[0.3em]
          & 0.15  & 2     & 100   & 100   & 100   & 100   & 100   & 100   & 100   & 100   & 100   & 100   & 100   & 100   & 100   & 100   & 100   & 100   & 100   & 100   & 100   & 100   & 99.9  & 100   & 100   & 100   & 100   & 100   & 100   & 100   & 100   & 100   & 100   & 100   & 100   & 99.9  & 100   & 100   & 100   & 99.3  & 99.1  & 98.8  & 99.0  & 99.5  \\
          &       & 4     & 100   & 100   & 100   & 100   & 100   & 100   & 100   & 100   & 100   & 100   & 100   & 100   & 100   & 100   & 100   & 100   & 100   & 100   & 100   & 100   & 100   & 100   & 100   & 100   & 100   & 100   & 100   & 100   & 100   & 100   & 100   & 100   & 99.8  & 99.8  & 99.5  & 100   & 99.8  & 99.5  & 98.7  & 97.8  & 97.7  & 98.8  \\
          &       & 6     & 100   & 100   & 100   & 100   & 100   & 100   & 100   & 100   & 100   & 100   & 100   & 100   & 100   & 100   & 100   & 100   & 100   & 100   & 100   & 100   & 100   & 100   & 100   & 100   & 100   & 100   & 100   & 100   & 100   & 100   & 100   & 100   & 99.9  & 99.7  & 99.1  & 100   & 100   & 99.4  & 98.6  & 97.8  & 98.1  & 98.5  \\
          &       & 8     & 100   & 100   & 100   & 100   & 100   & 100   & 100   & 100   & 100   & 100   & 100   & 100   & 100   & 100   & 100   & 100   & 100   & 100   & 100   & 100   & 100   & 100   & 100   & 100   & 100   & 100   & 100   & 100   & 100   & 100   & 100   & 100   & 100   & 99.7  & 99.2  & 100   & 100   & 99.4  & 98.7  & 98.1  & 97.3  & 98.5  \\[0.3em]
          & 0.40  & 2     & 100   & 100   & 100   & 100   & 100   & 100   & 100   & 100   & 100   & 100   & 100   & 100   & 100   & 100   & 100   & 100   & 100   & 100   & 100   & 100   & 100   & 100   & 100   & 100   & 100   & 100   & 100   & 100   & 100   & 100   & 100   & 100   & 100   & 100   & 99.9  & 100   & 99.8  & 99.3  & 97.9  & 97.3  & 97.6  & 98.7  \\
          &       & 4     & 100   & 100   & 100   & 100   & 100   & 100   & 100   & 100   & 100   & 100   & 100   & 100   & 100   & 100   & 100   & 100   & 100   & 100   & 100   & 100   & 100   & 100   & 100   & 100   & 100   & 100   & 100   & 100   & 100   & 100   & 100   & 100   & 100   & 99.7  & 99.4  & 99.8  & 99.8  & 98.9  & 97.5  & 95.3  & 94.9  & 96.6  \\
          &       & 6     & 100   & 100   & 100   & 100   & 100   & 100   & 100   & 100   & 100   & 100   & 100   & 100   & 100   & 100   & 100   & 100   & 100   & 100   & 100   & 100   & 100   & 100   & 100   & 100   & 100   & 100   & 100   & 100   & 100   & 100   & 100   & 100   & 100   & 99.7  & 98.8  & 99.8  & 99.8  & 98.5  & 97.3  & 94.2  & 93.3  & 95.5  \\
          &       & 8     & 100   & 100   & 100   & 100   & 100   & 100   & 100   & 100   & 100   & 100   & 100   & 100   & 100   & 100   & 100   & 100   & 100   & 100   & 100   & 100   & 100   & 100   & 100   & 100   & 100   & 100   & 100   & 100   & 100   & 100   & 100   & 100   & 100   & 99.7  & 98.4  & 99.8  & 99.7  & 98.9  & 96.6  & 92.7  & 91.5  & 94.5  \\[0.3em]
          & 1.20  & 2     & 100   & 100   & 100   & 100   & 100   & 100   & 100   & 100   & 100   & 100   & 100   & 100   & 100   & 100   & 100   & 100   & 100   & 100   & 100   & 100   & 100   & 100   & 100   & 100   & 100   & 100   & 100   & 100   & 100   & 100   & 100   & 100   & 100   & 100   & 99.9  & 99.7  & 99.5  & 98.9  & 96.9  & 95.2  & 94.3  & 95.8  \\
          &       & 4     & 100   & 100   & 100   & 100   & 100   & 100   & 100   & 100   & 100   & 100   & 100   & 100   & 100   & 100   & 100   & 100   & 100   & 100   & 100   & 100   & 100   & 100   & 100   & 100   & 100   & 100   & 100   & 100   & 100   & 100   & 100   & 100   & 100   & 99.8  & 98.7  & 98.5  & 98.3  & 97.4  & 92.9  & 87.0  & 84.6  & 88.8  \\
          &       & 6     & 100   & 100   & 100   & 100   & 100   & 100   & 100   & 100   & 100   & 100   & 100   & 100   & 100   & 100   & 100   & 100   & 100   & 100   & 100   & 100   & 100   & 100   & 100   & 100   & 100   & 100   & 100   & 100   & 100   & 100   & 100   & 100   & 100   & 99.5  & 97.4  & 98.0  & 97.5  & 96.1  & 89.5  & 79.0  & 76.0  & 82.5  \\
          &       & 8     & 100   & 100   & 100   & 100   & 100   & 100   & 100   & 100   & 100   & 100   & 100   & 100   & 100   & 100   & 100   & 100   & 100   & 100   & 100   & 100   & 100   & 100   & 100   & 100   & 100   & 100   & 100   & 100   & 100   & 100   & 100   & 100   & 100   & 98.8  & 95.5  & 96.6  & 96.0  & 94.5  & 86.5  & 72.4  & 68.5  & 77.6  \\
    \end{tabular}%
    }
  \label{tab:power_cbn}%
\end{sidewaystable}%
%\end{landscape}

\subsection{Power curve}
\label{sec:curve}
In this subsection, we perturb Models 1--3 so that the new sequence is not a MDS and present  power curves. For given constant $a\in\{0,0.5,1,1.5,2,2.5\}$, the model settings are as follows:
\begin{itemize}[leftmargin=1.8cm]
  \item[Model 1'.] Let $\bx_t$ follow Model 1 and $\by_t=\bx_t+a\exp(-2\bx_{t-1}^2)$.
   \item[Model 2'.] Let $\bx_t$ follow Model 2 and $\by_t=\bx_t+a\cos(\boldsymbol\varepsilon_{t-1}\circ\boldsymbol{\sigma}_{t-1})$, where $\boldsymbol \varepsilon_{t-1}$ and $\boldsymbol\sigma_{t-1}$ are specified in Model 2.
  \item[Model 3'.] Let $\bx_t$ follow Model 3 and $\by_t=\bx_t+a\log(\bx_{t-2}^2)$.
\end{itemize}

\begin{figure}
  \centering
  \includegraphics[scale=0.85]{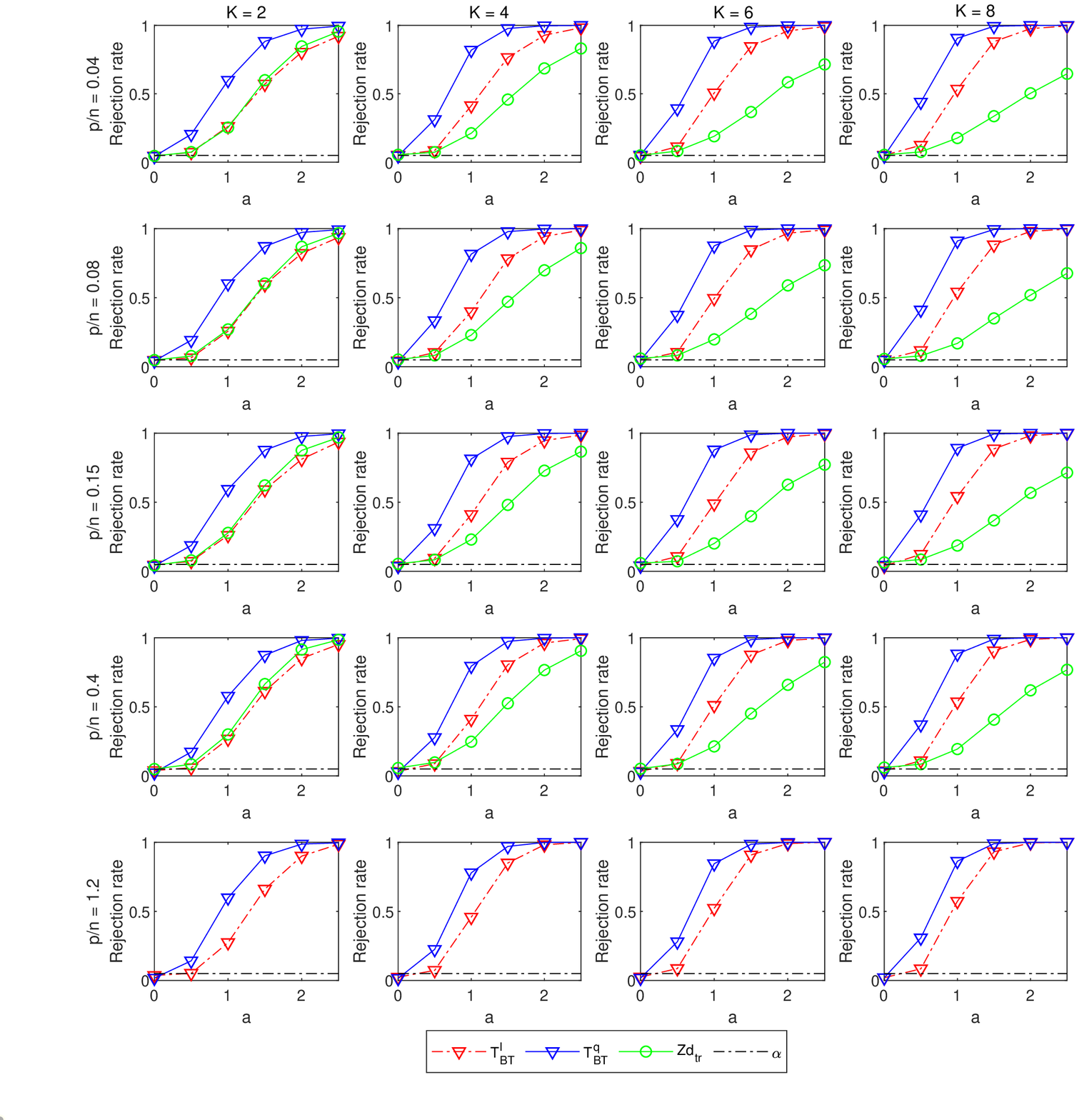}\\
  \caption{Empirical sizes and powers of $T_{\rm BT}^l$, $T_{\rm BT}^q$ and $Zd_{\rm tr}$ for Model 1' at the nominal level $\alpha=0.05$, where the sample size $n=100$.}
  \label{fig:curve1}
\end{figure}

\begin{figure}
  \centering
  \includegraphics[scale=0.85]{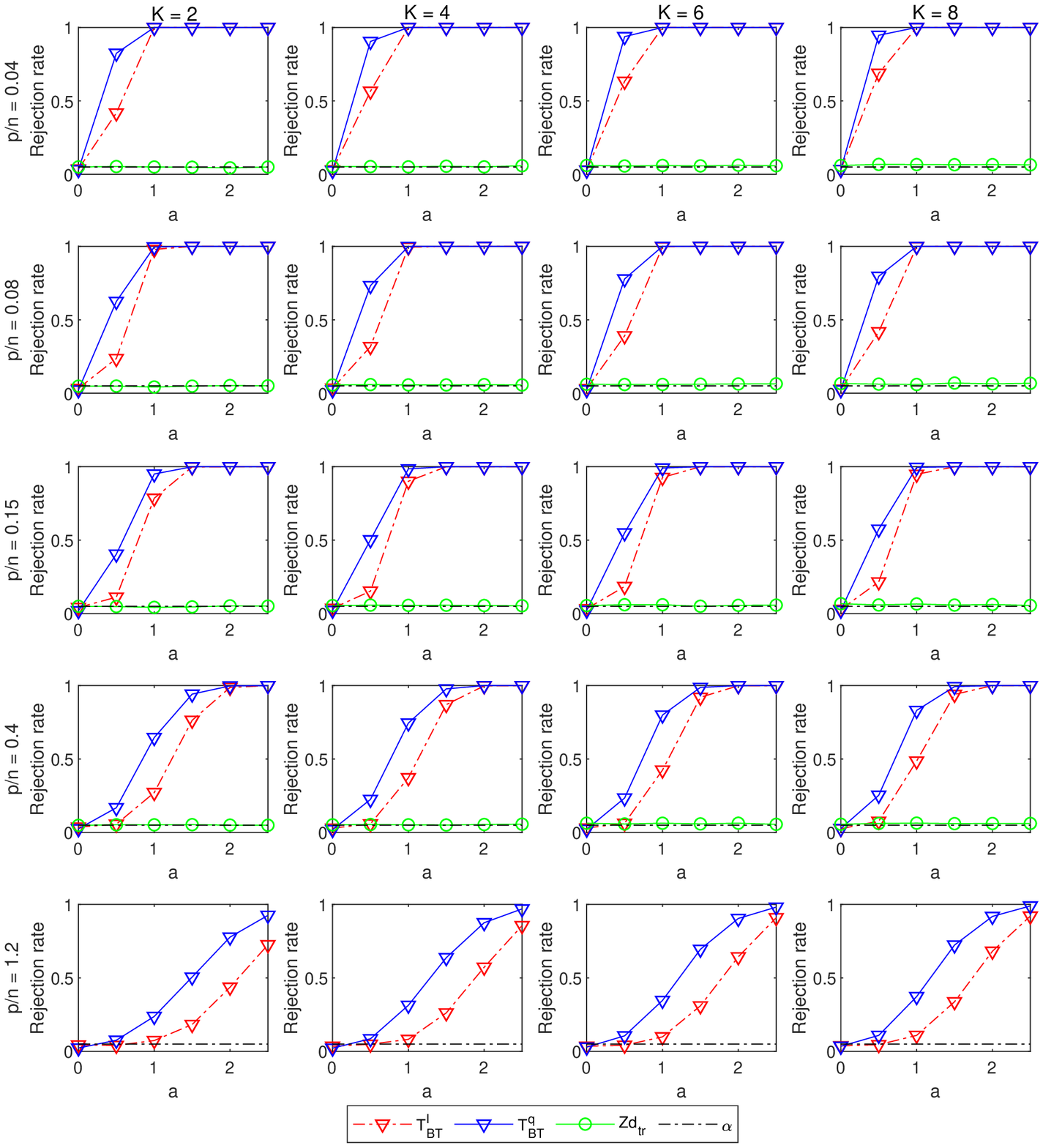}\\
  \caption{Empirical sizes and powers of $T_{\rm BT}^l$, $T_{\rm BT}^q$ and $Zd_{\rm tr}$ for Model 2' at the nominal level $\alpha=0.05$, where the sample size $n=100$.}
  \label{fig:curve2}
\end{figure}

\begin{figure}
  \centering
  \includegraphics[scale=0.85]{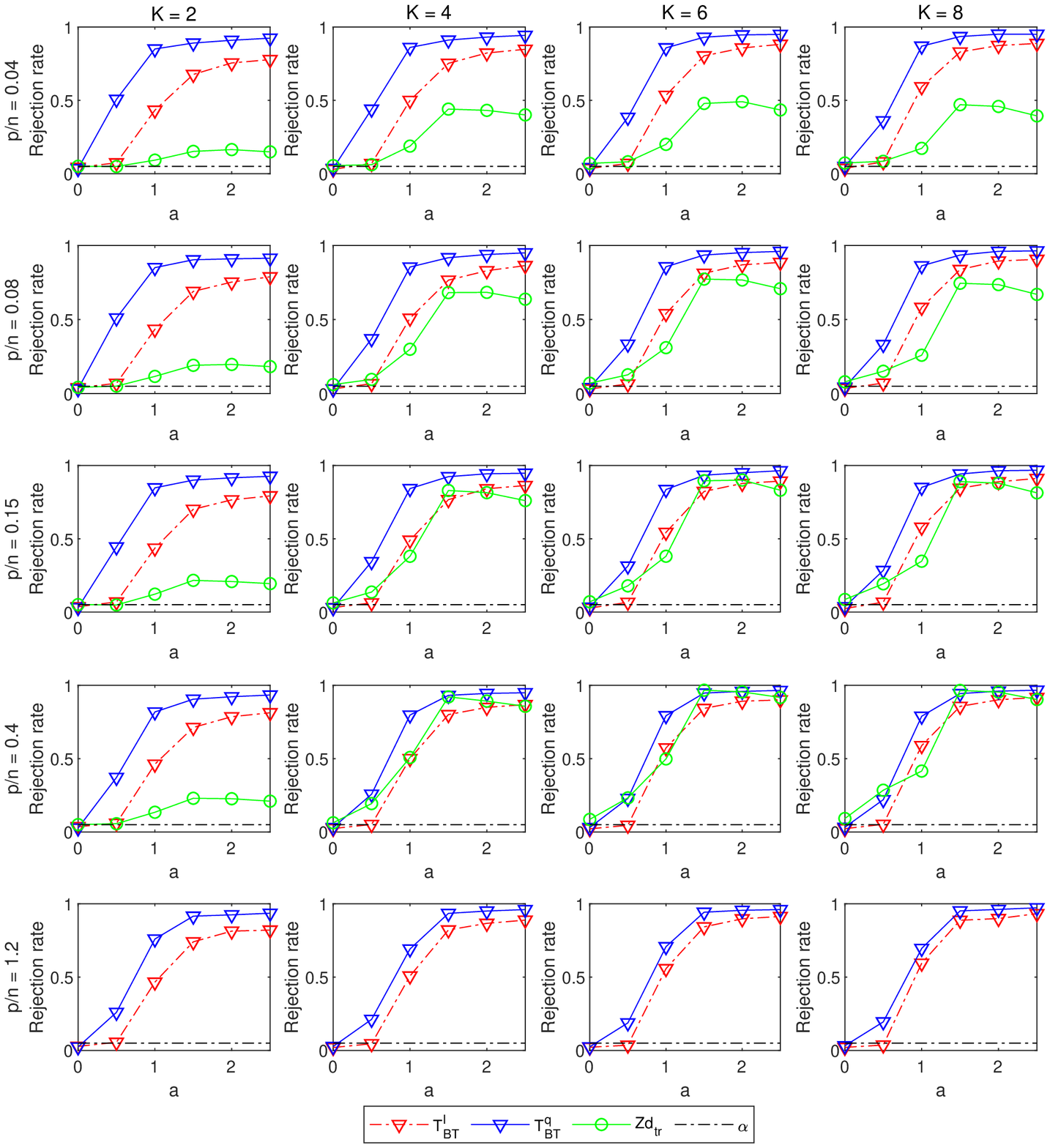}\\
  \caption{Empirical sizes and powers of $T_{\rm BT}^l$, $T_{\rm BT}^q$ and $Zd_{\rm tr}$ for Model 3' at the nominal level $\alpha=0.05$, where the sample size $n=100$.}
  \label{fig:curve3}
\end{figure}

We aim to test  whether $\{\by_t\}_{t\in\mathbb{Z}}$ defined in Models 1'--3' is a MDS. When $a=0$, $\by_t=\bx_t$ and Models 1'--3' become Models 1--3, respectively, which follow the null hypothesis.
Figures \ref{fig:curve1}--\ref{fig:curve3} display the empirical sizes and powers of our proposed tests ($T_{\rm BT}^l, T_{\rm BT}^q$) and \cite{HLZ2017}'s test ($Zd_{\rm tr}$) when the sample size $n=100$. Notice that $Zd_{\rm tr}$ is feasible when $p<n$. Thus when $p/n=1.2$, there is no power curve for $Zd_{\rm tr}$. As seen from  Figure \ref{fig:curve1},  our tests and \cite{HLZ2017}'s test control the empirical sizes well under the null hypothesis with $a=0$ and the empirical powers increase for larger values of the distance parameter $a$. But our tests outperform \cite{HLZ2017}'s test especially for large $K$. In Figure \ref{fig:curve2}, \cite{HLZ2017}'s test almost cannot detect the alternative hypotheses, but our tests still work well. This is presumably due to the inability of their test to capture nonlinear serial dependence. Based on Figure \ref{fig:curve3}, similar phenomenon is observed that the empirical powers increase as the distance $a$ grows.
Somewhat counter-intuitively, the empirical powers of $Zd_{\rm tr}$ decrease when $a$ increases from $2$ to $2.5$, which means the power is non-monotonic.
In addition, comparing the results of our tests for two maps, we find that the test based on linear and quadratic map is more powerful than the test only based on linear map for the three models. This should not be surprising. Since the alternatives in the three models are nonlinear transformations, the linear and quadratic map can capture both linear and nonlinear dependence. Generally speaking, both of the two maps perform well in the three models.

\section{Real data analysis}
\label{sec:real}

In this section, we apply our proposed tests to a real dataset, which collects weekly closing prices from 17 September 2004 to 26 December 2008 for 394 stocks. The returns of the stocks are obtained by the log difference of the data. And the sample size $n$ for the returns is 223.
These stocks can be classified into 9 major sectors, which consist of materials (22 stocks), real estate (25 stocks), utilities (26 stocks), consumer staples (30 stocks), healthcare (55 stocks), industrials (56 stocks), financials (58 stocks), IT (60 stocks), and consumer discretionary (62 stocks). Here we examine the validity of the martingale difference hypothesis within each sector and for all stocks using our tests and the ones proposed in \cite{HLZ2017}. Note that neither $Z_{{\rm tr}}$ nor
$Z_{{\rm det}}$ is applicable here, since $p<\sqrt{n}$ is violated for each sector. Hence we only present the results of $Zd_{\rm tr}$ for each sector, as it is not usable when we apply to all stock returns. Denote by $\bx_t$ the returns of these stocks at time $t$. Financial theory usually assumes the stock prices follow geometric Brownian Motion which implies $\mathbb{E}(\bx_t)=\bzero$ under the efficient markets hypothesis. We can propose the test statistic $T_{{\rm mean}}=|n^{-1/2}\sum_{t=1}^n\bx_t|_\infty$ for the null hypothesis $H_0:\mathbb{E}(\bx_t)=\bzero$. Using the method given in Section 4.1 of \cite{CCW2020} with three kernels (QS, PR, BT) to estimate the associated long-run covariance matrix, the associated p-values for such null hypothesis are 0.759, 0.749 and 0.753, respectively, which means there is no strong evidence against the zero-mean assumption of $\bx_t$ in our real data.

Table~\ref{tab:stock} reports the p-values of $Zd_{\rm tr}$ and our tests with assuming $\mathbb{E}(\bx_t)=\bzero$ and without assuming $\mathbb{E}(\bx_t)=\bzero$.
%Table~\ref{tab:stock} shows the p-values of six versions of our test and $Zd_{\rm tr}$ for each sector and all stocks combined.
It appears that there is no strong evidence against the martingale difference hypothesis based on all tests, except for a marginally significant p-value of $Zd_{\rm tr}$ when $K=2$ for the sector of consumer staples. Generally speaking, the martingale difference hypothesis is expected to hold for the weekly returns data, so in a sense both our tests and $Zd_{\rm tr}$ help confirming this property. For the same map, the use of different kernels  do not seem to affect the p-values much, indicating the insensitivity of our results with respect to the kernel. For this particular dataset, the use of linear and quadratic maps also  produces p-values that are not far away from the use of linear maps alone, for most sectors. The p-values corresponding to $Zd_{\rm tr}$ seem to monotonically decrease as $K$ goes down from $8$ to $2$ for all sectors, an interesting phenomenon worthy of
some theoretical investigation.
In addition, the results of our tests with assuming $\mathbb{E}(\bx_t)=\bzero$ and without assuming $\mathbb{E}(\bx_t)=\bzero$ are quite similar, which is consistent with the aforementioned conclusion that $\mathbb{E}(\bx_t)$ is not significantly different from zero.
Overall, our tests are preferred to the ones proposed in \cite{HLZ2017} due to the fact that they can be used regardless of whether the dimension $p$ exceeds the sample size $n$.

\begin{table}[htbp]
  \centering
  \caption{P-values of our tests and Hong et al.'s test for the weekly stock returns.}
  \vspace{-2mm}
    \resizebox{!}{8cm}{
    \begin{tabular}{rcp{0.5cm}|cccccc|cccccc|c}
    %\toprule
          &       &       & \multicolumn{6}{c|}{MDS test}                 & \multicolumn{6}{c|}{general MDS test}         & Hong et al.'s test \\[0.3em]
    \multicolumn{1}{l}{Sectors} & \multicolumn{1}{c}{$p$} & \multicolumn{1}{l|}{$K$} &   $T_{\rm QS}^l$  & $T_{\rm PR}^l$  & $T_{\rm BT}^l$  &  $T_{\rm QS}^q$  & $T_{\rm PR}^q$  & $T_{\rm BT}^q$    &       $T_{\rm QS}^l$  & $T_{\rm PR}^l$  & $T_{\rm BT}^l$  &  $T_{\rm QS}^q$  & $T_{\rm PR}^q$  & $T_{\rm BT}^q$   & $Zd_{\rm tr}$ \\[0.5em]
    \hline
    \multicolumn{1}{l}{Joint test} & 394   & 2     & 0.334  & 0.308  & 0.311  & 0.403  & 0.393  & 0.364  & 0.322  & 0.303  & 0.292  & 0.371  & 0.349  & 0.357  & NA \\
          &       & 4     & 0.523  & 0.517  & 0.546  & 0.440  & 0.462  & 0.443  & 0.524  & 0.508  & 0.520  & 0.465  & 0.465  & 0.450  & NA \\
          &       & 6     & 0.558  & 0.543  & 0.570  & 0.522  & 0.526  & 0.499  & 0.574  & 0.513  & 0.589  & 0.506  & 0.510  & 0.512  & NA \\
          &       & 8     & 0.612  & 0.602  & 0.643  & 0.553  & 0.538  & 0.521  & 0.639  & 0.600  & 0.643  & 0.518  & 0.527  & 0.546  & NA \\[0.3em]
    \hline
    \multicolumn{1}{l}{Materials} & 22    & 2     & 0.687  & 0.676  & 0.723  & 0.723  & 0.691  & 0.687  & 0.707  & 0.697  & 0.696  & 0.719  & 0.701  & 0.696  & 0.500  \\
          &       & 4     & 0.741  & 0.729  & 0.772  & 0.749  & 0.726  & 0.763  & 0.749  & 0.732  & 0.776  & 0.753  & 0.730  & 0.779  & 0.759  \\
          &       & 6     & 0.582  & 0.563  & 0.608  & 0.586  & 0.566  & 0.607  & 0.602  & 0.565  & 0.609  & 0.595  & 0.572  & 0.594  & 0.844  \\
          &       & 8     & 0.513  & 0.510  & 0.562  & 0.511  & 0.513  & 0.559  & 0.530  & 0.519  & 0.542  & 0.520  & 0.533  & 0.546  & 0.879  \\[0.3em]
    \multicolumn{1}{l}{Real estate} & 25    & 2     & 0.610  & 0.610  & 0.603  & 0.629  & 0.591  & 0.646  & 0.622  & 0.603  & 0.614  & 0.610  & 0.600  & 0.643  & 0.305  \\
          &       & 4     & 0.537  & 0.518  & 0.594  & 0.578  & 0.535  & 0.599  & 0.570  & 0.512  & 0.612  & 0.554  & 0.541  & 0.600  & 0.385  \\
          &       & 6     & 0.464  & 0.440  & 0.491  & 0.475  & 0.440  & 0.488  & 0.474  & 0.454  & 0.498  & 0.476  & 0.450  & 0.475  & 0.510  \\
          &       & 8     & 0.457  & 0.421  & 0.472  & 0.465  & 0.447  & 0.472  & 0.467  & 0.450  & 0.489  & 0.458  & 0.418  & 0.489  & 0.612  \\[0.3em]
    \multicolumn{1}{l}{Utilities} & 26    & 2     & 0.710  & 0.683  & 0.721  & 0.706  & 0.687  & 0.707  & 0.679  & 0.659  & 0.698  & 0.716  & 0.691  & 0.674  & 0.166  \\
          &       & 4     & 0.755  & 0.744  & 0.766  & 0.761  & 0.737  & 0.800  & 0.750  & 0.745  & 0.773  & 0.746  & 0.743  & 0.792  & 0.173  \\
          &       & 6     & 0.756  & 0.736  & 0.782  & 0.761  & 0.736  & 0.770  & 0.755  & 0.747  & 0.757  & 0.752  & 0.732  & 0.777  & 0.171  \\
          &       & 8     & 0.565  & 0.561  & 0.579  & 0.577  & 0.545  & 0.569  & 0.561  & 0.564  & 0.588  & 0.594  & 0.560  & 0.587  & 0.193  \\[0.3em]
    \multicolumn{1}{l}{Consumer} & 30    & 2     & 0.804  & 0.803  & 0.838  & 0.650  & 0.648  & 0.685  & 0.804  & 0.786  & 0.843  & 0.683  & 0.650  & 0.713  & 0.042  \\
    \multicolumn{1}{l}{staples} &       & 4     & 0.411  & 0.400  & 0.458  & 0.417  & 0.394  & 0.421  & 0.420  & 0.426  & 0.438  & 0.393  & 0.404  & 0.444  & 0.170  \\
          &       & 6     & 0.446  & 0.466  & 0.462  & 0.358  & 0.346  & 0.385  & 0.446  & 0.436  & 0.501  & 0.380  & 0.393  & 0.380  & 0.226  \\
          &       & 8     & 0.498  & 0.516  & 0.520  & 0.412  & 0.406  & 0.422  & 0.504  & 0.493  & 0.518  & 0.387  & 0.415  & 0.409  & 0.291  \\[0.3em]
    \multicolumn{1}{l}{Healthcare} & 55    & 2     & 0.835  & 0.808  & 0.846  & 0.813  & 0.816  & 0.855  & 0.803  & 0.794  & 0.838  & 0.809  & 0.796  & 0.848  & 0.131  \\
          &       & 4     & 0.611  & 0.603  & 0.618  & 0.549  & 0.543  & 0.544  & 0.590  & 0.626  & 0.590  & 0.537  & 0.542  & 0.559  & 0.172  \\
          &       & 6     & 0.636  & 0.626  & 0.665  & 0.592  & 0.578  & 0.591  & 0.625  & 0.616  & 0.642  & 0.597  & 0.616  & 0.605  & 0.188  \\
          &       & 8     & 0.661  & 0.641  & 0.657  & 0.615  & 0.621  & 0.636  & 0.626  & 0.626  & 0.656  & 0.618  & 0.595  & 0.614  & 0.351  \\[0.3em]
    \multicolumn{1}{l}{Industrials} & 56    & 2     & 0.588  & 0.541  & 0.595  & 0.579  & 0.547  & 0.603  & 0.575  & 0.549  & 0.588  & 0.553  & 0.547  & 0.590  & 0.365  \\
          &       & 4     & 0.642  & 0.640  & 0.677  & 0.665  & 0.617  & 0.696  & 0.676  & 0.625  & 0.697  & 0.670  & 0.657  & 0.686  & 0.485  \\
          &       & 6     & 0.637  & 0.630  & 0.676  & 0.650  & 0.626  & 0.665  & 0.666  & 0.629  & 0.678  & 0.639  & 0.637  & 0.692  & 0.573  \\
          &       & 8     & 0.697  & 0.698  & 0.739  & 0.694  & 0.680  & 0.730  & 0.706  & 0.692  & 0.742  & 0.705  & 0.683  & 0.723  & 0.631  \\[0.3em]
    \multicolumn{1}{l}{Financials} & 58    & 2     & 0.675  & 0.641  & 0.676  & 0.265  & 0.254  & 0.244  & 0.677  & 0.656  & 0.675  & 0.273  & 0.268  & 0.250  & 0.148  \\
          &       & 4     & 0.715  & 0.704  & 0.726  & 0.360  & 0.379  & 0.347  & 0.719  & 0.703  & 0.734  & 0.367  & 0.361  & 0.362  & 0.290  \\
          &       & 6     & 0.710  & 0.708  & 0.724  & 0.429  & 0.441  & 0.416  & 0.706  & 0.674  & 0.730  & 0.429  & 0.422  & 0.406  & 0.370  \\
          &       & 8     & 0.740  & 0.715  & 0.763  & 0.485  & 0.497  & 0.478  & 0.737  & 0.728  & 0.739  & 0.486  & 0.501  & 0.472  & 0.498  \\[0.3em]
    \multicolumn{1}{l}{IT} & 60    & 2     & 0.276  & 0.293  & 0.293  & 0.296  & 0.292  & 0.307  & 0.295  & 0.277  & 0.306  & 0.283  & 0.267  & 0.288  & 0.121  \\
          &       & 4     & 0.550  & 0.541  & 0.586  & 0.531  & 0.537  & 0.595  & 0.551  & 0.545  & 0.569  & 0.532  & 0.541  & 0.590  & 0.299  \\
          &       & 6     & 0.610  & 0.599  & 0.615  & 0.611  & 0.577  & 0.623  & 0.593  & 0.566  & 0.634  & 0.583  & 0.583  & 0.610  & 0.454  \\
          &       & 8     & 0.637  & 0.588  & 0.636  & 0.622  & 0.583  & 0.613  & 0.624  & 0.599  & 0.619  & 0.596  & 0.586  & 0.629  & 0.629  \\[0.3em]
    \multicolumn{1}{l}{Consumer} & 62    & 2     & 0.273  & 0.273  & 0.264  & 0.286  & 0.316  & 0.303  & 0.267  & 0.274  & 0.260  & 0.318  & 0.306  & 0.308  & 0.407  \\
    \multicolumn{1}{l}{discretionary} &       & 4     & 0.350  & 0.344  & 0.351  & 0.366  & 0.359  & 0.377  & 0.355  & 0.363  & 0.355  & 0.384  & 0.335  & 0.362  & 0.648  \\
          &       & 6     & 0.372  & 0.342  & 0.385  & 0.407  & 0.393  & 0.409  & 0.358  & 0.363  & 0.360  & 0.387  & 0.395  & 0.390  & 0.800  \\
          &       & 8     & 0.377  & 0.360  & 0.359  & 0.405  & 0.401  & 0.405  & 0.358  & 0.351  & 0.378  & 0.406  & 0.391  & 0.403  & 0.888  \\
    %\bottomrule
    \end{tabular}%
    }
  \label{tab:stock}%
\end{table}%

\section{Discussion}\label{sec:conc}

In this paper, we propose a new martingale difference test that captures nonlinear serial dependence and works in the high-dimensional environment, as motivated by the
increasing availability of high-dimensional nonlinear time series from economics and finance.  Under mild moment and weak temporal dependence assumptions, we establish the validity of  Gaussian approximation and provide a simulation-based approach for critical values. In addition to its built-in capability of
accommodating both low and high dimensions, our test also has a number of appealing features such as being robust to conditional moments of unknown forms and strong/weak cross-series dependence. From our numerical simulations and a real data analysis, we observe quite encouraging finite sample performance.  Therefore we feel confident to recommend its use by the practitioners when there is a need to assess the martingale difference hypothesis for econometric/financial time series of moderate or high dimension.

%Testing the martingale difference hypothesis can be viewed as a key component in  specification testing of time series models; see \cite{WZS2021} for a recent paper and references therein. As a natural next step, we may consider the application of our proposed method in model diagnosis problems. However, such extension is very nontrivial in high dimensional settings. See Section~\ref{sec:modelcheck} in the supplement for more discussions.

%plan to consider an extension of our test to test the specification of VAR$(m)$ models when the dimension  $p$ is high. This is very nontrivial as we have to take into account the estimation effect that arises  in the estimation of high-dimensional parameters, and possible bias if the LASSO-type regularization  is used under sparsity assumption on the VAR coefficient matrix [\cite{BM2015}].

In the literature, testing quantile/directional predictability has been studied for low-dimensional time series; see  \cite{HLOW2016}. It would be also interesting to extend their test to the high-dimensional setting.
A sound data-driven bandwidth choice in our simulation-based approach for generating the critical values merits additional research, especially from a testing-optimal viewpoint.
We leave these topics for future investigation.

%\section*{Acknowledgements}
%Chang was supported in part by the National Natural Science Foundation of China (grant nos.~71991472, 72125008 and 11871401), and the Center of Statistical Research at Southwestern University of Finance and Economics. Jiang was supported in part by
%the National Natural Science Foundation
%of China (grant nos. 71991472 and 12001442). We would like to thank the editor, the associate editor and two reviewers for their constructive suggestions which led to substantial improvements.

%\appendix
\section{Technical proofs}\label{sec:pfs}
In this section, we provide the detailed proofs for all theoretical results stated in the paper, and also introduce necessary lemmas and propositions with proofs.
Throughout this section, we use $C$ to denote a generic positive finite constant that does not depend on $(p,d,n,K)$ and may be different in different uses. For two sequences of positive numbers $\{a_n\}$ and $\{b_n\}$, we write $a_n\lesssim b_n$ or $b_n\gtrsim a_n$ if $\limsup_{n\rightarrow\infty}a_n/b_n\leqslant c_0$ for some positive constant $c_0$. We write $a_n\asymp b_n$ if $a_n\lesssim b_n$ and $b_n\lesssim a_n$ hold simultaneously. We write $a_n\ll b_n$ or $b_n\gg a_n $ if $\limsup_{n\rightarrow\infty}a_n/b_n=0$.  For a countable set $\mathcal{F}$, we use $|\cF|$ to denote the cardinality of $\cF$.

Write
$\bu := ( u_1,\ldots, u_{Kpd} )^{\T} =(\hat\bgamma_1^{\T},\ldots,\hat\bgamma_{K}^{\T})^{\T}$ with $\hat{\bgamma}_j=(n-j)^{-1}\sum_{t=1}^{n-j} {\rm vec}\{\bphi(\bx_{t})\bx_{t+j}^{\T}\}$ for any $j\in[K]$. Let $\tilde n=n-K$. Recall $\be_t=([{\rm vec} \{\bphi(\bx_t)\bx_{t+1}^{\T}\}]^\T,\ldots, [{\rm vec} \{ \bphi(\bx_t)\bx_{t+K}^{\T}\}]^\T )^{\T}$. Since $\{\bx_t\}$ is an $\alpha$-mixing process satisfying Condition \ref{cond.mix}, we know the newly defined process $\{\be_t\}$ is also $\alpha$-mixing with the $\alpha$-mixing coefficients $\{\tilde{\alpha}_K(k)\}_{k\geqslant1}$ satisfying
\begin{align}\label{eq:newalphamixing0}
\tilde \alpha_K(k)\leqslant C_3\exp (-C_4|k-K|_{+}^{\tau_2})\,,
\end{align}
where the positive constants $\tau_2$, $C_3$ and $C_4$ are specified in Condition 2.
Write $\bar \be :=(\bar{\eta}_1,\ldots,\bar{\eta}_{Kpd})^\T= \tilde n^{-1}\sum_{t=1}^{\tilde n}\be_t$.  For each $j\in[K]$, define
$Z_j =n\max_{\ell\in\mathcal{L}_j}  u_{\ell}^2 $ and $\tilde Z_j = \tilde n \max_{\ell\in\mathcal{L}_j} \bar\eta_\ell^2$ with $\mathcal{L}_j:=\{(j-1)pd+1,\ldots,jpd\}$. Then the test statistic can be written as
$
T_n=n\sum_{j=1}^{K}|\hat\bgamma_j|_{\infty}^2= \sum_{j=1}^K Z_j$. Furthermore, we let
$
\tilde T_n := \sum_{j=1}^K \tilde Z_j$.

\subsection{A key proposition}
Let $\{\bz_t\}_{t=1}^{n}$ be a $d_{z}$-dimensional dependent sequence with $\E(\bz_t)=\boldsymbol 0$ for any $t\in[n]$. Define $\bs_{n,z}=n^{-1/2}\sum_{t=1}^n \bz_t$ and $\bXi=\mbox{Var}(n^{-1/2}\sum_{t=1}^n \bz_t)$. Write $\bz_t=(z_{t,1},\ldots,z_{t,d_z})^\T$. We assume $\{\bz_t\}_{t=1}^n$ satisfy the following three assumptions:
\begin{itemize}
  \item[AS1.] There exist universal constants $b_1>1$, $b_2>0$ and $r_1 \in (0,1]$ such that
      $
      \sup_{t\in[n]}\sup_{j\in[d_{z}]}\bP(|z_{t,j}|>u) \leqslant b_1\exp(-b_2 u^{r_1} )$
      for any $u>0$.
  \item[AS2.] There exist universal constants $a_1>1$, $a_2>0$ and $r_2\in(0,1]$ such that the $\alpha$-mixing coefficients of the sequence $\{\bz_t\}_{t=1}^n$, denoted by $\{\alpha_z(k)\}_{k\geqslant1}$, satisfying
$
  \alpha_z(k) \leqslant a_1 \exp(-a_2|k-m|_{+}^{r_2})$
  for any $k\geqslant1$ and some $m=m(n)>0$, where $m=o(n)$ may diverge with $n$.
  \item[AS3.] There exists a universal constant $c>0$ such that $\E(|n^{-1/2}\sum_{t=1}^{n} z_{t,j}|^{2}) \geqslant c$ for any $j\in[d_{z}]$.
\end{itemize}
Let $\bs_{n,y}\sim \mathcal{N}(\bzero,\bXi)$ be independent of $\cZ_n=\{\bz_1,\ldots,\bz_n\}$. Define
\begin{align}\label{varrho}
  \varrho_n := \sup_{\bu\in\R^{d_z},\nu\in[0,1]} \big| \bP(\sqrt{\nu}\bs_{n,z}+\sqrt{1-\nu}\bs_{n,y}\leqslant\bu) - \bP(\bs_{n,y}\leqslant\bu) \big|\,.
\end{align} \cite{CCW2020} gives an upper bound for $\varrho_n$ when $m$ is a fixed constant. Proposition \ref{lem.approx} presents a more general result that allows $m$ to diverge with $n$, whose proof is presented in the supplementary material.

\begin{proposition}\label{lem.approx}
Assume $d_{z}\geqslant n^{\varpi}$ for some sufficiently small constant $\varpi>0$. Under {\rm AS1--AS3}, it holds that
\begin{align*}
  \varrho_n\lesssim \frac{m^{1/3}(\log d_z)^{2/3}}{n^{1/9}}\{m^{1/6}(\log d_z)^{1/2}+m^{1/3}+(\log d_z)^{1/(3r_2)}\}
\end{align*}
provided that $\log d_{z}\ll\min\{m^{3r/(6+2r)}n^{7r/(18+6r)}, m^{-3r_1/(6+2r_1)}n^{7r_1/(18+6r_1)}, n^{r_2/(9-3r_2)}\}$
with $m\lesssim n^{1/9}(\log n)^{1/3}$, where $r=r_1r_2/(r_1+r_2)$ and $r_1$ and $r_2$ are specified in {\rm AS1} and {\rm AS2}, respectively.
\end{proposition}

\newtheorem{lemma}{Lemma}
\setcounter{lemma}{0}
\renewcommand{\thelemma}{L\arabic{lemma}}

Proposition \ref{lem.approx} requires that $m$ involved in Assumption AS2 cannot diverge faster than $n^{1/9}(\log n)^{1/3}$. The proof of Proposition 3 is based on the widely used ``large-and-small-blocks" technique in time series analysis. The key step for the proof of Proposition \ref{lem.approx} is to establish the associated Gaussian approximation result for the partial sum over the large blocks, see Lemma \ref{lem.varrho2} in the supplementary material. The restrictions on $\log d_z$ given in Proposition \ref{lem.approx} are derived from the conditions of Lemma \ref{lem.varrho2} with suitable selections of the lengths of large and small blocks. In the proofs of Propositions \ref{prop.sigma} and \ref{lem.approx}, and Theorem \ref{thm.H1}, we need the following lemma whose proof is given in the supplementary material.

\begin{lemma}\label{lem.bern}
Under {\rm AS1--AS3},
it holds that
\begin{align}\label{bern.inq}
  \max_{0\leqslant a\leqslant n-q}\max_{j\in[d_z]}\bP\bigg(\max_{k\in[q]} \bigg|\sum_{t=a+1}^{a+k} z_{t,j}\bigg| \geqslant x\bigg) \lesssim
  & ~ \exp(-Cq^{-1}m^{-1}x^2) + qx^{-1}\exp(-Cx^r) \nonumber\\
  &  +qx^{-1}\exp(-Cm^{-r_1} x^{r_1})
\end{align}
for any $x>0$ and $m\leqslant q\leqslant n$, where $r=r_1r_2/(r_1+r_2)$.
\end{lemma}

\subsection{Proof of Proposition 1}\label{sec:b.1}
Recall $
T_n=\sum_{j=1}^K Z_j$ and
$
\tilde T_n := \sum_{j=1}^K \tilde Z_j$. To construct Proposition 1, we need the following lemma whose proof is given in the supplementary material.
\begin{lemma}\label{lem.eta}
Assume Conditions {\rm 1--3} hold. Let $\tau=\tau_1\tau_2/(\tau_1+\tau_2)$. If $\log(Kpd)=o(n^{\tau/2})$ and $K^{\tau_1}\log(Kpd)=o(n^{\tau_1/2})$, then
\begin{align*}
|T_n-\tilde T_n| \lesssim \frac{K^{3/2}\{\log(Kpd)\}^{1/2}}{{n}^{1/2}}\max[\{\log(Kpd)\}^{1/\tau}, K\{\log(Kpd)\}^{1/\tau_1}]
\end{align*} with probability at least $1-C(Kpd)^{-1}$ under $H_0$.
\end{lemma}

Recall $\bar\be=\tilde{n}^{-1}\sum_{t=1}^{\tilde{n}}\be_t$ and $G_K = \sum_{j=1}^K \max_{\ell\in\mathcal{L}_j} |g_\ell|^2$ with $\bg=(g_1,\ldots,g_{Kpd})^{\T} \sim \calN(\bzero, \bSigma_{n,K})$ where $\bSigma_{n,K}=\tilde{n}\E\{(\bar\be-\bmu)(\bar\be-\bmu)^{\T} \}$ and  $\bmu =\tilde{n}^{-1}\sum_{t=1}^{\tilde{n}}\mathbb{E}(\be_t)$. Under $H_0$, we have $\bmu=\bzero$. Thus $\bSigma_{n,K}=\tilde{n}\E(\bar\be\bar\be^{\T})$.
 Define
$\bv := (v_1,\ldots, v_{Kpd})^{\T}=  \tilde n^{1/2} \bar\be$.
Our proof includes two steps:
(i) using Proposition \ref{lem.approx} to show $\sup_{x>0}|\bP(\tilde T_n\leqslant x) - \bP(G_K\leqslant x) |=o(1)$, and
(ii) using Lemma \ref{lem.eta} to show $\sup_{x>0}|\bP(T_n\leqslant x) - \bP(G_K\leqslant x) |=o(1)$.

\bigskip
\textbf{Step 1.}
For any $j_1,\ldots,j_K\in[pd]$ and $x>0$, let $\mathcal A_{j_1,\ldots,j_K}(x)=\{\bb \in\R^{Kpd}: \bb_{S_{j_1,\ldots,j_K}}^{\T} \bb_{S_{j_1,\ldots,j_K}}\leqslant x\}$ with  $S_{j_1,\ldots,j_K}=\{j_1,j_2+pd,\ldots,j_K+(K-1)pd\}$. Define $\mathcal A(x;K) = \bigcap_{j_1=1}^{pd} \cdots \bigcap_{j_K=1}^{pd} \mathcal A_{j_1,\ldots,j_K}(x)$. We then have $\{\tilde T_n\leqslant x\}=\{\bv\in\mathcal A(x;K)\}$ and $\{G_K\leqslant x\}=\{\bg\in\mathcal A(x;K)\}$. Note that the set $\mathcal A_{j_1,\ldots,j_K}(x)$ is convex that only depends on the components in $S_{j_1,\ldots,j_K}$. For a generic integer $q\geqslant2$, denote by $\bS^{q-1}$ the $q$-dimensional unit sphere. We can reformulate $\mathcal A_{j_1,\ldots,j_K}(x)$ as follows:
\begin{align*}
  \mathcal A_{j_1,\ldots,j_K}(x)=\bigcap_{\ba\in\{\ba\in\bS^{Kpd-1}:\ba_{S_{j_1,\ldots,j_K}}\in\bS^{K-1}\}} \{\bb\in\R^{Kpd}:\ba^{\T} \bb \leqslant \sqrt{x}\} \,.
\end{align*}
Define $\cF
=\bigcup_{j_1=1}^{pd}\cdots\bigcup_{j_K=1}^{pd}\{\ba\in\bS^{Kpd-1}:\ba_{S_{j_1,\ldots,j_K}}\in\bS^{K-1}\}$. Then
$
  \mathcal A(x;K)=\bigcap_{\ba\in\cF}\{\bb\in\R^{Kpd}: \ba^{\T} \bb\leqslant\sqrt{x}\}$. For the unit sphere $\bS^{K-1}$ equipped with $|\cdot|_2$, it is well-known that its $\epsilon$-covering number $N_{\bS^{K-1},\epsilon}$ satisfies $\epsilon^{-K} \leqslant N_{\bS^{K-1},\epsilon}\leqslant (1+2\epsilon^{-1})^K$, see Lemma 5.2 of \cite{Vershynin2012}.
Let $\mathcal S_\epsilon$ be an $\epsilon$-net of $\bS^{K-1}$ with cardinality $N_{\bS^{K-1},\epsilon}$. Without loss of generality, we assume $\mathcal S_\epsilon \subset \bS^{K-1}$.
Then $\tilde{\mathcal S}_{\epsilon}^{(j_1,...,j_K)}:=\{\ba\in\bS^{Kpd-1}: \ba_{S_{j_1,\ldots,j_K}}\in\mathcal S_\epsilon \}$ provides an $\epsilon$-net of $\{\ba\in\bS^{Kpd-1}:\ba_{S_{j_1,\ldots,j_K}}\in\bS^{K-1}\}$ for any given $(j_1,\ldots,j_K)\in[pd]^K$, and $|\tilde{\mathcal{S}}_{\epsilon}^{(j_1,\ldots,j_K)}|=N_{\bS^{K-1},\epsilon}$. Furthermore, we know $\cF_\epsilon=\bigcup_{j_1=1}^{pd}\cdots\bigcup_{j_K=1}^{pd} \tilde{\mathcal S_\epsilon}^{(j_1,\ldots,j_K)}\subset\cF$ is an $\epsilon$-net of $\cF$ with $|\cF_\epsilon|$ satisfying
$
  \epsilon^{-K} \leqslant |\cF_\epsilon| \leqslant \{(2+\epsilon)\epsilon^{-1}pd\}^K$.
Recall
$  \mathcal A(x;K)=\bigcap_{\ba\in\cF}\{\bb\in\R^{Kpd}:\ba^{\T} \bb\leqslant\sqrt{x}\}$.
Define $A_1(x) = \bigcap_{\ba\in\cF_\epsilon} \{\bb\in\R^{Kpd}:\ba^{\T} \bb\leqslant(1-\epsilon)\sqrt{x}\} $ and $A_2(x) = \bigcap_{\ba\in\cF_\epsilon} \{\bb\in\R^{Kpd}:\ba^{\T} \bb\leqslant\sqrt{x}\}$. We can show that
$
  A_1(x) \subset \mathcal A(x;K) \subset A_2(x)$.
Define
\begin{align*}
  &\rho_{1,g}(x)  :=|\bP\{\bv\in A_1(x)\} - \bP\{\bg\in A_1(x)\}|\vee |\bP\{\bv\in A_2(x)\} - \bP\{\bg\in A_2(x)\} | \,,\\
  &\rho_{2,g}(x)  :=|\bP\{\bg\in A_2(x)\} - \bP\{\bg\in A_1(x)\}|\,.
\end{align*}
It then holds that
\begin{align*}
  \bP\{\bv\in\mathcal A(x;K)\}
  \leqslant&~ \bP\{\bv\in A_2(x)\}
                     \leqslant \bP\{\bg\in A_2(x)\} + \rho_{1,g}(x) \\
  \leqslant&~ \bP\{\bg\in A_1(x)\} + \rho_{2,g}(x) + \rho_{1,g}(x) \\
  \leqslant&~ \bP\{\bg\in\mathcal A(x;K)\} +  \rho_{1,g}(x) + \rho_{2,g}(x) \,.
\end{align*}
Analogously, we also have $\bP\{\bv\in\mathcal A(x;K)\} \geqslant \bP\{\bg\in\mathcal A(x;K)\} - \rho_{1,g}(x) - \rho_{2,g}(x)$. Hence, we have
\begin{align}\label{ks.v}
  |\bP\{\bv\in\mathcal A(x;K)\} - \bP\{\bg\in\mathcal A(x;K)\}| \leqslant \rho_{1,g}(x)+\rho_{2,g}(x) \,.
\end{align}
We set $\epsilon=n^{-1}$ throughout the following arguments. Then $|\cF_\epsilon|\geqslant n^K$.
Due to $\tau_2\in(0,1]$, it holds that $K\lesssim (\log|\cF_\epsilon|)^{1/\tau_2}$. Note that $K\lesssim n^{1/9}(\log n)^{1/3}$. By Proposition \ref{lem.approx} with $m=K$, $d_z\lesssim (npd)^K$ and $(r_1,r_2)=(\tau_1,\tau_2)$,  we have
\begin{align*}
\sup_{x> 0}  \rho_{1,g}(x)
= &~  \sup_{x> 0} \bigg|\bP\bigg(\max_{\ba\in\cF_\epsilon}\ba^{\T} \bv \leqslant x\bigg)
        - \bP\bigg(\max_{\ba\in\cF_\epsilon}\ba^{\T}\bg\leqslant x \bigg) \bigg| \\
 \lesssim&~ n^{-1/9}K^{5/3}\{\log(npd)\}^{7/6} + n^{-1/9}K^{(1+3\tau_2)/(3\tau_2)}\{\log(npd)\}^{(1+2\tau_2)/(3\tau_2)}
  \,,
\end{align*}
provided that
$\log(npd)\ll\min\{K^{(\tau-6)/(6+2\tau)}n^{7\tau/(18+6\tau)}, K^{-(6+5\tau_1)/(6+2\tau_1)}n^{7\tau_1/(18+6\tau_1)}, K^{-1}n^{\tau_2/(9-3\tau_2)}\}$.
To make $\sup_{x>0}\rho_{1,g}(x)=o(1)$, we need to require $\log(npd)\ll \min\{n^{2/21}K^{-10/7}, n^{\tau_2/(3+6\tau_2)}K^{-(1+3\tau_2)/(1+2\tau_2)}\}$. Notice that
$
  \rho_{2,g}(x)
= \bP\{(1-\epsilon)\sqrt{x}
< \max_{\ba\in\cF_\epsilon}\ba^{\T}\bg \leqslant \sqrt{x}\}$. If $x\leqslant K^3\{\log(npd)\}^{3}$, by Nazarov's inequality (Lemma A.1, \citealp{CCK2017}), we have
$
  \rho_{2,g}(x) \leqslant C\epsilon\sqrt{x\log |\cF_\epsilon|}
\lesssim n^{-1}K^2\{\log(npd)\}^2$.
If $x>K^3\{\log(npd)\}^3$, by Markov inequality, we have
\begin{align*}
  \rho_{2,g}(x) \leqslant \bP\bigg\{(1-\epsilon)\sqrt{x} \leqslant \max_{\ba\in\cF_\epsilon} \ba^{\T} \bg \bigg\}
\leqslant \frac{\E(\max_{\ba\in\cF_\epsilon}|\ba^{\T}\bg|)} {(1-\epsilon)K^{3/2}\{\log(npd)\}^{3/2}}\lesssim \{\log(npd)\}^{-1} \,,
\end{align*}
where the last step is based on Lemma 7.4 in \cite{FSZ2018}. Hence, $\sup_{x>0}\rho_{2,g}(x)=o(1)$ if $\log(npd)\ll \min\{n^{2/21}K^{-10/7}, n^{\tau_2/(3+6\tau_2)}K^{-(1+3\tau_2)/(1+2\tau_2)}\}$. Due to
$|\bP(\tilde T_n\leqslant x) - \bP(G_K\leqslant x) |
=|\bP\{\bv\in\mathcal A(x;K)\} - \bP\{\bg\in\mathcal A(x;K)\}|$, \eqref{ks.v} implies
$$
  \sup_{x>0}|\bP(\tilde T_n\leqslant x) - \bP(G_K\leqslant x) |= o(1)$$ 
  provided that
$\log(npd)\ll\min\{K^{(\tau-6)/(6+2\tau)}n^{7\tau/(18+6\tau)},  K^{-(6+5\tau_1)/(6+2\tau_1)}n^{7\tau_1/(18+6\tau_1)}, K^{-1}n^{\tau_2/(9-3\tau_2)}, \\ K^{-(1+3\tau_2)/(1+2\tau_2)}n^{\tau_2/(3+6\tau_2)},
K^{-10/7}n^{2/21}\}$.

\textbf{Step 2.}
For any  $\zeta>0$, we have
\begin{align}\label{ks.Tn}
  \sup_{x>0}|\bP( T_n\leqslant x) - \bP(G_K\leqslant x)|
   \leqslant&~ \sup_{x>0} |\bP(\tilde T_n\leqslant x) - \bP(G_K\leqslant x ) |
   + \bP( |T_n - \tilde T_n| > \zeta) \notag\\
   &  + \sup_{x>0}\bP(x-\zeta < G_K \leqslant x+\zeta) \,.
\end{align}
Note that $K=o(n)$. Selecting
$\zeta=CK^{3/2}\{\log(npd)\}^{1/2}n^{-1/2}\max[\{\log(npd)\}^{1/\tau}, K\{\log(npd)\}^{1/\tau_1}]$ for some sufficiently large constant $C>0$, Lemma \ref{lem.eta} yields
$
  \bP( |T_n - \tilde T_n| > \zeta)= o(1)$. In the sequel, we will consider $\mathbb{P}(x-\zeta<G_K\leqslant x+\zeta)$ under the scenarios $x\leqslant \zeta$ and $x>\zeta$, respectively. Notice that $(1,0,\ldots,0)^\T\in\cF$ and $(-1,0,\ldots,0)^\T\in\cF$. Recall that $\bg=(g_1,\ldots,g_{Kpd})^\T\sim\mathcal{N}(\bzero,\bSigma_{n,K})$ and $\{G_K\leqslant x\}=\{\max_{\ba\in\cF}\ba^\T\bg\leqslant\sqrt{x}\}$ for any $x>0$. Then we have
\begin{align}
  \sup_{x\leqslant\zeta}\bP(x-\zeta < G_K \leqslant x+\zeta)
  \leqslant&~ \sup_{x\leqslant\zeta}\bP( G_K \leqslant x+\zeta)=\sup_{x\leqslant\zeta}\bP\bigg(\max_{\ba\in\cF}\ba^\T\bg\leqslant \sqrt{x+\zeta}\bigg) \notag\\
  \leqslant&~ \sup_{x\leqslant\zeta}\bP(-\sqrt{x+\zeta}\leqslant g_1\leqslant\sqrt{x+\zeta})\lesssim \sqrt{\zeta} \,,\label{eq:ks.Tn1}
\end{align}
where the last step is due to the anti-concentration inequality of normal random variable.
For any $x>\zeta$, it holds that
\begin{align*}
  \bP(x-\zeta <  G_K \leqslant x+\zeta)
   =&~ \bP( G_K \leqslant x+\zeta) - \bP( G_K \leqslant x-\zeta) \\
   \leqslant&~ \bP\bigg(\max_{\ba\in\cF_{\epsilon}}\ba^{\T} \bg \leqslant \sqrt{x+\zeta} \bigg)
     - \bP\bigg\{ \max_{\ba\in\cF_{\epsilon}}\ba^{\T} \bg \leqslant (1-\epsilon)\sqrt{x-\zeta} \bigg\} \\
   \leqslant&~ \bP\bigg(\max_{\ba\in\cF_{\epsilon}}\ba^{\T} \bg \leqslant \sqrt{x}+\sqrt{\zeta} \bigg)
     - \bP\bigg\{ \max_{\ba\in\cF_{\epsilon}}\ba^{\T} \bg \leqslant (1-\epsilon)(\sqrt{x}-\sqrt{\zeta})\bigg\} \\
   \leqslant&~ \bP\bigg\{(1-\epsilon)(\sqrt{x}-\sqrt{\zeta})< \max_{\ba\in\cF_{\epsilon}}\ba^{\T} \bg \leqslant (1-\epsilon)\sqrt{x}\bigg\} \\
   & + \bP\bigg\{(1-\epsilon)\sqrt{x} < \max_{\ba\in\cF_{\epsilon}}\ba^{\T} \bg \leqslant \sqrt{x} + \sqrt{\zeta}\bigg\} \,.
\end{align*}
Recall $|\cF_\epsilon| \leqslant \{(2+\epsilon)\epsilon^{-1}pd\}^K$ with $\epsilon=n^{-1}$. By Nazarov's inequality, we have
$  \sup_{x>\zeta}\bP\{(1-\epsilon)(\sqrt{x}-\sqrt{\zeta}) < \max_{\ba\in\cF_{\epsilon}}\ba^{\T} \bg \leqslant (1-\epsilon)\sqrt{x}\}
  \lesssim \sqrt{\zeta K\log(npd)}$ and $\sup_{x>\zeta}\bP(\sqrt{x} < \max_{\ba\in\cF_{\epsilon}}\ba^{\T} \bg \leqslant \sqrt{x} + \sqrt{\zeta})\lesssim \sqrt{\zeta K\log(npd)}$.
Due to
$  \bP\{(1-\epsilon)\sqrt{x} < \max_{\ba\in\cF_{\epsilon}}\ba^{\T} \bg \leqslant \sqrt{x} + \sqrt{\zeta}\}
   =\rho_{2,g}(x) + \bP(\sqrt{x} < \max_{\ba\in\cF_{\epsilon}}\ba^{\T} \bg \leqslant \sqrt{x} + \sqrt{\zeta})$,  together with \eqref{eq:ks.Tn1}, we have
\begin{align*}
  \sup_{x>0}\bP(x-\zeta <  G_K \leqslant x+\zeta) \lesssim \sup_{x>0}\rho_{2,g}(x)+\sqrt{\zeta K\log(npd)}=o(1)+\sqrt{\zeta K\log(npd)}\,.
\end{align*}
If
$\log(npd)\ll \min\{K^{-5\tau/(3\tau+2)}n^{\tau/(3\tau+2)},K^{-7\tau_1/(3\tau_1+2)}n^{\tau_1/(3\tau_1+2)}\}$, then  $\zeta K\log(npd)=o(1)$. By \eqref{ks.Tn}, to make
$
  \sup_{x>0}|\bP( T_n\leqslant x) - \bP(G_K\leqslant x)|=o(1)$,
we need to require $K\lesssim n^{1/9}(\log n)^{1/3}$ and
\begin{align*}
 &\log(npd) \ll \begin{cases}
   K^{-(6-\tau)/(6+2\tau)}n^{7\tau/(18+6\tau)}\,,\\
    K^{-(6+5\tau_1)/(6+2\tau_1)}n^{7\tau_1/(18+6\tau_1)}\,, \\
    K^{-1}n^{\tau_2/(9-3\tau_2)}\,, \\
    K^{-10/7}n^{2/21}\,, \\
    K^{-(1+3\tau_2)/(1+2\tau_2)}n^{\tau_2/(3+6\tau_2)}\,, \\
    K^{-5\tau/(3\tau+2)}n^{\tau/(3\tau+2)}\,, \\
    K^{-7\tau_1/(3\tau_1+2)}n^{\tau_1/(3\tau_1+2)}\,.\\
 \end{cases}
\end{align*}
Due to $\log(npd)\rightarrow\infty$ as $n\rightarrow\infty$, $K$ should satisfy the restriction
$
  K \ll n^{f_1(\tau_1,\tau_2)}$ with $f_1(\tau_1,\tau_2)$ specified in \eqref{eq:f1}. If $K=O(n^\delta)$ for some constant $0\leqslant\delta<f_1(\tau_1,\tau_2)$, there exists a constant $c>0$ depending on $(\tau_1,\tau_2,\delta)$ such that $
  \sup_{x>0}|\bP( T_n\leqslant x) - \bP(G_K\leqslant x)|=o(1)$ provided that $\log(pd)\ll n^c$. $\hfill\Box$

\subsection{Proof of Proposition 2}\label{sec:sigma}
Write $\bmu=(\mu_1,\ldots,\mu_{Kpd})^\T =\tilde{n}^{-1}\sum_{t=1}^{\tilde{n}}\mathbb{E}(\be_t)$.
Define
$\bSigma_{n,K}^* = \sum_{j=-\tilde n+1}^{\tilde n-1} \mathcal K(j/b_n) \bH_j$,
where $\bH_j=\tilde{n}^{-1} \sum_{t=j+1}^{\tilde n}\E\{(\be_t-\bmu)(\be_{t-j}-\bmu)^{\T}\}$ if $j\geqslant0$ and $\bH_j=\tilde{n}^{-1} \sum_{t=-j+1}^{\tilde n}\E\{(\be_{t+j}-\bmu)(\be_t-\bmu)^{\T}\}$ if $j<0$.
By the triangle inequality, we have
\begin{align*}
  |\widehat{\bSigma}_{n,K}-\bSigma_{n,K}|_\infty \leqslant
  |\widehat{\bSigma}_{n,K}-\bSigma_{n,K}^*|_\infty + |\bSigma_{n,K}^*-\bSigma_{n,K}|_\infty \,.
\end{align*}
Let $\tau_*=(\tau_1\tau_2)/(\tau_1+2\tau_2)$.
As we will show later in Sections \ref{subsec.sigma1} and \ref{subsec.sigma2},
$|\bSigma_{n,K}^*-\bSigma_{n,K}|_\infty  \lesssim n^{-\rho}K^{2}$, and \begin{align*}
|\widehat\bSigma_{n,K}-\bSigma_{n,K}^*|_\infty
= &~O_{\p}\bigg[
  \frac{\{\log(npd)\}^{(2+\tau_1\vartheta-\tau_1)/(2\tau_1\vartheta-\tau_1)}} {n^{(2\rho+\vartheta-1-3\rho\vartheta)/(2\vartheta-1)}}\bigg]\\
  &+O_{\p}\bigg[ \frac{\{\log(npd)\}^{2/\tau_1}}{n^{(\rho+\vartheta-2\rho\vartheta-1)/\vartheta}}\bigg]
  +O_{\p}\bigg[ \frac{\{\log(npd)\}^{1/\tau_*}}{n^{1-\rho}} \bigg]
\end{align*}
provided that $K\lesssim n^{(2\rho\vartheta+1-2\rho)/(2\vartheta-1)}\{\log(npd)\}^{(4-\tau_1)/(2\tau_1\vartheta-\tau_1)} \wedge
n^{(1-\rho+\rho\vartheta)/\vartheta}$. Therefore, 
$
  K^3\{\log(npd)\}^2|\widehat\bSigma_{n,K} - \bSigma_{n,K}|_\infty  =o_\p(1)$
provided that  $0<\rho < (\vartheta-1)/(3\vartheta-2)$ and
\begin{align*}
\log(npd) \ll\begin{cases}
   K^{-5/2}n^{\rho/2}\,, \\
   \{K^{-(6\vartheta-3)} n^{(2\rho+\vartheta-1-3\rho\vartheta)}\}^{\tau_1/(2+5\tau_1\vartheta-3\tau_1)}\,,  \\
   \{K^{-3\vartheta}n^{\rho+\vartheta-2\rho\vartheta-1}\}^{\tau_1/(2\vartheta+2\tau_1\vartheta)}\,,  \\
   K^{-3\tau_*/(1+2\tau_*)}n^{(\tau_*-\rho\tau_*)/(1+2\tau_*)}\,.
\end{cases}
\end{align*}
Due to $\log(npd)\rightarrow\infty$ as $n\rightarrow\infty$, $K$ should satisfy the restriction
$
  K \ll n^{f_2(\rho,\vartheta)}$ with $f_2(\rho,\vartheta)$ specified in \eqref{eq:f2}. If $K=O(n^{\delta})$ for some constant $0\leqslant\delta<f_2(\rho,\vartheta)$, there exists a constant $c>0$ depending on $(\tau_1,\tau_2,\rho,\vartheta,\delta)$ such that $
  K^3\{\log(npd)\}^2|\widehat\bSigma_{n,K} - \bSigma_{n,K}|_\infty=o_{\p}(1)$ provided that $\log(pd)\ll n^c$. $\hfill\Box$

\subsubsection{Convergence rate of $|\widehat{\bSigma}_{n,K}-\bSigma_{n,K}^*|_\infty$.}\label{subsec.sigma1} 
Without lose of generality, we can assume $\bmu=\bzero$.
Recall that
$\widehat{\bSigma}_{n,K} = \sum_{j=-\tilde n+1}^{\tilde n-1} \mathcal K(j/b_n) \widehat{\bH}_j$, where  $\widehat \bH_j=\tilde{n}^{-1}\sum_{t=j+1}^{\tilde{n}}(\be_t-\bar{\be})(\be_{t-j}-\bar{\be})^{\T}$ if $j\geqslant0$,  $\widehat \bH_j=\tilde{n}^{-1}\sum_{t=-j+1}^{\tilde{n}}({\be}_{t+j}-\bar{\be})({\be}_{t}-\bar{\be})^{\T}$ otherwise, $\tilde{n}=n-K$ and
 $\bar{{\be}}=\tilde{n}^{-1}\sum_{t=1}^{\tilde{n}} {\be}_t$. By the triangle inequality, it holds that
\begin{align*}
   \bigg|\sum_{j=0}^{\tilde n-1}\mathcal K\bigg(\frac{j}{b_n}\bigg)
(\widehat{\bH}_j-\bH_j)\bigg|_\infty
 \leqslant&~ \underbrace{\bigg|\sum_{j=0}^{\tilde n-1}\mathcal K\bigg(\frac{j}{b_n}\bigg) \bigg[\frac{1}{\tilde{n}}\sum_{t=j+1}^{\tilde{n}}\{\be_t\be_{t-j}^\T - \E(\be_t\be_{t-j}^\T)\}\bigg] \bigg|_\infty}_{{\rm I}} \\
 &+ \underbrace{\bigg|\sum_{j=0}^{\tilde n-1}\mathcal K\bigg(\frac{j}{b_n}\bigg)
 \bigg(\frac{1}{\tilde{n}}\sum_{t=j+1}^{\tilde{n}}\be_t \bigg)\bar{\be}^{\T} \bigg|_\infty}_{\rm II}+ \underbrace{\bigg|\sum_{j=0}^{\tilde n-1}\mathcal K\bigg(\frac{j}{b_n}\bigg)
 \bar{\be}\bigg(\frac{1}{\tilde{n}}\sum_{t=j+1}^{\tilde{n}}\be_{t-j}\bigg)^{\T}\bigg|_\infty}_{\rm III} \\
 &+ \underbrace{\bigg|\sum_{j=0}^{\tilde n-1}\bigg(\frac{\tilde{n}-j}{\tilde{n}}\bigg) \mathcal K\bigg(\frac{j}{b_n}\bigg)\bar{\be}^{\otimes2}\bigg|_\infty}_{\rm IV} \,.
\end{align*}
In the sequel, we will specify the convergence rate of ${\rm I}$, ${\rm II}$, ${\rm III}$ and ${\rm IV}$ respectively.
Recall $\be_t=(\eta_{t,1},\ldots,\eta_{t,Kpd})^\T$.

\textbf{Convergence rate of ${\rm I}$.}
 Given $\ell_1,\ell_2\in[Kpd]$, we define $\psi_{t,j}=\eta_{t+j,\ell_1}\eta_{t,\ell_2} -\E(\eta_{t+j,\ell_1}\eta_{t,\ell_2})$. For any $M=o(n)\rightarrow\infty$ satisfying $M\gtrsim K$ and $b_n=o(M)$, we have
\begin{align}\label{sigma.trun}
&~ \bP\bigg(\bigg|\sum_{j=0}^{\tilde{n}-1}\calK\bigg(\frac{j}{b_n}\bigg) \bigg[{\frac{1}{\tilde{n}}\sum_{t=j+1}^{\tilde{n}} \{\eta_{t,\ell_1}\eta_{t-j,\ell_2}}-\E(\eta_{t,\ell_1}\eta_{t-j,\ell_2})\}\bigg]\bigg|>x\bigg)\notag\\
\leqslant&~ \bP\bigg\{ \sum_{j=0}^{M} \bigg| \calK\bigg(\frac{j}{b_n}\bigg)\bigg|\, \bigg|\frac{1}{\tilde{n}}\sum_{t=1}^{\tilde{n}-j} \psi_{t,j} \bigg| >\frac{x}{2}\bigg\}
+ \bP\bigg\{ \sum_{j=M+1}^{\tilde{n}-1} \bigg| \calK\bigg(\frac{j}{b_n}\bigg)\bigg|\,\bigg|\frac{1}{\tilde{n}}\sum_{t=1}^{\tilde{n}-j} \psi_{t,j} \bigg| >\frac{x}{2}\bigg\}
\end{align}
for any $x>0$. Lemma 2 of \cite{CTW2013} yields
$
\max_{0\leqslant j\leqslant \tilde{n}-1}\max_{t\in[\tilde{n}-j]}\bP(|\psi_{t,j}|>x)\leqslant C\exp(-Cx^{\tau_1/2})$
for any $x>0$. By Condition 4 and $b_n\asymp n^\rho$ for some $\rho\in(0,1)$, we have
$\sum_{j=M+1}^{\tilde{n}-1}\mathcal{K}(j/b_n)\lesssim \sum_{j=M+1}^{\tilde{n}-1}(j/b_n)^{-\vartheta}
\lesssim n^{\rho\vartheta} M^{1-\vartheta}$.
Analogous to Lemma 4 of \cite{CQYZ2018}, we can show that
\begin{align}\label{sigma.tail}
\bP\bigg\{\sum_{j=M+1}^{\tilde{n}-1}
   \bigg|\calK\bigg(\frac{j}{b_n}\bigg)\bigg| \bigg|\frac{1}{\tilde{n}}\sum_{t=1}^{\tilde{n}-j}\psi_{t,j}\bigg|>\frac{x}{2} \bigg\}
\leqslant&~ \sum_{j=M+1}^{\tilde{n}-1}
    \bP\bigg(\bigg|\frac{1}{\tilde{n}}\sum_{t=1}^{\tilde{n}-j}\psi_{t,j}\bigg| >\frac{CM^{\vartheta-1}x}{n^{\rho\vartheta}}\bigg)\notag\\
\leqslant&~ \sum_{j=M+1}^{\tilde{n}-1}\sum_{t=1}^{\tilde{n}-j}
   \bP\bigg(|\psi_{t,j}|
   >\frac{CM^{\vartheta-1}x}{n^{\rho\vartheta}}\bigg)\\
\leqslant&~ Cn^2\exp\bigg\{ -\frac{CM^{\tau_1(\vartheta-1)/2}x^{\tau_1/2}}{n^{\rho\vartheta\tau_1/2}}\bigg\}  \notag
\end{align}
for any $x>0$. %We next specify the convergence rate of $\bP\{\sum_{j=0}^{M}|\mathcal{K}({j}/{b_n})| |\tilde{n}^{-1}\sum_{t=1}^{\tilde{n}-j}\psi_{t,j}|>x/2\}$.
Write $D_n=\sum_{j=0}^M|\mathcal{K}(j/b_n)|$ and $\tau_*=(\tau_1\tau_2)/(\tau_1+2\tau_2)$.
It is easy to see $D_n\lesssim b_n\asymp n^\rho$. For each given $j$, we observe that $\{\psi_{t,j}\}$ is also an $\alpha$-mixing sequence and its $\alpha$-mixing coefficients $\tilde{\alpha}_{\psi_{t,j}}(k)\leqslant \tilde{\alpha}_K(|k-j|_+)\leqslant C_3\exp(-C_4|k-j-K|_+^{\tau_2})$, where $\tilde{\alpha}_K(\cdot)$ is the $\alpha$-mixing coefficients of the process $\{\be_t\}$ defined in \eqref{eq:newalphamixing0}. By Bonferroni inequality and Lemma \ref{lem.bern} with $q=\tilde{n}-j$, $m=j+K$, $r_1=\tau_1/2$, $r_2=\tau_2$ and $r=\tau_*$, we have
\begin{align*}
&~\bP\bigg\{\sum_{j=0}^{M}
   \bigg|\calK\bigg(\frac{j}{b_n}\bigg)\bigg| \bigg|\frac{1}{\tilde{n}}\sum_{t=1}^{\tilde{n}-j}\psi_{t,j}\bigg|>\frac{x}{2} \bigg\}
   \leqslant \sum_{j=0}^{M} \bP\bigg(\bigg|\frac{1}{\tilde{n}}\sum_{t=1}^{\tilde{n}-j}\psi_{t,j}\bigg|>\frac{x}{2D_n}\bigg) \\
%\lesssim&~ \sum_{j=0}^M\exp\bigg(-\frac{Cn^{1-2\rho}x^2}{ j+K}\bigg)
%   + \frac{n^\rho}{x} \sum_{j=0}^M\bigg[ \exp\{-Cn^{(1-\rho)\tau^*}x^{\tau^*}\}
%   +\exp\bigg\{-\frac{Cn^{(1-\rho)\tau_1/2}x^{\tau_1/2}}{{(j+K)}^{\tau_1/2}}\bigg\} \bigg] \\
\lesssim&~ M\exp\bigg(-\frac{Cn^{1-2\rho}x^2}{M}\bigg)
   + \frac{M n^\rho}{x} \bigg[ \exp\{-Cn^{(1-\rho)\tau_*}x^{\tau_*}\}
   +\exp\bigg\{-\frac{Cn^{(1-\rho)\tau_1/2}x^{\tau_1/2}}{M^{\tau_1/2}}\bigg\} \bigg]
\end{align*}
for any $x>0$.
%Write
%\begin{align*}
%{\rm I}=\bigg|\sum_{j=0}^{\tilde n-1}\mathcal K\bigg(\frac{j}{b_n}\bigg)
%   \bigg[\frac{1}{\tilde n} \sum_{t=j+1}^{\tilde n} \{\be_t\be_{t-j}^{\T} - \E(\be_t\be_{t-j}^{\T})\} \bigg] \bigg|_\infty\,.
%   \end{align*}
Together with \eqref{sigma.trun} and \eqref{sigma.tail}, it holds that
\begin{align*}
\bP({\rm I}>x)\lesssim&~\sum_{\ell_1\in[Kpd]}\sum_{\ell_2\in[Kpd]} \bP\bigg(\bigg|\sum_{j=0}^{\tilde{n}-1}\calK\bigg(\frac{j}{b_n}\bigg) \bigg[{\frac{1}{\tilde{n}}\sum_{t=j+1}^{\tilde{n}} \{\eta_{t,\ell_1}\eta_{t-j,\ell_2}}-\E(\eta_{t,\ell_1}\eta_{t-j,\ell_2})\}\bigg]\bigg|>x\bigg) \\
\lesssim&~ (Kpd
n)^2\exp\bigg\{ -\frac{CM^{\tau_1(\vartheta-1)/2}x^{\tau_1/2}}{n^{\rho\vartheta\tau_1/2}}\bigg\}
+M (Kpd)^2\exp\bigg(-\frac{Cn^{1-2\rho}x^2}{M}\bigg) \\
  &~ + \frac{M n^\rho (Kpd)^2}{x} \bigg[ \exp\{-Cn^{(1-\rho)\tau_*}x^{\tau_*}\}
   +\exp\bigg\{-\frac{Cn^{(1-\rho)\tau_1/2}x^{\tau_1/2}}{M^{\tau_1/2}}\bigg\} \bigg]
\end{align*}
for any $x>0$, which implies that
\begin{align}
{\rm I}
   =&~O_\p\bigg[
  \frac{n^{\rho\vartheta}\{\log(npd)\}^{2/\tau_1}}{M^{\vartheta-1}}\bigg]+
  O_\p\bigg[\frac{M^{1/2}\{\log(npd)\}^{1/2}}{n^{(1-2\rho)/2}}\bigg]\notag\\
  &+O_\p\bigg[\frac{M\{\log(npd)\}^{2/\tau_1}}{n^{1-\rho}}\bigg]+
  O_\p\bigg[\frac{\{\log(npd)\}^{1/\tau_*}}{n^{1-\rho}}
  \bigg]\,.\label{eq:Iconvrate}
\end{align}
To make ${\rm I}$ converge as fast as possible, we need to specify the optimal $M$ in \eqref{eq:Iconvrate}.
If $ \log(npd) \leqslant n^{(1-\rho)(\vartheta-1)\tau_1/\{\vartheta(4-\tau_1)\}}$, with selecting
$  M\asymp n^{(2\rho\vartheta+1-2\rho)/(2\vartheta-1)}\{\log(npd)\}^{(4-\tau_1)/(2\tau_1\vartheta-\tau_1)}$,
we have
\begin{align*}
  {\rm I}=O_\p\bigg\{\bigg[
  \frac{\{\log(npd)\}^{(2+\tau_1\vartheta-\tau_1)/\tau_1}} {n^{2\rho+\vartheta-1-3\rho\vartheta}}
   \bigg]^{1/(2\vartheta-1)}\bigg\}+O_\p\bigg[\frac{\{\log(npd)\}^{1/\tau_*}}{n^{1-\rho}} \bigg]\,.
\end{align*}
If $ \log(npd) > n^{(1-\rho)(\vartheta-1)\tau_1/\{\vartheta(4-\tau_1)\}}$, with selecting
$  M \asymp n^{(1-\rho+\rho\vartheta)/\vartheta}$, we have
\begin{align*}
{\rm I}=O_\p\bigg[
  \frac{\{\log(npd)\}^{2/\tau_1}}{n^{(\rho+\vartheta-2\rho\vartheta-1)/\vartheta}}
  \bigg]+O_\p\bigg[\frac{\{\log(npd)\}^{1/\tau_*}}{n^{1-\rho}} \bigg] \,.
\end{align*}
Therefore, we can conclude that
\begin{align*}
 {\rm I}
=&~ O_\p\bigg\{\bigg[
  \frac{\{\log(npd)\}^{(2+\tau_1\vartheta-\tau_1)/\tau_1}} {n^{2\rho+\vartheta-1-3\rho\vartheta}}
   \bigg]^{1/(2\vartheta-1)}\bigg\}+O_\p\bigg[
  \frac{\{\log(npd)\}^{2/\tau_1}}{n^{(\rho+\vartheta-2\rho\vartheta-1)/\vartheta}}
  \bigg]+O_\p\bigg[\frac{\{\log(npd)\}^{1/\tau_*}}{n^{1-\rho}} \bigg]
\end{align*}
provided that $K\lesssim n^{(2\rho\vartheta+1-2\rho)/(2\vartheta-1)}\{\log(npd)\}^{(4-\tau_1)/(2\tau_1\vartheta-\tau_1)} \wedge
n^{(1-\rho+\rho\vartheta)/\vartheta}$.

\textbf{Convergence rates of ${\rm II}$ and ${\rm III}$.} Given $\ell_1,\ell_2\in[Kpd]$, write
\begin{align*}
{\rm II}(\ell_1,\ell_2)=\bigg|\sum_{j=0}^{\tilde n-1}\mathcal K\bigg(\frac{j}{b_n}\bigg)\bigg(\frac{1}{\tilde{n}}\sum_{t=j+1}^{\tilde{n}}\eta_{t,\ell_1}\bigg) \bar{\eta}_{\ell_2}\bigg|\,.
 \end{align*}
By Bonferroni inequality and the triangle inequality, it holds that
\begin{align*}
 \bP\{{\rm II}(\ell_1,\ell_2)>x\}\leqslant &~
 \underbrace{\bP\bigg[\sum_{j=0}^{\tilde n-1}\bigg|\mathcal K\bigg(\frac{j}{b_n}\bigg)\bigg|
 \bigg|\frac{1}{\tilde{n}}\sum_{t=j+1}^{\tilde{n}}\{\eta_{t,\ell_1}-\E(\eta_{t,\ell_1})\}\bigg||\bar{\eta}_{\ell_2}| > \frac{x}{2}\bigg]}_{{\rm II}_{1,\ell_1,\ell_2}(x)} \\
 & + \underbrace{\bP\bigg[\sum_{j=0}^{\tilde n-1}\bigg|\mathcal K\bigg(\frac{j}{b_n}\bigg)\bigg|
 \bigg|\frac{1}{\tilde{n}}\sum_{t=j+1}^{\tilde{n}}\E(\eta_{t,\ell_1})\bigg||\bar{\eta}_{\ell_2}| > \frac{x}{2}\bigg]}_{{\rm II}_{2,\ell_1,\ell_2}(x)}
\end{align*}
for any $x>0$. Note that $\sum_{j=0}^{\tilde{n}-1}|\mathcal{K}(j/b_n)|\lesssim b_n\asymp n^\rho $.
 By Bonferroni inequality, the triangle inequality and Lemma \ref{lem.bern}, we have
\begin{align*}
 {\rm II}_{1,\ell_1,\ell_2}(x)
\leqslant&~ \sum_{j=0}^{\tilde{n}-1}\bP\bigg[ \bigg|\frac{1}{\tilde{n}}\sum_{t=j+1}^{\tilde{n}}\{\eta_{t,\ell_1}-\E(\eta_{t,\ell_1})\}\bigg||\bar{\eta}_{\ell_2}| > \frac{Cx}{n^\rho}\bigg] \\
\leqslant&~  \sum_{j=0}^{\tilde{n}-1}\bP\bigg[ \bigg|\frac{1}{\tilde{n}}\sum_{t=j+1}^{\tilde{n}}\{\eta_{t,\ell_1}-\E(\eta_{t,\ell_1})\}\bigg| > \frac{Cx^{1/2}}{n^{\rho/2}}\bigg]
+ n\bP\bigg(|\bar{\eta}_{\ell_2}| > \frac{Cx^{1/2}}{n^{\rho/2}}\bigg) \\
\lesssim&~ n\exp\bigg(-\frac{Cn^{1-\rho}x}{K}\bigg) + \frac{n^{\rho/2+1}}{x^{1/2}}
\bigg[\exp\{-Cn^{\tau(2-\rho)/2}x^{\tau/2}\} +\exp\bigg\{-\frac{Cn^{\tau_1(2-\rho)/2}x^{\tau_1/2}}{K^{\tau_1}}{}\bigg\} \bigg]
\end{align*}
for any $x>0$, where $\tau=\tau_1\tau_2/(\tau_1+\tau_2)$. Condition 1 yields that $\sup_{t\in[\tilde{n}]}\sup_{\ell\in[Kpd]}\E(|\eta_{t,\ell}|)\leqslant C$. Analogously, it holds that
\begin{align*}
  &{\rm II}_{2,\ell_1,\ell_2}(x)\lesssim n \exp\bigg(-\frac{Cn^{1-2\rho}x^2}{K}\bigg)
+\frac{n^{1+\rho}}{x}\bigg[\exp\{-Cn^{(1-\rho)\tau}x^\tau\} +\exp\bigg\{-\frac{Cn^{(1-\rho)\tau_1}x^{\tau_1}}{K^{\tau_1}}\bigg\}\bigg]
\end{align*}
for any $x>0$. Therefore,  by Bonferroni inequality, we have
\begin{align*}
\bP({\rm II}>x) \leqslant&~\sum_{\ell_1,\ell_2\in[Kpd]}\bP\{{\rm II}(\ell_1,\ell_2)>x\}\\
 \lesssim&~ n(Kpd)^2\bigg\{\exp\bigg(-\frac{Cn^{1-\rho}x}{K}\bigg)
 +\exp\bigg(-\frac{Cn^{1-2\rho}x^2}{K}\bigg) \bigg\} \\
 &+ \frac{n^{1+\rho}(Kpd)^2}{x}\bigg[\exp\{-Cn^{(1-\rho)\tau}x^\tau\} +\exp\bigg\{-\frac{Cn^{(1-\rho)\tau_1}x^{\tau_1}}{K^{\tau_1}}\bigg\}\bigg] \\
 & + \frac{n^{\rho/2+1}(Kpd)^2}{x^{1/2}}
\bigg[\exp\{-Cn^{\tau(2-\rho)/2}x^{\tau/2}\} +\exp\bigg\{-\frac{Cn^{\tau_1(2-\rho)/2}x^{\tau_1/2}}{K^{\tau_1}}{}\bigg\} \bigg]
\end{align*}
for any $x>0$, which implies that
\begin{align*}
  {\rm II}
  =&~O_{\rm p}\bigg\{\frac{K\log(npd)}{n^{1-\rho}}\bigg\}
  +O_{\rm p}\bigg[\frac{K^{1/2}\{\log(npd)\}^{1/2}}{n^{(1-2\rho)/2}}\bigg]
  +O_{\rm p}\bigg[\frac{\{\log(npd)\}^{1/\tau}}{n^{1-\rho}}\bigg]
  +O_{\rm p}\bigg[\frac{\{\log(npd)\}^{2/\tau}}{n^{2-\rho}}\bigg] \\
  & +O_{\rm p}\bigg[\frac{K\{\log(npd)\}^{1/\tau_1}}{n^{1-\rho}}\bigg]
  +O_{\rm p}\bigg[\frac{K^2\{\log(npd)\}^{2/\tau_1}}{n^{2-\rho}}\bigg] \,.
\end{align*}
Note that $M\gtrsim K$, $\tau_1\in(0,1]$ and $\tau_*<\tau$ in \eqref{eq:Iconvrate} and $K=o(n)$. Then
\begin{align*}
 {\rm II}
=&~ O_\p\bigg\{\bigg[
  \frac{\{\log(npd)\}^{(2+\tau_1\vartheta-\tau_1)/\tau_1}} {n^{2\rho+\vartheta-1-3\rho\vartheta}}
   \bigg]^{1/(2\vartheta-1)}\bigg\}+O_\p\bigg[
  \frac{\{\log(npd)\}^{2/\tau_1}}{n^{(\rho+\vartheta-2\rho\vartheta-1)/\vartheta}}
  \bigg]+O_\p\bigg[\frac{\{\log(npd)\}^{1/\tau_*}}{n^{1-\rho}} \bigg]
\end{align*}
provided that $K\lesssim n^{(2\rho\vartheta+1-2\rho)/(2\vartheta-1)}\{\log(npd)\}^{(4-\tau_1)/(2\tau_1\vartheta-\tau_1)} \wedge
n^{(1-\rho+\rho\vartheta)/\vartheta}$. Similarly, we also have
\begin{align*}
 {\rm III}
=&~ O_\p\bigg\{\bigg[
  \frac{\{\log(npd)\}^{(2+\tau_1\vartheta-\tau_1)/\tau_1}} {n^{2\rho+\vartheta-1-3\rho\vartheta}}
   \bigg]^{1/(2\vartheta-1)}\bigg\}+O_\p\bigg[
  \frac{\{\log(npd)\}^{2/\tau_1}}{n^{(\rho+\vartheta-2\rho\vartheta-1)/\vartheta}}
  \bigg]+O_\p\bigg[\frac{\{\log(npd)\}^{1/\tau_*}}{n^{1-\rho}} \bigg]
\end{align*}
provided that $K\lesssim n^{(2\rho\vartheta+1-2\rho)/(2\vartheta-1)}\{\log(npd)\}^{(4-\tau_1)/(2\tau_1\vartheta-\tau_1)} \wedge
n^{(1-\rho+\rho\vartheta)/\vartheta}$.

\textbf{Convergence rate of ${\rm IV}$.} Given $\ell_1,\ell_2\in[Kpd]$, write
\begin{align*}
{\rm IV}(\ell_1,\ell_2)=\bigg|\sum_{j=0}^{\tilde n-1}\bigg(\frac{\tilde{n}-j}{\tilde{n}}\bigg)\mathcal K\bigg(\frac{j}{b_n}\bigg)
 \bar\eta_{\ell_1}\bar\eta_{\ell_2} \bigg|\,.
 \end{align*}
By Bonferroni inequality and the triangle inequality, it holds that
\begin{align*}
\bP\{{\rm IV}(\ell_1,\ell_2)>x \}
\leqslant &~\bP\bigg\{\sum_{j=0}^{\tilde n-1}\bigg|\mathcal K\bigg(\frac{j}{b_n}\bigg)\bigg|
 |\bar\eta_{\ell_1}| |\bar\eta_{\ell_2}| > x\bigg\}
 \end{align*}
for any $x>0$. Identical to the arguments for deriving the upper bound of ${\rm II}_{1,\ell_1,\ell_2}(x)$, we know the same upper bound also holds for $\bP\{{\rm IV}(\ell_1,\ell_2)>x\}$.  Hence, we have
\begin{align*}
 {\rm IV}
=&~ O_\p\bigg\{\bigg[
  \frac{\{\log(npd)\}^{(2+\tau_1\vartheta-\tau_1)/\tau_1}} {n^{2\rho+\vartheta-1-3\rho\vartheta}}
   \bigg]^{1/(2\vartheta-1)}\bigg\}+O_\p\bigg[
  \frac{\{\log(npd)\}^{2/\tau_1}}{n^{(\rho+\vartheta-2\rho\vartheta-1)/\vartheta}}
  \bigg]+O_\p\bigg[\frac{\{\log(npd)\}^{1/\tau_*}}{n^{1-\rho}} \bigg]
\end{align*}
provided that $K\lesssim n^{(2\rho\vartheta+1-2\rho)/(2\vartheta-1)}\{\log(npd)\}^{(4-\tau_1)/(2\tau_1\vartheta-\tau_1)} \wedge
n^{(1-\rho+\rho\vartheta)/\vartheta}$.

Therefore, we can conclude that
\begin{align*}
  \bigg|\sum_{j=0}^{\tilde n-1}\mathcal K\bigg(\frac{j}{b_n}\bigg)
(\widehat{\bH}_j-\bH_j)\bigg|_\infty\leqslant&~{\rm I}+{\rm II}+{\rm III}+{\rm IV}\\
=&~ O_\p\bigg\{\bigg[
  \frac{\{\log(npd)\}^{(2+\tau_1\vartheta-\tau_1)/\tau_1}} {n^{2\rho+\vartheta-1-3\rho\vartheta}}
   \bigg]^{1/(2\vartheta-1)}\bigg\}\\
   &+O_\p\bigg[
  \frac{\{\log(npd)\}^{2/\tau_1}}{n^{(\rho+\vartheta-2\rho\vartheta-1)/\vartheta}}
  \bigg]+O_\p\bigg[\frac{\{\log(npd)\}^{1/\tau_*}}{n^{1-\rho}} \bigg]\,.
\end{align*}
 Identically, we can also show
\begin{align*}
  \bigg|\sum_{j=-\tilde{n}+1}^{-1}\mathcal K\bigg(\frac{j}{b_n}\bigg)
(\widehat{\bH}_j-\bH_j)\bigg|_\infty
=&~ O_\p\bigg\{\bigg[
  \frac{\{\log(npd)\}^{(2+\tau_1\vartheta-\tau_1)/\tau_1}} {n^{2\rho+\vartheta-1-3\rho\vartheta}}
   \bigg]^{1/(2\vartheta-1)}\bigg\}\\
   &+O_\p\bigg[
  \frac{\{\log(npd)\}^{2/\tau_1}}{n^{(\rho+\vartheta-2\rho\vartheta-1)/\vartheta}}
  \bigg]+O_\p\bigg[\frac{\{\log(npd)\}^{1/\tau_*}}{n^{1-\rho}} \bigg]\,.
\end{align*}
Hence, we have
\begin{align*}
|\widehat\bSigma_{n,K}-\bSigma_{n,K}^*|_\infty
= &~O_{\p}\bigg[
  \frac{\{\log(npd)\}^{(2+\tau_1\vartheta-\tau_1)/(2\tau_1\vartheta-\tau_1)}} {n^{(2\rho+\vartheta-1-3\rho\vartheta)/(2\vartheta-1)}}\bigg]\\
  &+O_{\p}\bigg[ \frac{\{\log(npd)\}^{2/\tau_1}}{n^{(\rho+\vartheta-2\rho\vartheta-1)/\vartheta}}\bigg]
  +O_{\p}\bigg[ \frac{\{\log(npd)\}^{1/\tau_*}}{n^{1-\rho}} \bigg]
\end{align*}
provided that $K\lesssim n^{(2\rho\vartheta+1-2\rho)/(2\vartheta-1)}\{\log(npd)\}^{(4-\tau_1)/(2\tau_1\vartheta-\tau_1)} \wedge
n^{(1-\rho+\rho\vartheta)/\vartheta}$.  $\hfill\Box$

\subsubsection{Convergence rate of $|\bSigma_{n,K}^*-\bSigma_{n,K}|_\infty$.}\label{subsec.sigma2}
Note that $\bSigma_{n,K} = \tilde{n}\E\{(\bar\be-\bmu)(\bar\be-\bmu)^\T)$,
$\bH_j=\tilde{n}^{-1} \sum_{t=j+1}^{\tilde n}\E\{(\be_t-\bmu)(\be_{t-j}-\bmu)^{\T}\}$ if $j\geqslant0$ and $\bH_j=\tilde{n}^{-1} \sum_{t=-j+1}^{\tilde n}\E\{(\be_{t+j}-\bmu)(\be_t-\bmu)^{\T}\}$ if $j<0$,
where $\bar\be=\tilde{n}^{-1}\sum_{t=1}^{\tilde{n}}\be_t$, $\bmu=\tilde{n}^{-1}\sum_{t=1}^{\tilde{n}}\mathbb{E}(\be_t)$ and $\be_t=(\eta_{t,1},\ldots,\eta_{t,Kpd})^\T$.
%Under $H_0$, we have $\bmu=\bzero$, which implies that $\bH_j=\tilde{n}^{-1} \sum_{t=j+1}^{\tilde n}\E(\be_t\be_{t-j}^{\T})$ if $j\geqslant0$ and $\bH_j=\tilde{n}^{-1} \sum_{t=-j+1}^{\tilde n}\E(\be_{t+j}\be_t^{\T})$ if $j<0$.
We write $\bSigma_{n,K}=\{\sigma_{n,K}(\ell_1,\ell_2)\}_{(Kpd)\times(Kpd)}$, $\bH_j=\{H_j(\ell_1,\ell_2)\}_{(Kpd)\times(Kpd)}$ and $\mathring\eta_{t,\ell}=\eta_{t,\ell}-\E(\eta_{t,\ell})$. For any $\ell_1,\ell_2\in[Kpd]$,  it holds that
\begin{align*}
 \sigma_{n,K}(\ell_1,\ell_2)
   =&~\tilde n\E\bigg\{\bigg(\frac{1}{\tilde n}\sum_{t=1}^{\tilde n}\mathring\eta_{t,\ell_1})\bigg) \bigg(\frac{1}{\tilde n}\sum_{t=1}^{\tilde n}\mathring\eta_{t,\ell_2}\bigg)\bigg\}  \\
   =&~\frac{1}{\tilde n} \sum_{t=1}^{\tilde n}\E(\mathring\eta_{t,\ell_1}\mathring\eta_{t,\ell_2})
   + \frac{1}{\tilde n} \sum_{t_1=1}^{\tilde n-1}\sum_{j=1}^{\tilde n-t_1} \E(\mathring\eta_{t_1,\ell_1}\mathring\eta_{t_1+j,\ell_2})
   + \frac{1}{\tilde n} \sum_{t_2=1}^{\tilde n-1}\sum_{j=1}^{\tilde n-t_2} \E(\mathring\eta_{t_2+j,\ell_1}\mathring\eta_{t_2,\ell_2})
   \\
   =&~H_{0}(\ell_1,\ell_2) +  \sum_{j=1}^{\tilde n-1} H_{-j}(\ell_1,\ell_2) +  \sum_{j=1}^{\tilde n-1} H_{j}(\ell_1,\ell_2)  \,.
\end{align*}
By Davydov's inequality,
$|H_{j}(\ell_1,\ell_2)| \leqslant \tilde n^{-1}\sum_{t=j+1}^{\tilde n}|\E(\mathring\eta_{t,\ell_1}\mathring\eta_{t-j,\ell_2})| \lesssim \tilde n^{-1}(\tilde n-j)\exp(-C|j-K|_+^{\tau_2})$ for any $j\geqslant 1$.
This bound also holds for $|H_{-j}(\ell_1,\ell_2)|$ with $j\geqslant 1$.
Observe that $  \bSigma_{n,K}^* := \{\sigma_{n,K}^*(\ell_1,\ell_2)\}_{(Kpd)\times(Kpd)}=\sum_{j=-\tilde n+1}^{\tilde n-1} \mathcal K(j/b_n) \bH_j$ and $\mathcal K(\cdot)$ is symmetric with $\mathcal K(0)=1$.
By the triangle inequality and Condition 4, %we have
\begin{align*}
 |\sigma_{n,K}^*(\ell_1,\ell_2) - \sigma_{n,K}(\ell_1,\ell_2)|
   \leqslant&~ \sum_{j=1}^{\tilde n-1}\bigg|\mathcal K\bigg(\frac{j}{b_n}\bigg)-1\bigg|\big\{|H_{j}(\ell_1,\ell_2)|
   +|H_{-j}(\ell_1,\ell_2)|\big\} \\
   \lesssim&~ \sum_{j=1}^{\tilde n-1}\frac{j(\tilde n-j)}{b_n\tilde n} \exp(-C|j-K|_+^{\tau_2}) \\
   \lesssim&~ \frac{1}{b_n} \bigg[ \sum_{j=1}^{K} j
   + \sum_{j=K+1}^{\tilde n-1}j\exp\{-C(j-K)^{\tau_2}\} \bigg]  \\
   \lesssim&~ b_n^{-1}K^{2} \,.
\end{align*}
Thus $|\bSigma_{n,K}^*-\bSigma_{n,K}|_\infty  \lesssim b_n^{-1}K^{2}$.    $\hfill\Box$

\subsection{Proof of Theorem 1}
%Let $\tilde{n}=n-K$ and $\bar{{\be}}=(\bar\eta_1,\ldots,\bar\eta_{Kpd})^\T =\tilde{n}^{-1}\sum_{t=1}^{\tilde{n}} {\be}_t$.
%Recall $\bSigma_{n,K} = \tilde{n}\E(\bar\be\bar\be^\T)$
%and $\widehat{\bSigma}_{n,K} = \sum_{j=-\tilde n+1}^{\tilde n-1} \mathcal K(j/b_n) \widehat{\bH}_j$, where $\widehat \bH_j=\tilde{n}^{-1}\sum_{t=j+1}^{\tilde{n}}(\be_t-\bar{\be})(\be_{t-j}-\bar{\be})^{\T}$ if $j\geqslant0$ and  $\widehat \bH_j=\tilde{n}^{-1}\sum_{t=-j+1}^{\tilde{n}}({\be}_{t+j}-\bar{\be})({\be}_{t}-\bar{\be})^{\T}$ otherwise.

Recall $G_K = \sum_{j=1}^{K} \max_{\ell\in\mathcal{L}_j} |g_\ell|^2$ and
$\hat{G}_K = \sum_{j=1}^{K} \max_{\ell\in\mathcal{L}_j} |\hat{g}_\ell|^2$ with $ \bg=({g}_1,\ldots,{g}_{Kpd})^\T \sim \calN(\bzero,\bSigma_{n,K})$ and $ \hat{\bg}=(\hat{g}_1,\ldots,\hat{g}_{Kpd})^\T \sim \calN(\bzero,\widehat\bSigma_{n,K})$. As shown in Proposition 1, $\sup_{x>0}|\bP(T_n\leqslant x)-\bP(G_K\leqslant x)|=o(1)$. Write $\mathcal{X}_n=\{\bx_1,\ldots,\bx_n\}$. To construct Theorem 1, it suffices to show $\sup_{x>0} |\bP(G_K\leqslant x) - \bP(\hat G_K\leqslant x\,|\,\mathcal{X}_n) | =o(1)$.
Recall $\rho_{2,g}(x)=|\bP\{\bg\in A_2(x)\} - \bP\{\bg\in A_1(x)\}|$ for $A_1(x)$ and $A_2(x)$ defined in Section \ref{sec:b.1}. Here we also define
$$
  \rho_{3,g}(x)  := |\bP\{\bg\in A_1(x)\} - \bP\{\hat\bg\in A_1(x)\,|\,\mathcal{X}_n\}|\vee|\bP\{\bg\in A_2(x)\} - \bP\{\hat\bg\in A_2(x)\,|\,\mathcal{X}_n\}|\,.$$ 
  Identical to the result $\{G_K\leqslant x\}=\{\bg\in\mathcal{A}(x;K)\}$ stated in Section \ref{sec:b.1}, we also have $\{\hat{G}_K\leqslant x\}=\{\hat{\bg}\in\mathcal{A}(x;K)\}$ for any $x>0$, where $\mathcal{A}(x;K)$ is defined in Section \ref{sec:b.1}.
Then it holds that
\begin{align*}
  \bP(\hat G_K\leqslant x \,|\,\mathcal{X}_n)
   =&~ \bP\{\hat{\bg}\in \mathcal A(x;K)\,|\,\mathcal{X}_n\}\leqslant \bP\{\hat{\bg}\in A_2(x)\,|\,\mathcal{X}_n\}\\
   \leqslant&~ \bP\{\bg\in A_2(x)\} + \rho_{3,g}(x) \\
   \leqslant&~ \bP\{\bg\in A_1(x)\} + \rho_{2,g}(x) + \rho_{3,g}(x) \\
   \leqslant&~ \bP\{\bg\in\mathcal A(x;K)\} +  \rho_{2,g}(x) + \rho_{3,g}(x) \\
   \leqslant&~ \bP(G_K\leqslant x)+ \rho_{2,g}(x) + \rho_{3,g}(x)
\end{align*}
for any $x>0$. Similarly, we can obtain the reverse inequality. Notice that we have shown in Section \ref{sec:b.1} that $\sup_{x>0}\rho_{2,g}(x)=o(1)$. Therefore,
\begin{align*}
  \sup_{x>0}|\bP(G_K\leqslant x ) - \bP(\hat G_K\leqslant x\,|\,\mathcal{X}_n)| \leqslant \sup_{x>0}\rho_{2,g}(x) + \sup_{x>0}\rho_{3,g}(x)=o(1)+\sup_{x>0}\rho_{3,g}(x)\,.
\end{align*}
By Lemma 13 of \cite{CCW2020}, it holds that
\begin{align*}
&~\sup_{x>0} |\bP\{\bg\in A_1(x)\} - \bP\{\hat\bg\in A_1(x)\,|\,\mathcal{X}_n\}| \\
   =&~ \sup_{x>0} \bigg| \bP\bigg\{\max_{\ba\in\cF_\epsilon} \ba^{\T} \bg \leqslant (1-\epsilon)\sqrt{x} \bigg\}
     - \bP\bigg\{\max_{\ba\in\cF_\epsilon} \ba^{\T} \hat\bg \leqslant (1-\epsilon)\sqrt{x} \,|\,\mathcal{X}_n\bigg\} \bigg|   \\
   \lesssim&~ \Delta_n^{1/3} \big\{ K\log(npd)\big\}^{2/3}
\end{align*}
with $\Delta_n=\max_{\ba_1,\ba_2\in\cF} |  \ba_1^{\T} (\bSigma_{n,K}-\widehat \bSigma_{n,K}) \ba_2|$, where $\mathcal{F}$ is defined in Section \ref{sec:b.1}. Recall $|\ba|_0\leqslant K$ and $|\ba|_2=1$ for any $\ba\in\cF$. Thus,
$
|\ba_1^{\T}(\bSigma_{n,K}-\widehat\bSigma_{n,K})\ba_2|
\leqslant |\ba_1|_1 |\ba_2|_1 |\bSigma_{n,K}-\widehat\bSigma_{n,K}|_\infty
\leqslant K|\bSigma_{n,K}-\widehat\bSigma_{n,K}|_\infty$.  Then we have $\sup_{x>0} |\bP\{\bg\in A_1(x)\} - \bP\{\hat\bg\in A_1(x)\,|\,\mathcal{X}_n\}|\lesssim K|\bSigma_{n,K}-\widehat\bSigma_{n,K}|_\infty^{1/3}\{\log(npd)\}^{2/3}$. Analogously, we also have $\sup_{x>0} |\bP\{\bg\in A_2(x)\} - \bP\{\hat\bg\in A_2(x)\,|\,\mathcal{X}_n\}|\lesssim K|\bSigma_{n,K}-\widehat\bSigma_{n,K}|_\infty^{1/3}\{\log(npd)\}^{2/3}$. Hence,
\begin{align}
\sup_{x>0}|\bP(G_K\leqslant x ) - \bP(\hat G_K\leqslant x\,|\,\mathcal{X}_n)|\lesssim K|\bSigma_{n,K}-\widehat\bSigma_{n,K}|_\infty^{1/3}\{\log(npd)\}^{2/3}+o(1)\,.
\end{align}
By Proposition 2, we complete the proof. $\hfill\Box$

\subsection{Proof of Theorem 2}
Recall that $\mathcal{X}_n=\{\bx_1,\ldots,\bx_n\}$ and  $\hat{G}_K = \sum_{j=1}^{K} \max_{\ell\in\mathcal{L}_j} |\hat{g}_\ell|^2$ with $\hat{\bg}=(\hat{g}_1,\ldots,\hat{g}_{Kpd})^\T$. By Bonferroni inequality, we have
\begin{align*}
 \bP(\hat{G}_K>x\,|\,\mathcal{X}_n)
    \leqslant\sum_{j=1}^{K} \bP\bigg(\max_{\ell\in\mathcal{L}_j}|\hat{g}_\ell|^2 > \frac{x}{K} \,\bigg|\,\mathcal{X}_n \bigg)
   = \sum_{j=1}^{K} \bP\bigg(\max_{\ell\in\mathcal{L}_j}|\hat{g}_\ell| > \frac{x^{1/2}}{K^{1/2}} \,\bigg|\, \mathcal{X}_n \bigg)
\end{align*}
for any $x>0$. Since $\hat\bg\sim \calN(\bzero,\widehat{\bSigma}_{n,K})$ with $\widehat{\bSigma}_{n,K}=\{\hat{\sigma}_{n,K}(\ell_1,\ell_2)\}_{Kpd\times Kpd}$, then 
$$
  \E\bigg(\max_{\ell\in\mathcal{L}_j}|\hat{g}_\ell|\,\bigg|\,\mathcal{X}_n\bigg) \leqslant \big[1+\{2\log(pd)\}^{-1}\big]\{2\log(pd)\}^{1/2}\max_{\ell\in\mathcal{L}_j}\{\hat{\sigma}_{n,K}(\ell,\ell)\}^{1/2}$$ for any $j\in[K]$.
%where $\Sigma_{n,K,1},\ldots,\Sigma_{n,K,Kpd}$ are the elements in the diagonal of $\bSigma_{n,K}$.
Recall  $\bSigma_{n,K}=\{{\sigma}_{n,K}(\ell_1,\ell_2)\}_{Kpd\times Kpd}$ and $\varrho=\max_{\ell\in[Kpd]}\sigma_{n,K}(\ell,\ell)$. Define an event $$\cE_0(\nu)=\bigg\{\max_{\ell\in[Kpd]}\bigg|\frac{\hat{\sigma}_{n,K}(\ell,\ell)}{\sigma_{n,K}(\ell,\ell)} -1\bigg| \leqslant \nu \bigg\}\,,$$
where $\nu>0$ and $\nu\asymp \{K\log(pd)\}^{-1}$. As shown in Proposition 2, 
$
  \max_{\ell\in[Kpd]}|\hat\sigma_{n,K}(\ell,\ell)-\sigma_{n,K}(\ell,\ell)|
  =o_{\rm p}[K^{-3}\{\log(npd)\}^{-2}]=o_{\rm p}(\nu)$.
From Condition 3, we have $\min_{\ell\in[Kpd]}\sigma_{n,K}(\ell,\ell)\geqslant C$, where $C$ is a positive constant.
It holds that
\begin{align*}
  \max_{\ell\in[Kpd]}\bigg|\frac{\hat\sigma_{n,K}(\ell,\ell)}{\sigma_{n,K}(\ell,\ell)}-1\bigg|
  \leqslant \frac{\max_{\ell\in[Kpd]}|\hat\sigma_{n,K}(\ell,\ell)-\sigma_{n,K}(\ell,\ell)|} {\min_{\ell\in[Kpd]}\sigma_{n,K}(\ell,\ell)} =o_{\rm p}(\nu) \,.
\end{align*}
Thus $\bP\{\cE_0(\nu)^c\} \to 0$ as $n\rightarrow\infty$. Restricted on $\cE_0(\nu)$, it holds that
\begin{align*}
\max_{j\in[K]}\E\bigg(\max_{\ell\in\mathcal{L}_j}|\hat{g}_\ell| \,\Big|\, \mathcal{X}_n\bigg) \leqslant (1+\nu)^{1/2}\varrho^{1/2}\big[1+\{2\log(pd)\}^{-1}\big]\{2\log(pd)\}^{1/2} \,.
\end{align*}
By Borell inequality for Gaussian process,
%{\color{blue}(As shown in \cite{Borell1975},)}
it holds that
\begin{align*}
  \bP\bigg\{\max_{\ell\in\mathcal{L}_j}|\hat{g}_\ell| \geqslant \E\bigg(\max_{\ell\in\mathcal{L}_j}|\hat{g}_\ell| \,\Big|\,\mathcal{X}_n\bigg) + x \,\bigg|\, \mathcal{X}_n\bigg\}
\leqslant 2\exp\bigg\{-\frac{x^2}{2\max_{\ell\in\mathcal{L}_j}\hat{\sigma}_{n,K}(\ell,\ell)}\bigg\}
\end{align*}
for any $x>0$.
Let $x_*= K(1+\nu)\varrho([1+\{2\log(pd)\}^{-1}]\{2\log(pd)\}^{1/2} + \{2\log(4K/\alpha)\}^{1/2} )^2$. Restricted on $\cE_0(\nu)$, we have
\begin{align*}
	\frac{x_*^{1/2}}{K^{1/2}} \geqslant \max_{j\in[K]}\E\bigg(\max_{\ell\in\mathcal{L}_j}|\hat{g}_\ell| \,\bigg|\,\mathcal{X}_n\bigg) + (1+\nu)^{1/2}\varrho^{1/2}\bigg\{2\log\bigg(\frac{4K}{\alpha}\bigg)\bigg\}^{1/2} \,,
\end{align*}
%$
%  x^{1/2}_*K^{-1/2}
%  >\max_{j\in[K]}\E(\max_{\ell\in\mathcal{L}_j}|\hat{g}_\ell| \,|\,\mathcal{X}_n)$,
 which yields that
\begin{align*}
   \bP\{\hat{G}_K>x_*,
   \,\mathcal{E}_0(\nu)
     \,|\,\mathcal{X}_n\}
   %\leqslant&~ \sum_{j=1}^{K} \bP\bigg(\max_{\ell\in[pd]_j}|\hat{g}_\ell| > \frac{x^{1/2}_*}{K^{1/2}} \,\bigg|\,\mathcal{X}_n \bigg) \\
   \leqslant&~  \sum_{j=1}^{K} \bP\bigg\{\max_{\ell\in\mathcal{L}_j}|\hat{g}_\ell| - \E\bigg(\max_{\ell\in\mathcal{L}_j}|\hat{g}_\ell| \,\Big|\,\mathcal{X}_n\bigg)
   > \frac{x^{1/2}_*}{K^{1/2}} - \E\bigg(\max_{\ell\in\mathcal{L}_j}|\hat{g}_\ell| \,\Big|\,\mathcal{X}_n \bigg),
      \,\mathcal{E}_0(\nu)
    \,\bigg|\,\mathcal{X}_n \bigg\} \\
   \leqslant&~ \sum_{j=1}^{K} \bP\bigg[\max_{\ell\in\mathcal{L}_j}|\hat{g}_\ell| - \E\bigg(\max_{\ell\in\mathcal{L}_j}|\hat{g}_\ell| \,\Big|\,\mathcal{X}_n \bigg)
   > (1+\nu)^{1/2}\varrho^{1/2}\bigg\{2\log\bigg(\frac{
   	4K}{\alpha}\bigg)\bigg\}^{1/2},
    \,\mathcal{E}_0(\nu)
    \,\bigg|\,\mathcal{X}_n   \bigg] \\
   \leqslant&~  2K \exp\bigg\{-\frac{2(1+\nu)\varrho\log(4K/\alpha) }{2(1+\nu)\varrho} \bigg\}
   = \frac{\alpha}{2}\,.
\end{align*}
Since $\bP\{\mathcal{E}_0(\nu)^{\rm c}\,|\,\cX_n\}=o_{\rm p}(1)$, then  $\bP\{\mathcal{E}_0(\nu)^{\rm c}\,|\,\cX_n\}\leqslant \alpha/4$ with probability approaching one. Hence, $\bP(\hat{G}_K>x_*\,|\,\cX_n)\leqslant 5\alpha/6$ with probability approaching one.
Following the definition of $\hat{\rm cv}_\alpha$, it holds with probability approaching one that
\begin{align}\label{cv.bound}
  \hat{\rm cv}_\alpha \leqslant (1+\nu)K\varrho\lambda^2(K,p,d,\alpha)\big[1+\{2\log(pd)\}^{-1}\big]^2
\end{align}
with $\lambda(K,p,d,\alpha)=\{2\log(pd)\}^{1/2} +\{2\log(4K/\alpha)\}^{1/2}$.

We next specify the lower bound of $T_n$. Recall that $T_n=n\sum_{j=1}^K|\hat{\bgamma}_j|_\infty^2= \sum_{j=1}^{K} \max_{\ell\in\mathcal{L}_j} (n^{1/2}u_\ell)^2$, where $\bu=(u_1,\ldots,u_{Kpd})^\T=(\hat\bgamma_1^\T,\ldots,\hat\bgamma_K^\T)^\T$ with $\hat{\bgamma}_j=(n-j)^{-1}\sum_{t=1}^{n-j}{\rm vec}\{\bphi(\bx_t)\bx_{t+j}^\T\}$.
Let $\tilde\bu=(\tilde{u}_{1},\ldots,\tilde{u}_{Kpd})^\T=(\bgamma_1^\T,\ldots,\bgamma_K^\T)^\T$ with $\bgamma_j=(n-j)^{-1}\sum_{t=1}^{n-j}\E[{\rm vec}\{\bphi(\bx_t)\bx_{t+j}^\T\}]$. Define $\ell_j^*=\arg\max_{\ell\in\mathcal{L}_j}|\tilde{u}_\ell|$ for $j\in[K]$. By Cauchy-Schwarz inequality,  it holds that
\begin{align*}
  T_n
  =\sum_{j=1}^{K} \max_{\ell\in\mathcal{L}_j}(n^{1/2}u_\ell)^2
  \geqslant &~\sum_{j=1}^{K}(n^{1/2}u_{\ell_j^*})^2
    = \sum_{j=1}^{K} \big(n^{1/2}u_{\ell_j^*} - n^{1/2}\tilde{u}_{\ell_j^*} + n^{1/2}\tilde{u}_{\ell_j^*}\big)^2 \\
  =&~ n\sum_{j=1}^{K} (u_{\ell_j^*}-\tilde{u}_{\ell_j^*})^2 + n\sum_{j=1}^{K}\tilde{u}_{\ell_j^*}^2 + 2n\sum_{j=1}^{K}\tilde{u}_{\ell_j^*}(u_{\ell_j^*}-\tilde{u}_{\ell_j^*}) \\
  \geqslant&~ n\sum_{j=1}^{K} (u_{\ell_j^*}-\tilde{u}_{\ell_j^*})^2 + n\sum_{j=1}^{K}\tilde{u}_{\ell_j^*}^2 - 2n \bigg(\sum_{j=1}^{K}\tilde{u}_{\ell_j^*}^2\bigg)^{1/2} \bigg\{\sum_{j=1}^{K}(u_{\ell_j^*}-\tilde{u}_{\ell_j^*})^2\bigg\}^{1/2} \,.
\end{align*}
According to the definition of $\bu$ and $\tilde{\bu}$, we have $n^{1/2}(u_{\ell_j^*}-\tilde{u}_{\ell_j^*})=n^{1/2}(n-j)^{-1}\sum_{t=1}^{n-j}[\phi_{l_1^*}(\bx_t)x_{t+j,l_2^*}-\E\{\phi_{l_1^*}(\bx_t)x_{t+j,l_2^*}\}]$ for some $l_1^*\in[d]$ and $l_2^*\in[p]$. Note that $K\ll n^{1/7}$. By Bonferroni inequality and Lemma \ref{lem.bern}, it holds that for any $x>0$
\begin{align*}
\bP\bigg\{n\sum_{j=1}^{K}(u_{\ell_j^*}-\tilde{u}_{\ell_j^*})^2 > x\bigg\}
\leqslant&~\sum_{j=1}^K \bP\bigg(\frac{n^{1/2}}{n-j} \bigg|\sum_{t=1}^{n-j} [\phi_{l_1^*}(\bx_t)x_{t+j,l_2^*}-\E\{\phi_{l_1^*}(\bx_t)x_{t+j,l_2^*}\}]\bigg| > \frac{x^{1/2}}{K^{1/2}}\bigg) \\
\lesssim&~ \frac{n^{1/2}K^{3/2}}{x^{1/2}}\bigg\{\exp\bigg(-\frac{Cn^{\tau/2}x^{\tau/2}}{K^{\tau/2}}\bigg)
+\exp\bigg(-\frac{Cn^{\tau_1/2}x^{\tau_1/2}}{K^{3\tau_1/2}}\bigg) \bigg\}\\
&+K\exp\bigg(-\frac{Cx}{K^{2}}\bigg)
\end{align*}
with $\tau=\tau_1\tau_2/(\tau_1+\tau_2)$, which implies that $n\sum_{j=1}^{K}(u_{\ell_j^*}-\tilde{u}_{\ell_j^*})^2=O_{\rm p}(K^2\log K )$.
%Based on Condition \ref{cond.tail}, we have $\max_{\ell\in[Kpd]}|\tilde{u}_\ell|\leqslant C$ for some positive constant $C$. Thus it holds that
%% provided that $K=o(n^\kappa)$ where $\kappa\in(0,1)$.
%\begin{align*}
%  &~\bP\bigg\{n\sum_{j=1}^{K}\tilde{u}_{\ell_j^*}(u_{\ell_j^*}-\tilde{u}_{\ell_j^*})>x\bigg\}
%  \lesssim\sum_{j=1}^K\bP\bigg\{|n(u_{\ell_j^*}-\tilde{u}_{\ell_j^*})|>\frac{Cx}{K}\bigg\} \\
%  \lesssim&~ K\exp(-\frac{Cx^2}{nK^3})
%\end{align*}
%for any $x>0$.
%
Choose $u>0$ such that $(1+\nu)^{1/2}[1+\{2\log(pd)\}^{-1}+u]=1+\epsilon_n$ for some $\epsilon_n>0$.
Due to $\sum_{j=1}^{K}\tilde{u}_{\ell_j^*}^2 \geqslant n^{-1}K\varrho\lambda^2(K,p,d,\alpha) (1+\epsilon_n)^2$
and $n\sum_{j=1}^{K}(u_{\ell_j^*}-\tilde{u}_{\ell_j^*})^2=O_{\rm p}(K^2\log K)$,
%restricted on $\cE_0(\nu)$,
by \eqref{cv.bound}, it holds with probability approaching one that
\begin{align*}
  T_n \geqslant
  &~ n\sum_{j=1}^{K} (u_{\ell_j^*}-\tilde{u}_{\ell_j^*})^2+(1+\nu)K\varrho\lambda^2(K,p,d,\alpha)[1+\{2\log(pd)\}^{-1}+u]^2
  \notag\\
  &- O_{\rm p} \big\{ K^{3/2}(\log K)^{1/2}\varrho^{1/2}\lambda(K,p,d,\alpha)(1+\epsilon_n)\}  \notag\\
  >
  &~ (1+\nu)K\varrho\lambda^2(K,p,d,\alpha)\big[1+\{2\log(pd)\}^{-1}\big]^2
  + 2K\varrho\lambda^2(K,p,d,\alpha)u \notag\\
  &- O_{\rm p} \big\{ K^{3/2}(\log K)^{1/2}\varrho^{1/2}\lambda(K,p,d,\alpha)(1+\epsilon_n)\} \notag\\
  >&~ \hat{\rm cv}_\alpha  + 2K\varrho\lambda^2(K,p,d,\alpha)u- O_{\rm p} \big\{ K^{3/2}(\log K)^{1/2}\varrho^{1/2}\lambda(K,p,d,\alpha)  \big\} \,.
\end{align*}
Notice that $\epsilon_n\rightarrow 0$ and $\varrho\lambda^2(K,p,d,\alpha)K^{-1}(\log K)^{-1}\epsilon_n^2\rightarrow\infty$.
Then it holds that
\begin{align*}
  \epsilon_n\gg \frac{K^{1/2}(\log K)^{1/2}}{\{\log(pd)\}^{1/2}+(\log K)^{1/2}}
  \gg \frac{1}{\log(pd)}\gtrsim\frac{1}{K\log(pd)} =\nu\,,
\end{align*}
which implies that $u\asymp \epsilon_n$. It yields that $K\varrho\lambda^2(K,p,d,\alpha)u \gg  K^{3/2}(\log K)^{1/2}\varrho^{1/2}\lambda(K,p,d,\alpha)$ and $K\varrho\lambda^2(K,p,d,\alpha)u\rightarrow\infty$.
 Therefore,
 %restricted on $\cE_0(\nu)$,
 we have
$  T_n-\hat{\rm cv}_\alpha >  K\varrho\lambda^2(K,p,d,\alpha)u$ with probability approaching one.
%Due to $\bP\{\cE_0(\nu)^c\} \to 0$ as $n\rightarrow\infty$,
%then
Hence,
$
  \bP_{H_1}(T_n>\hat{\rm cv}_\alpha)\to 1$ as $n\rightarrow\infty$.
$\hfill\Box$

\singlespacing
\small
%\bibliographystyle{agsm}
%\bibpunct{(}{)}{,}{a}{}{;}
%\bibliography{HDMDT_bib}

%%%%%%%%%%%%%%%%%%%%%%%%%%%%%%%%%%%%%%%%%%%%%%%%%%%%%%%%%%%%%%%%%%%%%%%%%%%%%%

\clearpage

 \def\spacingset#1{\renewcommand{\baselinestretch}%
 {#1}\small\normalsize} \spacingset{1.6}

\if1\blind
{
  \begin{center}
    {\LARGE\bf  Supplementary Materials for ``Testing the Martingale Difference  Hypothesis in High Dimension'' by Jinyuan Chang, Qing Jiang and Xiaofeng Shao}
  \end{center}

} \fi

\if0\blind
{
  \begin{center}
    {\LARGE\bf Supplementary Materials for ``Testing the Martingale Difference  Hypothesis in High Dimension''}
\end{center}

} \fi

\setcounter{equation}{0}

\renewcommand{\theequation}{S.\arabic{equation}}

\onehalfspacing
\appendix
\setcounter{page}{1}
\medskip

In the supplementary material, we provide the detailed proofs for some theoretical results stated in the paper, and also report some additional simulation results and comparisons. Specifically, in Sections S.1--S.3, we include the detailed proofs of Lemma L1, Proposition 3 and Lemma L2, respectively. Section S.4 describes a comparison of the computational costs for our proposed tests and \cite{HLZ2017}'s tests. Simulation comparisons between our proposed tests and three multivariate white noise tests are shown  in Section S.5. In Section S.6, we present some additional simulation results for both  Quadratic Spectral kernel and Parzen kernel in the examination of the influence of the data driven bandwidth.

Throughout the supplementary material, we use $C$ to denote a generic positive finite constant that does not depend on $(p,d,n,K)$ and may be different in different uses. For two sequences of positive numbers $\{a_n\}$ and $\{b_n\}$, we write $a_n\lesssim b_n$ or $b_n\gtrsim a_n$ if $\limsup_{n\rightarrow\infty}a_n/b_n\leqslant c_0$ for some positive constant $c_0$. We write $a_n\asymp b_n$ if $a_n\lesssim b_n$ and $b_n\lesssim a_n$ hold simultaneously. We write $a_n\ll b_n$ or $b_n\gg a_n $ if $\limsup_{n\rightarrow\infty}a_n/b_n=0$. For a generic $q$-dimensional vector $\ba=(a_1,\ldots,a_q)^\T$ and a given index set $\mathcal{L}\subset[q]$, denote by $\ba_{\mathcal{L}}$ the subvector of $\ba$ collecting the components indexed by $\mathcal{L}$. For a countable set $\mathcal{F}$, we use $|\cF|$ to denote the cardinality of $\cF$.

Write
$\bu := ( u_1,\ldots, u_{Kpd} )^{\T} =(\hat\bgamma_1^{\T},\ldots,\hat\bgamma_{K}^{\T})^{\T}$ with $\hat{\bgamma}_j=(n-j)^{-1}\sum_{t=1}^{n-j} {\rm vec}\{\bphi(\bx_{t})\bx_{t+j}^{\T}\}$ for any $j\in[K]$. Let $\tilde n=n-K$. Recall $\be_t=([{\rm vec} \{\bphi(\bx_t)\bx_{t+1}^{\T}\}]^\T,\ldots, [{\rm vec} \{ \bphi(\bx_t)\bx_{t+K}^{\T}\}]^\T )^{\T}$. Since $\{\bx_t\}$ is an $\alpha$-mixing process, we know the newly defined process $\{\be_t\}$ is also $\alpha$-mixing with the $\alpha$-mixing coefficients $\{\tilde{\alpha}_K(k)\}_{k\geqslant1}$ satisfying
\begin{align}\label{eq:newalphamixing}
\tilde \alpha_K(k)\leqslant C_3\exp (-C_4|k-K|_{+}^{\tau_2})\,,
\end{align}
where the positive constants $\tau_2$, $C_3$ and $C_4$ are specified in Condition 2.
Write $\bar \be :=(\bar{\eta}_1,\ldots,\bar{\eta}_{Kpd})^\T= \tilde n^{-1}\sum_{t=1}^{\tilde n}\be_t$.  For each $j\in[K]$, define
$Z_j =n\max_{\ell\in\mathcal{L}_j}  u_{\ell}^2 $ and $\tilde Z_j = \tilde n \max_{\ell\in\mathcal{L}_j} \bar\eta_\ell^2$ with $\mathcal{L}_j:=\{(j-1)pd+1,\ldots,jpd\}$. Then the test statistic can be written as
$
T_n=n\sum_{j=1}^{K}|\hat\bgamma_j|_{\infty}^2= \sum_{j=1}^K Z_j$. Furthermore, we let
$
\tilde T_n := \sum_{j=1}^K \tilde Z_j$.

\renewcommand{\thesection}{S.\arabic{section}}
\setcounter{table}{0}
\renewcommand{\thetable}{S\arabic{table}}

\section{Proof of Lemma \ref{lem.bern}}\label{sec:pflembern}
Write $a+[q]=\{a+1,\ldots,a+q\}$.
Recall $\mathbb{E}(z_{t,j})=0$ for any $t\in a+[q]$ and $j\in[d_z]$. Write $S_{k,j}=\sum_{t=a+1}^{a+k} z_{t,j}$. From {\rm AS1}, we have $
 \max_{t\in a+[q]}\max_{j\in[d_z]}\Var(z_{t,j})$ is uniformly bounded away from infinity. In the sequel, we will use the Fuk-Nagaev inequality to bound the tail probability of $\max_{k\in[q]}|S_{k,j}|$. Define
$
 \alpha_z^{-1}(u)=\sum_{k=1}^\infty I\{ u<\alpha_z(k) \}$
 for any $u\in(0,1]$ with $\alpha_z(k)$ specified in AS2, and
$
Q_j(u)=\sup_{t\in a+[q]}Q_{t,j}(u)$
with $Q_{t,j}(u)=\inf\{x>0: \bP(|z_{t,j}|>x)\leqslant u\}$ for any $u\in[0,1]$.

Notice that $\alpha_z(k) \leqslant a_1 \exp(-a_2|k-m|_{+}^{r_2})$ for any $k\geqslant0$. It then holds that $\alpha_z^{-1}(u) \leqslant m+ a_2^{-1/r_2}\{\log(a_1u^{-1})\}^{1/r_2}$.
By AS1, we have $\max_{t\in a+[q]}\max_{j\in[d_z]}\bP[|z_{t,j}|>b_2^{-1/r_1} \{\log(b_1u^{-1})\}^{1/r_1} ] \leqslant u$, which implies $Q_{t,j}(u)\leqslant b_2^{-1/r_1} \{\log(b_1u^{-1})\}^{1/r_1}$ for any $t\in a+[q]$. Thus,
$
  Q_j(u)\leqslant b_2^{-1/r_1} \{\log(b_1u^{-1})\}^{1/r_1}$.
Define $
R_j(u) = \alpha_z^{-1}(u)Q_j(u)$
for any $u\in(0,1]$. Based on the upper bounds of $\alpha_z^{-1}(u)$ and $Q_j(u)$ given above, we have $
  R_j(u)\leqslant c_1 \{\log(c_2u^{-1})\}^{1/r} + c_1m\{\log(c_2u^{-1})\}^{1/r_1}$ with $c_1=b_2^{-1/r_1}\max(a_2^{-1/r_2},1)$, $c_2=\max(a_1,b_1)$ and $r=r_1r_2/(r_1+r_2)$.
Since $R_j(u)$ is a right-continuous and non-increasing function, its inverse function
\begin{align*}
  H_j(x)=R_j^{-1}(x)&=\inf\{u>0: R_j(u)\leqslant x\}\\
&\leqslant \inf[u>0: c_1\{\log(c_2u^{-1})\}^{1/r} + c_1m\{\log(c_2u^{-1})\}^{1/r_1} \leqslant x ] \\
&\leqslant \tilde{c}_1\exp(-\tilde{c}_2x^r)+\tilde{c}_1\exp(-\tilde{c}_2m^{-r_1}x^{r_1})
\end{align*}
for any $x>0$ with $\tilde{c}_1=c_2$ and $\tilde{c}_2=\min\{(2c_1)^{-r},(2c_1)^{-r_1}\}$. Let $\tilde{u}=\tilde{c}_1\exp(-\tilde{c}_2x^r)+\tilde{c}_1\exp(-\tilde{c}_2m^{-r_1}x^{r_1})$. Therefore,
\begin{align*}
  \int_0^{H_j(x)}Q_j(u)\,\md u
  \leqslant&~ \int_0^{\tilde u} b_2^{-1/r_1} \{\log(b_1u^{-1})\}^{1/r_1}\,\md u
   \lesssim\int_{\log(b_1\tilde u^{-1})}^{+\infty} y^{1/r_1}e^{-y}\,\md y \\
  %\lesssim&~ \int_{\log(b_1\tilde{u}^{-1})}^\infty  (y^{1/r_1}-r_1^{-1} y^{1/r_1-1})e^{-y}\,\md y \\
  \lesssim&~ b_1^{-1}\tilde{u}\{\log(b_1\tilde{u}^{-1})\}^{1/r_1}
  \lesssim \tilde{u}^{1/2}\{\tilde{u}^{r_1/2}\log(b_1\tilde{u}^{-1})\}^{1/r_1}\,.
\end{align*}
As $x\rightarrow+\infty$, we have $\tilde{u}\rightarrow0^+$ which implies $\tilde u^{r_1/2} \log(b_1\tilde u^{-1})\to 0^+$. Hence, there exists a uniform constant $\varepsilon>0$ such that $\tilde u^{r_1/2} \log(b_1\tilde u^{-1})<\varepsilon$ for any $x>0$. By the definition of $\tilde{u}$, we have
\begin{align}\label{int.Q}
  \int_0^{H_j(x)}Q_j(u)\,\md u \lesssim\exp(-Cx^r)+\exp(-Cm^{-r_1}x^{r_1}) \,.
\end{align}
For any $j\in[d_z]$, it follows from Davydov's inequality that
\begin{align*}
  \sigma_{q,j}^2
:=&~ \sum_{t_1=a+1}^{a+q}\sum_{t_2=a+1}^{a+q}|\Cov(z_{t_1,j},z_{t_2,j})|\\
%\lesssim  \sum_{t_1=1}^n\sum_{t_2=1}^n \exp( -a_2||t_1-t_2|-m|_{+}^{r_2} ) \\
\lesssim&~ \sum_{t=a+1}^{a+q} 1 + \sum_{t_1>t_2}\exp(-a_2|t_1-t_2-m|_{+}^{r_2})\\
\lesssim&~ q + \sum_{j=1}^m (q-j) + \sum_{k=1}^{q-m-1}(q-m-k)\exp(-a_2k^{r_2})
     \lesssim qm \,,
\end{align*}
where the last inequality is due to $1\leqslant m\leqslant q$. By the Fuk-Nagaev inequality \cite[Theorem 6.2]{Rio2013} and \eqref{int.Q}, we have
\begin{align*}
  \bP\bigg(\max_{k\in[q]}|S_{k,j}| \geqslant 4\lambda\bigg)
   \leqslant&~   4\bigg(1+\frac{\lambda^2}{\delta \sigma_{q,j}^2} \bigg)^{-\delta/2} + 4q\lambda^{-1}\int_{0}^{H_j(\lambda/\delta)} Q_j(u) \,\md u  \notag\\
   \lesssim&~  \bigg(1+\frac{\lambda^2}{C\delta qm} \bigg)^{-\delta/2}
   + q\lambda^{-1}\big\{\exp(-C\lambda^{r}\delta^{-r})
   + \exp(-Cm^{-r_1}\lambda^{r_1}\delta^{-r_1}) \big\}
%   \leqslant&~  \bigg(1+\frac{\lambda^2}{C\delta nm} \bigg)^{-\delta/2}
 %  + n\lambda^{-1}\exp(-C\lambda^{r})
 %  + n\lambda^{-1}\exp(-Cm^{-r_1}\lambda^{r_1}) \,.
\end{align*}
for any positive $\lambda$ and any $\delta\geqslant 1$.
Notice that $(1+x^{-1} )^{-x}\to e^{-1}$ as $x\to\infty$. With a sufficiently large constant $\delta$,  we can conclude that
$
  \bP(\max_{k\in [q]} |S_{k,j} | \geqslant 4\lambda) \lesssim \exp(-Cq^{-1}m^{-1}\lambda^2) + q\lambda^{-1}\exp(-C\lambda^r) +q\lambda^{-1}\exp(-Cm^{-r_1} \lambda^{r_1})$
for any $\lambda>0$. We complete the proof of Lemma \ref{lem.bern}. $\hfill\Box$

\section{Proof of Proposition \ref{lem.approx}}
%To prove Proposition \ref{lem.approx}, we need the following auxiliary lemma.

Let $B=o(n)$  be a positive integer that will diverge with $n$. We first decompose the sequence $\{1,\ldots, n\}$ to $L+1$ blocks with $L=\lfloor  n/B \rfloor$: $\cG_\ell = \{(\ell-1)B+1,\ldots,\ell B\}$ for $\ell\in[L]$ and $\cG_{L+1}=\{LB+1,\ldots,n\}$, where $\lfloor \cdot \rfloor$ is the integer truncation operator. Let $b> s$ be two nonnegative integers such that $B=b+s$ with $s>m$ and $s=o(b)$. We then decompose each $\cG_\ell$ $(\ell\in[L])$ to a ``large'' block $\cB_\ell$ with length $b$ and a ``small'' block $\cS_\ell$ with length $s$: $\cB_\ell = \{(\ell-1)B+1,\ldots,(\ell-1)B+b\}$ and $\cS_\ell=\{(\ell-1)B+b+1,\ldots,\ell B\}$.
Define $\tilde \bz_\ell = b^{-1/2} \sum_{t\in\cB_\ell} \bz_t$ and
$\check{\bz}_\ell = s^{-1/2} \sum_{t\in\cS_\ell} \bz_t$  for each  $\ell\in[L]$. Set $\cS_{L+1} = \cG_{L+1}$ and $\check \bz_{L+1} = (n - LB)^{-1/2}\sum_{t\in\cS_{L+1}}\bz_t$.  Let $\{ \by_t\}_{t=1}^n$ be a sequence of independent normal random vectors with mean zero, where the covariance of $\by_t$ $(t\in\cB_\ell)$ is $\E(\tilde \bz_\ell \tilde \bz_\ell^\T)$. For each $\ell\in[L]$, define $\tilde \by_\ell=b^{-1/2}\sum_{t\in\cB_\ell} \by_t$. Write $\bs_{n,z}^{(1)}=L^{-1/2}\sum_{\ell=1}^L \tilde\bz_\ell$ and $\bs_{n,y}^{(1)}=L^{-1/2}\sum_{\ell=1}^L \tilde \by_\ell$. Furthermore, define
\begin{align}
  \bar{\varrho}_n:=&~\sup_{\bu\in\R^{d_{z}},\nu\in[0,1]} \big| \bP\{\sqrt{\nu}\bs_{n,z} + \sqrt{1-\nu}\bs_{n,y}^{(1)} \leqslant \bu\} - \bP\{\bs_{n,y}^{(1)}\leqslant\bu\} \big|\,.\label{varrho2}
\end{align}
Notice that $\bs_{n,z}$ is independent of $\bs_{n,y}$. As shown in \cite{CCW2020a}, it holds that
$
\varrho_n \leqslant \bar{\varrho}_n + 2\sup_{\bu\in\R^{d_{z}}}|\bP\{\bs_{n,y}^{(1)}\leqslant\bu \} - \bP(\bs_{n,y}\leqslant\bu)|$.
Recall $\bs_{n,y}^{(1)} \sim \mathcal{N}(0,\bXi_1)$ and $\bs_{n,y} \sim \mathcal{N}(\bzero,\bXi)$ with
$
\bXi_1=(Lb)^{-1}\sum_{\ell=1}^L \E\{(\sum_{t\in\cB_\ell}\bz_t) ^{\otimes2}\}$ and
$
\bXi={n}^{-1}\E\{(\sum_{t=1}^n\bz_t)^{\otimes2}\}$. To construct Proposition \ref{lem.approx}, we need the following two lemmas whose proof are given in Sections \ref{subsec.lemcov} and \ref{sec:pflavarrho2}, respectively.

\begin{lemma}\label{lem.covest}
Under {\rm AS1--AS2}, it holds that
$
|\bXi_1-\bXi|_\infty \lesssim m(sb^{-1} + bn^{-1})$.
\end{lemma}

\begin{lemma}\label{lem.varrho2}
Assume {\rm AS1--AS3} hold and $d_z\geqslant n^{\varpi}$ for some constant $\varpi>0$. Let $r=r_1r_2/(r_1+r_2)$. If $s= m+C(\log d_{z})^{1/r_2}$ for some sufficiently large $C>0$,
$ms\ll b \ll  n^{1/2}$ and $b\gtrsim n^{1/4}(\log d_z)^{-1/4} \max\{m^{3/4}, \\ (\log d_z)^{3/(4r_2)}\} $, it holds that
$
\bar{\varrho}_n\lesssim m^{1/2}L^{-1/6}(\log d_{z})^{7/6}$
provided that
$\log d_{z}\ll\min\{m^{3r/(6+2r)}b^{3r/(6+2r)} L^{r/(3+r)},\\L^{2/5}, m^{-3r_1/(6+2r_1)}b^{3r_1/(6+2r_1)}L^{r_1/(3+r_1)}\}$.
\end{lemma}

By Lemma 3 in \cite{CCW2020a}, we have
$
\sup_{\bu\in\R^{d_{z}}}|\bP\{\bs_{n,y}^{(1)}\leqslant\bu \} - \bP(\bs_{n,y}\leqslant\bu)| \lesssim |\bXi_1-\bXi|_\infty^{1/3}(\log d_{z})^{2/3}$. Note that $L\asymp nb^{-1}$, $b=o(n^{1/2})$  and $s=m+C(\log d_z)^{1/r_2}$. With selecting $b\asymp n^{1/3}$, by Lemmas \ref{lem.covest} and \ref{lem.varrho2}, it holds that
\begin{align*}
  \varrho_n
  \lesssim&~\frac{m^{1/2}b^{1/6}(\log d_z)^{7/6}}{n^{1/6}}
  + \frac{m^{2/3}(\log d_z)^{2/3}}{b^{1/3}} + \frac{m^{1/3}(\log d_z)^{(1+2r_2)/(3r_2)}}{b^{1/3}} \\
  \lesssim&~ \frac{m^{1/3}(\log d_z)^{2/3}}{n^{1/9}}\{m^{1/6}(\log d_z)^{1/2}+m^{1/3}+(\log d_z)^{1/(3r_2)}\}
  \,.
   \end{align*}
If $\log d_z=o\{n^{r_2/(9-3r_2)}\}$ and $m\lesssim n^{1/9}(\log n)^{1/3}$, we have $b\gtrsim n^{1/4}(\log d_z)^{-1/4}\max\{m^{3/4},  (\log d_z)^{3/(4r_2)}\}$ holds automatically. Therefore,
\begin{align*}
  \varrho_n\lesssim \frac{m^{1/3}(\log d_z)^{2/3}}{n^{1/9}}\{m^{1/6}(\log d_z)^{1/2}+m^{1/3}+(\log d_z)^{1/(3r_2)}\}
\end{align*}
provided that $\log d_{z}\ll\min\{m^{3r/(6+2r)}n^{7r/(18+6r)}, m^{-3r_1/(6+2r_1)}n^{7r_1/(18+6r_1)}, n^{r_2/(9-3r_2)}\}$ with $m\lesssim n^{1/9}(\log n)^{1/3}$. We complete the proof of Proposition \ref{lem.approx}. $\hfill\Box$

\subsection{Proof of Lemma \ref{lem.covest}}\label{subsec.lemcov}
Recall $\tilde \bz_\ell = b^{-1/2} \sum_{t\in\cB_\ell} \bz_t$ and
$\check{\bz}_\ell = s^{-1/2} \sum_{t\in\cS_\ell} \bz_t$  for $\ell\in[L]$, and $\check \bz_{L+1} = (n - LB)^{-1/2}\sum_{t\in\cS_{L+1}}\bz_t$. We adopt the convention $\cB_{L+1}=\emptyset$ and set $\tilde{\bz}_{L+1}=\bzero$. Write $\tilde{\bz}_\ell=(\tilde{z}_{\ell,1},\ldots,\tilde{z}_{\ell,d_z})^\T$ and $\check{\bz}_\ell=(\check{z}_{\ell,1},\ldots,\check{z}_{\ell,d_z})^\T$. Define $\tilde{a}_\ell=b^{1/2}$ for any $\ell\in[L+1]$, $\check{a}_\ell=s^{1/2}$ for any $\ell\in[L]$, and $\check{a}_{L+1}=(n-LB)^{1/2}$. Then
\begin{align*}
|\bXi_1-\bXi|_\infty =&~\max_{j_1,j_2\in[d_{z}]}\bigg| \frac{1}{Lb}\sum_{\ell=1}^L\tilde{a}_\ell^2\E(\tilde{z}_{\ell,j_1}\tilde{z}_{\ell,j_2}) - \frac{1}{n}\E\bigg\{\bigg(\sum_{t=1}^n z_{t,j_1}\bigg)\bigg(\sum_{t=1}^n z_{t,j_2}\bigg) \bigg\} \bigg| \\
\leqslant&~\max_{j_1,j_2\in[d_z]}\underbrace{\bigg| \frac{1}{Lb}\sum_{\ell=1}^L  \tilde{a}_{\ell}^2\E(\tilde{z}_{\ell,j_1}\tilde{z}_{\ell,j_2})
- \frac{1}{n}\sum_{\ell=1}^L \tilde{a}_{\ell}^2\E(\tilde{z}_{\ell,j_1}\tilde{z}_{\ell,j_2}) \bigg|}_{R_1(j_1,j_2)}\\
&+\max_{j_1,j_2\in[d_{z}]}\underbrace{\bigg| \frac{1}{n}\sum_{\ell=1}^L\tilde{a}_\ell^2\E(\tilde{z}_{\ell,j_1}\tilde{z}_{\ell,j_2}) - \frac{1}{n}\E\bigg\{\bigg(\sum_{t=1}^n z_{t,j_1}\bigg)\bigg(\sum_{t=1}^n z_{t,j_2}\bigg) \bigg\} \bigg|}_{R_2(j_1,j_2)}\,.
\end{align*}
As we will specify in Sections \ref{sec:R1j1j2} and \ref{sec:R2j1j2}, it holds that $\max_{j_1,j_2\in[d_z]}R_1(j_1,j_2)\lesssim m(sb^{-1} + bn^{-1})$ and $\max_{j_1,j_2\in[d_z]}R_2(j_1,j_2)\lesssim m(sb^{-1} + bn^{-1})$, above inequalities yield $|\bXi_1-\bXi|_\infty\lesssim m(sb^{-1} + bn^{-1})$.

\subsubsection{Convergence rate of $\max_{j_1,j_2\in[d_z]}R_1(j_1,j_2)$}\label{sec:R1j1j2}

For any $j_1,j_2\in[d_z]$, by the triangle inequality and Cauchy-Schwarz inequality, we have
\begin{align}\label{eq:R1}
R_1(j_1,j_2)\leqslant &~\frac{n-Lb}{nLb}\sum_{\ell=1}^L\sum_{t\in\cB_\ell} \{\E( z_{t,j_1}^2)\}^{1/2} \{\E(z_{t,j_2}^2)\}^{1/2}
+\frac{n-Lb}{nLb}\sum_{\ell=1}^L \sum_{t_1,t_2\in\cB_\ell, t_1\neq t_2}|\E( z_{t_1,j_1}z_{t_2,j_2})|\notag \\
%\lesssim &~ \frac{n-Lb}{n} + \frac{n-Lb}{nLb}\sum_{\ell=1}^L \sum_{t_1\neq t_2}\big|\E( z_{t_1,j_1}z_{t_2,j_2})\big|\\
\lesssim &~\frac{s}{b}+\frac{b}{n}+\frac{n-Lb}{nLb}\sum_{\ell=1}^L \sum_{t_1,t_2\in\cB_{\ell}, t_1\neq t_2}|\E( z_{t_1,j_1}z_{t_2,j_2})| \,,
\end{align}
where the last step is due to the facts $\max_{t\in[n]}\max_{j\in[d_z]}\E(z_{t,j}^2)\lesssim1$, $n-Lb< Ls+B$ and $L\asymp nb^{-1}$. Note that $m<s$. Based on Davydov's inequality, it holds that
\begin{align*}
&~\frac{n-Lb}{nLb}\sum_{\ell=1}^L \sum_{t_1,t_2\in\cB_{\ell},t_1> t_2}|\E( z_{t_1,j_1}z_{t_2,j_2})| \\
%\lesssim &~ \frac{n-Lb}{nLb} \sum_{\ell=1}^L \sum_{t_1=(\ell-1)B+2}^{(\ell-1)B+b} \sum_{t_2=(\ell-1)B+1}^{t_1-1} \exp(-C|t_1-t_2-m|_+^{r_2}) \\
\lesssim &~ \frac{n-Lb}{nLb} \sum_{\ell=1}^L \sum_{k=1}^{b-1} (b-k) \exp(-C|k-m|_+^{r_2}) \\
\lesssim &~ \frac{n-Lb}{nb}\sum_{k=1}^m (b-k) + \frac{n-Lb}{nb}\sum_{k=m+1}^{b-1}(b-k)\exp\{-C(k-m)^{r_2}\} \\
\lesssim &~ m(sb^{-1} + bn^{-1}) + sb^{-1}+bn^{-1} \lesssim m(sb^{-1} + bn^{-1}) \,.
\end{align*}
Analogously, we also have $(n-Lb)(nLb)^{-1}\sum_{\ell=1}^L \sum_{t_1,t_2\in\cB_{\ell},t_1< t_2}|\E( z_{t_1,j_1}z_{t_2,j_2})|\lesssim m(sb^{-1} + bn^{-1})$. Thus,
$
\max_{j_1,j_2\in[d_z]}R_1(j_1,j_2)
\lesssim m(sb^{-1} + bn^{-1})$.

\subsubsection{Convergence rate of $\max_{j_1,j_2\in[d_z]}R_2(j_1,j_2)$}\label{sec:R2j1j2}

For any given $j_1,j_2\in[d_z]$, it holds that
\begin{align}\label{sigma}
R_2(j_1,j_2)\leqslant&~\underbrace{\bigg|\frac{1}{n}\sum_{\ell=1}^{L+1}\tilde{a}_\ell\check{a}_\ell\E(\tilde{z}_{\ell,j_1}\check{z}_{\ell,j_2})\bigg|}_{R_{2,1}(j_1,j_2)}
+\underbrace{\bigg|\frac{1}{n}\sum_{\ell=1}^{L+1}\check{a}_\ell\tilde{a}_\ell\E(\check{z}_{\ell,j_1}\tilde{z}_{\ell,j_2})\bigg|}_{R_{2,2}(j_1,j_2)}
+ \underbrace{\bigg|\frac{1}{n}\sum_{\ell=1}^{L+1}\check{a}_\ell^2\E(\check{z}_{\ell,j_1}\check{z}_{\ell,j_2})\bigg|}_{R_{2,3}(j_1,j_2)} \notag \\
&+ \underbrace{\bigg|\frac{1}{n}\sum_{\ell_1\neq\ell_2}\tilde{a}_{\ell_1}\tilde{a}_{\ell_2}\E(\tilde{z}_{\ell_1,j_1}\tilde{z}_{\ell_2,j_2})\bigg|}_{R_{2,4}(j_1,j_2)}
+ \underbrace{\bigg|\frac{1}{n}\sum_{\ell_1\neq \ell_2}\tilde{a}_{\ell_1}\check{a}_{\ell_2}\E(\tilde{z}_{\ell_1,j_1}\check{z}_{\ell_2,j_2})\bigg|}_{R_{2,5}(j_1,j_2)}\\
&+\underbrace{\bigg|\frac{1}{n}\sum_{\ell_1\neq \ell_2}\check{a}_{\ell_1}\tilde{a}_{\ell_2}\E(\check{z}_{\ell_1,j_1}\tilde{z}_{\ell_2,j_2})\bigg|}_{R_{2,6}(j_1,j_2)}
+ \underbrace{\bigg|\frac{1}{n}\sum_{\ell_1\neq \ell_2}\check{a}_{\ell_1}\check{a}_{\ell_2}\E(\check{z}_{\ell_1,j_1}\check{z}_{\ell_2,j_2})\bigg|}_{R_{2,7}(j_1,j_2)} \,.   \nonumber
\end{align}
As shown later, we have $\max_{j_1,j_2\in[d_z]}R_{2,1}(j_1,j_2)=\max_{j_1,j_2\in[d_z]}R_{2,2}(j_1,j_2)\lesssim msb^{-1}$,
$\max_{j_1,j_2\in[d_z]}R_{2,3}(j_1,j_2)\\ \lesssim m(sb^{-1} + bn^{-1})$,~~
$\max_{j_1,j_2\in[d_z]}R_{2,4}(j_1,j_2)\lesssim o(b^{-1})$, ~~
$\max_{j_1,j_2\in[d_z]}R_{2,5}(j_1,j_2)=\max_{j_1,j_2\in[d_z]}R_{2,6}(j_1,j_2)\lesssim bn^{-1}+msb^{-1}$ and $\max_{j_1,j_2\in[d_z]}R_{2,7}(j_1,j_2)  \lesssim msn^{-1}+o(b^{-1})$, which implies $\max_{j_1,j_2\in[d_z]}R_2(j_1,j_2)\lesssim m(sb^{-1} + bn^{-1})$.

{\bf Convergence rate of $\max_{j_1,j_2\in[d_z]}R_{2,1}(j_1,j_2)$.} Due to $m<s$, by the triangle inequality and Davydov's inequality, it holds that
\begin{align*}
R_{2,1}(j_1,j_2)
   \leqslant&~
\frac{1}{n}\sum_{\ell=1}^{L+1}\sum_{t_1\in\cB_\ell}\sum_{t_2\in\cS_\ell} |\E(z_{t_1,j_1}z_{t_2,j_2})|
 \lesssim  \frac{1}{n}\sum_{\ell=1}^{L+1} \sum_{t_1\in\cB_\ell}\sum_{t_2\in\cS_\ell} \exp(-C|t_2-t_1-m|_+^{r_2}) \\
 \lesssim &~ \frac{1}{n}\sum_{\ell=1}^{L+1}\sum_{k=1}^{s}k\exp(-C|k-m|_+^{r_2})
 +\frac{1}{n}\sum_{\ell=1}^{L+1}\sum_{k=s+1}^{b} s\exp(-C|k-m|_+^{r_2}) \\
 &+ \frac{1}{n}\sum_{\ell=1}^{L+1}\sum_{k=b+1}^{B-1}(B-k)\exp(-C|k-m|_+^{r_2}) \,.
\end{align*}
Then
$\max_{j_1,j_2\in[d_z]}R_{2,1}(j_1,j_2)
\lesssim b^{-1}\sum_{k=1}^{m}k +{b}^{-1}\sum_{k=m+1}^s k\exp\{-C(k-m)^{r_2}\} +  {s}{b}^{-1}\sum_{k=s+1}^b \exp\{-C (k-m)^{r_2}\}
\lesssim  m^2b^{-1} + sb^{-1}$.

{\bf Convergence rate of $\max_{j_1,j_2\in[d_z]}R_{2,3}(j_1,j_2)$.} Applying the triangle inequality and Cauchy-Schwarz inequality again, it holds that
\begin{align*}
R_{2,3}(j_1,j_2)
%\lesssim &~\frac{1}{n}\sum_{\ell=1}^{L}\sum_{t_1,t_2\in\cS_\ell}|\E(z_{t_1,j_1}z_{t_2,j_2})| +\frac{1}{n}\sum_{t_1,t_2\in\cS_{L+1}}|\E(z_{t_1,j_1}z_{t_2,j_2})| \\
\lesssim &~\frac{1}{n}\sum_{\ell=1}^L\sum_{t\in\cS_\ell}\{\E (z_{t,j_1}^2)\}^{1/2}\{\E(z_{t,j_2}^2)\}^{1/2} +\frac{1}{n}\sum_{\ell=1}^L\sum_{t_1,t_2\in\cS_\ell,t_1\neq t_2}|\E(z_{t_1,j_1}z_{t_2,j_2})| \\
&~ + \frac{1}{n}\sum_{t\in\cS_{L+1}}\{\E(z_{t,j_1}^2)\}^{1/2}\{\E(z_{t,j_2}^2)\}^{1/2} + \frac{1}{n}\sum_{t_1,t_2\in\cS_{L+1},t_1\neq t_2}|\E(z_{t_1,j_1}z_{t_2,j_2})| \\
\lesssim &~ \frac{s}{b}+\frac{n-LB}{n} +\frac{1}{n}\sum_{\ell=1}^L\sum_{t_1,t_2\in\cS_\ell,t_1\neq t_2}|\E(z_{t_1,j_1}z_{t_2,j_2})| + \frac{1}{n}\sum_{t_1,t_2\in\cS_{L+1},t_1\neq t_2}|\E(z_{t_1,j_1}z_{t_2,j_2})| \,.
\end{align*}
By Davydov's inequality,
$
{n}^{-1}\sum_{\ell=1}^L\sum_{t_1,t_2\in\cS_\ell,t_1> t_2}|\E(z_{t_1,j_1}z_{t_2,j_2})|
\lesssim {n}^{-1} \sum_{\ell=1}^L\sum_{t_1,t_2\in\cS_\ell, t_1>t_2}\exp(-C|t_1-t_2-m|_+^{r_2})
\lesssim{b}^{-1}\sum_{k=1}^{m-1}(s-k)+{b}^{-1}\sum_{k=m}^{s-1}(s-k) \exp\{-C(k-m)^{r_2}\}\lesssim msb^{-1}+sb^{-1}\lesssim msb^{-1}$.
Analogously, we have  $n^{-1}\sum_{\ell=1}^L\sum_{t_1,t_2\in\cS_\ell,t_1< t_2}|\E(z_{t_1,j_1}z_{t_2,j_2})|\lesssim  msb^{-1}$. Thus
$
n^{-1}\sum_{\ell=1}^L\sum_{t_1,t_2\in\cS_\ell,t_1\neq t_2}|\E(z_{t_1,j_1}z_{t_2,j_2})|\lesssim msb^{-1}$.
Next, if $n-LB\leqslant m$, we have
${n}^{-1}\sum_{t_1,t_2\in\cS_{L+1},t_1\neq t_2}|\E(z_{t_1,j_1}z_{t_2,j_2})|
\lesssim n^{-1}\sum_{k=1}^{n-LB-1}(n-LB-k)\lesssim(n-LB)^2n^{-1}$.
If $n-LB>m$, we have ${n}^{-1}\sum_{t_1,t_2\in\cS_{L+1},t_1\neq t_2}|\E(z_{t_1,j_1}z_{t_2,j_2})|
\lesssim n^{-1}\sum_{k=1}^{m-1}(n-LB-k)+n^{-1}\sum_{k=m}^{n-LB-1}(n-LB-k)\exp\{-C(k-m)^{r_2}\}\lesssim(n-LB)mn^{-1}$.
Thus we can conclude that
${n}^{-1}\sum_{t_1,t_2\in\cS_{L+1},t_1\neq t_2}|\E(z_{t_1,j_1}z_{t_2,j_2})|
\lesssim\{(n-LB)\wedge m\}(n-LB)n^{-1}$.
Combining the above results, we obtain that
$
\max_{j_1,j_2\in[d_z]}R_{2,3}(j_1,j_2)
\lesssim msb^{-1} + \{(n-LB)\wedge m\}(n-LB)n^{-1}\lesssim m(sb^{-1}+bn^{-1})$.

{\bf Convergence rate of $\max_{j_1,j_2\in[d_z]}R_{2,4}(j_1,j_2)$.}
Applying the triangle inequality, it holds that
\begin{align*}
R_{2,4}(j_1,j_2)
\leqslant \bigg| \frac{1}{n}\sum_{\ell_1>\ell_2}\tilde{a}_{\ell_1}\tilde{a}_{\ell_2}\E(\tilde{z}_{\ell_1,j_1}\tilde{z}_{\ell_2,j_2})\bigg|
+ \bigg| \frac{1}{n}\sum_{\ell_1<\ell_2}\tilde{a}_{\ell_1}\tilde{a}_{\ell_2}\E(\tilde{z}_{\ell_1,j_1}\tilde{z}_{\ell_2,j_2})\bigg| \,.
\end{align*}
Due to $m<s$ and $k-m\geqslant -b+1-s=-B+1$ for any $-b+1\leqslant k\leqslant b-1$, by the triangle inequality and Davydov's inequality, we have
%\begin{align}\label{eq:longrunestbd1}
%\bigg| \frac{1}{n}\sum_{\ell_1>\ell_2}\tilde{a}_{\ell_1}\tilde{a}_{\ell_2}\E(\tilde{z}_{\ell_1,j_1}\tilde{z}_{\ell_2,j_2})\bigg|
%%\leqslant&~\frac{1}{n}\sum_{\ell_1>\ell_2}\sum_{t_1\in\cB_{\ell_1}}\sum_{t_2\in\cB_{\ell_2}} |\E(z_{t_1,j_1}z_{t_2,j_2})|\notag\\
%%\lesssim&~\frac{1}{n}\sum_{\ell_1>\ell_2}\sum_{t_1\in\cB_{\ell_1}}\sum_{t_2\in\cB_{\ell_2}}\exp(-C|t_1-t_2-m|_+^{r_2}) \\
%\lesssim&~ \frac{1}{n} \sum_{j=1}^{L-1} \sum_{k=-b+1}^{b-1}(L-j)(b-|k|)\exp(-C|jB+k-m|_+^{r_2}) \,.
%   \end{align}
%Due to $m<s$ and $k-m\geqslant -b+1-s=-B+1$ for any $-b+1\leqslant k\leqslant b-1$,  then \eqref{eq:longrunestbd1} yields
 \begin{align}\label{eq:longrunestbd2}
 \bigg| \frac{1}{n}\sum_{\ell_1>\ell_2}\tilde{a}_{\ell_1}\tilde{a}_{\ell_2}\E(\tilde{z}_{\ell_1,j_1}\tilde{z}_{\ell_2,j_2})\bigg|\lesssim&~\frac{1}{n}\sum_{j=1}^{L-1}\sum_{k=-b+1}^{b-1}(L-j)(b-|k|)\exp\{-C(jB+k-m)^{r_2}\}\,.
  \end{align}
Notice that there exist a sufficiently large constant $\delta>3$ and a positive constant $C_\delta$ depending on $\delta$ such that $\exp(-Cx^{r_2})\leqslant C_\delta x^{-\delta}$ for any $x\geqslant1$. For any $j=1$ and $m\neq s$, or $j\geqslant2$, we have that
\begin{align}\label{eq:longrunestsum}
&~\sum_{k=-b+1}^{b-1}(b-|k|)\exp\{-C(jB+k-m)^{r_2}\}\notag \\
\lesssim&~\sum_{k=-b+1}^0(b+k)\exp\{-C(jB+k-m)^{r_2}\}\lesssim \sum_{k=0}^{b-1}\frac{b-k}{(jB-k-m)^{\delta}} \notag\\
=&~\sum_{k=0}^{b-1}\frac{1}{(jB-k-m)^{\delta-1}}-(jB-b-m)\sum_{k=0}^{b-1} \frac{1}{(jB-k-m)^{\delta}} \\
\leqslant&~\int_{jB-b-m}^{jB-m}\frac{{\rm d}x}{x^{\delta-1}}-(jB-b-m)\int_{jB-b+1-m}^{jB-m+1}\frac{{\rm d}x}{x^\delta}\lesssim \frac{1}{(jB-b-m)^{\delta-2}}\,, \notag
\end{align}
which implies
\begin{align*}
&~\frac{1}{n}\sum_{j=1}^{L-1}\sum_{k=-b+1}^{b-1}(L-j)(b-|k|)\exp\{-C(jB+k-m)^{r_2}\}\\
\lesssim&~\frac{1}{b}\sum_{k=-b+1}^{b-1}(b-|k|)\exp\{-C(B+k-m)^{r_2}\}+\frac{1}{n}\sum_{j=2}^{L-1}\frac{L-j}{(jB-b-m)^{\delta-2}}\\
\lesssim&~\frac{1}{b(s-m+1)^{\delta-2}}+\frac{1}{n}\sum_{j=2}^{L-1}\frac{L}{\{(j-1)B\}^{\delta-2}} \lesssim\frac{1}{b(s-m+1)^{\delta-2}}+\frac{L}{nB^{\delta-2}}\\
\lesssim&~\frac{1}{b(s-m+1)^{\delta-2}}\,.
\end{align*}
By \eqref{eq:longrunestbd2}, it holds that $\max_{j_1,j_2\in[d_z]}|{n}^{-1}\sum_{\ell_1>\ell_2}\tilde{a}_{\ell_1}\tilde{a}_{\ell_2}\E(\tilde{z}_{\ell_1,j_1}\tilde{z}_{\ell_2,j_2})|\lesssim (s-m+1)^{2-\delta}b^{-1}$.
Analogously, we also have $
\max_{j_1,j_2\in[d_z]}|{n}^{-1}\sum_{\ell_1<\ell_2}\tilde{a}_{\ell_1}\tilde{a}_{\ell_2}\E(\tilde{z}_{\ell_1,j_1}\tilde{z}_{\ell_2,j_2})| \lesssim(s-m+1)^{2-\delta}b^{-1}$. Therefore, we have
$\max_{j_1,j_2\in[d_z]}R_{2,4}(j_1,j_2) \lesssim o(b^{-1})$.

{\bf Convergence rate of $\max_{j_1,j_2\in[d_z]}R_{2,5}(j_1,j_2)$.}
Notice that $\tilde{\bz}_{L+1}=\bzero$. Then
\begin{align*}
R_{2,5}(j_1,j_2)
\leqslant&~\bigg|\frac{1}{n}\sum_{\ell_2<\ell_1\leqslant L}\tilde{a}_{\ell_1}\check{a}_{\ell_2} \E(\tilde{z}_{\ell_1,j_1}\check{z}_{\ell_2,j_2})\bigg|
+ \bigg|\frac{1}{n}\sum_{\ell_1<\ell_2\leqslant L+1}\tilde{a}_{\ell_1}\check{a}_{\ell_2} \E(\tilde{z}_{\ell_1,j_1}\check{z}_{\ell_2,j_2})\bigg|\\
\leqslant&~\underbrace{\bigg|\frac{1}{n}\sum_{\ell_2<\ell_1\leqslant L}\tilde{a}_{\ell_1}\check{a}_{\ell_2} \E(\tilde{z}_{\ell_1,j_1}\check{z}_{\ell_2,j_2})\bigg|}_{R_{2,5,1}(j_1,j_2)}
+ \underbrace{\bigg|\frac{1}{n}\sum_{\ell_1<\ell_2\leqslant L}\tilde{a}_{\ell_1}\check{a}_{\ell_2} \E(\tilde{z}_{\ell_1,j_1}\check{z}_{\ell_2,j_2})\bigg|}_{R_{2,5,2}(j_1,j_2)}\\
&+\underbrace{\bigg|\frac{1}{n}\sum_{\ell_1=1}^L\tilde{a}_{\ell_1}\check{a}_{L+1} \E(\tilde{z}_{\ell_1,j_1}\check{z}_{L+1,j_2})\bigg|}_{R_{2,5,3}(j_1,j_2)}\,.
\end{align*}
Applying the triangle inequality and Davydov's inequality again, it holds that
\begin{align}\label{eq:R25bd1}
R_{2,5,1}(j_1,j_2)
%\lesssim&~\frac{1}{n}\sum_{\ell_2<\ell_1\leqslant L}\sum_{t_1\in\cB_{\ell_1}}\sum_{t_2\in\cS_{\ell_2}}\exp(-C|t_1-t_2-m|_+^{r_2}) \notag\\
\lesssim&~ \frac{1}{n} \sum_{j=1}^{L-1}(L-j)\sum_{k=-s+1}^{0}(s-|k|)\exp(-C|jB-b+k-m|_+^{r_2})\notag \\
&+ \frac{1}{n} \sum_{j=1}^{L-1}(L-j)\sum_{k=1}^{b-s}s\exp(-C|jB-b+k-m|_+^{r_2}) \\
&+\frac{1}{n} \sum_{j=1}^{L-1}(L-j) \sum_{k=b-s+1}^{b-1}(b-k)\exp(-C|jB-b+k-m|_+^{r_2})\notag \\
\lesssim&~ \frac{1}{n} \sum_{j=1}^{L-1}(L-j)\sum_{k=-s+1}^{0}(s-|k|)\exp(-C|jB-b+k-m|_+^{r_2}) \notag\\
&+ \frac{1}{n} \sum_{j=1}^{L-1}(L-j)\sum_{k=1}^{b-s}s\exp(-C|jB-b+k-m|_+^{r_2}) \notag\,.
\end{align}
Since $m<s$, we know $jB-b+k-m\geqslant 2B-b-s+1-m= B+1-m>0$ for any $j\geqslant 2$ and $-s+1\leqslant k\leqslant b-1$. Similar to \eqref{eq:longrunestsum}, for any $j\geqslant 2$, it holds that
$
\sum_{k=-s+1}^0(s-|k|) \exp\{-C(jB-b+k-m)^{r_2}\} \lesssim (jB-B-m)^{2-\delta}$,
which implies
$n^{-1}\sum_{j=2}^{L-1}(L-j)\sum_{k=-s+1}^0(s-|k|)\exp\{-C(jB-b+k-m)^{r_2}\}\lesssim B^{-1}(B-m)^{2-\delta}\ll sb^{-1}$.
Also, $n^{-1}\sum_{j=1}^{L-1}(L-j)\sum_{k=1}^{b-s}s\exp\{-C(jB-b+k-m)^{r_2}\}\lesssim sb^{-1}$. Therefore, by \eqref{eq:R25bd1}, we have
\begin{align}\label{eq:R25bd2}
\max_{j_1,j_2\in[d_z]}R_{2,5,1}(j_1,j_2)
\lesssim&~sb^{-1}+\frac{1}{b}\sum_{k=-s+1}^{0}(s-|k|)\exp(-C|B-b+k-m|_+^{r_2})\,.
%&+ \frac{1}{b}\sum_{k=1}^{b-s}s\exp(-C|B-b+k-m|_+^{r_2})\,.
\end{align}
Then we have $b^{-1}\sum_{k=-s+1}^0 (s-|k|)\exp(-C|B-b+k-m|_+^{r_2})=b^{-1}\sum_{k=0}^{s-m}(m+k) \exp(-Ck^{r_2})
+b^{-1}\sum_{k=s-m+1}^{s-1}(s-k)
\lesssim m^2b^{-1}$
%and $b^{-1}\sum_{k=1}^{b-s}s\exp\{-C(B-b+k-m)^{r_2}\}\lesssim sb^{-1}$.
 Thus, \eqref{eq:R25bd2} yields that
$
\max_{j_1,j_2\in[d_z]}R_{2,5,1}(j_1,j_2)
\lesssim m^2b^{-1}+sb^{-1}$. Similarly, we can also show $R_{2,5,2}(j_1,j_2)\lesssim m^2b^{-1}+sb^{-1}$. To bound $\max_{j_1,j_2\in[d_z]}R_{2,5,3}(j_1,j_2)$, we consider two cases: (i) $n-LB\geqslant b $ and (ii) $n-LB<b$.
In Case (i), it follows from the triangle inequality and Davydov's inequality that
\begin{align}\label{eq:R253bd2}
\max_{j_1,j_2\in[d_z]}R_{2,5,3}(j_1,j_2)
%\lesssim &~\frac{1}{n}\sum_{\ell_1=1}^{L}\sum_{t_1=(\ell_1-1)B+1}^{(\ell_1-1)B+b} \sum_{t_2=LB+1}^n \exp(-C|t_2-t_1-m|_+^{r_2}) \notag\\
\lesssim &~\frac{1}{n}\sum_{j=1}^L \sum_{k=1-b}^0(b+k)\exp(-C|jB+k-m|_+^{r_2})\notag\\
&+ \frac{1}{n}\sum_{j=1}^L\sum_{k=1}^{n-LB-b}b\exp\{-C(jB+k-m)^{r_2}\} \notag\\
& +\frac{1}{n}\sum_{j=1}^L\sum_{k=n-LB-b+1}^{n-LB-1}(n-LB-k)\exp\{-C(jB+k-m)^{r_2}\}\\
%\lesssim &~\frac{1}{n}\sum_{j=1}^L \sum_{k=1-b}^0(b+k)\exp(-C|jB+k-m|_+^{r_2})\notag\\
%&+ \frac{1}{n}\sum_{j=1}^L\sum_{k=1}^{n-LB-b}b\exp\{-C(jB+k-m)^{r_2}\} \notag\\
\lesssim &~\frac{b}{n}+\frac{1}{n}\sum_{j=1}^L \sum_{k=1-b}^0(b+k)\exp(-C|jB+k-m|_+^{r_2})\,,\notag
\end{align}
where the last step is due to the fact $\sum_{j=1}^L\sum_{k=1}^{n-LB-b}\exp\{-C(jB+k-m)^{r_2}\}\leqslant \sum_{j=1}^\infty \exp(-Cj^{r_2})<\infty$. Note that $n^{-1}\sum_{j=1}^L \sum_{k=1-b}^0(b+k)\exp(-C|jB+k-m|_+^{r_2})\leqslant n^{-1}\sum_{k=1-b}^0(b+k)\exp(-C|LB+k-m|_+^{r_2})+n^{-1}\sum_{j=1}^{L-1} \sum_{k=1-b}^{b-1}(L-j)(b-|k|)\exp(-C|jB+k-m|_+^{r_2})\lesssim bn^{-1}+n^{-1}\sum_{j=1}^{L-1} \sum_{k=1-b}^{b-1}(L-j)(b-|k|)\exp(-C|jB+k-m|_+^{r_2})$. As we have shown in the procedure for deriving the convergence rate of $\max_{j_1,j_2\in[d_z]}R_{2,4}(j_1,j_2)$, it holds that
$n^{-1}\sum_{j=1}^{L-1} \sum_{k=1-b}^{b-1}(L-j)(b-|k|)\exp(-C|jB+k-m|_+^{r_2})\lesssim(s-m+1)^{2-\delta}b^{-1}$, which implies that
$
\max_{j_1,j_2\in[d_z]}R_{2,5,3}(j_1,j_2)\lesssim
bn^{-1}+(s-m+1)^{2-\delta}b^{-1}$ in Case (i). In Case (ii), noting $n-LB-b<0$, analogous to \eqref{eq:R253bd2}, we have
\begin{align}\label{eq:R253bd3}
 \max_{j_1,j_2\in[d_z]}R_{2,5,3}(j_1,j_2)
\lesssim&~ \frac{1}{n}\sum_{j=1}^L\sum_{k=1-b}^{n-LB-b}(b+k)\exp(-C|jB+k-m|_+^{r_2}) \notag\\
&+\frac{1}{n}\sum_{j=1}^L\sum_{k=n-LB-b+1}^0 (n-LB)\exp(-C|jB+k-m|_+^{r_2})\\
\leqslant&~\frac{1}{n}\sum_{j=1}^L\sum_{k=1-b}^{0}(b+k)\exp(-C|jB+k-m|_+^{r_2}) \notag\\
\lesssim &~\frac{b}{n}+\frac{1}{(s-m+1)^{\delta-2}b}\notag
\end{align}
Combining the above results, we have
$\max_{j_1,j_2\in[d_z]}R_{2,5,3}(j_1,j_2)\lesssim bn^{-1}+(s-m+1)^{2-\delta}b^{-1}$
regardless of $n-LB\geqslant b$ or not. Therefore,
$
\max_{j_1,j_2\in[d_z]}R_{2,5}(j_1,j_2)\lesssim bn^{-1}+m^2b^{-1}+sb^{-1}\lesssim bn^{-1}+msb^{-1}$.

{\bf Convergence rate of $\max_{j_1,j_2\in[d_z]}R_{2,7}(j_1,j_2)$.} It holds that
\begin{align*}
\max_{j_1,j_2\in[d_z]}R_{2,7}(j_1,j_2)
\leqslant&~2\max_{j_1,j_2\in[d_z]}\bigg|\frac{1}{n}\sum_{\ell_1>\ell_2}\sum_{t_1\in\cS_{\ell_1}}\sum_{t_2\in\cS_{\ell_2}} \E(z_{t_1,j_1}z_{t_2,j_2})\bigg|\\
\lesssim&~\frac{1}{n}\sum_{\ell_2<\ell_1\leqslant L} \sum_{t_1\in\cS_{\ell_1}}\sum_{t_2\in\cS_{\ell_2}} \exp(-C|t_1-t_2-m|_+^{r_2})\\
&+ \frac{1}{n}\sum_{\ell_2=1}^L \sum_{t_2\in\cS_{\ell_2}} \sum_{t_1\in\cS_{L+1}} \exp(-C|t_1-t_2-m|_+^{r_2}) \\
=&~ \frac{1}{n} \sum_{j=1}^{L-1} \sum_{k=-s+1}^{s-1}(L-j)(s-|k|)\exp(-C|jB+k-m|_+^{r_2})\\
&+ \frac{1}{n}\sum_{\ell_2=1}^L \sum_{t_2\in\cS_{\ell_2}} \sum_{t_1\in\cS_{L+1}} \exp(-C|t_1-t_2-m|_+^{r_2}) \\
\lesssim &~\frac{1}{b(b-m)^{\delta-2}} + \underbrace{\frac{1}{n}\sum_{\ell_2=1}^L \sum_{t_2\in\cS_{\ell_2}} \sum_{t_1\in\cS_{L+1}} \exp(-C|t_1-t_2-m|_+^{r_2})}_{\Delta}
\end{align*}
where the last step follows from the similar arguments used to bound \eqref{eq:longrunestbd2}. In the sequel, we consider the second term $\Delta$ on the right-hand side of above inequality in two cases: (i) $n-LB\geqslant s$ and (ii) $n-LB<s$. In Case (i), due to $jB-b+k-m\geqslant 2B-b+1-s-m>0$ for any $j\geqslant 2$ and $1-s\leqslant k\leqslant n-LB-1$, and $jB-b+k-m>0$ for any $j\geqslant 1$ and $1\leqslant k\leqslant n-LB-1$ similar to \eqref{eq:R253bd2}, it holds that
\begin{align*}
\Delta
\lesssim&~ \frac{1}{n}\sum_{j=2}^L \sum_{k=1-s}^0 (s+k)\exp\{-C(jB-b+k-m)^{r_2}\}
+\frac{1}{n}\sum_{k=1-s}^0 (s+k)\exp(-C|s+k-m|_+^{r_2})\\
&+ \frac{1}{n}\sum_{j=1}^L\sum_{k=1}^{n-LB-s} s\exp\{-C(jB-b+k-m)^{r_2}\} \\
\lesssim&~ \frac{s}{n}
+\frac{1}{n}\sum_{k=0}^{s-1} (s-k)\exp(-C|s-k-m|_+^{r_2}) \,.
%+\frac{1}{n}\sum_{k=1}^{n-LB-s}s\exp(-C|s+k-m|_+^{r_2})\,.
\end{align*}
Due to $m< s$, we have
$n^{-1}\sum_{k=0}^{s-1} (s-k)\exp(-C|s-k-m|_+^{r_2})
% +n^{-1}\sum_{k=1}^{n-LB-s}s\exp(-C|s+k-m|_+^{r_2})
= n^{-1}\sum_{k=s-m}^{s-1}(s-k)
 +n^{-1}\sum_{k=0}^{s-m-1}(s-k)\exp\{-C(s-k-m)^{r_2}\}
% +n^{-1}\sum_{k=1}^{n-LB-s}s\exp\{-C(s+k-m)^{r_2}\}
\lesssim m^2n^{-1}$.
Thus, $\Delta\lesssim msn^{-1}$ in Case (i). In Case (ii), due to $jB-b+k-m\geqslant 2B-b+1-s-m>0$ for any $j\geqslant 2$ and $1-s\leqslant k\leqslant n-LB-1$, similar to \eqref{eq:R253bd3}, it holds that
\begin{align*}
\Delta
\lesssim &~ \frac{1}{n}\sum_{j=1}^L\sum_{k=1-s}^{0} (s+k)\exp(-C|jB-b+k-m|_+^{r_2})\\
\lesssim&~ \frac{s}{n}+\frac{1}{n}\sum_{k=1-s}^{0} (s+k)\exp(-C|s+k-m|_+^{r_2})\lesssim\frac{s}{n}+\frac{ms}{n}\lesssim\frac{ms}{n}\,,
\end{align*}
where the third step is based on the result $n^{-1}\sum_{k=0}^{s-1} (s-k)\exp(-C|s-k-m|_+^{r_2})\lesssim msn^{-1}$ that has shown in Case (i). Thus $\max_{j_1,j_2\in[d_z]}R_{2,7}(j_1,j_2)\lesssim msn^{-1}+o(b^{-1})$. $\hfill\Box$

\subsection{Proof of  Lemma \ref{lem.varrho2}}\label{sec:pflavarrho2}
%\begin{lemma}\label{lem.In}
%Assume (A.1)--(A.3) hold, then
%\begin{align}\label{In}
% \sup_{\bu\in\R^{d_z},\nu\in[0,1]} \big|\E(\cI_n)\big|
%  &\lesssim  \phi\sqrt{L} \max_{\ell\in[L]} \E\Big\{ \max_{j\in[d_z]} \big| \E(\tilde z_{\ell,j}\,|\,\cF_{-\ell}) \big| \Big\} \nonumber\\
%  & ~~~ + \, \phi^2(\log d_z) \max_{\ell\in[L]} \E\Big[ \max_{k,j\in[d_z]} \big| \E\big\{ \tilde z_{\ell,k}\tilde z_{\ell,j} - \E(\tilde z_{\ell,k}\tilde z_{\ell,j})\,|\,\cF_{-\ell} \big\} \big| \Big] \\
%   &  ~~~ + \, \frac{\phi^3 (\log d_z)^2}{\sqrt{L}} \max_{\ell\in[L]} \E \Big[ \max_{j\in[d_z]} \big| \E\{ |\tilde z_{\ell,j}|^3 - \E(|\tilde z_{\ell,j}|^3 ) \,|\,\cF_{-\ell} \} \big| \Big]  \nonumber\\
%   &  ~~~ + \, \frac{\phi^3(\log d_z)^2}{\sqrt{L}} \big\{ \phi^{-1}(\log d_z)^{1/2} + \varrho_n^{(1)} + M_{\tilde \bz}(\phi) + M_{\tilde \by}(\phi) \big\} \,,   \nonumber
%\end{align}
%where $\phi=\beta^{-1}\log d_z$ and $\cF_{-\ell}$ is the $\sigma$-field generated by $\{\tilde \bz_s\}_{s\neq\ell}$.
%\end{lemma}
% This lemma is directly obtained from Appendix A of \cite{CCW2020}.

Recall $\bs_{n,z}^{(1)}=L^{-1/2}\sum_{\ell=1}^L \tilde\bz_\ell$ and $\bs_{n,y}^{(1)}=L^{-1/2}\sum_{\ell=1}^L \tilde \by_\ell$. To construct Lemma \ref{lem.varrho2}, we need Lemma \ref{lem.bern} and the following lemma whose proofs are given in Sections \ref{sec:pflembern} and \ref{sec:pflem.varrho1}, respectively.

%\newtheorem{lem}{Lemma}
%\setcounter{lem}{0}
%\renewcommand{\thelem}{L\arabic{lem}}

%\begin{lem}\label{lem.bern}
%Under {\rm AS1--AS3},
%it holds that
%\begin{align}\label{bern.inq}
%  \max_{0\leqslant a\leqslant n-q}\max_{j\in[d_z]}\bP\bigg(\max_{k\in[q]} \bigg|\sum_{t=a+1}^{a+k} z_{t,j}\bigg| \geqslant x\bigg) \lesssim
%  & ~ \exp(-Cq^{-1}m^{-1}x^2) + qx^{-1}\exp(-Cx^r) \nonumber\\
%  &  +qx^{-1}\exp(-Cm^{-r_1} x^{r_1})
%\end{align}
%for any $x>0$ and $m\leqslant q\leqslant n$, where $r=r_1r_2/(r_1+r_2)$.
%\end{lem}	

\begin{lemma}\label{lem.varrho1}
Assume {\rm AS1--AS3} hold and $d_z\geqslant n^\varpi$ for some $\varpi>0$. Let  $r=r_1r_2/(r_1+r_2)$ and $\eta=m^{1/2}L^{1/3} (\log d_z)^{-1/3}$. If $s\geqslant m+C(\log d_{z})^{1/r_2}$ for some sufficiently large $C>0$ and $msb^{-1}+mbn^{-1}\ll 1$, it holds that
\begin{align*}
\check{\varrho}_n:=&~\sup_{\bu\in\R^{d_{z}},\nu\in[0,1]} \big| \bP\{\sqrt{\nu}\bs_{n,z}^{(1)} + \sqrt{1-\nu}\bs_{n,y}^{(1)} \leqslant \bu\} - \bP\{\bs_{n,y}^{(1)}\leqslant\bu\} \big|\lesssim\frac{m^{1/2}(\log d_z)^{7/6}}{L^{1/6}}
\end{align*}
provided that
$\log d_{z}\ll\min\{L^{2/5}, m^{3r/(6+2r)}b^{3r/(6+2r)}L^{r/(3+r)}, m^{-3r_1/(6+2r_1)}b^{3r_1/(6+2r_1)}L^{r_1/(3+r_1)}\}$ with $m \ll n^{2/3}b^{1/3}(\log n)^{-(6+2r_1)/(3r_1)}$.
\end{lemma}

Now we begin to prove Lemma \ref{lem.varrho2}. Notice that
\begin{align*}
  \bs_{n,z}=&~ \bs_{n,z}^{(1)} + \frac{1}{\sqrt{n}} \sum_{\ell=1}^{L+1} \sum_{t\in\cS_\ell} \bz_t + \bigg( \frac{1}{\sqrt{n}} - \frac{1}{\sqrt{Lb}}\bigg) \sum_{\ell=1}^L \sum_{t\in\cB_\ell} \bz_t=:\bs_{n,z}^{(1)} + \bdel_n \,.
\end{align*}
Given some $D_n>0$, define an event $\cE=\{|\bdel_n|_\infty \leqslant D_n \}$. Then we have
\begin{align*}
   &\bP\{\sqrt{\nu}\bs_{n,z}  +  \sqrt{1-\nu}\bs_{n,y}^{(1)}  \leqslant \bu\}    - \bP\{\bs_{n,y}^{(1)}\leqslant\bu\} \\
   &~~~~~\leqslant\bP\{\sqrt{\nu}\bs_{n,z}^{(1)} + \sqrt{1-\nu}\bs_{n,y}^{(1)} \leqslant \bu - \sqrt{\nu}\bdel_n, ~ \cE\} - \bP\{\bs_{n,y}^{(1)}\leqslant\bu\} + \bP(\cE^c) \\
   &~~~~~\leqslant\bP\{\sqrt{\nu}\bs_{n,z}^{(1)} + \sqrt{1-\nu}\bs_{n,y}^{(1)} \leqslant \bu + \sqrt{\nu}D_n\}
   - \bP\{\bs_{n,y}^{(1)}\leqslant\bu +  \sqrt{\nu}D_n \}  \\
   &~~~~~~~~ + \bP\{\bs_{n,y}^{(1)}\leqslant\bu +  \sqrt{\nu}D_n \} - \bP\{\bs_{n,y}^{(1)}\leqslant\bu \}+ \bP(\cE^c) \\
   &~~~~~\leqslant\check{\varrho}_n + \bP\{\bs_{n,y}^{(1)}\leqslant\bu +  \sqrt{\nu}D_n \} - \bP\{\bs_{n,y}^{(1)}\leqslant\bu \} + \bP(\cE^c)\,.
\end{align*}
Similarly, we also have
$
\bP\{\sqrt{\nu}\bs_{n,z}  +  \sqrt{1-\nu}\bs_{n,y}^{(1)}  \leqslant \bu\}    - \bP\{\bs_{n,y}^{(1)}\leqslant\bu\}
\geqslant - \check{\varrho}_n +
\bP\{\bs_{n,y}^{(1)}\leqslant\bu -  \sqrt{\nu}D_n \} - \bP\{\bs_{n,y}^{(1)}\leqslant\bu \} - \bP(\cE^c)$.
According to {\rm AS3} and Lemma \ref{lem.covest}, we have $\Xi_{1,j,j}\geqslant c$ for any $j\in[d_z]$ provided that $msb^{-1}+mbn^{-1}=o(1)$, where $\Xi_{1,j,j}$ is the $j$th diagonal element of $\bXi_1$. And the following discussion is restricted on $msb^{-1}+mbn^{-1}=o(1)$.
Then by
Nazarov's inequality (Lemma A.1, \citealp{CCK2017}), it holds that
$
  \sup_{\bu\in\R^{d_z}}|\bP\{ \bs_{n,y}^{(1)} \leqslant \bu+\sqrt{\nu}D_n \}
   - \bP\{\bs_{n,y}^{(1)} \leqslant \bu\}|\lesssim \sqrt{\nu} D_n (\log d_z)^{1/2}$ and $\sup_{\bu\in\R^{d_z}}|\bP\{ \bs_{n,y}^{(1)} \leqslant \bu-\sqrt{\nu}D_n \}
   - \bP\{\bs_{n,y}^{(1)} \leqslant \bu\}|\lesssim \sqrt{\nu} D_n (\log d_z)^{1/2}$.
Thus, with selecting $D_n=C m^{1/2}L^{-1/6}(\log d_z)^{2/3}$ for some sufficiently large $C>0$, $\bar{\varrho}_n$ defined as \eqref{varrho2} satisfies
\begin{align}\label{rho2.inq2}
  \bar{\varrho}_n \lesssim \check{\varrho}_n + D_n(\log d_z)^{1/2} + \bP(\cE^c)\lesssim \frac{m^{1/2}(\log d_z)^{7/6}}{L^{1/6}}+\bP(\cE^c) \,,
\end{align}
where the last step is obtained by Lemma \ref{lem.varrho1}. By Bonferroni inequality, it follows that
\begin{align*}
  \bP(\cE^c)\leqslant&~ \bP\bigg(\bigg| \sum_{\ell=1}^{L+1} \sum_{t\in\cS_\ell} \bz_t  \bigg|_\infty >\frac{D_n\sqrt{n}}{2}\bigg)
       + \bP\bigg\{  \bigg| \sum_{\ell=1}^L \sum_{t\in\cB_\ell} \bz_t\bigg|_\infty  >\frac{D_n\sqrt{nLb}}{2(\sqrt{n}-\sqrt{Lb})}\bigg\} \\
   \leqslant&~ \sum_{j=1}^{d_z}  \bP\bigg(\bigg| \sum_{\ell=1}^{L+1} \sum_{t\in\cS_\ell}z_{t,j}  \bigg| >\frac{D_n\sqrt{n}}{2}\bigg)
       + \sum_{j=1}^{d_z} \bP\bigg\{  \bigg| \sum_{\ell=1}^L \sum_{t\in\cB_\ell}z_{t,j}\bigg|  >\frac{D_n\sqrt{nLb}}{2(\sqrt{n}-\sqrt{Lb})}\bigg\}  \,.
\end{align*}
Note that, for each fixed $j\in[d_z]$, $\{z_{t,j}\}_{t\in\cB_\ell, \, \ell\in[L]}$ and $\{z_{t,j}\}_{t\in\cS_\ell,\ell\in[L+1]}$ are both $\alpha$-mixing processes with $\alpha$-mixing coefficients $\{\alpha_z(k)\}_{k\geqslant1}$. Due to $Lb\asymp n$,  identical to Lemma \ref{lem.bern}, we can show
\begin{align*}
&\max_{j\in[d_z]} \bP\bigg(\bigg| \sum_{\ell=1}^{L+1} \sum_{t\in\cS_\ell} z_{t,j}  \bigg| >\frac{D_n\sqrt{n}}{2}\bigg)+\max_{j\in[d_z]} \bP\bigg\{  \bigg| \sum_{\ell=1}^L \sum_{t\in\cB_\ell} z_{t,j}\bigg|  > \frac{D_n\sqrt{nLb}}{2(\sqrt{n}-\sqrt{Lb})} \bigg\}\\
&~~~~~~\lesssim\exp\{-C(n-Lb)^{-1}m^{-1} nD_n^2\}+ \frac{n-Lb}{n^{1/2}D_n}\big\{\exp(-Cn^{r/2}D_n^{r})+\exp(-Cm^{-r_1}n^{r_1/2}D_n^{r_1})\big\} \,.
\end{align*}
Recall $D_n=C m^{1/2}L^{-1/6}(\log d_z)^{2/3}$, $n-Lb\leqslant Ls+b+s$ and $s=o(b)$. Then it holds that
\begin{align*}
  \bP(\cE^c)\lesssim&~
  \underbrace{d_z\exp\{-CnL^{-1/3}(\log d_z)^{4/3}(Ls+b)^{-1}\}}_{{\rm I}_1}\\
  &+\underbrace{\frac{d_z(Ls+b)}{n^{1/2}m^{1/2}L^{-1/6}(\log d_z)^{2/3}}\exp\{-Cn^{r/2}m^{r/2}L^{-r/6}(\log d_z)^{2r/3}\}}_{{\rm I}_2} \\
  &+\underbrace{\frac{d_z(Ls+b)}{n^{1/2}m^{1/2}L^{-1/6}(\log d_z)^{2/3}} \exp\{-Cm^{-r_1/2}n^{r_1/2}L^{-r_1/6}(\log d_z)^{2r_1/3}\}}_{{\rm I}_3} \,.
\end{align*}
Since $n\asymp Lb$, $d_z\geqslant n^\varpi$ for some $\varpi>0$ and $\log d_{z}\ll\min\{L^{2/5}, m^{3r/(6+2r)}b^{3r/(6+2r)}L^{r/(3+r)}, m^{-3r_1/(6+2r_1)} \\ b^{3r_1/(6+2r_1)}L^{r_1/(3+r_1)}\}$, we have ${\rm I}_2\lesssim\exp\{- Cm^{r/2}b^{r/2}L^{r/3}(\log d_z)^{-r/3}\}\lesssim m^{1/2}L^{-1/6}(\log d_z)^{7/6}$ and ${\rm I}_3\lesssim \exp\{-Cm^{-r_1/2}b^{r_1/2}L^{r_1/3}(\log d_z)^{-r_1/3}\}\lesssim m^{1/2}L^{-1/6}(\log d_z)^{7/6}$.
Note that $b=o(n^{1/2})$. It follows that
$
  {\rm I}_1 \lesssim d_z\exp\{-CnL^{-4/3}s^{-1}(\log d_z)^{4/3}\} \asymp \exp\{\log d_z-Cn^{-1/3}b^{4/3}s^{-1}(\log d_z)^{4/3}\}$.
Recall $s=m+C(\log d_z)^{1/r_2}$. To make ${\rm I}_1=o(1)$, we need to impose the restriction $b\geqslant C\max\{n^{1/4}m^{3/4} (\log d_z)^{-1/4}, n^{1/4}(\log d_z)^{(3-r_2)/(4r_2)}\}$ for some sufficiently large $C>0$. Under such restriction, ${\rm I}_1\lesssim \exp(-C\log d_z)\lesssim m^{1/2}L^{-1/6}(\log d_z)^{7/6}$. Thus we have $\bP(\cE^c)\lesssim m^{1/2}L^{-1/6}(\log d_z)^{7/6}$.
Due to that $m<b=o(n^{1/2})$, the restriction $m \ll n^{2/3}b^{1/3}(\log n)^{-(6+2r_1)/(3r_1)}$ required in Lemma \ref{lem.varrho1} holds automatically.
From \eqref{rho2.inq2}, We complete the proof of Lemma \ref{lem.varrho2}. $\hfill\Box$

\subsection{Proof of Lemma \ref{lem.varrho1}}\label{sec:pflem.varrho1}
Let $\cW_n=\{\bw_1,\ldots,\bw_n\}$ be a copy of $\cY_n=\{\by_1,\ldots,\by_n\}$. Assume $\cZ_n=\{\bz_1,\ldots,\bz_n\}$, $\cY_n$ and $\cW_n$ are independent.
Recall $\bs_{n,y}^{(1)}=L^{-1/2}\sum_{\ell=1}^L \tilde \by_\ell$ with $\tilde \by_\ell=b^{-1/2}\sum_{t\in\cB_\ell} \by_t$, where $\by_t\sim\mathcal{N}\{\bzero,\E(\tilde \bz_\ell \tilde \bz_\ell^\T)\}$ for any $t\in\cB_\ell$. We can define $\bs_{n,w}^{(1)}=L^{-1/2}\sum_{\ell=1}^L \tilde \bw_\ell$ with $\tilde \bw_\ell=b^{-1/2}\sum_{t\in\cB_\ell} \bw_t$.
Then $\check{\varrho}_n$ defined in Lemma \ref{lem.varrho1} can be rewritten as
\begin{align*}
  \check{\varrho}_n=\sup_{\bu\in\R^{d_z},\nu\in[0,1]} \big| \bP\{\sqrt{\nu}\bs_{n,z}^{(1)} + \sqrt{1-\nu}\bs_{n,y}^{(1)} \leqslant \bu\} - \bP\{\bs_{n,w}^{(1)}\leqslant\bu\} \big|\,.
\end{align*}
Let $\beta:=\phi\log d_z$. For a given $\bu=(u_1,\ldots,u_{d_z})^\T\in\R^{d_z}$, define
$
F_\beta(\bv) = \beta^{-1}\log[ \sum_{j=1}^{d_z} \exp\{ \beta(v_j-u_j)\}]
$
for any $\bv=(v_1,\ldots,v_{d_z})^\T\in\R^{d_z}$, which satisfies the property
$
  0\leqslant F_\beta(\bv)-\max_{j\in[d_z]}(v_j-u_j) \leqslant \beta^{-1}\log d_z=\phi^{-1}$  for any $\bv=(v_1,\ldots,v_{d_z})^\T\in\R^{d_z}$. Select a thrice continuously differentiable function $g_0: \R\to[0,1]$ whose derivatives up to the third order are all bounded such that $g_0(t)=1$ for $t\leqslant 0$ and $g_0(t)=0$ for $t\geqslant 1$. Define $g(t)=g_0(\phi t)$ for any $t\in\R$, and
$q(\bv)=g\{F_\beta(\bv)\}$ for any $\bv\in\R^{d_z}$. Define
$
\cI_n:=q\{ \sqrt{\nu}\bs_{n,z}^{(1)} + \sqrt{1-\nu} \bs_{n,y}^{(1)} \} - q\{\bs_{n,w}^{(1)}\}$
and $\bv_n:=\sqrt{\nu}\bs_{n,z}^{(1)} + \sqrt{1-\nu} \bs_{n,y}^{(1)}$.
Then we have
\begin{align*}
  \bP(\bv_n\leqslant \bu-\phi^{-1})
   \leqslant&~ \bP\{F_\beta(\bv_n) \leqslant 0\}\leqslant\E\{q(\bv_n)\} = \E[ q\{\bs_{n,w}^{(1)} \}] + \E(\cI_n) \\
   \leqslant&~ \bP[F_\beta\{\bs_{n,w}^{(1)}\} \leqslant\phi^{-1}] + \E(\cI_n) \\
   \leqslant&~ \bP\{\bs_{n,w}^{(1)}\leqslant \bu+\phi^{-1}\} + |\E(\cI_n)| \\
   \leqslant&~ \bP\{\bs_{n,w}^{(1)}\leqslant \bu-\phi^{-1}\} + C\phi^{-1}(\log d_z)^{1/2} + |\E(\cI_n)| \,,
\end{align*}
where the first to the fourth steps follow from the properties of $F_\beta(\bv)$ and $g(t)$, and the last inequality is based on Nazarov's inequality and {\rm AS3} provided that $msb^{-1}+mbn^{-1}=o(1)$. Likewise we have
$\bP(\bv_n\leqslant \bu-\phi^{-1}) \geqslant \bP\{\bs_{n,w}^{(1)}\leqslant \bu-\phi^{-1}\} - C\phi^{-1}(\log d_z)^{1/2} - |\E(\cI_n)|$.
Then
\begin{align*}
  \check{\varrho}_n  \lesssim \phi^{-1}(\log d_z)^{1/2} + \sup_{\bu\in\R^{d_z},\nu\in[0,1]}|\E(\cI_n)| \,.
\end{align*}
Define
\begin{align}
  M_{\tilde{z}}(u) =&~ \max_{\ell\in[L]} \E[|\tilde \bz_\ell|^3_\infty I\{ |\tilde \bz_\ell|_\infty > \sqrt{L}(4u\log d_{z})^{-1} \} ] \label{eq:mtildez} \\
  M_{\tilde{y}}(u) =&~ \max_{\ell\in[L]} \E[|\tilde \by_\ell|^3_\infty I\{ |\tilde \by_\ell|_\infty > \sqrt{L} (4u\log d_{z})^{-1} \} ] \label{eq:mtildey}
\end{align}
for any $u>0$. Recall $\tilde{\bz}_{\ell}=(\tilde{z}_{\ell,1},\ldots,\tilde{z}_{\ell,d_z})^\T$.
%Identical to (A.1)\footnote{Need to check equation number in final version} in \cite{CCW2020a},
As shown in Section \ref{sec:In}, we have
\begin{align}\label{eq:In}
  \sup_{\bu\in\R^{d_z},\nu\in[0,1]} |\E(\cI_n)|
  \lesssim&~  \phi\sqrt{L} \max_{\ell\in[L]} \E\bigg\{ \max_{j\in[d_z]} \big| \E(\tilde z_{\ell,j}\,|\,\cF_{-\ell}) \big| \bigg\} \notag\\
   & +  \phi^2(\log d_z) \max_{\ell\in[L]} \E\bigg[ \max_{k,j\in[d_z]} \big| \E\big\{ \tilde z_{\ell,k}\tilde z_{\ell,j} - \E(\tilde z_{\ell,k}\tilde z_{\ell,j})\,|\,\cF_{-\ell} \big\} \big| \bigg]  \\
   & +  \frac{\phi^3(\log d_z)^2}{\sqrt{L}} \max_{\ell\in[L]} \E \bigg[ \max_{j\in[d_z]} \big| \E\big\{ |\tilde z_{\ell,j}|^3 - \E(|\tilde z_{\ell,j}|^3 )\,|\,\cF_{-\ell} \big\} \big| \bigg]  \notag\\
   &  + \frac{m^{3/2}\phi^3(\log d_z)^2}{\sqrt{L}} \big\{ \phi^{-1}(\log d_z)^{1/2} + \check{\varrho}_n \big\} \notag\\
   & + \frac{\phi^3(\log d_z)^2}{\sqrt{L}} \big\{ M_{\tilde z}(\phi) + M_{\tilde y}(\phi) \big\} \,, \notag
\end{align}
where $\cF_{-\ell}$ is the $\sigma$-filed generated by $\{\tilde \bz_s:s\neq\ell\}$.
With selecting $\phi = CL^{1/6}m^{-1/2}(\log d_z)^{-2/3}$ for some sufficiently small $C>0$, we have
\begin{align}\label{rho1.inq}
  \check{\varrho}_n
   \lesssim&~ \frac{m^{1/2}(\log d_z)^{7/6}}{L^{1/6}} + \frac{1}{m^{3/2}}M_{\tilde z} \bigg\{ \frac{C L^{1/6}}{m^{1/2}(\log d_z)^{2/3}} \bigg\} +  \frac{1}{m^{3/2}}M_{\tilde y}\bigg\{ \frac{C L^{1/6}}{m^{1/2}(\log d_z)^{2/3}} \bigg\} \notag\\
   & +  \frac{L^{2/3}}{m^{1/2}(\log d_z)^{2/3}} \max_{\ell\in[L]}  \E\bigg\{ \max_{j\in[d_z]} \big| \E(\tilde z_{\ell,j}\,|\,\cF_{-\ell}) \big| \bigg\} \\
   & +  \frac{L^{1/3}}{m(\log d_z)^{1/3}} \max_{\ell\in[L]} \E\bigg[ \max_{k,j\in[d_z]} \big| \E\big\{ \tilde z_{\ell,k}\tilde z_{\ell,j} - \E(\tilde z_{\ell,k}\tilde z_{\ell,j})\,|\,\cF_{-\ell} \big\} \big| \bigg]  \notag\\
   & + \frac{1}{m^{3/2}} \max_{\ell\in[L]} \E \bigg[ \max_{j\in[d_z]} \big| \E\big\{ |\tilde z_{\ell,j}|^3 - \E(|\tilde z_{\ell,j}|^3 )\,|\,\cF_{-\ell} \big\} \big| \bigg] \,.  \notag
\end{align}
In the sequel, we will specify the convergence rate of each term on the right-hand side of \eqref{rho1.inq}.

We first consider $M_{\tilde{y}}\{Cm^{-1/2}L^{1/6}(\log d_z)^{-2/3}\}$ and $M_{\tilde{z}}\{Cm^{-1/2}L^{1/6}(\log d_z)^{-2/3}\}$.
To simplify the notations, let $\eta=m^{1/2}L^{1/3}(\log d_z)^{-1/3}$.
Notice that $\tilde{\by}_\ell\sim\mathcal{N}\{\bzero,\mathbb{E}(\tilde{\bz}_\ell\tilde{\bz}_{\ell}^\T)\}$. Denote by $\sigma_{\ell,j,j}^2$ the $(j,j)$th element of $\mathbb{E}(\tilde{\bz}_\ell\tilde{\bz}_{\ell}^\T)$. It then holds that for sufficiently large $\eta$,
\begin{align*}
  \E\bigg\{ |\tilde \by_\ell|^3_\infty I\bigg( |\tilde \by_\ell|_\infty > \frac{\eta}{4C} \bigg) \bigg\}
  =&~\bigg(\frac{\eta}{4C}\bigg)^3 \bP\bigg( |\tilde \by_\ell|_\infty > \frac{\eta}{4C}\bigg)
  + 3\int_{\frac{\eta}{4C}}^\infty u^2\bP( |\tilde \by_\ell|_\infty > u)\,\md u \\
  \lesssim&~\eta^2d_z\exp\bigg(-\frac{\eta^2}{32C^2\max_{j\in[d_z]}\sigma_{\ell,j,j}^2}\bigg)\max_{j\in[d_z]}\sigma_{\ell,j,j}\\
  &+d_z\exp\bigg(-\frac{\eta^2}{32C^2\max_{j\in[d_z]}\sigma_{\ell,j,j}^2}\bigg)\max_{j\in[d_z]}\sigma_{\ell,j,j}^3\,.
  \end{align*}
By Lemma \ref{lem.bern}, we have $\max_{\ell\in[L]}\max_{j\in[d_z]}\sigma_{\ell,j,j}\lesssim m^{1/2}$. Then if $\log d_z = o(L^{2/5})$, it holds that
\begin{align}\label{My.inq}
  M_{\tilde y} \bigg\{\frac{CL^{1/6}}{m^{1/2}(\log d_z)^{2/3}} \bigg\}
    \lesssim d_zm^{1/2}(\eta^2+m)\exp(-Cm^{-1}\eta^2)\lesssim \frac{m^{3/2}(\log d_z)^{7/6}}{L^{1/6}} \,.
\end{align}
By Bonferroni inequality and Lemma \ref{lem.bern}, it holds that
$\bP( |\tilde\bz_\ell |_\infty > u )
    \lesssim d_z \exp(-Cm^{-1}u^2)
        + d_z b^{1/2}u^{-1} \\ \exp(-Cb^{r/2}u^r)+  d_zb^{1/2}u^{-1} \exp(-Cm^{-r_1}b^{r_1/2}u^{r_1})$
for any $u>0$. Noticing that $b>m$ and $d_z\geqslant n^\varpi$ for some $\varpi>0$, analogous to \eqref{My.inq}, we can also show that
$
  M_{\tilde z}\{Cm^{-1/2}L^{1/6}(\log d_z)^{-2/3}\}\lesssim m^{3/2}L^{-1/6}(\log d_z)^{7/6}$
provided that $\log d_{z}\ll\min\{L^{2/5}, m^{3r/(6+2r)}b^{3r/(6+2r)}L^{r/(3+r)}, m^{-3r_1/(6+2r_1)}b^{3r_1/(6+2r_1)}L^{r_1/(3+r_1)}\}$.

Next we consider the other three terms on the right-hand side of \eqref{rho1.inq}. Given some $D_{1n}\to\infty$, we have
$
  \E(\tilde z_{\ell,j} \,|\, \cF_{-\ell} )
  =\E\{\tilde z_{\ell,j} I (|\tilde z_{\ell,j}|\leqslant D_{1n} ) \,|\, \cF_{-\ell} \} - \E\{\tilde z_{\ell,j} I (|\tilde z_{\ell,j}|\leqslant D_{1n} )  \} +  \E\{\tilde z_{\ell,j} I (|\tilde z_{\ell,j}| > D_{1n} ) \,|\, \cF_{-\ell} \big\} - \E\{\tilde z_{\ell,j} I (|\tilde z_{\ell,j}| > D_{1n} )  \}$,
which implies that
\begin{align}\label{m1.inq}
  \E\{ |\E(\tilde z_{\ell,j} \,|\, \cF_{-\ell} )| \}
   \leqslant&~ \E\big[ \big|\E\{\tilde z_{\ell,j} I (|\tilde z_{\ell,j}|\leqslant D_{1n} ) \,|\, \cF_{-\ell} \} - \E\{\tilde z_{\ell,j} I (|\tilde z_{\ell,j}|\leqslant D_{1n} ) \} \big| \big] \notag\\
   & +  2\E\{ |\tilde z_{\ell,j} | I (|\tilde z_{\ell,j}|>D_{1n} ) \} \,.
\end{align}
It follows from Lemma \ref{lem.bern} that
\begin{align}\label{m12.inq}
  \E\{ |\tilde z_{\ell,j} | I (|\tilde z_{\ell,j}|>D_{1n} ) \}
   =&~ D_{1n}\bP(|\tilde z_{\ell,j}|>D_{1n} ) + \int_{D_{1n}}^\infty \bP(|\tilde z_{\ell,j}|>u ) \,\md u  \notag \\
   \lesssim&~ D_{1n}\exp(-Cm^{-1}D_{1n}^2)
   + b^{1/2}\exp(-Cb^{r/2}D_{1n}^r)  \\
   & +  b^{1/2}\exp(-Cm^{-r_1}b^{r_1/2}D_{1n}^{r_1})   \notag
\end{align}
provided that $m\ll D_{1n}^2\wedge b^{1/2}D_{1n}$.
On the other hand, due to $\E\{\tilde z_{\ell,j} I (|\tilde z_{\ell,j}|\leqslant D_{1n} )\,|\, \cF_{-\ell} \}
   =-D_{1n}+D_{1n}\bP(|\tilde z_{\ell,j}|>D_{1n} \,|\, \cF_{-\ell} )
  -2D_{1n}\bP(\tilde z_{\ell,j}>D_{1n} \,|\, \cF_{-\ell}) + \int_{-D_{1n}}^{D_{1n}}\bP( \tilde z_{\ell,j} >u \,|\, \cF_{-\ell})\,\md u$
and
$
  \E\{\tilde z_{\ell,j} I (|\tilde z_{\ell,j}|\leqslant D_{1n} )  \}
    = -D_{1n}+D_{1n}\bP(|\tilde z_{\ell,j}|>D_{1n} )
  -2D_{1n}\bP(\tilde z_{\ell,j}>D_{1n} )  + \int_{-D_{1n}}^{D_{1n}}\bP( \tilde z_{\ell,j} >u )\,\md u$, we have
$
\E[|\E\{\tilde z_{\ell,j} I (|\tilde z_{\ell,j}|\leqslant D_{1n}) \,|\, \cF_{-\ell} \}  - \E\{\tilde z_{\ell,j} I (|\tilde z_{\ell,j}|\leqslant D_{1n})  \}|] \leqslant6D_{1n}\bP(|\tilde z_{\ell,j}|>D_{1n} )
  + \int_{-D_{1n}}^{D_{1n}}\E\{ | \bP( \tilde z_{\ell,j} >u \,|\, \cF_{-\ell}\big) - \bP( \tilde z_{\ell,j} >u  ) | \}\,\md u$. By Lemma \ref{lem.bern}, $\bP(|\tilde{z}_{\ell,j}|>D_{1n})\lesssim \exp(-Cm^{-1}D_{1n}^2)+b^{1/2}D_{1n}^{-1}\exp(-Cb^{r/2}D_{1n}^r)+b^{1/2}D_{1n}^{-1}\exp(-Cm^{-r_1}b^{r_1/2}D_{1n}^{r_1})$.
According to Equation (1.10c) of \cite{Rio2013}, we have
$
\int_{-D_{1n}}^{D_{1n}}\E\{| \bP( \tilde z_{\ell,j} >u \,|\, \cF_{-\ell}) - \bP( \tilde z_{\ell,j} >u  ) | \}\,\md u \lesssim D_{1n}\alpha_z(s)$.
Therefore,
\begin{align*}
  &\max_{\ell\in[L],j\in[d_z]}\E\big[ \big|\E\{\tilde z_{\ell,j} I (|\tilde z_{\ell,j}|\leqslant D_{1n} ) \,|\, \cF_{-\ell} \} - \E\{\tilde z_{\ell,j} I (|\tilde z_{\ell,j}|\leqslant D_{1n} )  \} \big| \big]  \\
  &~~~~~\lesssim D_{1n}\exp(-Cm^{-1}D_{1n}^2) +b^{1/2}\exp(-Cb^{r/2}D_{1n}^r) +b^{1/2}\exp(-Cm^{-r_1}b^{r_1/2}D_{1n}^{r_1}) + D_{1n}\alpha_z(s) \,.
\end{align*}
Together with \eqref{m12.inq}, it follows from \eqref{m1.inq} that
\begin{align*}
  \max_{\ell\in[L]} \E\bigg\{\max_{j\in[d_z]} |\E(\tilde z_{\ell,j} \,|\, \cF_{-\ell} )|\bigg\}
  \leqslant&~ \max_{\ell\in[L]} \sum_{j=1}^{d_z} \E\big\{|\E(\tilde z_{\ell,j} \,|\, \cF_{-\ell})|\big\} \\
  \lesssim&~ d_z D_{1n}\exp(-Cm^{-1}D_{1n}^2) + d_z b^{1/2}\exp(-Cb^{r/2}D_{1n}^r) \\
  & + d_z b^{1/2}\exp(-Cm^{-r_1}b^{r_1/2}D_{1n}^{r_1}) + d_z D_{1n}\exp(-C|s-m|_+^{r_2})  \,.
\end{align*}
Select $D_{1n}=C' \eta=C'm^{1/2}L^{1/3}(\log d_z)^{-1/3}$ and $s\geqslant m+C'\{\log(nd_{z})\}^{1/r_2}$ for sufficiently large $C'>0$.
Due to $\log d_{z}\ll\min\{L^{2/5}, m^{3r/(6+2r)}b^{3r/(6+2r)}L^{r/(3+r)}, m^{-3r_1/(6+2r_1)}b^{3r_1/(6+2r_1)}L^{r_1/(3+r_1)}\}$,
then we have $\max_{\ell\in[L]} \E\{\max_{j\in[d_z]} |\E(\tilde z_{\ell,j} \,|\, \cF_{-\ell} )|\} \lesssim m^{7/4}\eta^{-5/2}$,
which implies that
\begin{align}\label{M1}
 \frac{L^{2/3}}{m^{1/2}(\log d_z)^{2/3}}\max_{\ell\in[L]} \E\bigg\{\max_{j\in[d_z]} |\E(\tilde z_{\ell,j} \,|\, \cF_{-\ell} )|\bigg\}
 \lesssim m^{1/4}\eta^{-1/2} \lesssim \frac{m^{1/2}(\log d_z)^{7/6}}{L^{1/6}} \,.
\end{align}
By Lemma 2 of \cite{CTW2013} and Lemma \ref{lem.bern}, we have
\begin{align*}
   \max_{\ell\in[L]} \max_{k,j\in[d_z]} \bP\big\{ |\tilde z_{\ell,k} \tilde z_{\ell,j} - \E(\tilde z_{\ell,k} \tilde z_{\ell,j}) | > u \big\}
   \lesssim&~  \exp(-Cm^{-1}u) + b^{1/2}u^{-1/2}\exp(-Cb^{r/2}u^{r/2})  \\
   & +  b^{1/2}u^{-1/2}\exp(-Cm^{-r_1}b^{r_1/2}u^{r_1/2}) \\
   \max_{\ell\in[L]}\max_{j\in[d_z]} \bP\big\{\big| |\tilde z_{\ell,j}|^3 - \E(|\tilde z_{\ell,j}|^3) \big| >u  \big\}
   \lesssim&~  \exp(-Cm^{-1}u^{2/3}) + b^{1/2}u^{-1/3}\exp(-Cb^{r/2}u^{r/3})  \\
   & +  b^{1/2}u^{-1/3}\exp(-Cm^{-r_1}b^{r_1/2}u^{r_1/3})
\end{align*}
for any $u>0$. Repeating the same arguments for deriving \eqref{M1}, we can also show
\begin{align*}
 &\frac{L^{1/3}}{m(\log d_z)^{1/3}} \max_{\ell\in[L]} \E\bigg[ \max_{k,j\in[d_z]} \big| \E\{ \tilde z_{\ell,k}\tilde z_{\ell,j}  - \E(\tilde z_{\ell,k}\tilde z_{\ell,j}) \,|\, \cF_{-\ell} \} \big| \bigg]
  \lesssim \frac{m^{1/2}(\log d_z)^{7/6}}{L^{1/6}}  \,,\\
 &~~~~~~~~~~\frac{1}{m^{3/2}}\max_{\ell\in[L]} \E\bigg[ \max_{j\in[d_z]} \big| \E\{ |\tilde z_{\ell,j}|^3  - \E(|\tilde z_{\ell,j}|^3) \,|\, \cF_{-\ell} \} \big| \bigg]
  \lesssim \frac{m^{1/2}(\log d_z)^{7/6}}{L^{1/6}}
\end{align*}
provided that $\log d_{z}\ll\min\{L^{2/5}, m^{3r/(6+2r)}b^{3r/(6+2r)}L^{r/(3+r)}, m^{-3r_1/(6+2r_1)}b^{3r_1/(6+2r_1)}L^{r_1/(3+r_1)}\}$. Due to $d_z\geqslant n^\varpi$ for some $\varpi>0$, we also need to require
$
  \log n \ll m^{-3r_1/(6+2r_1)}b^{3r_1/(6+2r_1)}L^{r_1/(3+r_1)}$, which is equivalent to $
  m \ll n^{2/3}b^{1/3}(\log n)^{-(6+2r_1)/(3r_1)}$.
 We complete the proof of Lemma \ref{lem.varrho1}. $\hfill\Box$

\subsection{Proof of \eqref{eq:In}}\label{sec:In}
Define $\mathring\bz(t)=\sum_{\ell=1}^L\mathring\bz_\ell(t)$, where $\mathring\bz_\ell(t)=L^{-1/2}\{\sqrt{t}(\sqrt{\nu}\tilde{\bz}_\ell+\sqrt{1-\nu}\tilde{\by}_\ell)+\sqrt{1-t}\tilde{\bw}_\ell\}$.
Then $\mathring\bz(1)=\sqrt{\nu}\bs_{n,z}^{(1)}+\sqrt{1-\nu}\bs_{n,y}^{(1)}$ and $\mathring\bz(0)=\bs_{n,w}^{(1)}$.
Let $\mathring\bz^{(-\ell)}(t)=\mathring\bz(t)-\mathring\bz_\ell(t)$ and
$\dot{\bz}_\ell(t):=L^{-1/2}\{t^{-1/2}(\sqrt{\nu}\tilde{\bz}_\ell+\sqrt{1-\nu}\tilde{\by}_\ell) -(1-t)^{-1/2}\tilde{\bw}_\ell\}=\{\dot{z}_{\ell,1}(t),\ldots,\dot{z}_{\ell,p}(t)\}^{\T}$.
For brevity of notation, we write $\partial_jq(\bv)=\partial q(\bv)/\partial v_j$, $\partial_{jk}q(\bv)=\partial^2 q(\bv)/\partial v_j\partial v_k$ and $\partial_{jkl}q(\bv)=\partial^3 q(\bv)/\partial v_j\partial v_k\partial v_l$.
Note that $\cI_n=q\{\sqrt{\nu}\bs_{n,z}^{(1)}+\sqrt{1-\nu}\bs_{n,y}^{(1)}\}-q\{\bs_{n,w}^{(1)}\}=\int_0^1\,\md q\{\mathring\bz(t)\}=2^{-1}\sum_{j=1}^{d_z}\sum_{\ell=1}^L\int_0^1\partial_jq\{\mathring\bz(t)\}  \dot{z}_{\ell,j}(t)\,\md t$. By Taylor expansion,
\begin{align*}
  2\E(\cI_n)
  =&~\underbrace{\sum_{j=1}^{d_z}\sum_{\ell=1}^L\int_0^1\E[\partial_jq\{\mathring\bz^{(-\ell)}(t)\}\dot{z}_{\ell,j}(t)]\,\md t}_{\rm I}
   + \underbrace{\sum_{j,k=1}^{d_z}\sum_{\ell=1}^L\int_0^1\E[\partial_{jk}q\{\mathring\bz^{(-\ell)}(t)\} \mathring{z}_{\ell,k}(t)\dot{z}_{\ell,j}(t)]\,\md t}_{\rm II} \\
  &~ + \underbrace{\sum_{j,k,l=1}^{d_z}\sum_{\ell=1}^L\int_0^1\int_0^1(1-\tau)\E[\partial_{jkl}q\{\mathring{\bz}^{(-\ell)}(t)+\tau\mathring\bz_\ell(t)\} \mathring{z}_{\ell,k}(t)\mathring{z}_{\ell,l}(t)\dot{z}_{\ell,j}(t)]\,\md\tau\md t}_{\rm III} \,.
\end{align*}
In the sequel, we will bound the three terms ${\rm I}$, ${\rm II}$ and ${\rm III}$, respectively. To simplify the notation and without causing much confusion, we write $\mathring\bz_{\ell}(t)$, $\mathring\bz^{(-\ell)}(t)$, $\mathring{z}_{\ell,j}(t)$ and $\dot{z}_{\ell,j}(t)$
as $\mathring\bz_\ell$, $\mathring\bz^{(-\ell)}$, $\mathring{z}_{\ell,j}$ and $\dot{z}_{\ell,j}$, respectively, for any $\ell\in[L]$ and $j\in[d_z]$.

For $\textrm{I}$, notice that
$
L^{1/2}\partial_jq\{\mathring\bz^{(-\ell)}\}\dot{z}_{\ell,j}=\nu^{1/2}t^{-1/2}\tilde{z}_{\ell,j}\partial_jq\{\mathring\bz^{(-\ell)}\} +(1-\nu)^{1/2}t^{-1/2}\tilde{y}_{\ell,j}\partial_jq\{\mathring\bz^{(-\ell)}\} -(1-t)^{-1/2}\tilde{w}_{\ell,j}\partial_jq\{\mathring\bz^{(-\ell)}\}$.
Since $\tilde{\by}_\ell$ is independent of $\{\tilde{\bz}_\ell\}_{\ell=1}^L$, $\{\tilde{\bw}_\ell\}_{\ell=1}^L$ and $\{\tilde{\by}_s\}_{s\neq \ell}$, we know $\mathbb{E}[\partial_jq\{\mathring\bz^{(-\ell)}\}\tilde{y}_{\ell,j}]=0$, which implies
$
\sum_{j=1}^{d_z}\sum_{\ell=1}^L\mathbb{E}[\partial_jq\{\mathring\bz^{(-\ell)}\}\tilde{y}_{\ell,j}]=0$.
Analogously, we also have $
\sum_{j=1}^{d_z}\sum_{\ell=1}^L\mathbb{E}[\partial_jq\{\mathring\bz^{(-\ell)}\}\tilde{w}_{\ell,j}]\\=0$. Then
$
{\rm I}=\nu^{1/2}L^{-1/2}\sum_{j=1}^{d_z}\sum_{\ell=1}^L\int_0^1t^{-1/2}\mathbb{E}[\tilde{z}_{\ell,j}\partial_jq\{\mathring\bz^{(-\ell)}\}]\,{\rm d}t$.
Let $\mathcal{F}_{-\ell}^*$ be the $\sigma$-filed generated by $\{\tilde{\bz}_s,\tilde{\by}_s,\tilde{\bw}_s\}_{s\neq \ell}$.
Note that $\mathbb{E}[\tilde{z}_{\ell,j}\partial_jq\{\mathring\bz^{(-\ell)}\}] =\mathbb{E}[\partial_jq\{\mathring\bz^{(-\ell)}\}\mathbb{E}(\tilde{z}_{\ell,j}\,|\,\mathcal{F}_{-\ell}^*)]$ and $\sum_{j=1}^{d_z}|\partial_jq(\bv)|\leqslant C\phi$ for any $\bv\in\mathbb{R}^{d_z}$. Recall $\mathcal{F}_{-\ell}$ is the $\sigma$-filed generated by $\{\tilde{\bz}_s\}_{s\neq \ell}$. Since $\tilde{\bz}_\ell$ is independent of $\{\tilde{\by}_s,\tilde{\bw}_s\}_{s\neq \ell}$, we have
\begin{align}\label{eq:I}
|\textrm{I}|
\lesssim&\,\,\frac{1}{\sqrt{L}}\sum_{\ell=1}^L\sum_{j=1}^{d_z}\int_0^1\frac{1}{\sqrt{t}} \mathbb{E}\bigg[|\partial_jq\{\mathring\bz^{(-\ell)}\}|\max_{j\in[d_z]}|\mathbb{E}(\tilde{z}_{\ell,j}\,|\,\mathcal{F}_{-\ell}^*)|\bigg]\,{\rm d}t\notag\\
\lesssim&\,\,\phi \sqrt{L}\max_{\ell\in[L]}\mathbb{E}\bigg\{\max_{j\in[d_z]}|\mathbb{E}(\tilde{z}_{\ell,j}\,|\,\mathcal{F}_{-\ell}^*)|\bigg\}=\phi \sqrt{L}\max_{\ell\in[L]}\mathbb{E}\bigg\{\max_{j\in[d_z]}|\mathbb{E}(\tilde{z}_{\ell,j}\,|\,\mathcal{F}_{-\ell})|\bigg\}\,.
\end{align}
For $\textrm{II}$,
since $\tilde{\by}_\ell$ is independent of $\tilde{\bw}_\ell$, and $(\tilde{\by}_\ell,\tilde{\bw}_\ell)$ is independent of $\{\tilde{\bz}_\ell\}_{\ell=1}^L$ and $\{\tilde{\by}_s,\tilde{\bw}_s\}_{s\neq\ell}$,  we have
$
\mathbb{E}[\partial_{jk}q\{\mathring\bz^{(-\ell)}\}\mathring{z}_{\ell,k}\dot{z}_{\ell,j}]=\nu L^{-1}\mathbb{E}[\partial_{jk}q\{\mathring\bz^{(-\ell)}\}\tilde{z}_{\ell,k}\tilde{z}_{\ell,j}] +(1-\nu)L^{-1}\mathbb{E}[\partial_{jk}q\{\mathring\bz^{(-\ell)}\}]\mathbb{E}(\tilde{y}_{\ell,j}\tilde{y}_{\ell,k}) -L^{-1}\mathbb{E}[\partial_{jk}q\{\mathring\bz^{(-\ell)}\}]\\ \mathbb{E}(\tilde{w}_{\ell,j}\tilde{w}_{\ell,k})$.
Due to the fact $\mathbb{E}(\tilde{z}_{\ell,j}\tilde{z}_{\ell,k})=\mathbb{E}(\tilde{y}_{\ell,j}\tilde{y}_{\ell,k})=\mathbb{E}(\tilde{w}_{\ell,j}\tilde{w}_{\ell,k})$, it holds that
$
\mathbb{E}[\partial_{jk}q\{\mathring\bz^{(-\ell)}\}\mathring{z}_{\ell,k}\dot{z}_{\ell,j}]=\nu L^{-1}\mathbb{E}[\partial_{jk}q\{\mathring\bz^{(-\ell)}\}\mathbb{E}\{\tilde{z}_{\ell,k}\tilde{z}_{\ell,j} -\mathbb{E}(\tilde{z}_{\ell,k}\tilde{z}_{\ell,j})\,|\,\mathcal{F}_{-\ell}^*\}]$.
Lemmas A.5 and A.6 of \cite{CCK2013} show that $\sum_{j,k=1}^{d_z}|\partial_{jk}q(\bv)|\lesssim\phi^2+\phi\beta$ for any $\bv\in\mathbb{R}^{d_z}$. Then
\begin{align}\label{eq:II}
|\textrm{II}|\lesssim&\,\,(\phi^2+\phi\beta)\max_{\ell\in[L]}\mathbb{E}\bigg[\max_{k,j\in[d_z]} |\mathbb{E}\{\tilde{z}_{\ell,k}\tilde{z}_{\ell,j}-\mathbb{E}(\tilde{z}_{\ell,k}\tilde{z}_{\ell,j})\,|\,\mathcal{F}_{-\ell}^*\}|\bigg]\notag\\
\lesssim&\,\,\phi^2(\log d_z)\max_{\ell\in[L]}\mathbb{E}\bigg[\max_{k,j\in[d_z]} |\mathbb{E}\{\tilde{z}_{\ell,k}\tilde{z}_{\ell,j}-\mathbb{E}(\tilde{z}_{\ell,k}\tilde{z}_{\ell,j})\,|\,\mathcal{F}_{-\ell}\}|\bigg]\,.
\end{align}

To bound $\textrm{III}$, we first define
$
\chi_\ell=I\{\max_{j\in[d_z]}(|\tilde{z}_{\ell,j}|\vee|\tilde{y}_{\ell,j}|\vee|\tilde{w}_{\ell,j}|)\leqslant \sqrt{L}/(4\beta)\}$. Then
\begin{align*}
\textrm{III}=&\,\underbrace{\sum_{j,k,l=1}^{d_z}\sum_{\ell=1}^L\int_0^1\int_0^1(1-\tau)\mathbb{E}[\chi_\ell \partial_{jkl}q\{\mathring\bz^{(-\ell)}+\tau \mathring\bz_\ell\} \mathring{z}_{\ell,k}\mathring{z}_{\ell,l}\dot{z}_{\ell,j}]\,{\rm d}\tau{\rm d}t}_{\textrm{III}_1}\\
&+\underbrace{\sum_{j,k,l=1}^{d_z}\sum_{\ell=1}^L\int_0^1\int_0^1(1-\tau)\mathbb{E}[(1-\chi_\ell) \partial_{jkl}q\{\mathring\bz^{(-\ell)}+\tau \mathring\bz_\ell\} \mathring{z}_{\ell,k}\mathring{z}_{\ell,l}\dot{z}_{\ell,j}]\,{\rm d}\tau{\rm d}t}_{\textrm{III}_2}\,.
\end{align*}
Define
$
\kappa(t)=(\sqrt{t}\wedge\sqrt{1-t})^{-1}$ and $
h(\bv,t)=I\{-\phi^{-1}-t/\beta<\max_{j\in[d_z]}(v_j-u_j)\leqslant \phi^{-1}+t/\beta\}$
for any $\bv\in\mathbb{R}^{d_z}$ and $t>0$. Lemmas A.5 and A.6 of \cite{CCK2013} show that $\sum_{j,k,l=1}^{d_z}|\partial_{jkl}q(\bv)|\lesssim\phi\beta^2$ for any $\bv\in\mathbb{R}^{d_z}$. Then
\begin{align}\label{eq:III2alpha}
|{\rm III}_2|\lesssim \phi\beta^2\sum_{\ell=1}^L\int_0^1\mathbb{E}\bigg\{(1-\chi_\ell)\max_{j,k,l\in[d_z]} |\mathring{z}_{\ell,k}\mathring{z}_{\ell,l}\dot{z}_{\ell,j}|\bigg\}\,{\rm d}t\,.
\end{align}
Observe that
$
\max_{j,k,l\in[d_z]}|\mathring{z}_{\ell,k}\mathring{z}_{\ell,l}\dot{z}_{\ell,j}|\lesssim \kappa(t)L^{-3/2}(|\tilde{\bz}_{\ell}|_\infty^3\vee|\tilde{\by}_{\ell}|_\infty^3\vee|\tilde{\bw}_{\ell}|_\infty^3)$
and
$
1-\chi_\ell\leqslant I\{|\tilde{\bz}_{\ell}|_\infty>\sqrt{L}/(4\beta)\}+I\{|\tilde{\by}_{\ell}|_\infty>\sqrt{L}/(4\beta)\}+I\{|\tilde{\bw}_{\ell}|_\infty>\sqrt{L}/(4\beta)\}$.
Thus, by Chebyshev's association inequality \cite[Lemma B.1]{CCK2017}, it holds that
\begin{align}\label{eq:bdident}
\mathbb{E}\bigg\{(1-\chi_\ell)\max_{j,k,l\in[d_z]}|\mathring{z}_{\ell,k}\mathring{z}_{\ell,l}\dot{z}_{\ell,j}|\bigg\}\lesssim \frac{\kappa(t)}{L^{3/2}}\{M_{\tilde{z}}(\phi)+M_{\tilde{y}}(\phi)\}
\end{align}
with $M_{\tilde{z}}(\cdot)$ and $M_{\tilde{y}}(\cdot)$ defined in \eqref{eq:mtildez} and \eqref{eq:mtildey}, respectively,
which implies
\begin{align}\label{eq:III2}
|{\rm III}_2|\lesssim \frac{\phi\beta^2}{\sqrt{L}}\{M_{\tilde{z}}(\phi)+M_{\tilde{y}}(\phi)\}=\frac{\phi^3(\log d_z)^2}{\sqrt{L}}\{M_{\tilde{z}}(\phi)+M_{\tilde{y}}(\phi)\}\,.
\end{align}
If $\chi_\ell=1$, then
$
\max_{j\in[d_z]}|\mathring{z}_{\ell,j}|
\leqslant(4\beta)^{-1}\{\sqrt{t}(\sqrt{\nu}+\sqrt{1-\nu})+\sqrt{1-t}\}\leqslant(4\beta)^{-1}\sqrt{3}$
for any $t\in[0,1]$.
If $h\{\mathring\bz^{(-\ell)},1\}=0$, then $\max_{j\in[d_z]}\{\mathring{z}_{j}^{(-\ell)}-u_j\}\leqslant -\phi^{-1}-\beta^{-1}$ or $\max_{j\in[d_z]}\{\mathring{z}_{j}^{(-\ell)}-u_j\}>\phi^{-1}+\beta^{-1}$.
When $\max_{j\in[d_z]}\{\mathring{z}_{j}^{(-\ell)}-u_j\}\leqslant -\phi^{-1}-\beta^{-1}$ and $\chi_\ell=1$, we have
$
F_\beta\{\mathring\bz^{(-\ell)}+\tau\mathring\bz_\ell\}\leqslant -\beta^{-1}(1-\sqrt{3}/4)
$ for any $\tau\in[0,1]$, which implies that $q\{\mathring\bz^{(-\ell)}+\tau \mathring\bz_{\ell}\}=1$ for any $t\in[0,1]$ and $\tau\in[0,1]$.
When $\max_{j\in[d_z]}\{\mathring{z}_{j}^{(-\ell)}-u_j\}> \phi^{-1}+\beta^{-1}$ and $\chi_\ell=1$, $F_\beta\{\mathring\bz^{(-\ell)}+\tau\mathring\bz_{\ell}\}>\phi^{-1}+\beta^{-1}(1-\sqrt{3}/4)$ for any $\tau\in[0,1]$, which implies $q\{\mathring\bz^{(-\ell)}+\tau\mathring\bz_{\ell}\}=0$ for any $t\in[0,1]$ and $\tau\in[0,1]$.
Therefore, if $\chi_\ell=1$ and $h\{\mathring\bz^{(-\ell)},1\}=0$, we have $\partial_{jkl}q\{\mathring\bz^{(-\ell)}+\tau\mathring\bz_{\ell}\}=0$ for any $t\in[0,1], \tau\in[0,1]$ and $j,k,l\in[d_z]$.
Lemmas A.5 and A.6 of \cite{CCK2013} indicate that there exist $U_{jkl}(\bv)$ such that $|\partial_{jkl}q(\bv)|\leqslant U_{jkl}(\bv)$ and $\sum_{j,k,l=1}^{d_z}U_{jkl}(\bv)\lesssim\phi\beta^2$ for any $\bv\in\mathbb{R}^{d_z}$.
Then $\chi_\ell |\partial_{jkl}q\{\mathring\bz^{(-\ell)}+\tau\mathring\bz_{\ell}\}|
=\chi_\ell h\{\mathring\bz^{(-\ell)},1\} |\partial_{jkl}q\{\mathring\bz^{(-\ell)}+\tau\mathring\bz_{\ell}\}|
\lesssim\chi_\ell h\{\mathring\bz^{(-\ell)},1\} U_{jkl}\{\mathring\bz^{(-\ell)}\}\leqslant h\{\mathring\bz^{(-\ell)},1\} U_{jkl}\{\mathring\bz^{(-\ell)}\}$, which implies
\begin{align}\label{eq:III1}
|\textrm{III}_1|
\lesssim&\sum_{j,k,l=1}^{d_z}\sum_{\ell=1}^L\int_0^1\mathbb{E}[h\{\mathring\bz^{(-\ell)},1\} U_{jkl}\{\mathring\bz^{(-\ell)}\}|\mathring{z}_{\ell,k}\mathring{z}_{\ell,l}\dot{z}_{\ell,j}|]\,{\rm d}t\notag\\
\lesssim&\sum_{j,k,l=1}^{d_z}\sum_{\ell=1}^L\int_0^1\frac{\kappa(t)}{L^{3/2}}\mathbb{E}[h\{\mathring\bz^{(-\ell)},1\} U_{jkl}\{\mathring\bz^{(-\ell)}\}(|\tilde{z}_{\ell,k}|^3+|\tilde{y}_{\ell,k}|^3+|\tilde{w}_{\ell,k}|^3)]\,{\rm d}t\,.
\end{align}
In the sequel, we will show
\begin{align}\label{eq:alphatoprove1}
R:=&\,\sum_{j,k,l=1}^{d_z}\sum_{\ell=1}^L\int_0^1\frac{\kappa(t)}{L^{3/2}}\mathbb{E}[h\{\mathring\bz^{(-\ell)},1\} U_{jkl}\{\mathring\bz^{(-\ell)}\}(|\tilde{z}_{\ell,k}|^3+|\tilde{y}_{\ell,k}|^3+|\tilde{w}_{\ell,k}|^3)]\,{\rm d}t\notag\\
\lesssim&\,\,\frac{\phi^3(\log d_z)^2}{\sqrt{L}}\{m^{3/2}\phi^{-1}(\log d_z)^{1/2}+m^{3/2}\check{\varrho}_n +M_{\tilde{z}}(\phi)+M_{\tilde{y}}(\phi)\}\\
&~~~~~~+\frac{\phi^3(\log d_z)^2}{\sqrt{L}} \max_{\ell\in[L]}\mathbb{E}\bigg[\max_{k\in[d_z]}|\mathbb{E}\{|\tilde{z}_{\ell,k}|^3-\mathbb{E}(|\tilde{z}_{\ell,k}|^3)\,|\,\mathcal{F}_{-\ell}\}|\bigg]\,.\notag
 \end{align}
Recall that ${\rm III}={\rm III}_1+{\rm III}_2$. Together with (\ref{eq:III2}) and \eqref{eq:III1}, we have
\begin{align*}
|{\rm III}|\lesssim&\,\,\frac{\phi^3(\log d_z)^2}{\sqrt{L}}\{m^{3/2}\phi^{-1}(\log d_z)^{1/2} +m^{3/2}\check{\varrho}_n+M_{\tilde{z}}(\phi)+M_{\tilde{y}}(\phi)\}\\
&+\frac{\phi^3(\log d_z)^2}{\sqrt{L}} \max_{\ell\in[L]}\mathbb{E}\bigg[\max_{j\in[d_z]}|\mathbb{E}\{|\tilde{z}_{\ell,j}|^3-\mathbb{E}(|\tilde{z}_{\ell,j}|^3)\,|\,\mathcal{F}_{-\ell}\}|\bigg]\,.
\end{align*}
Due to $\mathbb{E}(\mathcal{T}_n)=2^{-1}({\rm I}+{\rm II}+{\rm III})$, together with (\ref{eq:I}) and (\ref{eq:II}), we have \eqref{eq:In}.

Now we begin to prove \eqref{eq:alphatoprove1}. Write $u_{\ell,k}=|\tilde{z}_{\ell,k}|^3+|\tilde{y}_{\ell,k}|^3+|\tilde{w}_{\ell,k}|^3-\mathbb{E}(|\tilde{z}_{\ell,k}|^3)-\mathbb{E}(|\tilde{y}_{\ell,k}|^3)-\mathbb{E}(|\tilde{w}_{\ell,k}|^3)$. Then
\begin{align}\label{eq:sd}
R=&\underbrace{\sum_{j,k,l=1}^{d_z}\sum_{\ell=1}^L\int_0^1\frac{\kappa(t)}{L^{3/2}}\mathbb{E}[\chi_\ell h\{\mathring\bz^{(-\ell)},1\} U_{jkl}\{\mathring\bz^{(-\ell)}\}] \mathbb{E}(|\tilde{z}_{\ell,k}|^3+|\tilde{y}_{\ell,k}|^3+|\tilde{w}_{\ell,k}|^3)\,{\rm d}t}_{R_1}\notag\\
&+\underbrace{\sum_{j,k,l=1}^{d_z}\sum_{\ell=1}^L\int_0^1\frac{\kappa(t)}{L^{3/2}}\mathbb{E}[(1-\chi_\ell) h\{\mathring\bz^{(-\ell)},1\} U_{jkl}\{\mathring\bz^{(-\ell)}\}] \mathbb{E}(|\tilde{z}_{\ell,k}|^3+|\tilde{y}_{\ell,k}|^3+|\tilde{w}_{\ell,k}|^3)\,{\rm d}t}_{R_2}\notag\\
&+\underbrace{\sum_{j,k,l=1}^{d_z}\sum_{\ell=1}^L\int_0^1\frac{\kappa(t)}{L^{3/2}}\mathbb{E}[h\{\mathring\bz^{(-\ell)},1\} U_{jkl}\{\mathring\bz^{(-\ell)}\}u_{\ell,k}]\,{\rm d}t}_{R_3}\,.
 \end{align}
Observe that if $\chi_\ell=1$ and $h(\mathring\bz,2)=0$, then $h\{\mathring\bz^{(-\ell)},1\}=0$.  By Lemma \ref{lem.bern}, $\max_{\ell\in[L]}\max_{j\in[d_z]}\E(|\tilde{z}_{\ell,j}|^3)\lesssim m^{3/2}$.
Due to $\sum_{j,k,l=1}^{d_z}U_{jkl}\{\bz^{(-\ell)}\}\lesssim\phi\beta^2$ for any $t\in[0,1]$ and $\ell\in[L]$, by the fact that $\mathbb{E}(|\tilde y_{\ell,j}|^3)\lesssim\{\mathbb{E}( |\tilde y_{\ell,j}|^2)\}^{3/2}=\{\mathbb{E}(|\tilde z_{\ell,j}|^2)\}^{3/2}\leqslant\mathbb{E}(|\tilde z_{\ell,j}|^3)$, we have
\begin{align}\label{eq:III111}
R_1\leqslant&\,\sum_{j,k,l=1}^{d_z}\sum_{\ell=1}^L\int_0^1\frac{\kappa(t)}{L^{3/2}}\mathbb{E}[\chi_\ell h(\mathring\bz,2)U_{jkl}\{\mathring\bz^{(-\ell)}\}]\mathbb{E}(|\tilde{z}_{\ell,k}|^3+2|\tilde{y}_{\ell,k}|^3)\,{\rm d}t\notag\\
\lesssim&\,\,\frac{\phi\beta^2}{L^{3/2}}\sum_{\ell=1}^L\int_0^1\kappa(t)\mathbb{E}\{h(\mathring\bz,2)\} \max_{k\in[d_z]}\mathbb{E}(|\tilde{z}_{\ell,k}|^3+|\tilde{y}_{\ell,k}|^3)\,{\rm d}t\\
\lesssim&\,\,\frac{\phi\beta^2m^{3/2}}{\sqrt{L}}\int_0^1\kappa(t)\mathbb{E}\{h(\mathring\bz,2)\}\,{\rm d}t\,.\notag
\end{align}
Recall that $\check{\varrho}_n=\sup_{\bu\in\mathbb{R}^{d_z},\nu\in[0,1]}|\mathbb{P}\{\sqrt{\nu}\bs_{n,z}^{(1)} +\sqrt{1-\nu}\bs_{n,y}^{(1)}\leqslant \bu\}-\mathbb{P}\{\bs_{n,y}^{(1)}\leqslant \bu\}|$ with $\bs_{n,z}^{(1)}=L^{-1/2}\sum_{\ell=1}^L\tilde{\bz}_\ell$ and $\bs_{n,y}^{(1)}=L^{-1/2}\sum_{\ell=1}^L\tilde{\by}_\ell$.
Since
$\mathring\bz=L^{-1/2}\sum_{\ell=1}^L\{\sqrt{t\nu}\tilde{\bz}_\ell+\sqrt{t(1-\nu)}\tilde{\by}_\ell +\sqrt{1-t}\tilde{\bw}_\ell\} \overset{d}{=}\sqrt{t\nu}\bs_{n,z}^{(1)}+\sqrt{1-t\nu}{\bs}_{n,y}^{(1)}$,
we have $\mathbb{P}(\mathring\bz\leqslant \bu+\phi^{-1}+2\beta^{-1})\leqslant \mathbb{P}\{\bs_{n,y}^{(1)}\leqslant \bu+\phi^{-1}+2\beta^{-1}\}+\check\varrho_n$ and $\mathbb{P}(\mathring\bz\leqslant \bu-\phi^{-1}-2\beta^{-1})\geqslant \mathbb{P}\{\bs_{n,y}^{(1)}\leqslant \bu-\phi^{-1}-2\beta^{-1}\}-\check\varrho_n$. Then
$
\mathbb{E}\{h(\mathring\bz,2)\}=\mathbb{P}(\mathring\bz\leqslant \bu+\phi^{-1}+2\beta^{-1})-\mathbb{P}(\mathring\bz\leqslant \bu-\phi^{-1}-2\beta^{-1})\leqslant\mathbb{P}\{\bs_{n,y}^{(1)}\leqslant \bu+\phi^{-1}+2\beta^{-1}\}-\mathbb{P}\{\bs_{n,y}^{(1)}\leqslant \bu-\phi^{-1}-2\beta^{-1}\}+2\check{\varrho}_n\lesssim\phi^{-1}(\log d_z)^{1/2}+\check\varrho_n$,
where the last step is based on Nazarov's inequality and {\rm AS3} provided that $msb^{-1}+mbn^{-1}=o(1)$. Therefore, by \eqref{eq:III111},
\begin{align}\label{eq:III11}
R_1
\lesssim\frac{\phi^3m^{3/2}(\log d_z)^2}{\sqrt{L}}\{\phi^{-1}(\log d_z)^{1/2}+\check\varrho_n\}\,.
\end{align}
On the other hand, it holds that
\begin{align}\label{eq:bdterm1}
R_2\lesssim&\,\,\frac{\phi\beta^2}{L^{3/2}}\sum_{\ell=1}^L\int_0^1\kappa(t)\mathbb{E}[(1-\chi_\ell) h\{\mathring\bz^{(-\ell)},1\}]\mathbb{E}\bigg\{\max_{k\in[d_z]}(|\tilde{z}_{\ell,k}|^3\vee|\tilde{y}_{\ell,k}|^3\vee|\tilde{w}_{\ell,k}|^3)\bigg\}\,{\rm d}t\notag\\
\leqslant&\,\,\frac{\phi\beta^2}{L^{3/2}}\sum_{\ell=1}^L\int_0^1\kappa(t)\mathbb{E}(1-\chi_\ell) \mathbb{E}\bigg\{\max_{k\in[d_z]}(|\tilde{z}_{\ell,k}|^3\vee|\tilde{y}_{\ell,k}|^3\vee|\tilde{w}_{\ell,k}|^3)\bigg\}\,{\rm d}t\\
\leqslant&\,\,\frac{\phi\beta^2}{L^{3/2}}\sum_{\ell=1}^L\int_0^1\kappa(t)\mathbb{E}\bigg\{(1-\chi_\ell) \max_{k\in[d_z]}(|\tilde{z}_{\ell,k}|^3\vee|\tilde{y}_{\ell,k}|^3\vee|\tilde{w}_{\ell,k}|^3)\bigg\}\,{\rm d}t\notag\\
\lesssim&\,\,\frac{\phi^3(\log d_z)^2}{\sqrt{L}}\{M_{\tilde{z}}(\phi)+M_{\tilde{y}}(\phi)\}\,,\notag
\end{align}
where the third step and last step are based on Chebyshev's association inequality \cite[Lemma B.1]{CCK2017}.
Recall $\mathcal{F}_{-\ell}^*$ is the $\sigma$-filed generated by $\{\tilde{\bz}_s,\tilde{\by}_s,\tilde{\bw}_s\}_{s\neq \ell}$. Since $(\tilde{\by}_\ell,\tilde{\bw}_\ell)$ are independent of $\{\tilde{\bz}_s,\tilde{\by}_s,\tilde{\bw}_s\}_{s\neq \ell}$, and $\{\tilde{\bz}_\ell\}_{\ell=1}^L$ is independent of $\{\tilde{\by}_s,\tilde{\bw}_s\}_{s\neq \ell}$, we have
\begin{align*}
R_3=&\sum_{j,k,l=1}^{d_z}\sum_{\ell=1}^L\int_0^1\frac{\kappa(t)}{L^{3/2}}\mathbb{E}[h\{\mathring\bz^{(-\ell)},1\} U_{jkl}\{\mathring\bz^{(-\ell)}\}\{|\tilde{z}_{\ell,k}|^3-\mathbb{E}(|\tilde{z}_{\ell,k}|^3)\}]\,{\rm d}t\\
=&\sum_{j,k,l=1}^{d_z}\sum_{\ell=1}^L\int_0^1\frac{\kappa(t)}{L^{3/2}}\mathbb{E}[h\{\mathring\bz^{(-\ell)},1\} U_{jkl}\{\mathring\bz^{(-\ell)}\}\mathbb{E}\{|\tilde{z}_{\ell,k}|^3-\mathbb{E}(|\tilde{z}_{\ell,k}|^3)\,|\,\mathcal{F}_{-\ell}^*\}]\,{\rm d}t\\
\lesssim&\,\,\frac{\phi^3(\log d_z)^2}{\sqrt{L}}\max_{\ell\in[L]} \mathbb{E}\bigg[\max_{k\in[d_z]}|\mathbb{E}\{|\tilde{z}_{\ell,k}|^3-\mathbb{E}(|\tilde{z}_{\ell,k}|^3)\,|\,\mathcal{F}_{-\ell}^*\}|\bigg]\\
=&\,\,\frac{\phi^3(\log d_z)^2}{\sqrt{L}} \mathbb{E}\bigg[\max_{k\in[d_z]}|\mathbb{E}\{|\tilde{z}_{\ell,k}|^3-\mathbb{E}(|\tilde{z}_{\ell,k}|^3)\,|\,\mathcal{F}_{-\ell}\}|\bigg]\,,
\end{align*}
where $\mathcal{F}_{-\ell}$ is the $\sigma$-field generated by $\{\tilde{\bz}_s\}_{s\neq\ell}$. Together with \eqref{eq:III11} and \eqref{eq:bdterm1},  (\ref{eq:sd}) implies \eqref{eq:alphatoprove1} holds. $\hfill\Box$

%%=============================================================================
%%=============================================================================

%%=============================================================================
%%=============================================================================

%\newtheorem{lem}{lemma}
%\setcounter{lem}{1}
%\renewcommand{\thelem}{L\arabic{lem}}

\section{Proof of Lemma \ref{lem.eta}}\label{sec:T}

%\begin{lem}\label{lem.eta}
%Assume Conditions {\rm 1--3} hold. Let $\tau=\tau_1\tau_2/(\tau_1+\tau_2)$. If $\log(Kpd)=o(n^{\tau/2})$ and $K^{\tau_1}\log(Kpd)=o(n^{\tau_1/2})$, it holds that
%\begin{align*}
%|T_n-\tilde T_n| \lesssim \frac{K^{3/2}\{\log(Kpd)\}^{1/2}}{{n}^{1/2}}\max[\{\log(Kpd)\}^{1/\tau}, K\{\log(Kpd)\}^{1/\tau_1}]
%\end{align*} with probability at least $1-C(Kpd)^{-1}$ under $H_0$.
%\end{lem}

Recall $T_n=\sum_{j=1}^K Z_j$ and $\tilde T_n := \sum_{j=1}^K \tilde Z_j$. Write $\bar{\be}=(\bar{\eta}_1,\ldots,\bar{\eta}_{Kpd})^\T=\tilde{n}^{-1}\sum_{t=1}^{\tilde{n}}\be_t$ with $\be_t=([{\rm vec} \{\bphi(\bx_t)\bx_{t+1}^{\T}\}]^\T,\ldots, [{\rm vec} \{ \bphi(\bx_t)\bx_{t+K}^{\T}\}]^\T )^{\T}$ and $\tilde{n}=n-K$. We first consider the convergence rate of $\max_{j\in[K]}|Z_j-\tilde Z_j|$. Due to $Z_j =n\max_{\ell\in\mathcal{L}_j}  u_{\ell}^2 $ and $\tilde Z_j = \tilde n \max_{\ell\in\mathcal{L}_j} \bar\eta_\ell^2$,
by the triangle inequality, it holds that
\begin{align}
  \max_{j\in[K]}|Z_j-\tilde Z_j|
  =&~ \max_{j\in[K]}\bigg|\max_{\ell\in\mathcal{L}_j}  (n^{1/2}u_{\ell})^2 - \max_{\ell\in\mathcal{L}_j} (\tilde n^{1/2}\bar\eta_\ell)^2 \bigg|\leqslant\max_{\ell\in[Kpd]} \big|(n^{1/2}u_{\ell})^2 -  (\tilde n^{1/2}\bar\eta_\ell)^2\big| \notag\\
 % \leqslant&~ \max_{\ell\in[Kpd]} |n^{1/2}u_{\ell} - \tilde n^{1/2}\bar\eta_\ell |
 %    \cdot \max_{\ell\in[Kpd]} |n^{1/2}u_{\ell} + \tilde n^{1/2}\bar\eta_\ell | \\
 % \leqslant&~ \max_{\ell\in[Kpd]} |n^{1/2}u_{\ell} - \tilde n^{1/2}\bar\eta_\ell |
 % \Big(\max_{\ell\in[Kpd]} |n^{1/2}u_{\ell} - \tilde n^{1/2}\bar\eta_\ell | + 2\max_{\ell\in[Kpd]}|\tilde n^{1/2}\bar\eta_\ell| \Big) \\
  \leqslant&~ \bigg(\max_{\ell\in[Kpd]} |n^{1/2}u_{\ell} - \tilde n^{1/2}\bar\eta_\ell |\bigg)^2 + 2\underbrace{\max_{\ell\in[Kpd]} |n^{1/2}u_{\ell} - \tilde n^{1/2}\bar\eta_\ell |}_{{\rm I}_1} \cdot\underbrace{\max_{\ell\in[Kpd]}|\tilde n^{1/2}\bar\eta_\ell|}_{{\rm I}_2}\,.\label{eq:diffz}
\end{align}
Note that $\bu := ( u_1,\ldots, u_{Kpd} )^{\T} =(\hat\bgamma_1^{\T},\ldots,\hat\bgamma_{K}^{\T})^{\T}$ with $\hat{\bgamma}_j=(n-j)^{-1}\sum_{t=1}^{n-j} {\rm vec}\{\bphi(\bx_{t})\bx_{t+j}^{\T}\}$ for any $j\in[K]$. We can formulate $\bu=n^{-1}\sum_{t=1}^{\tilde n} \be_t+\bR_n$, where $\bR_n=(R_{n,1},\ldots,R_{n,Kpd})^\T$ is the remainder term. Then
\begin{align*}
  {\rm I_1}
  \leqslant&~ (n^{1/2}-\tilde n^{1/2})\max_{\ell\in[Kpd]}|u_\ell|
  + \tilde n^{1/2} \max_{\ell\in[Kpd]}|u_\ell-\bar\eta_\ell| \\
  \leqslant&~
  \underbrace{\frac{K}{\sqrt{n}+\sqrt{n-K}} \max_{\ell\in[Kpd]} |u_\ell|}_{\rm I_{11}}
   + \underbrace{\frac{K\sqrt{n-K}}{n} \max_{\ell\in[Kpd]}|\bar\eta_\ell|}_{\rm I_{12}}
   + \underbrace{\sqrt{n-K} \max_{\ell\in[Kpd]} |R_{n,\ell}| }_{\rm I_{13}}\,.
\end{align*}
Recall $\bphi(\bx)=\{\phi_1(\bx),\ldots,\phi_d(\bx)\}^\T$ and $\bx_t=(x_{t,1},\ldots,x_{t,p})^\T$. Write $\be_t=(\eta_{t,1},\ldots,\eta_{Kpd})^\T$. For any given $\ell\in[Kpd]$, there exists a triple $(j,l_1,l_2)\in[K]\times[d]\times[p]$ such that $\eta_{t,\ell}=\phi_{l_1}(\bx_t)x_{t+j,l_2}$, which implies $u_\ell=(n-j)^{-1}\sum_{t=1}^{n-j}\eta_{t,\ell}$. According to Condition 1, we have
$ \max_{\ell\in[Kpd]} \bP(|\eta_{t,\ell}|>x)
\leqslant C\exp(-Cx^{\tau_1})$ for any $x>0$. We have $\mathbb{E}(\eta_{t,\ell})=0$ under $H_0$. Notice that $\{\be_t\}$ is an $\alpha$-mixing sequence with $\alpha$-mixing coefficients $\{\tilde{\alpha}_K(k)\}_{k\geqslant1}$ specified in \eqref{eq:newalphamixing}. For any $x>0$, applying Bonferroni inequality and Lemma \ref{lem.bern}, it holds that
\begin{align*}
& \bP({\rm I_{11}}>x)+\bP({\rm I_{12}}>x)   \\
&~~~~~~~~~~~\lesssim    Kpd \exp\bigg(-\frac{Cn^2x^2}{K^{3}}\bigg)
+ \frac{K^2pd}{\sqrt{n}x}\bigg\{ \exp\bigg(-\frac{Cn^{3\tau/2}x^{\tau}}{K^{\tau}}\bigg)
          + \exp\bigg(-\frac{Cn^{3\tau_1/2}x^{\tau_1}}{K^{2\tau_1}}\bigg)\bigg\}
\end{align*}
under $H_0$ provided that $K=o(n)$.
Analogously, we have
\begin{align*}
\bP({\rm I_{13}}>x)
  \lesssim&~    Kpd \max_{j\in[K]}\max_{\ell\in\mathcal{L}_j} \bP\bigg\{ \frac{j}{n(n-j)} \bigg|\sum_{t=1}^{n-j}\eta_{t,\ell}\bigg|> \frac{x}{2\sqrt{n-K}} \bigg\}\\
  & + Kpd\max_{j\in[K]}\max_{\ell\in\mathcal{L}_j}  \bP\bigg\{ \frac{1}{n} \bigg|\sum_{t=n-K+1}^{n-j}\eta_{t,\ell}\bigg|> \frac{x}{2\sqrt{n-K}} \bigg\} \\
  \lesssim&~  Kpd\exp\bigg(-\frac{Cnx^2}{K^2}\bigg)
+ \frac{K^2pd}{\sqrt{n}x}\bigg\{\exp(-Cn^{\tau/2}x^{\tau})+\exp\bigg(-\frac{Cn^{\tau_1/2}x^{\tau_1}}{K^{\tau_1}}\bigg)\bigg\}
\end{align*}
under $H_0$ provided that $K=o(n)$. Therefore,
\begin{align*}%\label{I1.inq}
  \bP({\rm I_1}>x)
  \lesssim&~   Kpd\exp\bigg(-\frac{Cnx^2}{K^2}\bigg)
+ \frac{K^2pd}{\sqrt{n}x}\bigg\{\exp(-Cn^{\tau/2}x^{\tau})+\exp\bigg(-\frac{Cn^{\tau_1/2}x^{\tau_1}}{K^{\tau_1}}\bigg)\bigg\}
\end{align*}
for any $x>0$ under $H_0$, which implies that ${\rm I}_1=O_\p(n^{-1/2}\max[\{\log(Kpd)\}^{1/\tau}, K\{\log(Kpd)\}^{1/\tau_1}])=o_\p(1)$ provided that $\log(Kpd)=o(n^{\tau/2})$ and $K^{\tau_1}\log(Kpd)=o(n^{\tau_1/2})$. More specifically, we have
$
{\rm I_1} \leqslant Cn^{-1/2}\max[\{\log(Kpd)\}^{1/\tau}, K\{\log(Kpd)\}^{1/\tau_1}]
$
 with probability at least $1-C(Kpd)^{-1}$. Following the same arguments, we can show
\begin{align*}
\bP({\rm I_2}>x)
   \lesssim  Kpd\exp\bigg(-\frac{Cx^2}{K}\bigg)
+ \frac{\sqrt{n}Kpd}{x}\bigg\{ \exp(-Cn^{\tau/2}x^{\tau})
    +  \exp\bigg(-\frac{Cn^{\tau_1/2}x^{\tau_1}}{K^{\tau_1}}\bigg)\bigg\}
\end{align*}
for any $x>0$ under $H_0$, provided that $K=o(n)$, which implies that
$
{\rm I_2}\leqslant CK^{1/2}\{\log(Kpd)\}^{1/2}$
 with probability at least $1-C(Kpd)^{-1}$. Hence, by \eqref{eq:diffz}, we have under $H_0$ that
\begin{align*}
\max_{j\in[K]}|Z_j-\tilde{Z}_j|
\lesssim \frac{K^{1/2}\{\log(Kpd)\}^{1/2}}{\sqrt{n}}\max[\{\log(Kpd)\}^{1/\tau}, K\{\log(Kpd)\}^{1/\tau_1}]\end{align*}
with probability at least $1-C(Kpd)^{-1}$. Since $ |T_n-\tilde T_n|\leqslant K\max_{j\in[K]}|Z_j-\tilde{Z}_j|$, then
\begin{align*}
  |T_n-\tilde T_n|
  \lesssim\frac{K^{3/2}\{\log(Kpd)\}^{1/2}}{\sqrt{n}}\max[\{\log(Kpd)\}^{1/\tau}, K\{\log(Kpd)\}^{1/\tau_1}]
\end{align*}
with probability at least $1-C(Kpd)^{-1}$. We complete the proof of Lemma \ref{lem.eta}.  $\hfill\Box$

\section{Computational costs}
\label{sec:cost}
In this section, we compare the computational costs for all the tests under examination.
Table \ref{tab:time} reports the average CPU time of a single trial (averaged over 100 trials) when the simulated data is generated from Model 1 with $n=100$, where Model 1 is i.i.d. normal sequence: $\bx_t\overset{{\rm i.i.d.}}{\sim} \mathcal{N}(\bzero, \bA)$ where $\bA=(a_{kl})_{p\times p}$ with $a_{kl}=0.995^{|k-l|}$ for any $k,l\in [p]$. The simulation is conducted on the Windows platform with Intel(R) Core(TM) i5-7500 CPU at 3.40GHz, using R software for our proposed method and  Matlab for the tests proposed in \cite{HLZ2017} (code available at: https://obl20.com/2019/09/02/). As stated in \cite{HLZ2017}, the test statistics $Z_{\rm tr}$ and $Z_{\rm det}$ require $p\leqslant \sqrt{n}$, and $Zd_{\rm tr}$ requires $p<n$. Thus we present 'NA' in the table when the dimensional constraints are violated.

According to Table \ref{tab:time}, the six versions of our proposed test cost less computational time as compared to  $Z_{\rm tr}$ and $Z_{\rm det}$ and the advantage is especially noticeable for larger $p$ and $K$. On the other hand, $Zd_{\rm tr}$ is computationally cheaper than ours when $p$ and $K$ are small. This is reasonable since the critical value for $Zd_{\rm tr}$ has closed form, whereas we perform 2000 bootstrap replications to obtain the critical values. But as $p$ and $K$ increase, the CPU time for $Zd_{\rm tr}$ grows much faster as compared to our tests. For example,  when $p/n=0.4$ and $K$ varies from $2$ to $8$, the time for $Zd_{\rm tr}$ increases by about  46 times, while by just 3-4 times for our tests. This implies that the computational complexity associated with $Zd_{\rm tr}$ is at least a quadratic function of $K$, as compared to the linear growth with respect to $K$ for our test. Similarly, for a fixed $K=2$ or $4$, when we increase $p$ from $4$ to $40$, we can see the quadratic/cubic growth rate with the computational time for $Zd_{\rm tr}$, but linear growth rate for our test.
When $p=120$ and $K=8$, we can see that our test takes more than 20 seconds for the linear and quadratic map, which can be substantially improved by performing parallel computing due to the bootstrap we used.

\begin{table}[htbp]
\footnotesize
  \centering
  \caption{The average CPU time (in seconds) for a single trial when the data is generated from the  Model 1 with $n=100$.}
    \begin{tabular}{cc|cccccc|ccc}
    $p$   & $K$      & $T_{\rm QS}^l$  & $T_{\rm PR}^l$  & $T_{\rm BT}^l$  & $T_{\rm QS}^q$  & $T_{\rm PR}^q$  & $T_{\rm BT}^q$    & $Z_{\rm tr}$   & $Z_{\rm det}$  & $Zd_{\rm tr}$ \\[0.3em]
    \hline
    4  & 2     & 0.020  & 0.021  & 0.019  & 0.022  & 0.022  & 0.021  & 0.026  & 0.020  & 0.002  \\
          & 4     & 0.022  & 0.022  & 0.023  & 0.025  & 0.024  & 0.024  & 0.030  & 0.029  & 0.010  \\
          & 6     & 0.023  & 0.023  & 0.022  & 0.027  & 0.026  & 0.026  & 0.046  & 0.047  & 0.027  \\
          & 8     & 0.023  & 0.023  & 0.023  & 0.030  & 0.029  & 0.029  & 0.070  & 0.068  & 0.049  \\[0.2em]
    8  & 2     & 0.026  & 0.026  & 0.024  & 0.032  & 0.031  & 0.031  & 0.146  & 0.148  & 0.004  \\
          & 4     & 0.031  & 0.032  & 0.031  & 0.045  & 0.045  & 0.045  & 0.162  & 0.165  & 0.023  \\
          & 6     & 0.040  & 0.037  & 0.037  & 0.057  & 0.057  & 0.056  & 0.200  & 0.198  & 0.062  \\
          & 8     & 0.046  & 0.043  & 0.042  & 0.069  & 0.068  & 0.068  & 0.258  & 0.258  & 0.120  \\[0.2em]
    15  & 2     & 0.044  & 0.044  & 0.042  & 0.065  & 0.065  & 0.065  & NA    & NA    & 0.010  \\
          & 4     & 0.067  & 0.064  & 0.064  & 0.118  & 0.118  & 0.120  & NA    & NA    & 0.071  \\
          & 6     & 0.090  & 0.087  & 0.087  & 0.171  & 0.171  & 0.173  & NA    & NA    & 0.188  \\
          & 8     & 0.117  & 0.115  & 0.114  & 0.220  & 0.221  & 0.221  & NA    & NA    & 0.355  \\[0.2em]
    40   & 2     & 0.209  & 0.206  & 0.206  & 0.399  & 0.396  & 0.396  & NA    & NA    & 1.214  \\
          & 4     & 0.396  & 0.391  & 0.392  & 0.767  & 0.773  & 0.768  & NA    & NA    & 10.579  \\
          & 6     & 0.575  & 0.570  & 0.574  & 1.150  & 1.142  & 1.138  & NA    & NA    & 29.412  \\
          & 8     & 0.756  & 0.750  & 0.747  & 1.521  & 1.524  & 1.521  & NA    & NA    & 56.532  \\[0.2em]
    120   & 2     & 1.815  & 1.790  & 1.797  & 3.800  & 3.801  & 3.789  & NA    & NA    & NA \\
          & 4     & 3.775  & 3.759  & 3.730  & 7.775  & 7.763  & 7.777  & NA    & NA    & NA \\
          & 6     & 5.707  & 5.707  & 5.707  & 12.221  & 12.166  & 12.145  & NA    & NA    & NA \\
          & 8     & 7.759  & 7.770  & 7.791  & 23.753  & 23.910  & 23.681  & NA    & NA    & NA \\
    \end{tabular}%
  \label{tab:time}%
\end{table}%

\section{Comparison with multivariate white noise tests}
In this section, we compare our proposed test statistics with three multivariate portmanteau tests with test statistics
 \begin{align*}
  Q_{\rm BP}=&~ n\sum_{k=1}^K{\rm tr}(C_k^{\T}C_0^{-1}C_k C_0^{-1})  \,, \\
  Q_{\rm HS}=&~ n^2\sum_{k=1}^K(n-k)\,{\rm tr}(C_k^{\T}C_0^{-1}C_k C_0^{-1})\,,  \\
  Q_{\rm LM}=&~ n\sum_{k=1}^K{\rm tr}(C_k^{\T}C_0^{-1}C_k C_0^{-1}) + p^2K(K+1)/(2n) \,,
 \end{align*}
where $C_k=n^{-1}\sum_{t=k+1}^n \bx_t\bx_{t-k}^{\T}$, for any $k=0,1,\ldots,K$. Here the test statistic $Q_{\rm BP}$ is first proposed by \cite{BP1970} for univariate time series and extended to multivariate time series by \cite{Hosking1980}. Then \cite{Hosking1980} also constructed a modified test statistic $Q_{\rm HS}$. The third test statistic $Q_{\rm LM}$ is from \cite{LM1981}. The simulation results are presented in the following Tables \ref{tab2:M1-M3} and \ref{tab2:M4-M6}, which are based on the simulation settings of Model 1--Model 6 in the main paper.

Table \ref{tab2:M1-M3} shows that all  tests have similar good performance when the dimension $p$ is small except for Model 3, e.g., $n=100$, $p=4$ and $n=300$, $p=12$. In general, our proposed test statistics perform much better than the portmanteau tests $Q_{\rm BP}$, $Q_{\rm HS}$ and $Q_{\rm LM}$, which fail badly to attain the nominal significance level as the dimension $p$ increases and can not be implemented when $p>n$. For Model 3, the portmanteau tests apparently cannot control the empirical sizes even when $p$ is small.

Table \ref{tab2:M4-M6} indicates that our proposed tests are more powerful than the portmanteau tests no matter when the dimension $p$ is small or large for Model 4 and Model 5. In the two models, the empirical powers of the portmanteau tests decrease so fast and even tend to zero, such as, $n=100$, $p=40$ or $n=300$, $p=120$. But for Model 6, their tests exhibit great power, which is probably due to the fact that the model implies strong linear serial dependence although it is a nonlinear model. This phenomenon is also observed  for Hong et al.'s test. Additionally, we see that $Q_{\rm HS}$ and $Q_{\rm LM}$ perform similarly, and outperform $Q_{\rm BP}$ when $p$ is large, which is consistent with the fact that the asymptotic approximations for $Q_{\rm HS}$ and $Q_{\rm LM}$ are more accurate than that for $Q_{\rm BP}$.

%\begin{landscape}
\begin{sidewaystable}[htbp]
\scriptsize
  \centering
  \caption{Empirical sizes ($\%$) of the tests $T_{\rm QS}^l$, $T_{\rm PR}^l$, $T_{\rm BT}^l$, $T_{\rm QS}^q$, $T_{\rm PR}^q$, $T_{\rm BT}^q$, $Q_{\rm BP}$, $Q_{\rm HS}$ and $Q_{\rm LM}$ for Models 1--3 at the 5\% nominal level. }
  \resizebox{!}{6cm}{
\begin{tabular}{ccc|ccccccccc| ccccccccc| ccccccccc}
          &       &       & \multicolumn{9}{c|}{Model 1}                                   & \multicolumn{9}{c|}{Model 2}                                     & \multicolumn{9}{c}{Model 3} \\[0.6em]
    $n$     & $p/n$   & $K$    & $T_{\rm QS}^l$  & $T_{\rm PR}^l$  & $T_{\rm BT}^l$  & $T_{\rm QS}^q$  & $T_{\rm PR}^q$  & $T_{\rm BT}^q$    & $Q_{\rm BP}$   & $Q_{\rm HS}$  & $Q_{\rm LM}$ & $T_{\rm QS}^l$  & $T_{\rm PR}^l$  & $T_{\rm BT}^l$  & $T_{\rm QS}^q$  & $T_{\rm PR}^q$  & $T_{\rm BT}^q$    & $Q_{\rm BP}$   & $Q_{\rm HS}$  & $Q_{\rm LM}$  & $T_{\rm QS}^l$  & $T_{\rm PR}^l$  & $T_{\rm BT}^l$  & $T_{\rm QS}^q$  & $T_{\rm PR}^q$  & $T_{\rm BT}^q$    & $Q_{\rm BP}$   & $Q_{\rm HS}$  & $Q_{\rm LM}$ \\[0.5em]
    \hline
    100   & 0.04  & 2     & 4.2   & 4.5   & 4.5   & 4.3   & 4.3   & 4.4   & 3.3   & 3.9   & 3.7   & 4.2   & 4.4   & 4.7   & 2.2   & 2.4   & 2.5   & 3.6   & 4.1   & 3.9   & 3.5   & 3.6   & 4.2   & 2.9   & 2.8   & 3.2   & 9.6   & 10.9  & 10.6  \\
          &       & 4     & 5.1   & 5.0   & 5.2   & 4.6   & 4.3   & 4.5   & 3.1   & 4.3   & 3.9   & 3.1   & 3.3   & 3.5   & 2.9   & 2.8   & 3.2   & 3.1   & 4.8   & 4.2   & 3.2   & 3.2   & 3.5   & 3.0   & 3.1   & 3.4   & 9.6   & 12.5  & 11.8  \\
          &       & 6     & 4.6   & 4.4   & 4.8   & 4.5   & 4.5   & 4.6   & 2.1   & 4.3   & 3.9   & 3.0   & 2.9   & 3.5   & 2.8   & 2.7   & 2.9   & 1.9   & 3.8   & 3.4   & 3.1   & 3.1   & 3.6   & 3.9   & 4.0   & 4.1   & 10.5  & 15.8  & 14.7  \\
          &       & 8     & 4.4   & 4.4   & 4.7   & 5.0   & 4.9   & 5.1   & 2.0   & 3.9   & 3.4   & 2.8   & 2.8   & 3.3   & 2.9   & 2.8   & 3.1   & 1.8   & 4.7   & 3.8   & 3.3   & 3.3   & 4.0   & 4.5   & 4.4   & 4.9   & 11.1  & 20.2  & 18.5  \\[0.3em]
          & 0.08  & 2     & 4.1   & 4.1   & 4.1   & 3.1   & 3.2   & 3.4   & 2.5   & 3.3   & 3.2   & 3.8   & 3.9   & 4.0   & 1.9   & 1.8   & 1.8   & 2.9   & 4.2   & 3.9   & 3.3   & 3.2   & 3.5   & 2.7   & 2.5   & 2.7   & 10.3  & 13.3  & 12.8  \\
\end{tabular}%
    }
  \label{tab2:M1-M3}%
\end{sidewaystable}%
%\end{landscape}

%\begin{landscape}
  \begin{sidewaystable}[htbp]
  \scriptsize
  \centering
  \caption{Empirical powers ($\%$) of the tests $T_{\rm QS}^l$, $T_{\rm PR}^l$, $T_{\rm BT}^l$, $T_{\rm QS}^q$, $T_{\rm PR}^q$, $T_{\rm BT}^q$, $Q_{\rm BP}$, $Q_{\rm HS}$ and $Q_{\rm LM}$  for Models 4--6 at the 5\% nominal level. }
    \resizebox{!}{6.3cm}{
    \begin{tabular}{ccc|ccccccccc| ccccccccc| ccccccccc}
          &       &       & \multicolumn{9}{c|}{Model 4}                           & \multicolumn{9}{c|}{Model 5}                           & \multicolumn{9}{c}{Model 6} \\[0.6em]
    $n$     & $p/n$   & $K$    & $T_{\rm QS}^l$  & $T_{\rm PR}^l$  & $T_{\rm BT}^l$  & $T_{\rm QS}^q$  & $T_{\rm PR}^q$  & $T_{\rm BT}^q$    & $Q_{\rm BP}$   & $Q_{\rm HS}$  & $Q_{\rm LM}$ & $T_{\rm QS}^l$  & $T_{\rm PR}^l$  & $T_{\rm BT}^l$  & $T_{\rm QS}^q$  & $T_{\rm PR}^q$  & $T_{\rm BT}^q$    &  $Q_{\rm BP}$   & $Q_{\rm HS}$  & $Q_{\rm LM}$  & $T_{\rm QS}^l$  & $T_{\rm PR}^l$  & $T_{\rm BT}^l$  & $T_{\rm QS}^q$  & $T_{\rm PR}^q$  & $T_{\rm BT}^q$    &  $Q_{\rm BP}$   & $Q_{\rm HS}$  & $Q_{\rm LM}$  \\[0.5em]
    \hline
    100   & 0.04  & 2     & 79.0  & 78.1  & 81.2  & 93.5  & 93.5  & 94.5  & 60.9  & 63.0  & 62.4  & 65.7  & 65.2  & 69.1  & 94.2  & 94.3  & 95.5  & 46.2  & 48.7  & 48.1  & 77.0  & 77.4  & 84.4  & 80.5  & 81.3  & 85.4  & 99.7  & 99.8  & 99.8  \\
          &       & 4     & 87.8  & 87.5  & 89.2  & 97.8  & 97.8  & 98.3  & 74.1  & 77.3  & 76.6  & 81.5  & 80.3  & 83.4  & 98.3  & 98.4  & 98.6  & 61.5  & 65.7  & 64.5  & 66.7  & 66.5  & 77.7  & 75.7  & 76.2  & 81.8  & 99.6  & 99.6  & 99.6  \\
          &       & 6     & 91.1  & 90.9  & 92.6  & 98.7  & 98.7  & 99.0  & 81.6  & 85.6  & 84.9  & 87.1  & 86.8  & 89.1  & 99.0  & 99.0  & 99.2  & 68.7  & 74.7  & 73.4  & 64.7  & 63.8  & 76.8  & 75.6  & 76.8  & 82.3  & 99.4  & 99.6  & 99.6  \\
          &       & 8     & 93.5  & 93.0  & 94.5  & 99.3  & 99.1  & 99.4  & 83.3  & 88.7  & 87.9  & 90.9  & 90.8  & 92.3  & 99.2  & 99.1  & 99.4  & 72.5  & 80.5  & 79.1  & 66.6  & 65.1  & 78.0  & 77.8  & 77.9  & 84.8  & 99.1  & 99.5  & 99.5  \\[0.3em]
          & 0.08  & 2     & 82.2  & 81.6  & 84.8  & 93.0  & 93.2  & 94.7  & 37.7  & 42.6  & 41.7  & 71.8  & 71.7  & 76.3  & 94.9  & 95.0  & 96.0  & 27.0  & 31.5  & 30.7  & 75.0  & 75.1  & 85.2  & 71.0  & 72.3  & 78.7  & 100   & 100   & 100  \\
\end{tabular}%
    }
  \label{tab2:M4-M6}%
\end{sidewaystable}%
%\end{landscape}

\section{Additional simulation results}\label{sec:addsimu}

%%%%%%%%%%%%%%%%%%%%%%%%%%%%%%%%%%%%%%  SIZE c*bn %%%%%%%%%%%%%%%%%%%%%%%%%%%%%%%%%%%%%%%%%
%\begin{landscape}
\begin{sidewaystable}[htbp]
\renewcommand\arraystretch{1.2}
\scriptsize
  \centering
  \caption{Empirical sizes ($\%$) of the tests $T_{\rm QS}^l$ and $T_{\rm QS}^q$ for Models 1--3 at the 5\% nominal level, where $c$ represents the constant which is multiplied by Andrews' bandwidth. }
  \resizebox{!}{5.7cm}{
    \begin{tabular}{ccc|ccccccc|ccccccc|ccccccc|ccccccc|ccccccc|ccccccc}
          &       &       & \multicolumn{7}{c|}{Model 1 with $T_{\rm QS}^l$}            & \multicolumn{7}{c|}{Model 1 with $T_{\rm QS}^q$}            & \multicolumn{7}{c|}{Model 2 with $T_{\rm QS}^l$}            & \multicolumn{7}{c|}{Model 2 with $T_{\rm QS}^q$}            & \multicolumn{7}{c|}{Model 3 with $T_{\rm QS}^l$}            & \multicolumn{7}{c}{Model 3 with $T_{\rm QS}^q$} \\[0.4em]
    \hline
          &       &       & \multicolumn{7}{c|}{$c$}   & \multicolumn{7}{c|}{$c$}   & \multicolumn{7}{c|}{$c$}        & \multicolumn{7}{c|}{$c$}   & \multicolumn{7}{c|}{$c$}    & \multicolumn{7}{c}{$c$} \\[0.2em]
    $n$     & $p/n$   & $K$     & $2^{-3}$ & $2^{-2}$ & $2^{-1}$ & $2^0$  & $2^1$  & $2^2$  & $2^3$  & $2^{-3}$ & $2^{-2}$ & $2^{-1}$ & $2^0$  & $2^1$  & $2^2$  & $2^3$  & $2^{-3}$ & $2^{-2}$ & $2^{-1}$ & $2^0$  & $2^1$  & $2^2$  & $2^3$  & $2^{-3}$ & $2^{-2}$ & $2^{-1}$ & $2^0$  & $2^1$  & $2^2$  & $2^3$  & $2^{-3}$ & $2^{-2}$ & $2^{-1}$ & $2^0$  & $2^1$  & $2^2$  & $2^3$  & $2^{-3}$ & $2^{-2}$ & $2^{-1}$ & $2^0$  & $2^1$  & $2^2$  & $2^3$ \\[0.3em]
    \hline
    100   & 0.04  & 2     & 4.6   & 5.0   & 4.6   & 4.4   & 5.4   & 6.3   & 10.9  & 4.1   & 4.4   & 4.0   & 3.9   & 4.4   & 4.8   & 9.2   & 4.2   & 4.3   & 4.0   & 3.8   & 3.6   & 3.9   & 5.7   & 3.0   & 2.9   & 2.5   & 2.3   & 2.3   & 2.9   & 4.8   & 4.2   & 4.0   & 3.4   & 3.2   & 3.7   & 3.7   & 5.1   & 4.1   & 3.7   & 2.7   & 2.5   & 2.9   & 3.0   & 4.6  \\
          &       & 4     & 5.1   & 5.0   & 4.4   & 4.4   & 4.1   & 5.1   & 7.2   & 5.0   & 4.4   & 4.3   & 4.3   & 3.4   & 4.7   & 6.7   & 4.4   & 4.5   & 3.5   & 3.4   & 2.3   & 1.8   & 3.1   & 2.9   & 3.6   & 3.0   & 3.0   & 2.1   & 1.9   & 2.6   & 4.4   & 4.6   & 3.7   & 3.1   & 2.1   & 2.2   & 2.6   & 3.9   & 4.0   & 3.7   & 2.9   & 3.0   & 2.6   & 3.8  \\
\end{tabular}%
    }
  \label{tab:size_cbn_qs}%
\end{sidewaystable}%
%\end{landscape}

%\begin{landscape}
\begin{sidewaystable}[htbp]
\renewcommand\arraystretch{1.2}
\scriptsize
  \centering
  \caption{Empirical sizes ($\%$) of the tests $T_{\rm PR}^l$ and $T_{\rm PR}^q$ for Models 1--3 at the 5\% nominal level, where $c$ represents the constant which is multiplied by Andrews' bandwidth.}
  \resizebox{!}{5.7cm}{
    \begin{tabular}{ccc|ccccccc|ccccccc|ccccccc|ccccccc|ccccccc|ccccccc}
          &       &       & \multicolumn{7}{c|}{Model 1 with $T_{\rm PR}^l$}            & \multicolumn{7}{c|}{Model 1 with $T_{\rm PR}^q$}            & \multicolumn{7}{c|}{Model 2 with $T_{\rm PR}^l$}            & \multicolumn{7}{c|}{Model 2 with $T_{\rm PR}^q$}            & \multicolumn{7}{c|}{Model 3 with $T_{\rm PR}^l$}            & \multicolumn{7}{c}{Model 3 with $T_{\rm PR}^q$} \\[0.4em]
    \hline
          &       &       & \multicolumn{7}{c|}{$c$}   & \multicolumn{7}{c|}{$c$}   & \multicolumn{7}{c|}{$c$}        & \multicolumn{7}{c|}{$c$}   & \multicolumn{7}{c|}{$c$}    & \multicolumn{7}{c}{$c$} \\[0.2em]
    $n$     & $p/n$   & $K$     & $2^{-3}$ & $2^{-2}$ & $2^{-1}$ & $2^0$  & $2^1$  & $2^2$  & $2^3$  & $2^{-3}$ & $2^{-2}$ & $2^{-1}$ & $2^0$  & $2^1$  & $2^2$  & $2^3$  & $2^{-3}$ & $2^{-2}$ & $2^{-1}$ & $2^0$  & $2^1$  & $2^2$  & $2^3$  & $2^{-3}$ & $2^{-2}$ & $2^{-1}$ & $2^0$  & $2^1$  & $2^2$  & $2^3$  & $2^{-3}$ & $2^{-2}$ & $2^{-1}$ & $2^0$  & $2^1$  & $2^2$  & $2^3$  & $2^{-3}$ & $2^{-2}$ & $2^{-1}$ & $2^0$  & $2^1$  & $2^2$  & $2^3$ \\[0.3em]
    \hline
    100   & 0.04  & 2     & 4.4   & 4.8   & 4.5   & 4.4   & 5.7   & 7.2   & 12.4  & 4.1   & 4.2   & 4.1   & 3.8   & 4.7   & 5.0   & 10.6  & 4.4   & 4.3   & 4.0   & 3.9   & 3.8   & 4.4   & 7.2   & 2.9   & 2.8   & 2.5   & 2.4   & 2.7   & 3.3   & 6.0   & 4.0   & 3.9   & 3.6   & 3.3   & 3.8   & 4.2   & 6.7   & 3.8   & 3.7   & 2.5   & 2.5   & 3.1   & 3.3   & 6.0  \\
          &       & 4     & 5.1   & 4.9   & 4.7   & 4.3   & 4.0   & 5.5   & 8.6   & 4.9   & 4.1   & 4.2   & 4.3   & 3.5   & 4.9   & 8.0   & 4.3   & 4.2   & 3.2   & 3.3   & 2.5   & 2.1   & 4.1   & 2.9   & 3.6   & 2.9   & 2.9   & 2.1   & 2.3   & 3.9   & 4.4   & 4.4   & 3.8   & 2.7   & 2.2   & 2.5   & 3.7   & 4.0   & 3.9   & 3.7   & 2.8   & 3.1   & 2.7   & 4.8  \\
\end{tabular}%
    }
  \label{tab:size_cbn_pr}%
\end{sidewaystable}%
%\end{landscape}

%%%%%%%%%%%%%%%%%%%%%%%%% power cbn %%%%%%%%%%%%%%%%%%%%%%%%%%%%%%%%%%%%%%%%%%%%%%
%\begin{landscape}
\begin{sidewaystable}[htbp]
\scriptsize
\renewcommand\arraystretch{1.2}
  \centering
  \caption{Empirical powers ($\%$) of the tests $T_{\rm QS}^l$ and $T_{\rm QS}^q$ for Models 4--6 at the 5\% nominal level, where $c$ represents the constant which is multiplied by Andrews' bandwidth. }
  \resizebox{!}{5.2cm}{
    \begin{tabular}{ccc|ccccccc|ccccccc|ccccccc|ccccccc|ccccccc|ccccccc}
          &       &       & \multicolumn{7}{c|}{Model 4 with $T_{\rm QS}^l$}            & \multicolumn{7}{c|}{Model 4 with $T_{\rm QS}^q$}            & \multicolumn{7}{c|}{Model 5 with $T_{\rm QS}^l$}            & \multicolumn{7}{c|}{Model 5 with $T_{\rm QS}^q$}            & \multicolumn{7}{c|}{Model 6 with $T_{\rm QS}^l$}            & \multicolumn{7}{c}{Model 6 with $T_{\rm QS}^q$} \\[0.4em]
    \hline
          &       &       & \multicolumn{7}{c|}{$c$}    & \multicolumn{7}{c|}{$c$}     & \multicolumn{7}{c|}{$c$}    & \multicolumn{7}{c|}{$c$}    & \multicolumn{7}{c|}{$c$}     & \multicolumn{7}{c}{$c$} \\[0.3em]
    $n$     & $p/n$   & $K$     & $2^{-3}$ & $2^{-2}$ & $2^{-1}$ & $2^0$  & $2^1$  & $2^2$  & $2^3$  & $2^{-3}$ & $2^{-2}$ & $2^{-1}$ & $2^0$  & $2^1$  & $2^2$  & $2^3$ & $2^{-3}$ & $2^{-2}$ & $2^{-1}$ & $2^0$  & $2^1$  & $2^2$  & $2^3$  & $2^{-3}$ & $2^{-2}$ & $2^{-1}$ & $2^0$  & $2^1$  & $2^2$  & $2^3$  & $2^{-3}$ & $2^{-2}$ & $2^{-1}$ & $2^0$  & $2^1$  & $2^2$  & $2^3$  & $2^{-3}$ & $2^{-2}$ & $2^{-1}$ & $2^0$  & $2^1$  & $2^2$  & $2^3$ \\[0.4em]
    \hline
    100   & 0.04  & 2     & 81.4  & 81.8  & 80.4  & 76.8  & 73.2  & 69.0  & 67.6  & 95.6  & 96.0  & 95.2  & 92.8  & 91.8  & 91.0  & 91.4  & 72.7  & 73.9  & 71.3  & 65.5  & 59.8  & 57.0  & 57.9  & 96.1  & 96.2  & 95.8  & 94.8  & 93.6  & 92.7  & 92.2  & 99.4  & 96.9  & 88.5  & 76.7  & 72.9  & 77.1  & 87.1  & 99.2  & 94.0  & 85.5  & 79.3  & 82.0  & 87.5  & 96.1  \\
          &       & 4     & 89.9  & 90.0  & 89.1  & 87.3  & 82.7  & 75.4  & 71.7  & 99.1  & 98.7  & 98.8  & 97.8  & 96.8  & 95.0  & 95.8  & 85.7  & 85.5  & 85.2  & 81.3  & 74.4  & 65.5  & 64.1  & 99.0  & 98.9  & 98.8  & 98.4  & 97.3  & 96.3  & 95.2  & 98.3  & 96.3  & 84.6  & 68.0  & 59.5  & 64.2  & 78.2  & 99.2  & 94.3  & 85.4  & 76.1  & 75.3  & 82.7  & 93.9  \\
\end{tabular}%
    }
  \label{tab:power_cbn_qs}%
\end{sidewaystable}%
%\end{landscape}

%\begin{landscape}
\begin{sidewaystable}[htbp]
\scriptsize
\renewcommand\arraystretch{1.2}
  \centering
  \caption{Empirical powers ($\%$) of the tests $T_{\rm PR}^l$ and $T_{\rm PR}^q$ for Models 4--6 at the 5\% nominal level, where $c$ represents the constant which is multiplied by Andrews' bandwidth. }
  \resizebox{!}{5.2cm}{
    \begin{tabular}{ccc|ccccccc|ccccccc|ccccccc|ccccccc|ccccccc|ccccccc}
          &       &       & \multicolumn{7}{c|}{Model 4 with $T_{\rm PR}^l$}            & \multicolumn{7}{c|}{Model 4 with $T_{\rm PR}^q$}            & \multicolumn{7}{c|}{Model 5 with $T_{\rm PR}^l$}            & \multicolumn{7}{c|}{Model 5 with $T_{\rm PR}^q$}            & \multicolumn{7}{c|}{Model 6 with $T_{\rm PR}^l$}            & \multicolumn{7}{c}{Model 6 with $T_{\rm PR}^q$} \\[0.4em]
    \hline
          &       &       & \multicolumn{7}{c|}{$c$}    & \multicolumn{7}{c|}{$c$}     & \multicolumn{7}{c|}{$c$}    & \multicolumn{7}{c|}{$c$}    & \multicolumn{7}{c|}{$c$}     & \multicolumn{7}{c}{$c$} \\[0.3em]
    $n$     & $p/n$   & $K$     & $2^{-3}$ & $2^{-2}$ & $2^{-1}$ & $2^0$  & $2^1$  & $2^2$  & $2^3$  & $2^{-3}$ & $2^{-2}$ & $2^{-1}$ & $2^0$  & $2^1$  & $2^2$  & $2^3$ & $2^{-3}$ & $2^{-2}$ & $2^{-1}$ & $2^0$  & $2^1$  & $2^2$  & $2^3$  & $2^{-3}$ & $2^{-2}$ & $2^{-1}$ & $2^0$  & $2^1$  & $2^2$  & $2^3$  & $2^{-3}$ & $2^{-2}$ & $2^{-1}$ & $2^0$  & $2^1$  & $2^2$  & $2^3$  & $2^{-3}$ & $2^{-2}$ & $2^{-1}$ & $2^0$  & $2^1$  & $2^2$  & $2^3$ \\[0.4em]
    \hline
    100   & 0.04  & 2     & 81.4  & 81.9  & 80.0  & 76.9  & 73.7  & 70.4  & 71.1  & 95.9  & 95.7  & 95.2  & 92.6  & 92.2  & 91.8  & 92.9  & 73.4  & 73.4  & 70.2  & 65.1  & 60.7  & 59.0  & 62.6  & 96.0  & 96.1  & 95.7  & 94.9  & 93.6  & 93.4  & 93.6  & 99.3  & 95.8  & 87.1  & 76.6  & 75.2  & 81.1  & 91.5  & 98.9  & 92.8  & 85.1  & 80.5  & 84.2  & 90.5  & 97.7  \\
          &       & 4     & 90.0  & 90.0  & 88.8  & 87.2  & 82.6  & 76.2  & 74.7  & 99.0  & 98.7  & 98.7  & 97.7  & 96.9  & 95.5  & 96.6  & 85.9  & 85.4  & 84.9  & 80.9  & 74.4  & 67.6  & 67.6  & 99.2  & 98.7  & 98.7  & 98.4  & 97.4  & 96.7  & 96.4  & 98.1  & 95.1  & 82.4  & 68.2  & 61.7  & 69.0  & 84.3  & 98.6  & 93.3  & 84.2  & 77.2  & 77.7  & 87.5  & 96.4  \\
\end{tabular}%
    }
  \label{tab:power_cbn_pr}%
\end{sidewaystable}%
%\end{landscape}

\singlespacing
\small
%\bibliographystyle{agsm}
%\bibpunct{(}{)}{,}{a}{}{;}
%\bibliography{HDMDT_bib}

\end{document}